\documentclass[letterpaper,12pt]{article}

\usepackage[left=2cm,right=2cm,top=3cm,bottom=3cm]{geometry}
\usepackage[onehalfspacing]{setspace}
\usepackage{comment}
\usepackage{float} % not to make figures float around.. use [H] option

\usepackage{colortbl}
\usepackage{subfig}
\usepackage{graphicx}
\usepackage[T1]{fontenc}

\usepackage{natbib} % avoid 'hyperref' problem due to
\usepackage{amsmath}
\usepackage{amssymb}
\usepackage{amsfonts}
\usepackage{amsthm}

\usepackage{newtxtext,newtxmath} % times new roman font (okay with bold)
\usepackage{bbm}
\usepackage{bm}

\usepackage{booktabs}
\usepackage{multirow}
\usepackage{makecell}
\usepackage{colortbl}
\usepackage{todonotes}
\usepackage{color}
\usepackage{hyperref} % enable hyperlink

\hypersetup{bookmarks,colorlinks,%
citecolor=blue,%
pdftex}

\def\ttheta{\tilde{\theta}}
\def\tphi{\tilde{\phi}}
\def\teps{\tilde{\eps}}
\def\trho{\tilde{\rho}}
\def\tnu{\tilde{\nu}}
\def\bt{\overline{t}}

\newcommand{\ul}[1]{\underline{#1}}
\newcommand{\ol}[1]{\overline{#1}}
\DeclareMathOperator*{\E}{\text{E}}

\DeclareMathOperator*{\var}{\textnormal{Var}}
\DeclareMathOperator*{\cov}{\textnormal{Cov}}
\newcommand{\eps}[0]{\ensuremath{\varepsilon}}

\newcommand{\tmu}[0]{\ensuremath{\tilde{\mu}}}

\newcommand{\parrow}[0]{\ensuremath{\stackrel{p}{\rightarrow}}}

\newcommand{\tcb}{\textcolor{blue}}
\newcommand{\tcr}{\textcolor{red}}

\newcommand{\tcg}{\textcolor{green}}

\newcommand{\ys}[1]{\todo[inline, backgroundcolor=blue!20!white]{#1}}

\newtheorem{prop}{Proposition}
\newenvironment{propbis}[1]
  {\renewcommand{\theprop}{\ref{#1}$'$}%
   \addtocounter{prop}{-1}%
   \begin{prop}}
  {\end{prop}}
\newtheorem{lemma}{Lemma}
\newtheorem{ass}{Assumption}
\newenvironment{assbis}[1]
  {\renewcommand{\theass}{\ref{#1}$'$}%
   \addtocounter{ass}{-1}%
   \begin{ass}}
  {\end{ass}}

\title{The Evolution of Unobserved Skill Returns in the U.S.: \\ A New Approach Using Panel Data\thanks{An earlier (and much different) version of this paper circulated under the title ``Wage Dynamics and Returns to Unobserved Skill."  For valuable comments on this line of research, we thank Pat Bayer, Magne Mogstad, Terry Moon, Fabrizio Perri, Gianluca Violante and Thomas Lemieux as well as numerous seminar and conference participants over the years.   
Lochner acknowledges generous support from SSHRC.  The views expressed here are those of the authors and do not necessarily reflect those of the Bank of Canada. Parts of this analysis were first performed using the SIPP Synthetic Beta (SSB) on the Synthetic Data Server housed at Cornell University, which is funded by NSF Grant \#SES-1042181. These data are for public use and may be accessed by researchers outside secure Census facilities. For more information, visit \href{https://www.census.gov/programs-surveys/sipp/guidance/sipp-synthetic-beta-data-product.html}{\url{https://www.census.gov/programs-surveys/sipp/guidance/sipp-synthetic-beta-data-product.html}}. Final results for this paper were obtained by Census Bureau staff using the authors' programs and the SIPP Completed Gold Standard Files. This does not imply endorsement by the Census Bureau of any methods, results, opinions, or views presented in this paper.}}

\author{\textbf{Lance Lochner}\\
University of Western Ontario\\
\and
\textbf{Youngmin Park}\\
Bank of Canada\\
\and
\textbf{Youngki Shin}\\
McMaster University
}

\date{\today}

\begin{document}

\maketitle
\pagenumbering{gobble} % no page number on the cover page
\maketitle

\begin{abstract} 
  Economists disagree about the factors driving the substantial increase in residual wage inequality in the US over the past few decades.  To identify changes in the returns to unobserved skills, we make a novel assumption about the dynamics of skills rather than about the stability of skill distributions across cohorts, as is standard.  We show that our assumption is supported by data on test score dynamics for older workers in the HRS. Using survey data from the PSID and administrative data from the IRS and SSA, we estimate that the returns to unobserved skills \textit{declined} substantially in the late-1980s and 1990s despite an increase in residual inequality.   Accounting for firm-specific pay differences yields similar results. 
  Extending our framework to consider occupational differences in returns to skill and multiple unobserved skills, we further show that skill returns display similar patterns for workers employed in each of cognitive, routine, and social occupations.  Finally, our results suggest that increasing skill dispersion, driven by rising skill volatility, explains most of the growth in residual wage inequality since the 1980s.
\end{abstract}

\newpage
\pagenumbering{arabic} % start page numbering

\section{Introduction}

\begin{figure}[h]
  \centering
  \includegraphics[width=0.45\columnwidth]{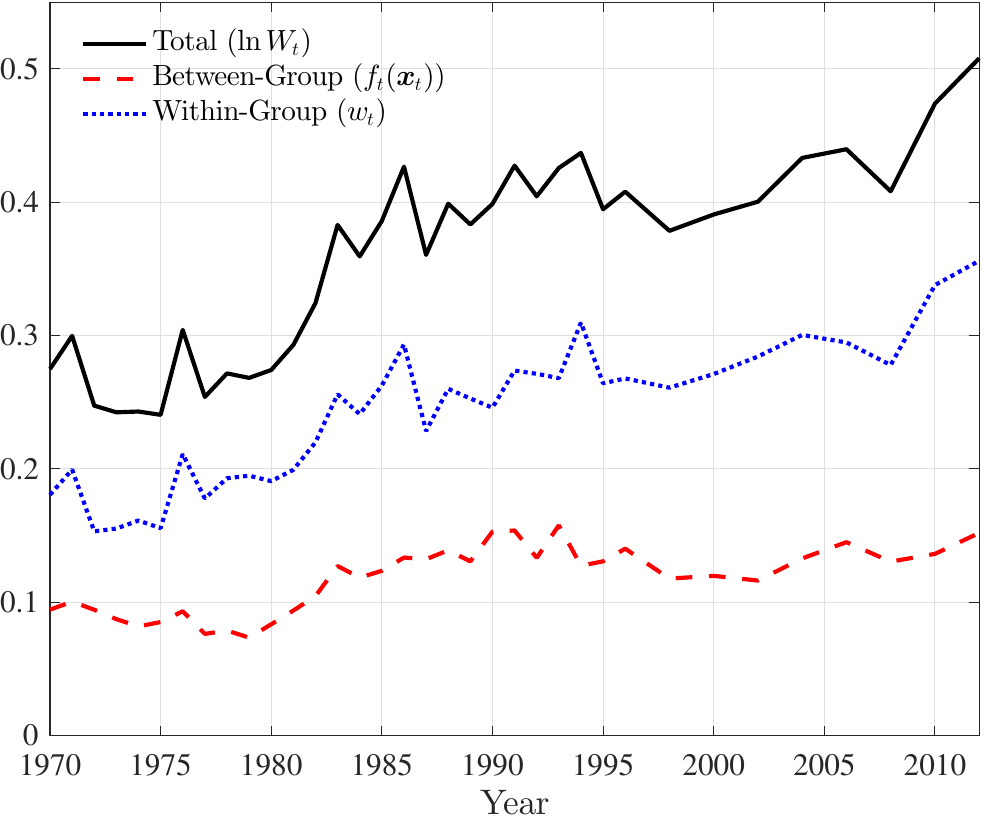}
\caption{Between- and Within-Group Variances of Log Wages}
    \label{fig: variance of hourly earnings}
\end{figure}

Wage inequality has risen considerably in the United States since the 1960s.  The long-term increases in wage differentials by education and experience are widely attributed to rising returns to skill \citep{bound_johnson_1992,katz_murphy_1992}.  In addition to these trends, wage inequality within narrowly defined groups (e.g.\ by race, education, and age/experience) also rose substantially. 
Figure~\ref{fig: variance of hourly earnings} reports these trends for men based on data from the Panel Study of Income Dynamics (PSID) used in this study.\footnote{In obtaining between- and within-group log wage variances, we condition log wages on potential experience, race/ethnicity, and 7 educational attainment categories, separately by year and college vs.\ non-college status. See Section~\ref{sec: PSID} for details.}

Since the seminal work of \cite{juhn_murphy_pierce_1993}, economists have often equated rising within-group, or residual, inequality with an increase in the returns to `unobserved' ability or skill \citep[see, e.g.,][]{card_lemieux_1996,katz_autor_1999,acemoglu_2002,autor_katz_kearney_2008}.  This interpretation, along with the rising returns to `observable' skills (i.e, education, experience),  motivated an enormous and still influential literature on skill-biased technical change (SBTC).\footnote{Many early theoretical studies aimed specifically to explain rising residual inequality and returns to unobserved ability/skill \citep[e.g.,][]{galor_tsiddon_1997, acemoglu_1999, caselli_1999, galor_moav_2000, violante_2002}.} More recent task-based models of the labor market also explore the influence of automation and globalization on wage and employment inequality between and within groups by altering the demand for both observed and unobserved skills \citep[e.g.,][]{autor_levy_murnane_2003,acemoglu_autor_2011, acemoglu_restrepo_2022, acemoglu_loebbing_2022}.

In an important challenge to the conventional view, \cite{lemieux_2006} demonstrates that the rise in residual inequality is at least partially explained by an increase in the variance of unmeasured skills resulting from composition changes in the labor market, especially in the late-1980s and 1990s, as the workforce shifted increasingly to older and more educated workers who exhibit greater within-group inequality. \cite{lemieux_2006} and \cite{gottschalk_moffitt_2009} further argue that increasing measurement error and short-term wage volatility have also contributed to rising residual inequality.
The extent to which the rise in residual inequality reflects an increase in returns to unobserved skills, growing unobserved skill inequality, or increased wage volatility unrelated to skills is critical to understanding both the economic causes and welfare consequences of rising inequality.  This paper develops a new approach for disentangling the importance of these distinct economic forces.
%\footnote{Economists have long recognized the importance of distinguishing between transitory and permanent income shocks for understanding inequality in consumption and welfare.  Several recent studies \citep[e.g.,][]{krueger_perri_2006, blundell_pistaferri_preston_2008, heathcote_storesletten_violante_2014} further show that conclusions about the nature and degree of consumption insurance (over time) depend critically on estimated correlations between consumption and income changes as well as the relative importance of transitory vs.\ permanent income shocks (over time).}

Several recent studies have turned to richer data to incorporate additional measures of skills or occupational tasks, directly estimating their effects on wages at different points in time.  Using the 1979 and 1997 Cohorts of the National Longitudinal Surveys of Youth (NLSY), \cite{castex_dechter_2014} estimate that the wage returns to cognitive achievement, as measured during adolescence by the Armed Forces Qualifying Test (AFQT), declined substantially between the late-1980s and late-2000s in the United States. 
\cite{deming_2017} confirms this finding but further estimates that the returns to social skills have risen across these two cohorts.\footnote{\cite{edin_et_al_2022} estimate relatively stable returns to cognitive skills and rising returns to a measure of teamwork and leadership skills in Sweden.}  Among others, \cite{autor_levy_murnane_2003} and \cite{autor_dorn_2013} document a decline in demand for middle-skill workers caused by the automation of routine tasks, which has led to a fall in the wages for workers in many  middle-skill relative to low- and high-skill occupations, dubbed `polarization'.  \cite{caines_hoffmann_kambourov_2017} instead argue that occupational task complexity has become a stronger determinant of wages in recent years, more so than routineness.

While efforts to better measure skills and job tasks have enriched our understanding of wage inequality, much of the cross-sectional variation in wages remains unexplained in these studies.  More importantly, difficult measurement challenges have led to strong (often implicit) assumptions on the evolution of skills over the lifecycle and across cohorts. For example, \cite{castex_dechter_2014} and \cite{deming_2017} examine the effects of pre-market skills on the wages of workers in their late-20s, thereby ignoring early-career lifecycle skill accumulation that may vary across workers and over time.  The vast majority of studies taking a task-based approach do not use individual-level data on skills or job tasks, implicitly assuming that worker skills and tasks within each occupation are time-invariant.  As a result, these studies attribute all time variation in wages across occupations to changes in the returns to skills/tasks.\footnote{%\cite{autor_handel_2013} find that person-level job tasks vary systematically across demographic groups within occupations, suggesting that worker skills and job tasks are likely to change within occupations as the workforce composition changes. 
Comparing skill requirements by occupation across editions of the Dictionary of Occupational Titles (DOT) and O*NET, \cite{cavounidis_et_al_2021} and \cite{cortes_jaimovich_siu_2023} document within-occupation changes in the skill/task content/requirements of jobs in the U.S. \ Using data with individual-level measures of job tasks, \cite{spitz-oener_2006} shows that most task changes in Germany over the 1980s and 1990s occurred within occupations.}

Studies of long-term changes in residual wage inequality or the returns to unobserved skills largely rely on repeated samples of cross-section data, making it difficult to distinguish changes in skill returns from changes in the distributions of skills.  As we show, panel data are naturally more useful.
Intuitively, if heterogeneity in skills is important, then workers earning a high wage one year should continue, on average, to earn a high wage many years later (even after the influence of transitory wage shocks has faded). As such, heterogeneity in unobserved skills implies that differences in wage residuals across workers should be predictive of long-term future residual differences, with those predicted differences growing (shrinking) as the returns to skill rise (fall) over time.
%Moreover, an increase in the return to unobserved skills should lead to divergence in average log wage residuals across workers with different initial wage residuals.  This is not what we observe during the late-1980s and 1990s.  

\begin{figure}[h]
  \centering
  \includegraphics[width=0.45\columnwidth]{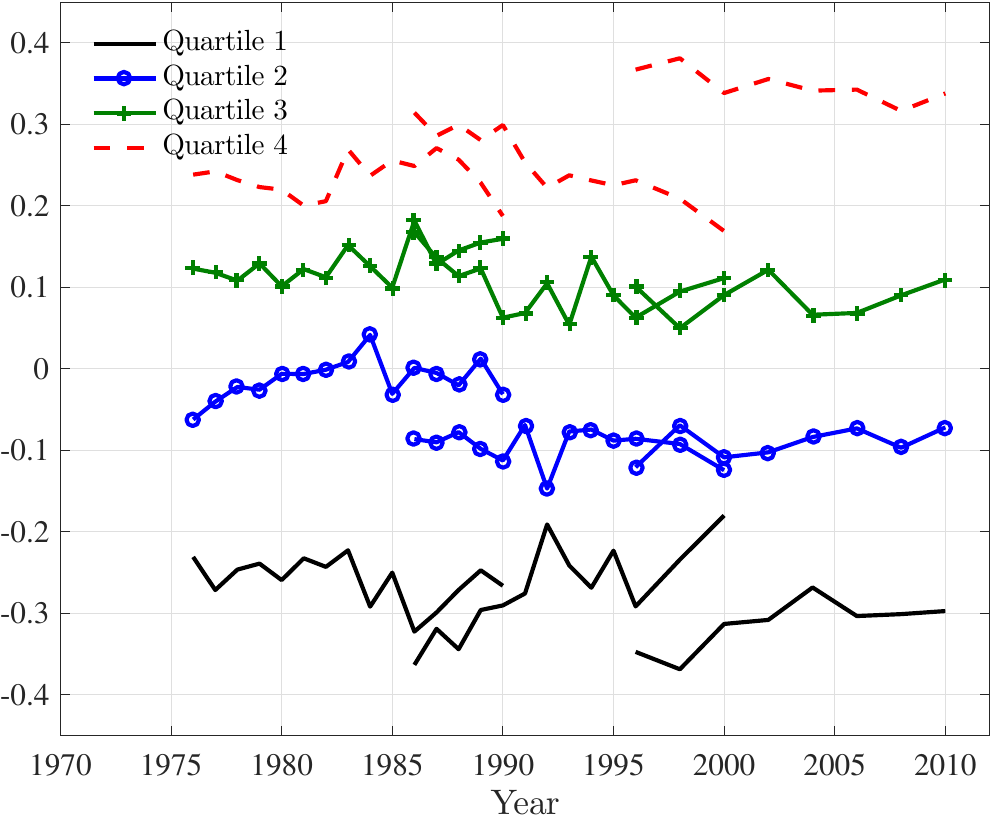}
\caption{Average Predicted Log Wage Residuals by Baseline Residual Quartile}
    \label{fig: resid quart pred}
\caption*{Notes: Each line reflects average log wage residuals in year $t$ conditional on the log wage residual quartile in a base year 6--20 years earlier.  Moving from left to right, the baseline years are 1970, 1980, and 1990.}
\end{figure}

Categorizing workers based on their log wage residual quartile in three different base years ($b=$ 1970, 1980, and 1990), Figure~\ref{fig: resid quart pred} reports their average residuals 6--20 years later (i.e., years $t=b+6,...,b+20$).   Consistent with an important role for unobserved skills, those with higher wage residuals in any given base year also earn more, on average, up to 20 years later.\footnote{Differences in levels across lines for any given quartile are due to differences in base year wage distributions.}
The strong convergence in average residuals (conditional on base-year quartiles) over the late-1980s and 1990s indicates that either the persistence of skill differences across workers fell (e.g., greater skill depreciation at the top of the skill distribution relative to the bottom) or that the returns to skills declined over this period.  This paper shows that the latter explanation is most consistent with a broad array of evidence: the returns to unobserved skill \textit{fell} sharply over the late-1980s and 1990s. 
By contrast, returns were more stable in earlier and later years.

We show that if unobserved skill growth is uncorrelated with sufficiently lagged skill levels and if non-skill wage shocks exhibit limited persistence, then a simple instrumental variable (IV) strategy can be used to estimate growth in skill returns over time.
While endogenous skill investments raise concerns about the skill-growth assumption for young workers, it is much more natural for older workers whose skill investments are likely to be negligible \citep{becker_1964, ben-porath_1967}.  Indeed, we show that panel data on cognitive test scores for older men in the Health and Retirement Study (HRS) support this assumption, as do several other specification tests based on the autocovariance structure for residuals. We also show that this assumption, along with the assumption of limited persistence in non-skill wage shocks, can be relaxed; although, estimation then requires a more general moment-based approach that exploits more of the autocovariance structure for log wage residuals.\footnote{The voluminous literature on earnings dynamics uses a similar moment-based estimation approach but focuses on a different set of questions from ours: identifying the changing importance of permanent vs.\ transitory shocks in earnings and the resulting implications for consumption and wealth inequality \citep[e.g.,][]{gottschalk_moffitt_1994, blundell_preston_1998, haider_2001,moffitt_gottschalk_2002, meghir_pistaferri_2004, bonhomme_robin_2010, heathcote_storesletten_violante_2010, heathcote_perri_violante_2010,  moffitt_gottschalk_2012, blundell_graber_mogstad_2015}.}
Once the returns to skill have been estimated (from, e.g., older workers), it is straightforward to estimate the distribution of unobserved skills, skill growth, and non-skill shocks over time.  Importantly, there is no need to observe independent measures of skills or what workers do on their jobs, enabling application of our approach in widely available panel data sets.

Using PSID data on log hourly wages, we estimate the evolution of returns to unobserved skills for American men from 1970 to 2012. Our main finding is that the returns to unobserved skills were relatively stable from 1970 to the mid-1980s, then \textit{fell} considerably through the late-1980s and 1990s, stabilizing thereafter. The drop in estimated returns reflects the sharp convergence in predicted wage residuals conditional on earlier differences as documented in Figure~\ref{fig: resid quart pred} and is robust to different estimation strategies and assumptions about the dynamics of skills and non-skill wage shocks. The decline in skill returns appears to be slightly stronger for non-college workers, consistent with the recent literature on polarization \citep{autor_levy_murnane_2003,acemoglu_autor_2011,autor_dorn_2013}.

The flip side of declining returns is that the variance of unobserved
skill has risen substantially since the early-1980s, driving most of the increase in residual wage  inequality.\footnote{The widening skill growth distributions within education and experience groups are not accounted for in the composition adjustments of \cite{lemieux_2006}.} Consistent with stability of AFQT distributions among teenagers across NLSY cohorts \citep{altonji_bharadwaj_lange_2012}, this increase is not driven by growth in the variance of early-career skill levels across cohorts. Instead, we find that the growing skill dispersion reflects an increase in the variance of idiosyncratic skill growth innovations, consistent with the notion of growing economic turbulence studied by \cite{ljungqvist_sargent_1998}.  We find little evidence of heterogeneity in systematic lifecycle skill growth, as studied by \cite{lillard_weiss_1979}, \cite{macurdy_1982}, \cite{baker_solon_2003}, \cite{guvenen_2009}, and \cite{moffitt_gottschalk_2012}.

A growing number of studies highlight differences in pay across firms and/or occupations, as well as the potential for different trends in the returns to heterogeneous skills (see, e.g., \cite{acemoglu_autor_2011}, \cite{sanders_taber_2012}, \cite{kline_2024}, and \cite{woessmann_2024} for recent reviews). We first show that heterogeneity in firm-specific pay cannot account for our estimated declines in skill returns. We then extend our analysis to consider occupation-specific wage schedules over a multi-dimensional skill vector.  We show that our IV estimator identifies a weighted-average of returns across different (unobserved) skills and use occupation-stayers to estimate the evolution of occupation-specific skill returns.  Based on the PSID, we estimate very similar long-run declines in the returns to skills within routine, cognitive, and social occupations.

Given recent concerns about differences in the dynamics of log earnings residuals between survey and administrative data \citep[see, e.g.,][]{sabelhaus_song_2010}, we reproduce key results using earnings measures from W-2 forms (collected by the the Internal Revenue Service, IRS) linked with several panels of the Survey of Income and Program Participation (SIPP). This analysis also indicates substantial long-run declines in average returns to unobserved skills; however, it suggests weaker (though not statistically different) declines in skill returns within cognitive relative to routine occupations since 1990.

This paper proceeds as follows. Section~\ref{sec: identification} describes our baseline assumptions used to identify and estimate the returns to skill over time using panel data on wages, contrasting these assumptions with those used in prior work \citep[e.g.][]{juhn_murphy_pierce_1993, lemieux_2006, castex_dechter_2014}.  We also test our main assumption on unobserved skill dynamics using cognitive test scores in the HRS.
Section~\ref{sec: evidence} describes the PSID data used for most of our empirical analysis and reports estimated returns to unobserved skill in the U.S. since the late-1970s. 
Section~\ref{sec: skill dist} discusses identification of unobserved skill distributions, separately from the distributions of non-skill shocks, and provides estimates of these distributions over time.  We also decompose the variance of skills into contributions from heterogeneity in initial skills and variation due to idiosyncratic lifecycle skill growth.
Section~\ref{sec: occ multiple skills} extends our analysis to account for differences across firms, occupations, and multiple skills. We confirm our PSID-based empirical findings with administrative data in Section~\ref{sec: GSF} before concluding in Section~\ref{sec: conclusion}.

\section{Identifying and Estimating the Returns to Unobserved Skills}\label{sec: identification}

We consider the following specification for log wages motivated by the literature on unobserved skills \cite[e.g.,][]{juhn_murphy_pierce_1993, card_lemieux_1994, chay_lee_2000, lemieux_2006}:
\begin{eqnarray}
  \ln W_{i,t} &=& f_t(\bm x_{i,t}) +w_{i,t} \label{eq: log wage spec} \\
  w_{i,t}  &=& \mu_t\theta_{i,t}+\varepsilon_{i,t}, \label{eq: resid}
\end{eqnarray}
where $W_{i,t}$ reflects wages for individual $i=1,...,N$ in period $t=\ul{t},...,\bt$, $f_t(\bm x_{i,t})$ reflects the time-varying influence of observed characteristics $\bm x_{i,t}$ (e.g.\ education, race, experience), and $w_{i,t}$ is the log wage ``residual'' satisfying $\E[\theta_{t}|\bm x_{t}]=\E[\eps_{t}|\bm x_{t}]=0$.\footnote{Let $z_t$ be a random variable and its realization for individual $i$ be $z_{i,t}$. Denote its cross-sectional first and second moments by $\E[z_t]$, $\var(z_t)$, and $\cov(z_t,z_{t'})$.} 
There are two ways to think about $\theta_t$: 
it may reflect variation in a homogenous skill (produced through schooling and other investments) conditional on educational attainment (and other factors in $\bm x_t$), or it may reflect an unobserved skill completely distinct from any skills produced through schooling.  Thus, $\mu_t$ could reflect returns to a single skill, with estimated ‘returns’ to schooling over time in $f_t(\bm x_t)$ capturing returns to that skill as well as differences in the average amount of skill produced through schooling.  Or, $\mu_t$ could reflect returns to a distinct skill not produced through education.  The latter is a more natural interpretation when estimated returns to education follow a different time path from $\mu_t$; although, such differences could  be explained by changes in how education produces skills.  We take the interpretation of \cite{juhn_murphy_pierce_1993} and most subsequent studies, treating $\theta_t$ as its own unique unobserved skill and $\mu_t$ as the return to that skill.

The residual $w_{i,t}$ reflects the contributions of unobserved skill (equivalently, worker productivity) $\theta_{i,t}$ and idiosyncratic non-skill shocks $\eps_{i,t}$, which may include measurement error.\footnote{\cite{chay_lee_2000}, \cite{card_lemieux_1994}, and \cite{lemieux_2006} consider the same log wage residual decomposition; however, they assume that the variances of skills within observable groups (e.g.\ education, experience, race) are time invariant.  Thus, their approaches only account for changes in the overall variance of unobserved skills due to changes in the composition of workers across observable types. In estimating the importance of these composition changes, \cite{lemieux_2006}  ignores any variation in the transitory component, $\eps_{i,t}$; although, he also provides a separate analysis documenting increases in log wage measurement error over time.  Our use of panel data facilitates a more general analysis.}
Note that average unobserved skill growth, which may vary by observable characteristics, is subsumed in changes in $f_t(\bm x_{i,t})$.\footnote{The assumption of separability between $\bm x_{i,t}$ and $\theta_{i,t}$ is both common and convenient, though not necessary. One can condition the analysis that follows on $\bm x_{i,t}$.  Our empirical analysis separately studies non-college and college educated workers.}
Individuals may come from different cohorts (i.e.\ different years of labor market entry), as discussed further below.
Our analysis focuses on the log wage residual of equation~\eqref{eq: resid} with the aim of identifying and estimating the returns to unobserved skill $\mu_t$ over time.\footnote{Equations~\eqref{eq: log wage spec} and \eqref{eq: resid} imply wage \textit{levels} that are non-linear in unobserved skill. As such, variation in $\mu_t$ over time is inconsistent with perfect substitutability across workers of different skill levels, since this would imply log wages functions that are additively separable in `prices' and skills.  Instead, this non-linearity is consistent with assignment and task-based models of the labor market \citep[see, e.g.,][]{sattinger_1993, costinot_vogel_2010,acemoglu_autor_2011}. See \cite{lochner_park_shin_2018} for assumptions on the production technology and distribution of skills and firm productivity that yield wage functions given by equations~\eqref{eq: log wage spec} and \eqref{eq: resid} in a standard assignment model.}  We also use the residual $w_{i,t}$ to identify and estimate the evolution of unobserved skill variation over time.

A few recent studies \citep[e.g.,][]{castex_dechter_2014, deming_2017} take advantage of skill measurements, or test scores, to aid in identification of the returns to skill.  To facilitate discussion of these  studies and to test our own assumptions, consider a (potentially) repeated skill measurement, $T_{i,t}$, in period $t$:
\begin{equation} \label{eq: T}
T_{i,t} = g_{t}(\bm x_{i,t}) +  \tau \theta_{i,t} + \eta_{i,t}.
\end{equation}
This specification allows test scores to vary with both observed factors and unobserved skills.\footnote{Note that $g_{j,t}(\cdot)$ may reflect differential measurement quality across groups or differences in skills across groups (e.g., total skills measured by the test may be given by $\tilde{g}_{j,t}(\bm x_t) + \theta_t$, in which case $g_{j,t}(\bm x_t) = \tau_j \tilde{g}_{j,t}(\bm x_t)$).}   We assume that unobserved skills have the same effect on scores for the same test regardless of when the test is taken (i.e., $\tau$ is time-invariant). We also assume throughout our analysis that test measurement errors are serially uncorrelated and are uncorrelated with other observed variables $\bm x_{t}$, unobserved skills $\theta_{t}$, and non-skill shocks $\eps_{t}$.\footnote{Specifically, we assume that %$\cov(\eta_{t},\eta_{t'})=0$ for all $t \neq t'$ and that $\E(\eta_{t}|\bm x_{t''}) =\cov(\eta_{t},\theta_{t'}|\bm x_{t''})=\cov(\eta_{t},\eps_{t'}|\bm x_{t''})=0$ for all $t,t',t''$.
 $\E[\eta_t|\bm x_t]=\cov(\eta_t,\theta_t|\bm x_t)=0$ for all $t$, $\cov(\eta_t,\theta_{t'})=\cov(\eta_t,\eps_{t'})=0$ for all $t,t'$, and $\cov(\eta_t,\eta_{t'})=0$ for all $t\neq t'$.} It is useful to define test score residuals, $\tilde{T}_{i,t} \equiv T_{i,t} - g_{t}(\bm x_{i,t}) = \tau \theta_{i,t} + \eta_{i,t}$.

\subsection{Prior Assumptions in the Literature}

We briefly consider the strategies and assumptions previously employed in the literature on skill returns.

\subsubsection{\cite{juhn_murphy_pierce_1993}}

By equating the increase in residual inequality with an increase in skill returns, $\mu_t$, \cite{juhn_murphy_pierce_1993} assume that the cross-sectional distributions of unobserved skills and non-skill shocks are time-invariant.  To ``test'' this assumption, they compare growth in the variance of residuals when following cohorts vs.\ experience groups over the period they examine (1963--1989).  Unfortunately, this comparison is not very informative about the evolution of skill distributions or returns over time.
To see why, let $c$ reflect a cohort's year of labor market entry and assume for simplicity that $\cov(\eps_t,\theta_t|c)=0$ and $\var(\eps_{t}|c)=\var(\eps_{t})$ for all $t\geq c$. Then,
\begin{eqnarray*}
  \big[\underbrace{\var(w_{t+\ell}|c)-\var(w_{t}|c)}_{\text{within-cohort}}\big]-  \big[\underbrace{\var(w_{t+\ell}|c+\ell)-\var(w_{t}|c)}_{\text{within-experience}}\big] &=&   \var(w_{t+\ell}|c) - \var(w_{t+\ell}|c+\ell) \\
& = &  \mu_{t+\ell}^2\left[\var(\theta_{t+\ell}|c) - \var(\theta_{t+\ell}|c+\ell)\right],
\end{eqnarray*}
which equals zero if the variance of skills (in period $t+\ell$) does not differ across cohorts.\footnote{This would arise if, for example, growth in the variance of skills accumulated via labor market experience was exactly offset by growth in the variance of initial skills across cohorts.}  Notice that $\var(\theta_{t+\ell}|c) = \var(\theta_{t+\ell}|c+\ell)$ for all $\ell \geq 0$ implies that the variance of skills within each period is the same across all cohorts, but it does not say anything about the evolution of skill variation or returns over time. As discussed further in Appendix~\ref{app: JMP}, equal within-cohort and within-experience growth in the variance of residuals is consistent with growth in skill returns, the variance of skill growth, or the variance of non-skill wage shocks.\footnote{Empirically, \cite{juhn_murphy_pierce_1993} estimate very similar within-experience and within-cohort growth in \textit{log wage} variation between 1964 and 1989; yet, their results suggest 20--30\% stronger within-cohort (relative to within-experience) growth in \textit{log wage residual} variation from 1970 to 1988.  These results are, therefore, consistent with stronger growth in unobserved skill variation over the lifecycle than across cohorts during the 1970s and 1980s, but they say little about changes in the population-wide distribution of unobserved skill over this period.}

\subsubsection{\cite{lemieux_2006}} \label{subsec: lemieux}

Assuming that the variance of skills conditional on observed characteristics is time-invariant, $\var(\theta_t|\bm x_t=\bm x) = \sigma^2(\bm x), \ \forall (t,\bm x)$, \cite{lemieux_2006} shows that the variance of skills increased over time due to compositional shifts in the labor market. To estimate changes in returns to unobserved skill over time, he implicitly ignores time variation in $\var(\eps_t)$ and re-weights $\var(w_{t}|\bm x_t)$ each year to account for composition shifts in $\bm x_{t}$.\footnote{A separate analysis in \cite{lemieux_2006} shows that measurement error in wages increased over time, at least in the widely used March Current Population Survey (CPS).}

Since Lemieux's assumption is not related to the dynamics of skill, it can be tested using repeated cross-section data with the same skill measurement over time. To see this, notice that equation \eqref{eq: T} implies
\begin{align*}
    \var(T_{t}|\bm x_t)=\E[\tilde{T}_{t}^2|\bm x_t]=\tau^2\var(\theta_t|\bm x_t) + \var(\eta_{t}|\bm x_t).
\end{align*}
Assuming the variance of test score measurement error, $\var(\eta_{t}|\bm x_t)$, does not change over time, time-invariance of $\var(\theta_t|\bm x_t)$ implies that $\var(T_{t}|\bm x_t)$ should also be constant over time.

Using data on men with 30--50 years of experience in the 1996--2018 HRS, we test whether the variance of cognitive memory scores conditional on $\bm x_t$ has changed over time by regressing the squared residual of memory scores on indicators for race, educational attainment, experience, and year.\footnote{The cognitive memory measure is a combination of immediate and delayed recall with raw scores ranging from 0 to 20. While the HRS contain other skill measures, they are either discrete (with very few values) or available for limited years.
The memory recall measure we use has a correlation of roughly 0.3 with other cognitive tests focused on math skills, roughly 0.25 with log wages, and 0.07 with log wage residuals.  See Appendix~\ref{app: HRS} for details on these measures and our HRS data.} If $\var(\theta_t|\bm x_t=\bm x)$ is time-invariant, then the coefficients on year indicators should all be equal.  Our results, shown in Appendix~\ref{app: Lemieux}, strongly reject time-invariance for highly experienced men in the HRS, indicating changes in within-group skill inequality over time.

\subsubsection{\cite{castex_dechter_2014} and \cite{deming_2017}}

A few recent studies incorporate direct skill measurements in estimating the returns to (traditionally unobserved) skills over time \citep[e.g.,][]{castex_dechter_2014, deming_2017}.
Regressing log wages of workers in their late-20s on adolescent skill measures for different cohorts, these studies identify changes in the effects of adolescent skills on adult earnings (10--15 years later), confounded by any changes in measurement reliability.  Even ignoring idiosyncratic errors in skill measurements, these estimates do not necessarily identify the evolution of returns to contemporaneous skills, $\mu_t$,  because they are confounded by any cross-cohort changes in the relationship between adolescent skills and adult skills.

In our context, OLS regression of log wage residuals in year $t+\ell$ on test residuals in year $t$ identifies
\[
 \hat{\beta}_{t,t+\ell}\parrow  \frac{\cov(w_{t+\ell},\tilde{T}_{t})}{\var(\tilde{T}_{t})}
  = \frac{\mu_{t+\ell}}{\tau} \underbrace{\left[1 + \frac{ \cov(\theta_{t+\ell}-\theta_t,\theta_{t})}{\var(\theta_{t})} \right]}_{\text{Skill Dynamics}} \underbrace{\left[\frac{\tau^2\var(\theta_{t})}{\tau^2\var(\theta_{t})+\var(\eta_{t})} \right]}_{\text{Test Reliability Ratio } (R_{t})},
\]
where $\ell \geq 0$ reflects the years between wage measurement and the time tests were administered.
Under ideal conditions, this regression identifies returns to skill in any period $t+\ell$ up to the test score scale: $\mu_{t+\ell}/\tau$. Unfortunately, the two terms in brackets complicate identification of skill return growth.  

Following a single cohort over time (i.e., varying $\ell$ for fixed $t$) will confound systematic heterogeneity in skill growth with changes in the returns to skill, as reflected in the ``Skill Dynamics'' term.\footnote{See \cite{murnane_willett_levy_1995} and \cite{cawley_heckman_vytlacil_2001} for efforts to sort out the rising importance of schooling vs.\ cognitive ability for earnings using the NLSY79.} 
Instead of following the same cohort over time, \cite{castex_dechter_2014}  compare estimates $\hat{\beta}_{t,t+\ell}$ and $\hat{\beta}_{t',t'+\ell}$ across the NLSY79 and NLSY97 cohorts where $t\approx 1980$ and $t'\approx 1997$, respectively.  They use the same cognitive test measure, AFQT, in both periods with wages reported roughly 10--15 years after the tests were administered.\footnote{The NLSY79 (NLSY97) surveyed youth born 1957--1964 (1980--1984) administering a battery of tests to all respondents. The AFQT tests measure math and reading skills and were administered in 1980 (NLSY79) and 1997 (NLSY97) for most respondents in the two cohorts.  In practice, the Armed Services Vocational Aptitude Battery (ASVAB) underlying the AFQT was taken via pencil and paper in the NLSY79, while it was administered in computer-adaptive form for the NLSY97.}
Comparing across cohorts, the ``Skill Dynamics'' term may differ due to cohort differences in the dynamics of skill accumulation between the year tests are taken ($t$ and $t'$) and the year wages are measured ($t+\ell$ and $t'+\ell$).
If the variance of test score measurement error is time-invariant, then the ``Test Reliability Ratio'', $R_{t}$, will be the same across cohorts if and only if $\var(\theta_{t}) = \var(\theta_{t'})$, which then implies that the variance of test score residuals should also be the same across cohorts.
For $\ell \geq 1$, $\hat{\beta}_{t',t'+\ell} /\hat{\beta}_{t,t+\ell}$ is unlikely to identify growth in unobserved skill returns, $\mu_{t'+\ell}/\mu_{t+\ell}$,  if (i) the process for unobserved skill dynamics (over the first $\ell$ years after tests are measured) differs across cohorts or (ii) ``initial'' skill distributions differ across cohorts.\footnote{If wages are observed during the same years skills are measured for both cohorts (i.e., $\ell=0$), then the ``Skill Dynamics" term equals one, and consistent estimation of growth in skill returns, $\mu_{t'}/\mu_{t}$, depends only on time-invariance of the test reliability across cohorts.}

Given modest changes in the distribution of AFQT scores across cohorts \citep{altonji_bharadwaj_lange_2012}, the ``Test Reliability Ratio'' term is likely to be very similar across NLSY cohorts.
By contrast, there are good reasons to think that skill dynamics during early-adulthood have changed.  For example, \cite{ashworth_et_al_2021} document increases in work experience throughout high school and college, coupled with a rise in time to college degree for the NLSY97 cohort. Additionally, Appendix~\ref{app: NLSY occ} documents substantial changes across NLSY cohorts in the types of occupational experience accumulated over ages 17--26.  Most notably, experience accumulated in sales positions nearly tripled, while experience in manager and professional positions increased by 23\% and 54\%, respectively.  Increases in management and professional experience were particularly strong at the high end of the AFQT distribution, while increases in sales and service experience were more uniform or concentrated at the low end.

Even ignoring these concerns, \cite{castex_dechter_2014} are only able to estimate changes in the returns to skill across two snapshots in time, from the late-1980s to around 2010.  These estimates, as well as similar estimates for AFQT by \cite{deming_2017}, suggest that the returns to math and reading skills \textit{fell} by roughly half over this 20-year period.  Our estimated returns to skill presented below imply a similar drop, indicating that much of the decline occurred during the late-1980s and 1990s with relative stability in the 2000s.

This discussion has assumed that the same skill measurement is used for both cohorts; otherwise, cross-cohort comparisons of $\hat{\beta}_{t,t+\ell}$ and $\hat{\beta}_{t',t'+\ell}$ would also be confounded by differences in the mapping between skills and their measurement (i.e., differences in $\tau$ across tests).  This is an additional challenge faced by \cite{deming_2017}, who aims to estimate changes in the return to social skills across NLSY cohorts. Unfortunately, he is forced to use different measures of social skills across cohorts, which means that his results cannot distinguish between differences in the ``strength'' of those measures vs.\ changes in skill returns, even if one is willing to assume that initial social skill distributions and their accumulation through high school, college, and early work-years remained the same across NLSY cohorts -- a questionable assumption given the aforementioned cross-cohort increases in experience working in service, professional, and management occupations.\footnote{\cite{deming_2017} normalizes his available measures of social skills to have a standard deviation of one in both cohorts before estimating $\hat{\beta}_{t,t+\ell}$ and $\hat{\beta}_{t',t'+\ell}$ (for social skills); however, this does not eliminate bias coming from differences in $\tau$ across measurements. See Appendix~\ref{app: CD} for details. \cite{edin_et_al_2022} take advantage of more consistent measures of cognitive and social/leadership skills across cohorts in Sweden, estimating modest reductions in returns to cognitive skills and increases in returns to social/leadership skills. While scores need not be re-normalized for each cohort, this analysis still relies on the assumption that early-career skill dynamics are identical across cohorts, as well as stability in measurement reliability ratios over time.}

\subsection{Identification using Panel Data on Wages} \label{subsec: returns identification}

Previous efforts to estimate returns to unobserved skills rely on assumptions about the stability of skill distributions  or early skill dynamics across cohorts.
Using panel data, we introduce a different approach based primarily on an assumption about lifecycle skill and wage dynamics.
Central to our approach is the classical idea of \cite{friedman_kuznets_1954} that earnings consist of a permanent component related to skills and a transitory component unrelated to skills. Although the transitory component, which may include measurement error, can be serially correlated, the correlation between transitory components far apart in time is likely to be negligible.\footnote{Also see \cite{carroll_1992} and \cite{moffitt_gottschalk_2011}, who make similar assumptions ensuring that ``long'' autocovariances for log earnings residuals reflect a permanent component.}  
We make the following assumption, which imposes restrictions on the lifecycle dynamics of skill, the interaction between skills and non-skill shocks, and the persistence of non-skill shocks.
%We begin with the following assumption.

\begin{ass}  \label{assum: gen mu ident}
For known $k\geq 1$ and for all $t-t' \geq k$:
(i) $\cov(\Delta \theta_t, \theta_{t'})=0$;
(ii) $\cov(\Delta \theta_t, \eps_{t'}) = 0$;
(iii) $\cov(\eps_t, \theta_{t'}) = 0$; and
(iv) $\cov(\eps_t,\eps_{t'})=0$.
\end{ass}
Condition (i) assumes that skill growth is uncorrelated with sufficiently lagged skill levels.  This allows for both permanent and transitory skill innovations.  Condition (ii) allows for non-skill shocks to influence skill growth in the short-term but not in the long-term.  For example, family illness or short-term work disruptions (including transitory firm-level productivity disruptions) may impact skill growth in the same year or even over the next $k-1$ years.  Condition (iii) is satisfied if skill levels are uncorrelated with non-skill shocks $k$ or more years later, while condition (iv) requires that non-skill shocks have limited persistence (e.g., they may follow an MA$(q)$ process where $q\leq k-1$). We discuss all of these conditions in greater detail below, empirically testing or relaxing those most central to identifying skill returns.

Our analysis assumes a sufficiently long panel with length satisfying $\bt - \ul{t} \geq k+1$.  Let $\Delta$ reflect the first-difference operator.

\begin{prop} \label{prop: gen mu ident}
Assumption~\ref{assum: gen mu ident} implies that for all $t-t' \geq k+1$, the following instrumental variable (IV) estimator identifies skill return growth rates:
\begin{equation} \label{eq: IV estimator}
\frac{\cov(\Delta w_t,w_{t'})}{\cov(w_{t-1},w_{t'})} =\frac{\Delta\mu_t}{\mu_{t-1}}.
\end{equation}
For $\bt - \ul{t} \geq k+1$ and normalizing $\mu_{t^*}=1$ for some $t^*\geq \ul{t}+k$, all other $\mu_{\ul{t}+k},\mu_{\ul{t}+k+1},...,\mu_{\bt}$ are identified.
\end{prop}

\noindent Proof: For all $t-t' \geq k$,
\begin{align}
\cov(w_t,w_{t'}) &= \cov(\mu_t \theta_t + \eps_t, \mu_{t'}\theta_{t'} + \eps_{t'}) \nonumber \\
    &= \mu_t\cov(\theta_t,\mu_{t'}\theta_{t'}+\eps_{t'}) + \mu_{t'}\cov(\eps_t,\theta_{t'}) + \cov(\eps_t,\eps_{t'}) \nonumber \\
    &= \mu_t\cov(\theta_t,\mu_{t'}\theta_{t'}+\eps_{t'})    & \text{[Assum \ref{assum: gen mu ident}(iii)--(iv)]} \nonumber  \\
    &= \mu_t\cov(\theta_{t'+k-1}+\Delta\theta_{t'+k} + \Delta\theta_{t'+k+1}+...+\Delta \theta_{t},\mu_{t'}\theta_{t'} + \eps_{t'}) \nonumber \\
    &= \mu_t\underbrace{[\mu_{t'}\cov(\theta_{t'+k-1},\theta_{t'}) + \cov(\theta_{t'+k-1},\eps_{t'})]}_{\equiv \Omega_{t'}}   & \text{[Assum \ref{assum: gen mu ident}(i)--(ii)].}  \label{eq: cov gen}
\end{align}
Thus, for  $t-t'\geq k+1$,
\[
\frac{\cov(\Delta w_t,w_{t'})}{\cov(w_{t-1},w_{t'})} =\frac{\Delta \mu_t\Omega_{t'}}{\mu_{t-1}\Omega_{t'}} = \frac{\Delta \mu_t}{\mu_{t-1}}.
\]
$\Box$

Proposition~\ref{prop: gen mu ident} shows that $\Delta \mu_t/\mu_{t-1}$ can be estimated by regressing $\Delta w_{i,t}$ on $w_{i,t-1}$ using sufficiently lagged $w_{i,t'}$ as an instrument.  This IV approach is intuitive, since wage residuals can be thought of as `noisy' measures of skill levels.
To further this line of reasoning,  follow the approach of \cite{holtz-eakin_newey_rosen_1988}, using $\theta_{i,t}= \theta_{i,t-1} +\Delta \theta_{i,t}$ and $\theta_{i,t-1}=(w_{i,t-1}-\eps_{i,t-1})/\mu_{t-1}$ to obtain an expression for $\Delta w_{i,t}$ in terms of $w_{i,t-1}$:
\begin{equation}
  \Delta w_{i,t} = \left[ \mu_t\left(\frac{w_{i,t-1}-\varepsilon_{i,t-1}}{\mu_{t-1}}+\Delta \theta_{i,t}\right)+\varepsilon_{i,t} \right] - w_{i,t-1}= \frac{\Delta \mu_t}{\mu_{t-1}} w_{i,t-1} + \left(\varepsilon_{i,t} - \frac{\mu_t}{\mu_{t-1}} \varepsilon_{i,t-1} + \mu_t \Delta \theta_{i,t}\right).  \label{eq:  regression}
\end{equation}
This suggests that lagged residuals $w_{i,t-1}$, much like a test score, might serve as a proxy for unobserved skills.  However, $w_{i,t-1}=\mu_{t-1}\theta_{i,t-1}+\varepsilon_{i,t-1}$ is a `noisy' measure of unobserved skill, so it is correlated with the error $\varepsilon_{i,t-1}$, as well as $\varepsilon_{i,t}$ if $\cov(\varepsilon_{t},\varepsilon_{t-1}) \neq 0$.  Simply
regressing $\Delta w_{i,t}$ on $w_{i,t-1}$ would, therefore, produce a biased
estimate of $\Delta \mu_t/\mu_{t-1}$. To address this problem, lagged wage
residuals from the distant past (i.e.\ any $w_{i,t'}$ for $t' \leq t-k-1$) can be used as instrumental variables in 2SLS estimation, since they are correlated with $w_{i,t-1}$ (through unobserved skills) but uncorrelated with $\varepsilon_{i,t-1}$, $\varepsilon_{i,t}$, and $\Delta \theta_{i,t}$ (under Assumption~\ref{assum: gen mu ident}).

In general, future wage residuals are not valid instruments in equations~\eqref{eq: IV estimator} or \eqref{eq:  regression}, because skill growth has lasting effects on future skills, generating a correlation between future  wage residuals and $\Delta \theta_{i,t}$.
This correlation biases the IV estimator (for $\Delta \mu_t/\mu_{t-1}$) and makes it challenging to estimate skill returns during early sample periods. Appendix~\ref{app: identify early mu} discusses conditions under which different cohorts may be used to eliminate the bias, enabling estimation of skill returns over the full sample period.  Given the lengthy period covered by many panel data sets and stronger identification requirements for early skill returns, we focus on identification and estimation for periods $t \geq \ul{t}+k$.

The evolution of returns to skill are also directly related to predicted differences in wages across workers given any prior differences.  Strengthening Assumption~\ref{assum: gen mu ident} to mean independence, $\E[\eps_t|\theta_{t'},\eps_{t'}]=\E[\Delta\theta_t|\theta_{t'},\eps_{t'}]=0$ for $t-t'\geq k$, implies that
\[
\E[w_t|w_{t'}] = \mu_t \underbrace{\left(\frac{w_{t'} - \E[\eps_{t'}|w_{t'}]}{\mu_{t'}} + \E[\theta_{t'+k}-\theta_{t'}| w_{t'}]\right)}_{\equiv \Psi_{t'}(w_{t'})}, \quad \text{for all $t \geq t'+k$}.
\]
Because wages are increasing in skills and skills are persistent, workers with a high wage in one period will also tend to have a high wage in the future, even after the influence of transitory non-skill shocks has disappeared.\footnote{This discussion assumes that $\Psi_{t'}(w_{t'})$ is an increasing function, as observed empirically.}

More importantly for our purposes, for any given differences in year $t'$ residuals across workers, long-term differences in expected future residuals, $\E[w_t|w_{t'}]$,  will increase (decrease) over time as the returns to skill $\mu_t$ increase (decrease):
\begin{equation} \label{eq: predicted wages}
\E[w_{t}|w_{t'} = w^H] - \E[w_{t}|w_{t'} = w^L] = \mu_t\big(\Psi_{t'}(w^H)- \Psi_{t'}(w^L)\big), \quad \text{for all $t \geq t'+k$}.
\end{equation}
Thus, the strong convergence in predicted future log wage residuals by prior residual quartiles over the late 1980s and 1990s shown in Figure~\ref{fig: resid quart pred} indicates a sharp decline in the returns to skill over those years.

In Sections~\ref{subsect: relax skill growth assum} and \ref{subsec: persistent eps}, we consider generalizations of Assumption~\ref{assum: gen mu ident} that relax conditions (i) and (iv) on the dynamics of skills and non-skill shocks.  Here, we briefly
make a few additional observations on identification of skill returns.

\paragraph{Transitory skill shocks.} Proposition~\ref{prop: gen mu ident} also applies if $\eps_{it}$ shocks are considered a component of skills, i.e., if $w_{i,t} = \mu_t (\theta_{i,t} + \eps_{i,t})$.\footnote{In this case, $\Omega_{t'} = \mu_{t'}[\cov(\theta_{t'+k-1},\theta_{t'}) + \cov(\theta_{t'+k-1},\eps_{t'})]$.} Whether transitory shocks are assumed to be related or unrelated to skills has no effect on identification or IV estimation of the returns to skill under Assumption~\ref{assum: gen mu ident}.  Conceptually, it seems natural to think that transitory wage innovations have little to do with skills, so we continue with residuals as defined in equation~\eqref{eq: resid}.

%\paragraph{Serially correlated non-skill shocks.} Our identification strategy has, thus far, relied on the assumption that non-skill shocks, $\eps_{t}$, become serially uncorrelated when observations are far enough apart.  This is not critical; although, identification is most transparent in this case.  Appendix~\ref{app: ARMA ident} shows identification of skill returns when the `transitory' component $\eps_{t}$ contains an autoregressive component, such that the serial correlation in non-skill shocks depreciates over time but never fully disappears. Section~\ref{subsec: persistent eps} shows that estimates assuming $\eps_t$ contains an AR(1) component are quite similar to those obtained under our baseline Assumption~\ref{assum: gen mu ident}.

\paragraph{Time-invariant skills.}  If skills are heterogeneous but time-invariant (i.e., $\theta_{i,t}=\theta_i$ with $\cov(\eps_t,\theta)=0$ for all $t$), then
\begin{equation} \label{eq: cov invariant theta}
\cov(w_t,w_{t'}) = \mu_t\mu_{t'}\var(\theta), \quad \text{for all $|t-t'| \geq k$.}
\end{equation}
In this case, $\Delta \mu_t/\mu_{t-1}$ could be identified and estimated using the IV estimator in equation~\eqref{eq: IV estimator} with sufficiently \textit{lagged} or \textit{future} log wage residuals (i.e., $w_{t'}$ satisfying $t'\leq t-k-1$ or $t' \geq t+k$) as instruments.  For panel length satisfying $\bt -\ul{t} \geq 2k$ and a single normalization (e.g., $\mu_{\ul{t}+k}=1$), all $\mu_{\ul{t}},...,\mu_{\bt}$ would be identified along with $\var(\theta)$.
Comparing IV estimates using past vs.\ future wage residuals as instruments, our empirical analysis below provides strong evidence against fixed  unobserved skills over the lifecycle.

\paragraph{Conditioning on observable subgroups.}  Assumption~\ref{assum: gen mu ident} can be modified so that all conditions (and results) hold for any observable subgroup, including specific cohort, age, or experience groups.  For example, it is natural to condition on older (or more experienced) workers for whom endogenous human capital investments are likely to be negligible \citep{becker_1964,ben-porath_1967}.\footnote{Appealing to \cite{becker_1964} and \cite{ben-porath_1967}, previous studies rely on the assumption of zero skill growth among older workers to identify the evolution of additively separable (log) skill prices \citep[e.g.][]{heckman_lochner_taber_1998, bowlus_robinson_2012} or the distribution of skill shocks \cite[e.g.][]{huggett_ventura_yaron_2011}. This assumption is stronger than needed in our context, where condition (i) of Assumption~\ref{assum: gen mu ident} only rules out persistent unobserved heterogeneity in skill growth among experienced workers.  Heterogeneity in skill growth based on observable characteristics is accounted for through $f_t(\bm x_t)$ in obtaining residuals.} Our empirical analysis below pays particular attention to experienced workers, estimating returns to skill based on this subgroup.

\subsection{Testing our Assumption on Skill Dynamics} \label{subsec: AR(1) skills test}

Even if endogenous skill investments become negligible as workers approach the end of their careers, skill growth rates may still be correlated with past skill levels for older workers due to other factors (e.g., skill depreciation may systematically differ across workers with different skill levels).  We examine this possibility using the same skill measure and sample of men with 30--50 years of experience in the 1996--2018  HRS data used earlier in Section~\ref{subsec: lemieux}.

We test whether $\cov(\Delta_2\theta_{t+2},\theta_{t-\ell})=0$ using the following moments:
\begin{equation} \label{eq: gamma  moment}
\E\left[(\Delta_2 \tilde{T}_{t+2}- \varrho \tilde{T}_{t})\tilde{T}_{t-\ell}\right]=0, \quad \text{for $\ell\geq k$,}
\end{equation}
where $\Delta_2$ reflects the two-period time difference given our use of biennial data from the HRS.  (These moments are consistent with 2SLS regression of $\Delta_2 \tilde{T}_{t+2}$ on $\tilde{T}_{t}$ using $\tilde{T}_{t-\ell}$ as an instrument.)
We test whether Assumption~\ref{assum: gen mu ident}(i) holds for
various $k$ values.  The first four columns of Table~\ref{tab: test Assum 1} test this assumption for $k=2$ by testing whether $\varrho=0$ when using instruments of lags $\ell \geq 2$. Columns 5 and 6 test the assumption for $k=4$ and $k=6$, respectively, using only longer lags as instruments.
Panel A of Table~\ref{tab: test Assum 1} reports GMM estimates of $\varrho$ using residualized memory recall scores.  Although we reject $\varrho=0$ at the 5\% significance level when instruments of lags $\ell\leq 4$ are used, the estimated $\varrho$ values are quite small.  For perspective, if skills follow a simple autoregressive process (i.e., $\theta_t = \rho \theta_{t-1} + \nu_t$ with $\cov(\theta_t,\nu_{t'})=0$ for all $t'\geq t+1$), then $\varrho =\rho^2-1$.  The reported estimates in columns 1--5 would all imply $\rho$ values of 0.97--1.02, very close to a random walk.  The last column of Table~\ref{tab: test Assum 1} reports an estimated $\varrho$ of 0.018 when using lags $\ell=6,8$.  This estimate is not significantly different from zero and suggests that Assumption~\ref{assum: gen mu ident}(i) is satisfied for $k=6$.  That is, skill growth rates $\Delta \theta_t$ are uncorrelated with skill levels (at least) 6 years earlier, $\theta_{t-6}$.

Panel~B of Table~\ref{tab: test Assum 1} reports estimates of $\varrho$ when also including lagged log wage residuals, $w_{t-\ell}$, as additional instruments. In this case, $\varrho=0$ implies that both conditions (i) and (ii) of Assumption~\ref{assum: gen mu ident} are satisfied for the relevant $k$.  These estimates are nearly identical to those using only lagged memory test score residuals as instruments in Panel~A, indicating that condition (ii) is likely to be satisfied.  Altogether, the estimates reported in Table~\ref{tab: test Assum 1} suggest that, for older men at least, conditions (i) and (ii) of Assumption~\ref{assum: gen mu ident} are satisfied for $k=6$, while violations of those conditions are quite modest for $k$ as small as 2.
%\footnote{See Appendix~\ref{app: HRS} for additional details and analogous 2SLS estimates of $\varrho$ using the same moments. }  
We conduct most of our analysis assuming that these conditions are satisfied for $k=6$; however, we reach very similar conclusions when relaxing condition (i) in Section~\ref{subsect: relax skill growth assum}.

\begin{table}[h]
\small
  \caption{GMM estimates of $\varrho$ in equation~\eqref{eq: gamma moment} using ($\tilde{T}_{i,t-\ell}$, $w_{i,t-\ell}$) as instruments}
  \label{tab: test Assum 1}
\centering
{
\def\sym#1{\ifmmode^{#1}\else\(^{#1}\)\fi}
\begin{tabular}{l*{6}{c}}
  \toprule
                    & $\ell=2$ &  $\ell=2, 4$ & $\ell=2, 4, 6$ & $\ell=2, 4, 6, 8$ & $\ell=4, 6, 8$ &$\ell=6, 8$  \\
  \midrule
\multicolumn{7}{l}{A. Instruments: $\tilde{T}_{i,t-\ell}$ }    \\
\quad Estimated $\varrho$                  &       0.045\sym{*}&      -0.040\sym{*}&      -0.029\sym{*}&      -0.031\sym{*}&      -0.057\sym{*}&       0.018       \\
                    &     (0.020)       &     (0.012)       &     (0.011)       &     (0.010)       &     (0.014)       &     (0.022)      \vspace{4pt}\\
\quad Implied $\rho = \sqrt{1+\varrho}$ & 1.022        &   0.980            & 0.985        & 0.984     & 0.971     & 1.009 \\
\addlinespace
\multicolumn{7}{l}{B. Instruments: $\tilde{T}_{i,t-\ell},w_{i,t-\ell}$ }    \\
\quad Estimated $\varrho$       &       0.044\sym{*}&      -0.040\sym{*}&      -0.030\sym{*}&      -0.031\sym{*}&      -0.059\sym{*}&       0.018       \\
                    &     (0.019)       &     (0.011)       &     (0.011)       &     (0.010)       &     (0.013)       &     (0.022)           \vspace{4pt}\\
\quad Implied $\rho = \sqrt{1+\varrho}$ & 1.022        & 0.980     & 0.985     & 0.984     & 0.970     & 1.009 \\
\bottomrule
\multicolumn{7}{l}{Notes: $\tilde{T}_{t}$ are residuals from regressions of word recall on experience, cohort, race, and} \\
\multicolumn{7}{l}{education dummies. $w_{t}$ are residuals from year-specific regressions on the same covariates.} \\
\multicolumn{7}{l}{Uses 1996--2018 HRS data for men ages 50--70 with 30--50 years of experience. Estimated} \\
\multicolumn{7}{l}{ via two-step optimal GMM with cluster-robust weighting matrix. $^*$ denotes significance at} \\
\multicolumn{7}{l}{0.05 level.} \\
\end{tabular}
}
\end{table}

%%%%%%%%%%%%%%%%%%%%%%%%%%%%%%%%%%%%%%%%%%%%%%%%%%%%%%%%%%%%%%%%%%%%%%%%%%%%%%%%%%%%%%%%%%%%%%%%%%%%%%%%%%%%%%%%%%%%%%

\section{New Evidence on Returns to Unobserved Skill} \label{sec: evidence}

Our primary objective is estimation of returns to unobserved skill over time.  We mainly exploit data from the PSID; however, we replicate key results using administrative data on earnings in Section~\ref{sec: GSF}.  This section briefly describes the PSID before turning to estimated returns to skill based on these data.

\subsection{Panel Study of Income Dynamics (PSID)}\label{sec: PSID}

The PSID is a longitudinal survey of a representative sample of
individuals and families in the U.S.\ beginning in 1968.  The survey
was conducted annually through 1997 and biennially since. We use data
collected from 1971 through 2013. Since earnings
were collected for the year prior to each survey, our analysis
studies hourly wages from 1970 to 2012.

Our sample is based on male heads of households from the core
(SRC) sample restricted to years these men were ages 16--64, had potential experience of 1--40 years,
reported positive wage and salary income, had positive hours worked, and
were not enrolled as students.  Our sample is 92\% white with
an average age of 39 years.  Roughly half of our sample completed more than 12 years of schooling, which we refer to as ``college workers''.
The wage measure used in our analysis divides annual earnings by annual hours worked, trimming
the top and bottom 1\% of all wages within year and college/non-college status by ten-year experience
cells. The resulting sample contains 3,766 men and 44,547 person-year
observations -- roughly 12 observations for each individual.
See Appendix~\ref{app: PSID} for further details.

Our analysis focuses on log wage residuals $w_{i,t}$ from equation \eqref{eq: log wage spec} after controlling for differences in educational attainment, race, and experience.  Specifically, we estimate $f_t(\bm x_{i,t})$ by year and college vs.\ non-college status from separate linear regressions of log hourly wages  on indicators for each year of potential experience, race/ethnicity, and 7 educational attainment categories, along with interactions between a cubic in experience and both race and education indicators.

\subsection{Underlying Trends}
As documented in Figure~\ref{fig: variance of hourly earnings}, log wage inequality has increased substantially since 1970, with particularly strong increases in the early-1980s and after 2000. The
evolution of residual inequality closely mirrors this pattern, explaining a larger share of the total variance than the between-group variance.  

Consistent with an important role for unobserved skills, Figure~\ref{fig: resid quart pred} shows that those with higher wage residuals in any given year also have higher wage residuals, on average, up to 20 years later.\footnote{For comparison, Appendix Figure~\ref{fig: resid quart actual} shows average log wage residuals for each quartile over time without conditioning on prior residual levels.}  The sharp convergence in log wage residuals (across quartiles of earlier residual levels) over the late-1980s and 1990s  indicates that the returns to skill fell over that period (see equation~\eqref{eq: predicted wages}), despite modest growth in residual inequality at the time.

\begin{figure}[!htbp]
    \centering
    \subfloat[All Men]{
      \includegraphics[width=0.45\columnwidth]{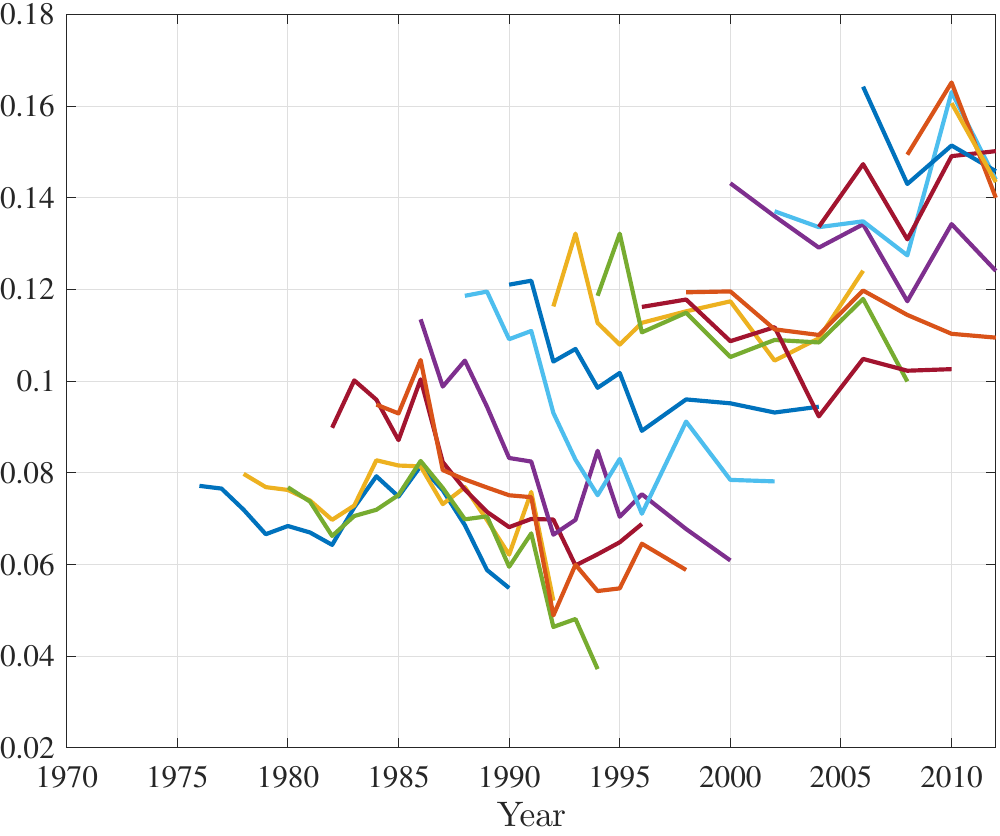}
}\quad
    \subfloat[Men with 15--30 Years of Experience in Year $b$]{
      \includegraphics[width=0.45\columnwidth]{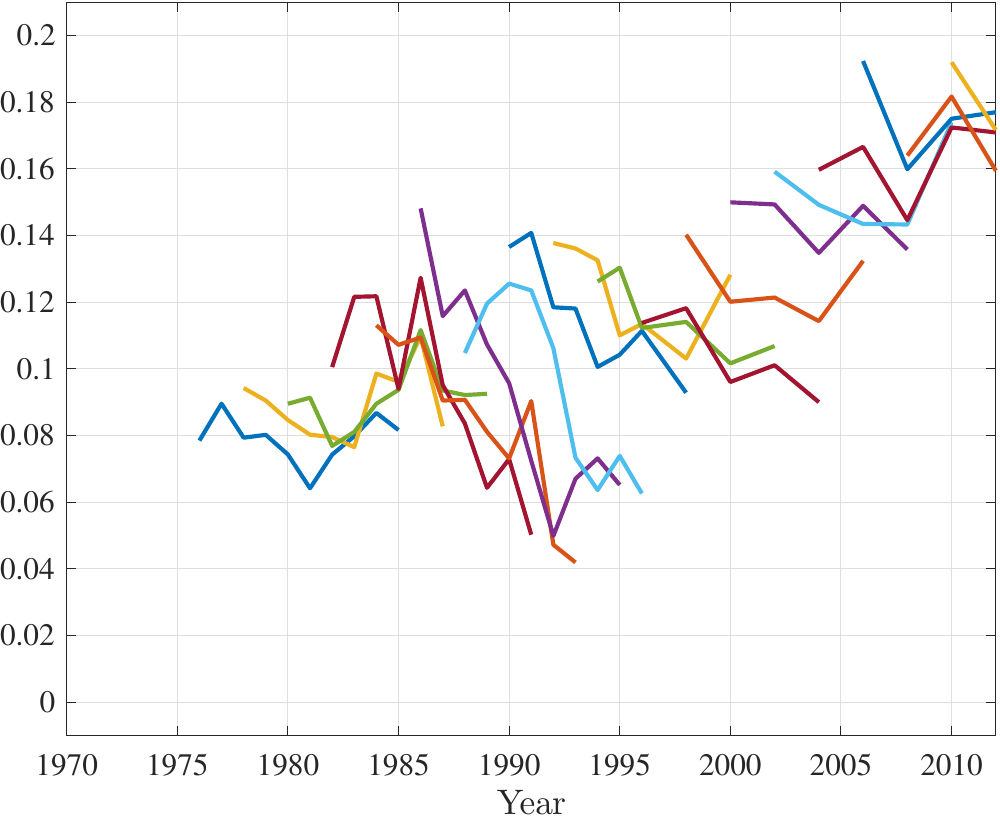}
    }
\caption{Autocovariances for Log Wage Residuals}
    \label{fig: PSID autocov}
\end{figure}

Section~\ref{subsec: returns identification} suggests that long autocovariances in wage residuals offer a more direct way to identify changes in the returns to unobserved skill. Figure~\ref{fig: PSID autocov}(a) reports $\cov(w_b,w_t)$ for $t=b+6,...,b+20$ with each line reporting autocovariances for a different `base' year $b$ and 15 subsequent years.\footnote{Appendix Figure~\ref{fig: PSID autocov no attrition} shows that sample attrition due to non-response or aging/retirement does not affect the autocovariance patterns documented here.}  For example, the leftmost line beginning in 1976 reflects autocovariances for $b=1970$ and values of $t$ ranging from 1976--1990. If systematic differences in unobserved skill growth are negligible and $t-b$ is large enough such that transitory shocks are uncorrelated, then $\cov(w_b,w_t) = \mu_t\Omega_b$  (see equation~\eqref{eq: cov gen}) and following each line over $t$ is directly informative about the evolution of $\mu_t$. 
The sharply declining autocovariances over the late-1980s and 1990s (regardless of base year) suggest that the returns to unobserved skill  fell substantially over this period, while the more stable autocovariances during earlier and later years are consistent with more modest changes in returns during those years.
As discussed in Section~\ref{sec: skill dist}, the upward-shifting lines beginning in the 1980s signal a period of rising skill dispersion.
Figure~\ref{fig: PSID autocov}(b) reveals very similar autocovariance patterns when restricting the sample to men with 21+ years of experience in years $t \geq b+6$.\footnote{Appendix Figure~\ref{fig: PSID autocov expr 1-14} shows qualitatively similar, though flatter, autocovariance patterns for less-experienced men.}

%As noted earlier, it is useful to focus on more experienced workers for whom differences in unobserved skill growth should be largely idiosyncratic due to diminished investment incentives \citep{becker_1964, ben-porath_1967}. Figure~\ref{fig: PSID autocov}(b) reveals very similar autocovariance patterns to Figure~\ref{fig: PSID autocov}(a) when restricting the sample to men with 15--30 years of experience as of baseline $b$ years (21+ years of experience in years $t \geq b+6$).\footnote{Appendix Figure~\ref{fig: PSID autocov expr 1-14} shows qualitatively similar, though flatter, autocovariance patterns for less-experienced men.

\begin{figure}[h]
  \centering
  \includegraphics[width=0.45\columnwidth]{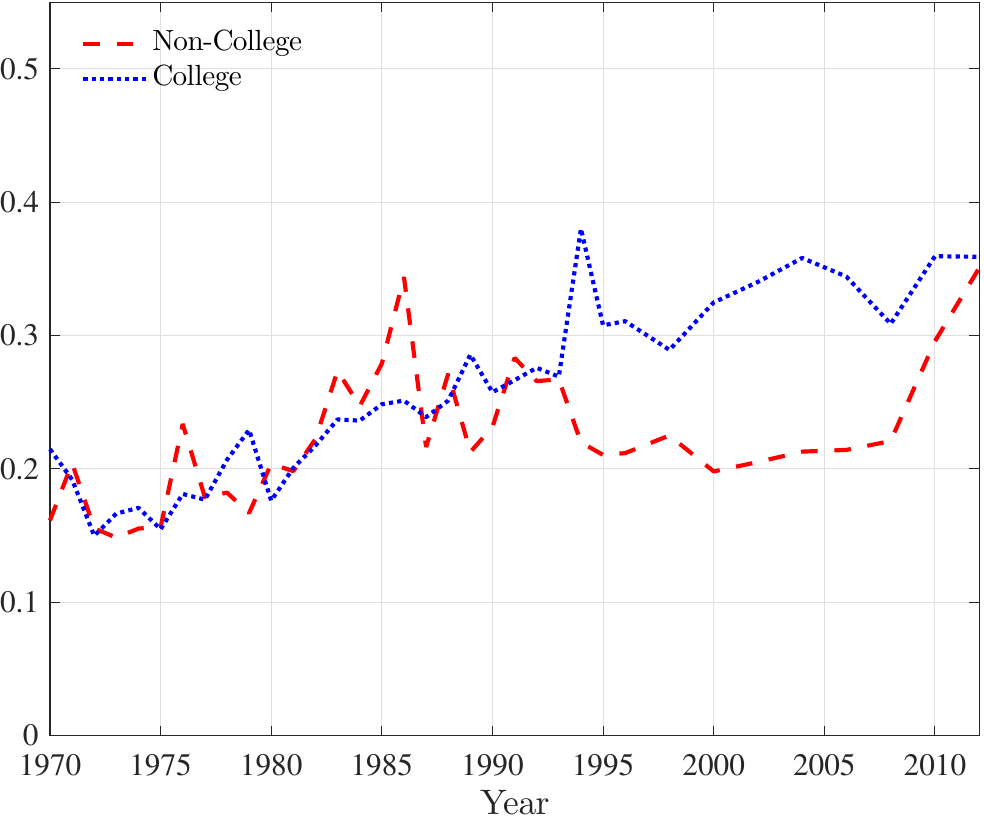}
\caption{Variance of Log Wage Residuals by Education}
    \label{fig: variance of residuals by sector}
\end{figure}

\begin{figure}%[h]
    \centering
    \subfloat[Non-College]{
      \includegraphics[width=0.45\columnwidth]{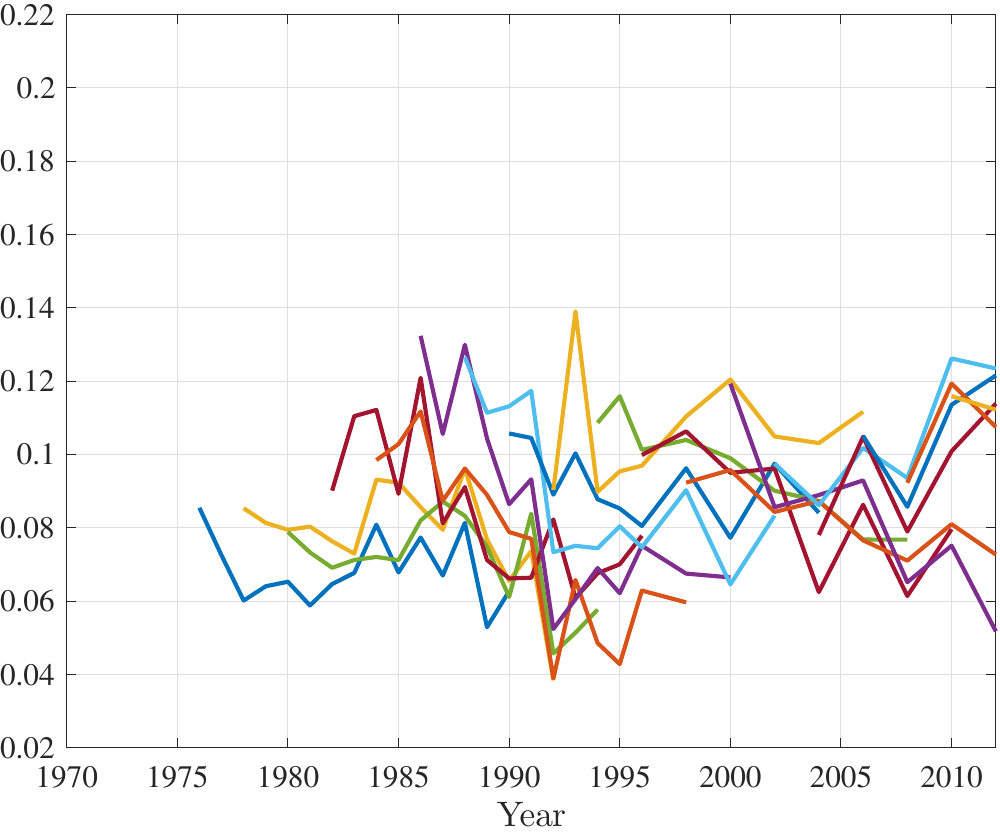}
}\quad
    \subfloat[College]{
      \includegraphics[width=0.45\columnwidth]{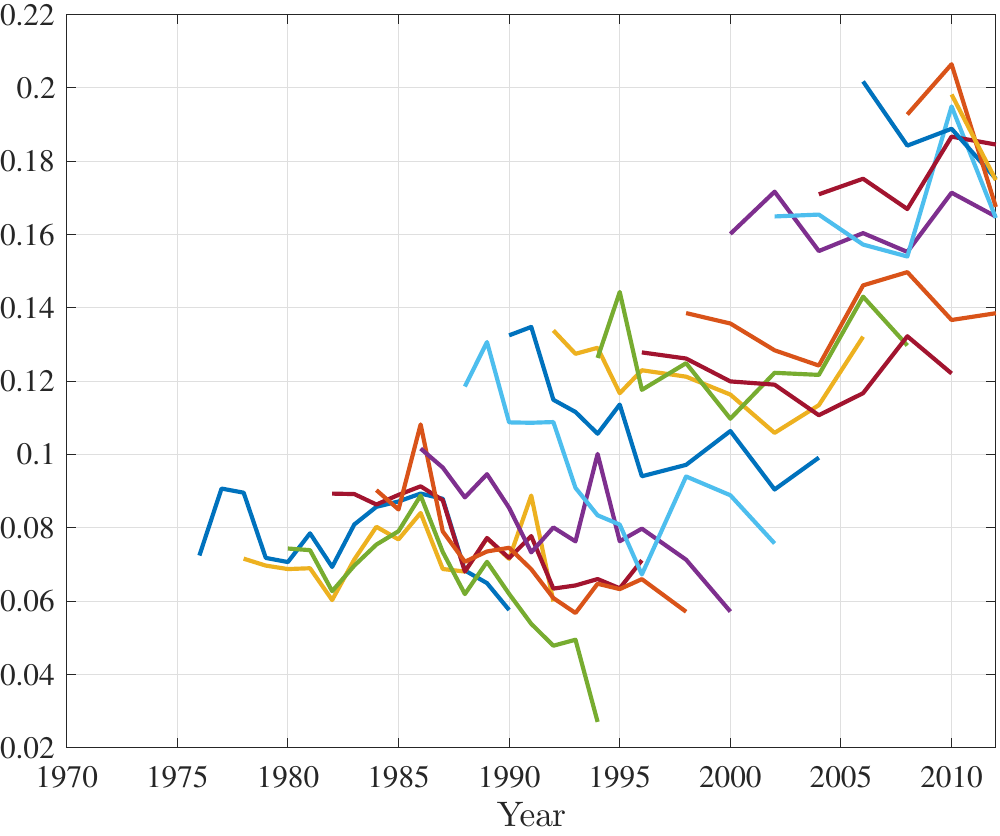}
}
\caption{Autocovariances for Log Wage Residuals by Education, All Experience levels}
    \label{fig: PSID autocov by educ}
\end{figure}

As emphasized by the literature on `polarization' in the U.S.\ labor market \citep{autor_levy_murnane_2003,autor_katz_kearney_2008,acemoglu_autor_2011,autor_dorn_2013}, wage inequality has evolved differently at the top and bottom of the wage distribution.
Figure~\ref{fig: variance of residuals by sector} shows that the rise in residual inequality over the early-1980s was stronger among non-college workers, before falling and then quickly stabilizing in the mid-1980s, while it continued to increase among college workers throughout our sample period. Do these trends reflect differences in the evolution of returns to skill by educational attainment, as is often assumed?
Figure~\ref{fig: PSID autocov by educ} shows that long-autocovariances declined sharply during the late-1980s and 1990s (given any base year) for both non-college- and college-educated men, indicating qualitatively similar declines in the returns to skill for both education groups over that period.\footnote{As discussed in Section~\ref{sec: skill dist}, the larger shift upward of the lines for college men (in Figure~\ref{fig: PSID autocov by educ}) implies a stronger increase in skill dispersion over the latter half of our sample period.}  These autocovariance patterns are central to our empirical approach and drive our main findings.

%Figure~\ref{fig: PSID autocov by educ} reports long autocovariances separately for non-college and college educated men.  The time patterns are qualitatively similar for both education groups with two noteworthy differences:  First, the autocovariance lines continue declining for non-college men throughout the early-2000s when they flatten out for college men.  This suggests that the returns to skill continued falling for non-college men several years after they stabilized for college men. Second, the lines generally shift upward over time, with particularly strong increases over the late-1990s and early-2000s for college men. As discussed in Section~\ref{sec: skill dist}, these shifts reflect rising skill dispersion, partially muted by declining skill returns.

\subsection{2SLS Estimation of Skill Returns}\label{sec: IV}

In this subsection, we directly estimate growth rates in the returns to unobserved skill based on the IV strategy described in Section \ref{subsec: returns identification}.
Because our data is only available every other year later in the sample period, we slightly modify the 2SLS approach based on equation~\eqref{eq:  regression} to estimate two-year growth rates,  $\Delta_2\mu_{t}/\mu_{t-2}$, based on the following:
\begin{equation}
\Delta_2w_{i,t} = \left(\frac{\Delta_2\mu_t}{\mu_{t-2}}\right) w_{i,t-2} + \left[\mu_t(\Delta \theta_{i,t-1}+\Delta \theta_{i,t}) + \varepsilon_{i,t} - \frac{\mu_t}{\mu_{t-2}} \varepsilon_{i,t-2}\right].  \label{eq: 2SLS regression}
\end{equation}
Under Assumption~\ref{assum: gen mu ident}, we can obtain consistent estimates of $\Delta_2\mu_{t}/\mu_{t-2}$ by estimating equation \eqref{eq: 2SLS regression} via 2SLS using lags $w_{i,t'}$ for $t' \leq t-k-2$ as instrumental variables.

Table~\ref{tab: iv1} reports 2SLS estimates of skill return growth rates
using equation \eqref{eq: 2SLS regression} for years $t$ covering
1979--1995, assuming that skill return growth rates are constant within
two- or three-year periods (i.e.\ 1979--1980, 1981--1983, ..., 1993--1995).
Assuming $k=6$, we use $(w_{i,t-8},w_{i,t-9})$ as instruments.  Table~\ref{tab: iv2} reports 2SLS estimates for the later years of the PSID ($t$ covering 1996--2012) when observations become biennial.\footnote{Estimates in Table~\ref{tab: iv2} assume two-year return growth rates are constant within each of the periods 1996--2000, 2002--2006, and 2008--2012, and use $(w_{i,t-8},w_{i,t-9})$ as instruments for 1996--2000 and $(w_{i,t-8},w_{i,t-10})$ thereafter.}  In all specifications, the instruments are `strong' with very large first-stage $F$-statistics.

\begin{figure}[h]
  \centering
    \subfloat[All Experience Levels]{
  \includegraphics[width=0.45\columnwidth]
  {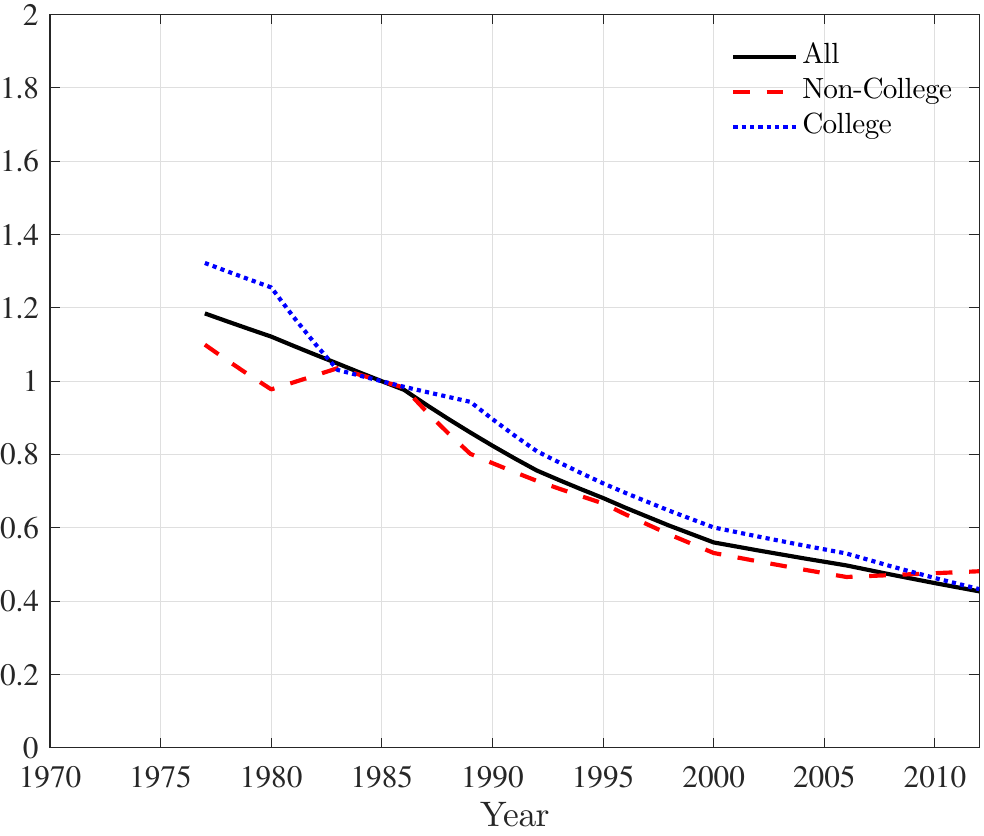}}
\quad
    \subfloat[21--40 Years of Experience in Year $t$]{
  \includegraphics[width=0.45\columnwidth]
  {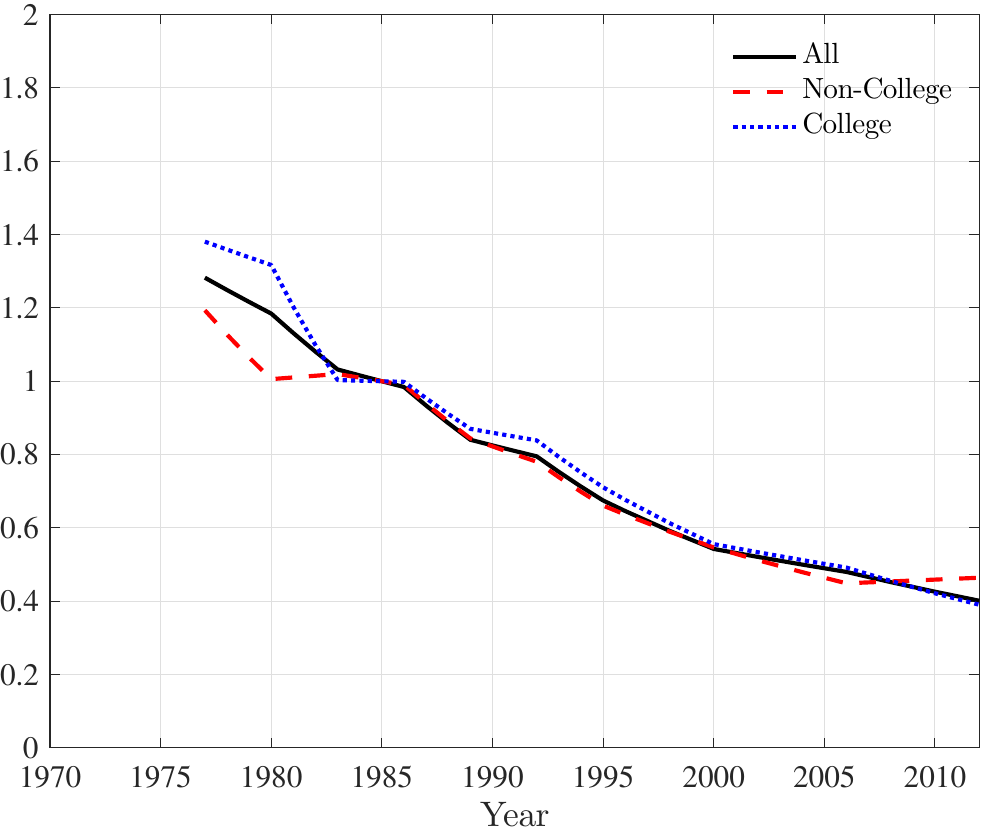}}
  \caption{$\mu_t$ Implied by 2SLS Estimates ($\mu_{1985}=1$)} \label{fig: IV mu}
\end{figure}

Panel A of Tables~\ref{tab: iv1} and \ref{tab: iv2} reports estimates for the full sample of men in the PSID, while panels~B and~C report separate estimates for non-college and college men.  Consistent with the autocovariances reported earlier, nearly all of these estimates are negative, with several statistically significant.  Appendix Tables~\ref{tab: iv1 old} and \ref{tab: iv2 old} report analogous results for the subsample of men with 21--40 years of experience (in year $t$) for whom we expect systematic heterogeneity in skill growth to be negligible. Figure~\ref{fig: IV mu} combines these estimates to trace out the implied paths for $\mu_t$ from 1979--2012, normalizing $\mu_{1985}=1$. Altogether, these results suggest that the returns to unobserved skill  \textit{declined} by roughly half since the mid-1980s, mirroring the substantial decline in returns to cognitive skills between the NLSY79 and NLSY97 cohorts estimated by \cite{castex_dechter_2014}. 
%Section~\ref{sec: occ multiple skills} considers the interpretation of these estimated return series when there are multiple unobserved skills whose returns may evolve differently over time.  We also estimate similar return patterns for men working in different occupation types.

\begin{table} \small
\def\sym#1{\ifmmode^{#1}\else\(^{#1}\)\fi}
\caption{2SLS estimates of $\Delta_2\mu_{t}/\mu_{t-2}$ for two- or three-year periods, 1979--1995}\label{tab: iv1} 
\center
\begin{tabular}{l*{6}{c}}
\hline
                    & 1979--1980 & {1981--1983}& {1984--1986}  & {1987--1989}    & {1990--1992}   &  {1993--1995}\\
 \hline
\multicolumn{7}{c}{\underline{A. All men}} \vspace{2pt}\\
$\Delta_2\mu_{t}/\mu_{t-2}$ &     -0.036         &     -0.044         &     -0.046         &     -0.081$^{*}$  &     -0.082$^{*}$  &     -0.067      \\
                    &   (0.045)         &    (0.038)         &    (0.038)         &    (0.034)         &    (0.035)         &    (0.035)        \\
\addlinespace
Observations         & 1,349        & 2,077        & 2,188         & 2,245        &  2,189         & 2,095         \\
1st stage \(F\)-Statistic&     163.09         &     191.61         &     114.85         &     209.42         &     227.13         &     286.96       \\
\\
\multicolumn{7}{c}{\underline{B. Non-college men}} \vspace{2pt}\\
$\Delta_2\mu_{t}/\mu_{t-2}$ &      -0.075         &      0.039         &     -0.035         &      -0.127$^*$  &     -0.062   &   -0.057 \\
                    &    (0.061)         &    (0.056)         &    (0.060)         &    (0.050)         &    (0.058)         &    (0.054)        \\
\addlinespace
Observations        & {740}         & {1,080}         & {997}         & {965}         &  {897}         & {851}               \\
1st stage \(F\)-Statistic&      81.85         &      85.23         &      39.48         &      98.34         &      92.27         &      91.33             \\
\\
\multicolumn{7}{c}{\underline{C. College men}} \vspace{2pt}\\
$\Delta_2\mu_{t}/\mu_{t-2}$&     -0.034         &      -0.123$^{*}$  &     -0.030     &     -0.028     &     -0.097$^{*}$  &     -0.074   \\
                    &    (0.061)         &    (0.048)         &    (0.049)         &    (0.047)         &    (0.047)      &    (0.046)    \\
\addlinespace
Observations      & {508}         & {884}         & {1,046}         & {1,109}         & {1,107}         & {1,242}              \\
1st stage \(F\)-Statistic&     100.95         &     115.03         &  123.38 & 97.29  &     122.42         &     208.04             \\
\hline
\multicolumn{7}{l}{Notes: Estimates from 2SLS regression of $\Delta_2 w_{i,t}$ on $w_{i,t-2}$ using instruments $(w_{i,t-8}, w_{i,t-9})$.  }\\
\multicolumn{7}{l}{$^*$ denotes significance at 0.05 level. }
\end{tabular}
\end{table}
%\clearpage

\begin{table} \small
\def\sym#1{\ifmmode^{#1}\else\(^{#1}\)\fi}
\caption{2SLS estimates of $\Delta_2\mu_{t}/\mu_{t-2}$ for five-year periods, 1996--2012}\label{tab: iv2}  
\center
\begin{tabular}{l*{4}{c}}
\hline
                    &\multicolumn{1}{c}{1996--2000}&\multicolumn{1}{c}{2002--2006}&\multicolumn{1}{c}{2008--2012}\\
\hline
\multicolumn{4}{c}{\underline{A. All men}} \vspace{2pt}\\
$\Delta_2\mu_{t}/\mu_{t-2}$ &       -0.075$^{*}$ &     -0.039         &     -0.050          \\
                    &          (0.025)         &    (0.028)         &    (0.027)        \\
\addlinespace
Observations     & {2,122}         & {2,129}         & {1,968}       \\
1st stage \(F\)-Statistic&      369.09         &     344.25         &     341.36          \\
\\
\multicolumn{4}{c}{\underline{B. Non-college men}} \vspace{2pt}\\
$\Delta_2\mu_{t}/\mu_{t-2}$ &       -0.087$^{*}$  &     -0.043         &      0.011            \\
   &    (0.043)         &    (0.047)         &    (0.075)           \\
\addlinespace
Observations    & {862}         & {826}         & {615}         \\
1st stage \(F\)-Statistic     &     121.44         &     142.56         &     104.92         \\
\\
\multicolumn{4}{c}{\underline{C. College men}} \vspace{2pt}\\
$\Delta_2\mu_{t}/\mu_{t-2}$&         -0.070$^*$  &     -0.041         &     -0.065$^*$   \\
                    &         (0.031)         &    (0.034)         &    (0.029)         \\
\addlinespace
Observations        & {1,252}         &{1,293}         & {1,141}         \\
1st stage \(F\)-Statistic &    260.47         &     218.64         & 229.40  \\
\hline
\multicolumn{4}{l}{Notes: Estimates from 2SLS regression of $\Delta_2 w_{i,t}$ on}\\
\multicolumn{4}{l}{ $w_{i,t-2}$ using instruments $(w_{t-8},w_{t-9})$ for 1996--2000 and}\\
\multicolumn{4}{l}{ $(w_{t-8},w_{t-10})$ for 2002--2006 and 2008--2012. $^*$ denotes } \\
\multicolumn{4}{l}{significance at 0.05 level. }
\end{tabular}
\end{table}

%\clearpage

Appendix \ref{app: GMM} shows that GMM estimates analogous to those in Tables~\ref{tab: iv1} and~\ref{tab: iv2} are very similar.\footnote{The GMM estimates exploit the same moments but use the optimal weighting matrix (allowing for heteroskedasticity and serial correlation within individuals). 
%As with those reported in Tables~\ref{tab: iv1} and \ref{tab: iv2}, these estimates require no assumptions about the variance of individual skill innovations $\Delta\theta_{i,t}$ (or non-skill shocks, $\varepsilon_{i,t}$) over time, across cohorts, or across experience groups.
}  More importantly, we test the validity of our lagged instruments (using Hansen $J$-tests), since the model is overidentified when using multiple instruments.  We cannot reject exogeneity of our instruments at conventional levels in any year, suggesting that Assumption~\ref{assum: gen mu ident} cannot be rejected (for $k=6$).  By contrast, we show that future residuals are invalid instruments (during most time periods), highlighting the importance of accounting for idiosyncratic variation in lifecycle skill growth.

We have, thus far, used a limited set of lagged residuals as instruments to keep the specifications similar across years and to allow estimation of skill return growth rates back to 1979. Rather than report several sets of 2SLS estimates with different instrument sets, we next employ minimum distance (MD) estimation to take advantage of all long autocovariances available in the data.
%The additional moments improve precision but yield similar estimates for the time sequence of $\mu_t$.

\subsection{Minimum Distance Estimation of Skill Returns using Long Autocovariances} \label{sec: MD long autocov}
We now explicitly incorporate cohorts, $c$, into our analysis.  Assuming all conditions in
Assumption~\ref{assum: gen mu ident} hold for each cohort, $\mu_t$ and $\Omega_{c,t'}$ are identified from long autocovariances as shown in equation~\eqref{eq: cov gen}.
Separately for non-college and college men, we estimate $\mu_t$ and $\Omega_{C,t'}$ for all $(t,t')$ satisfying
$t-t'\geq 6$ for men with 21--40 years of experience in year $t$.
Due to small sample sizes of single-year cohorts, we
consider 4 broad cohort groups denoted by $C$, where each cohort group consists of 10-year labor
market entry cohorts.\footnote{We estimate %$\Omega_{C,t'} \equiv \cov(w_t,w_{t'}|c \in C)/\mu_t$
$\Omega_{C,t'} \equiv \mu_{t'}\cov(\theta_{t'+k-1},\theta_{t'}|c\in C)+\cov(\theta_{t'+k-1},\eps_{t'}|c\in C)$ and make no effort to separately identify $\Omega_{c,t'}$ for each annual entry cohort.  Given Assumption~\ref{assum: gen mu ident}, this does not impose any assumptions on variation in $\Omega_{c,t'}$ across annual cohorts even for cohorts $c$ within broader cohort groups $C$. Requiring that all single-year cohorts in each cohort group have 21--40 years of experience in each year of $t$, we exclude older (1936--1941) and younger (1982--1991) cohorts due to limited variation in $t$.}  Table~\ref{tab: cohort groups} describes these cohort groups, parameters estimated, and autocovariances used in estimation. 
Altogether, we exploit 157 covariances and use equally weighted MD to estimate 63 parameters (normalizing $\mu_{1985}=1$) separately for non-college and college men.  See Appendix~\ref{app: MD desc} for further details on our MD estimation.

\begin{table}[!htbp]
  \centering\small
\caption{Cohort grouping}
\label{tab: cohort groups}
\begin{tabular}{cccccccccc}
  \toprule
\multirow{2}{*}{Cohort Group $C$}  &\multicolumn{6}{c}{Range}&\multicolumn{3}{c}{Number}\\
  \cmidrule(lr){2-7}  \cmidrule(lr){8-10}
  &\multicolumn{2}{c}{Cohort $c$}&  \multicolumn{2}{c}{Year $t'$}& \multicolumn{2}{c}{Year $t$}&
   $\Omega_{C,t'}$&$\mu_t$&$\cov(w_t,w_{t'}|C)$\\
  \midrule
1 & 1942 & 1951 & 1970 & 1976 & 1976 & 1982 & 7 & 7 & 28 \\
2 & 1952 & 1961 & 1976 & 1986 & 1982 & 1992 & 11 & 11 & 66 \\
3 & 1962 & 1971 & 1986 & 1996 & 1992 & 2002 & 11 & 8 & 42\\
4 & 1972 & 1981 & 1996 & 2006 & 2002 & 2012 & 6 & 6 & 21 \\
  \midrule
Total &&&&&&& 35 & 29 & 157  \\
\bottomrule
\multicolumn{10}{l}{Notes: Since $\mu_t$ is not cohort-specific, the total number of $\mu_t$ parameters does not equal the}\\
\multicolumn{10}{l}{sum for each cohort due to overlap in years across cohorts.  }\\
\end{tabular}
\end{table}

\begin{figure}%[h]
  \centering
  \subfloat[Non-College]{
    \includegraphics[width=0.45\columnwidth]
    {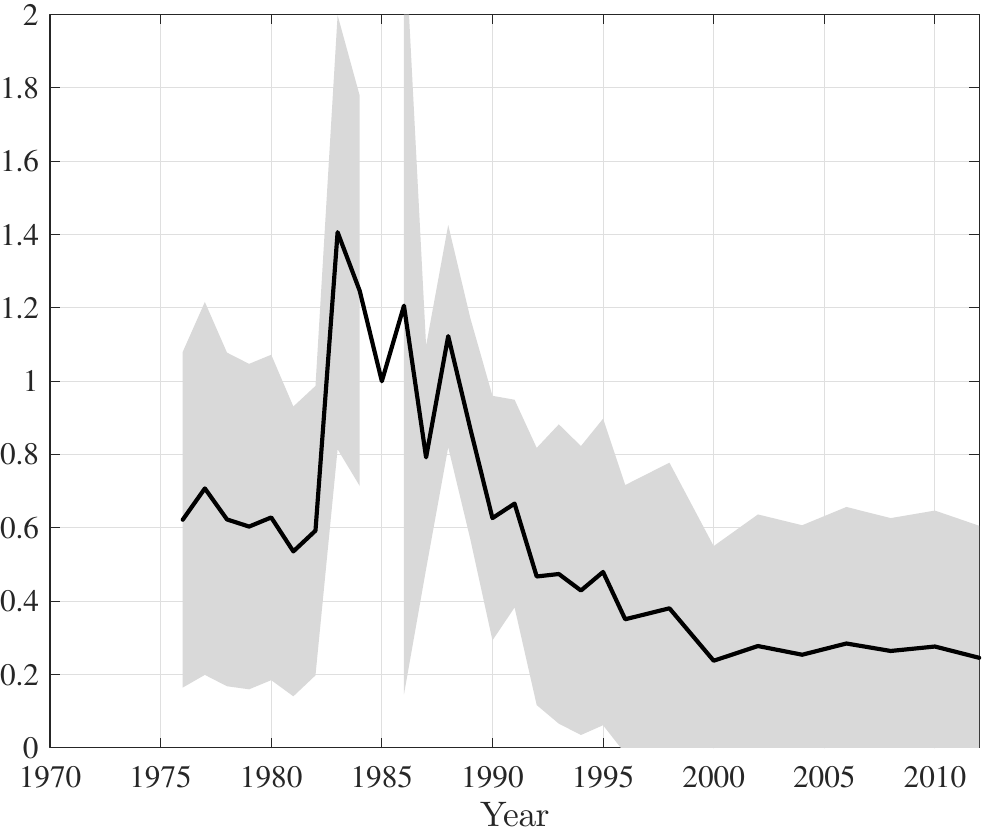}
  }\quad
  \subfloat[College]{
    \includegraphics[width=0.45\columnwidth]
    {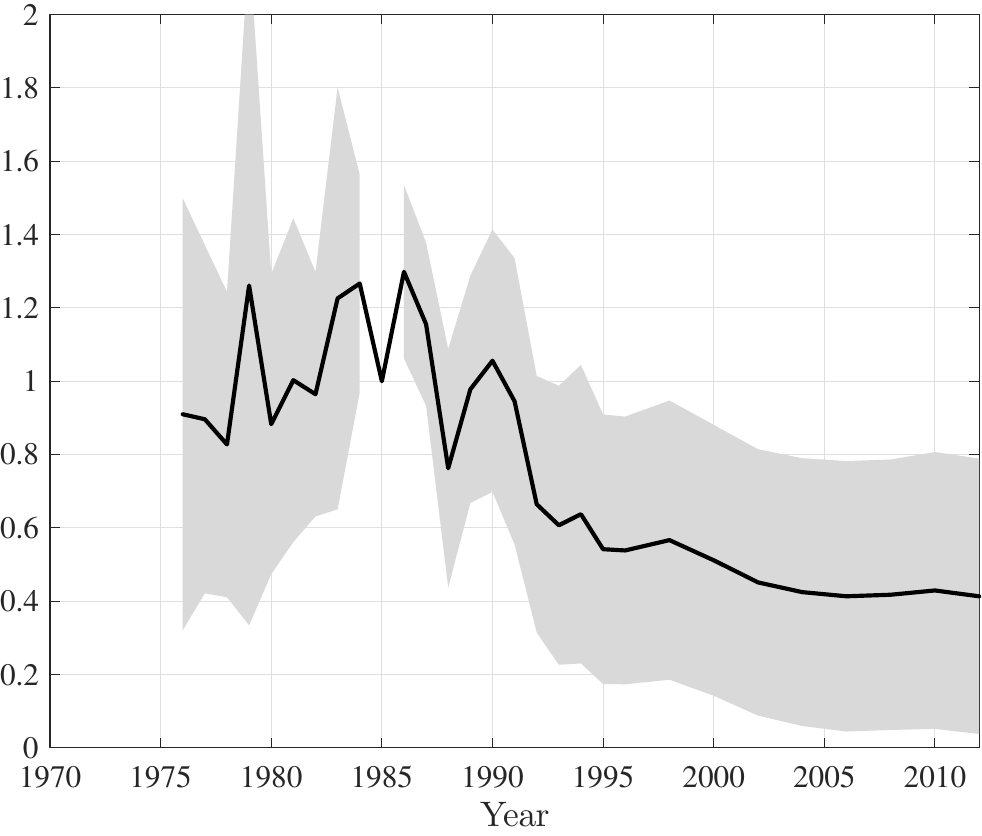}
  }
    \caption{$\mu_t$ implied by MD estimates using long autocovariances, 21--40 years of experience}
    \label{fig: MD long autocov old mu}
\end{figure}

\begin{figure}%[h]
  \centering
  \subfloat[Non-College]{
    \includegraphics[width=0.45\columnwidth]
    {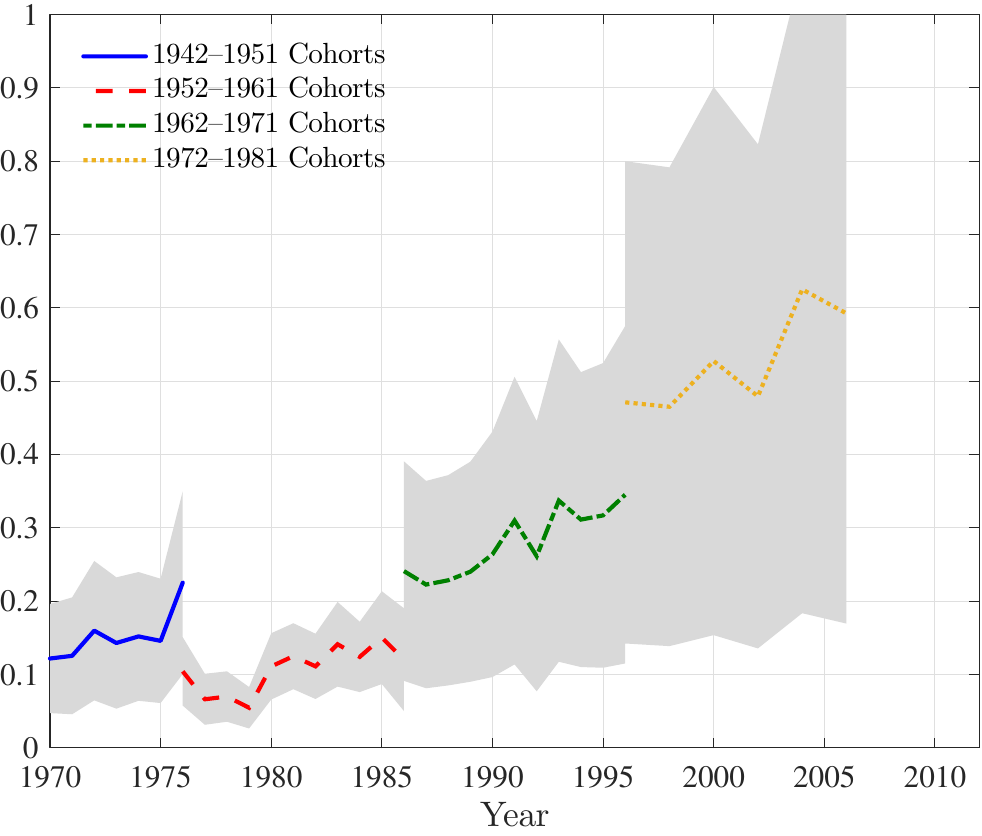}
  }\quad
  \subfloat[College]{
    \includegraphics[width=0.45\columnwidth]
    {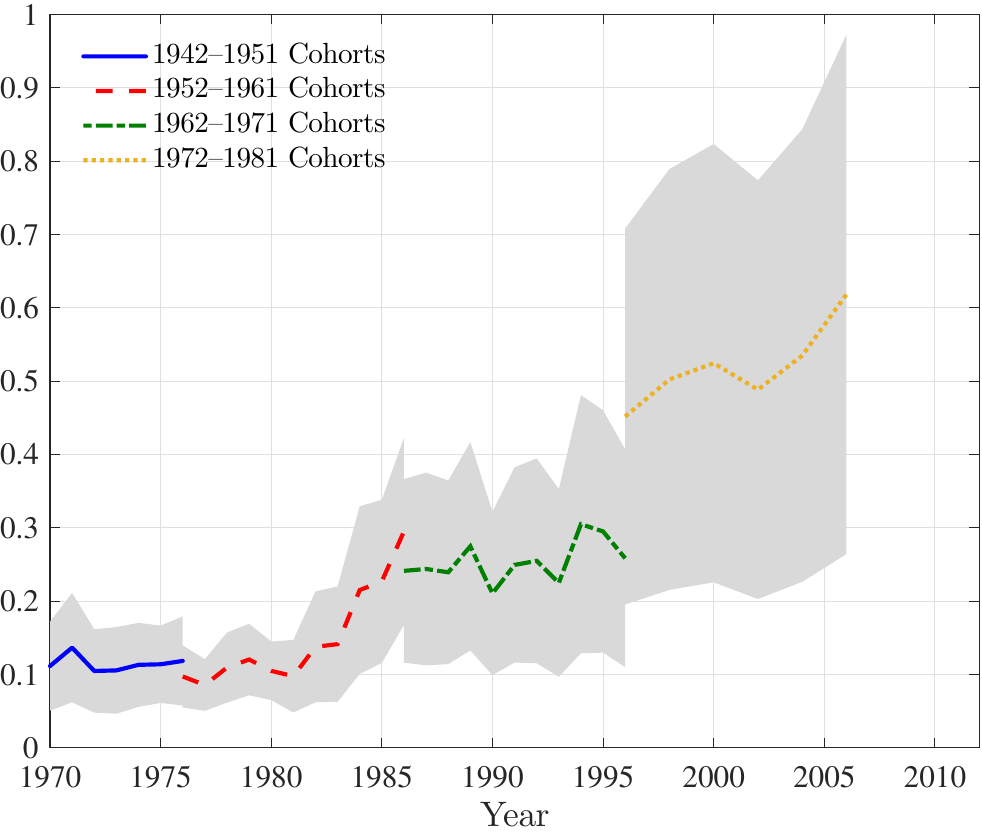}
  }
    \caption{$\Omega_{C,t'}$ implied by MD estimates using long autocovariances, 21--40 years of experience}
    \label{fig: MD long autocov old Omega}
\end{figure}

Figure~\ref{fig: MD long autocov old mu} reports MD estimates of $\mu_t$, while Figure~\ref{fig: MD long autocov old Omega} reports estimated $\Omega_{C,t'}$, both separately for non-college and college men.  (Shaded areas in these figures reflect 95\% confidence intervals.)  Like their 2SLS counterparts, MD estimates of $\mu_t$ indicate substantial declines (roughly 50--70\%) in the returns to skill over the late-1980s and 1990s. This contrasts sharply with the estimated rise in returns during the late-1970s and early-1980s.  While the additional autocovariances used in MD estimation (compared to 2SLS) improve precision, confidence intervals in Figure~\ref{fig: MD long autocov old mu} still admit the possibility that skill returns were relatively stable prior to 1985.  They also suggest that returns fell by at least 40\% for non-college men and 20\% for college men. Estimated $\Omega_{C,t'}$ profiles in Figure~\ref{fig: MD long autocov old Omega} show a strong upward trend beginning in the 1980s.  Under (mild) additional assumptions, we show in Section~\ref{sec: skill dist} that the $\Omega_{C,t'}$ trends indicate substantial growth in the variance of unobserved skills over time for the two most recent cohort groups.

%\clearpage

%%%%%%%%%%%%%%%%%%%%%%%%%%%%%%%%%%%%%%%%%%%%%%%%%%%%%%%%%%%%%%%%%%%%%%%%%%%%%%%%%%%%%%%%%%%%%%%%%%%%%%%%%%%%%%%

\subsection{Relaxing our Assumption on Skill Growth} \label{subsect: relax skill growth assum}

Our analysis, thus far, has relied on the assumption that skill growth is uncorrelated with sufficiently lagged skill levels. In this subsection, we consider two alternative specifications for skill dynamics that violate condition~(i) of Assumption~\ref{assum: gen mu ident}.  To simplify the discussion, it is useful to slightly strengthen conditions~(ii) and (iii) of Assumption~\ref{assum: gen mu ident} to $\cov(\theta_t,\eps_{t'})=0$ for all $t,t'$, while maintaining condition (iv) (i.e., limited persistence of non-skill shocks).  In this case, our IV estimator, using \textit{past} or \textit{future} residuals as instruments, converges to
\begin{equation} \label{eq: IV HIP}
 \gamma_{t,t'} \equiv  \frac{\cov(\Delta w_t,w_{t'})}{\cov(w_{t-1},w_{t'})}=\frac{\Delta \mu_t}{\mu_{t-1}}+\frac{\mu_t}{\mu_{t-1}}\frac{\cov(\Delta \theta_t,\theta_{t'})}{\cov(\theta_{t-1},\theta_{t'})},\quad \text{for } t'-t\geq k \text{ or }t-t'\geq k+1,
\end{equation}
where $\cov(\Delta \theta_t,\theta_{t'}) \neq 0$ would  bias estimates of skill return growth.

%%%%%%%%%%%%%%%%%%%%%%%%%%%%%%%%%%%%%%%%%%%%%%%%%%%%%%%%%%%%%%%%%%%%%%%%%%%%%%%%%%%%%%%%%%%%%%%%%%%%%%%%%%%%%%%

\subsubsection{Heterogeneity in Lifecycle Skill Growth} \label{subsect: HIP}

We begin by exploring the possibility that unobserved skill growth innovations are correlated over time as in the heterogeneous income profile (HIP) models estimated in, e.g., \cite{haider_2001}, \cite{baker_solon_2003}, \cite{guvenen_2009}, and \cite{moffitt_gottschalk_2012}.  We consider a more flexible process governing this skill growth heterogeneity, assuming
\begin{equation} \label{eq: HIP skill growth}
  \Delta\theta_{i,t}= \lambda_t(c_i)\delta_i+\nu_{i,t},
\end{equation}
where $\delta_i$ is a mean zero individual-specific lifecycle growth rate factor, and the $\lambda_t(c) \geq 0$ terms allow for variation in systematic skill growth across time and cohorts/experience.\footnote{We assume $\var(\delta|c)>0$, allowing for the possibility that $\lambda_t(c)=0$ for all $t$ and $c$ in the absence of heterogeneity in skill growth. Letting $\psi_i$ reflect the initial skill for an individual entering the labor market, the level of unobserved skill for individual $i$ from cohort $c_i$ in year $t$ can be written as
\begin{equation*} 
  \theta_{i,t}
=\psi_i + \Lambda_t(c_i) \delta_i + \sum_{j=0}^{t-c_i-1} \nu_{i,t-j},
\end{equation*}
where $\Lambda_t(c)\equiv \sum^{t-c-1}_{j=0}\lambda_{t-j}(c)$ reflects the accumulated influence of skill growth heterogeneity.}
This skill process generally violates condition (i) of Assumption~\ref{assum: gen mu ident} when $\lambda_t(c) > 0$, where the bias for IV estimator $\gamma_{t,t'}$ depends on
\begin{align*}
\cov(\Delta\theta_t,\theta_{t'})=\E\big[\cov(\Delta\theta_t,\theta_{t'}|c)\big]=\E\big[\lambda_t(c)\cov(\delta,\theta_{t'}|c)+ \mathbbm{1}(t'>t)\var(\nu_t|c)\big],
\end{align*}
the expectation is taken over cohorts, $c$, and $\mathbbm{1}(\cdot)$ is the indicator function.  This shows that $\gamma_{t,t'}$ estimates will be biased downward only when workers with higher skill growth rates, $\delta$, have lower skill levels, $\theta_{t'}$. This is only likely to be a concern for very young workers for whom initial skills may be negatively correlated with incentives to acquire new skills.  Hence, our focus on experienced workers makes it unlikely that the estimated declines in skill returns over the late-1980s and 1990s are explained by systematic lifecycle skill growth heterogeneity.

This model also offers testable predictions related to the year, $t'$,  from which we take $w_{t'}$ as an instrument. In the absence of HIP (i.e., $\lambda_t=0$ for all $t$), IV estimates should not vary with the year of lagged residuals (satisfying $t' \leq  t-k-1$) used as instruments nor with the year of future residuals (satisfying $t' \geq t+ k$) used as instruments; although, estimates will be greater when using any future (rather than any past) residuals as instruments if $\var(\nu_t)>0$. 
By contrast, HIP (i.e., $\lambda_t > 0$ for all $t$) implies that IV estimates will generally vary with the year of lagged or future residuals used as instruments.

Tables~\ref{tab: testing HIP noncoll} and \ref{tab: testing HIP coll} report GMM estimates of $\gamma_{t,t'}$ (using two-year differences) for all non-college and college men, respectively, using moments $\E[(\Delta_2 w_t - \gamma_{t,t'} w_{t-2})w_{t'}]=0$ with different residual leads and lags, $w_{t'}$, as instruments.  We highlight two patterns.  First, we cannot reject equality of estimates when using only lags of $t-8$ and $t-12$ as instruments (specification 1) or when using only leads of $t+6$ and $t+10$ (specification 2) as instruments.  Second, we reject equality of estimates when using lags ($t-8$) and leads ($t+6$) together as instruments (specification 3).  Together, these results provide no indication of systematic heterogeneity in unobserved skill growth.  Absent this heterogeneity, the larger $\gamma_{t,t'}$ estimates obtained when using leads as instruments imply an important role for idiosyncratic skill growth innovations (i.e., $\var(\nu_{t})>0$).\footnote{Exogeneity tests reported in Section~\ref{sec: IV} and Appendix~\ref{app: iv var of skill growth} also suggest that (i) estimated growth in returns does not vary significantly with the year of (sufficiently) lagged wages and (ii) estimated skill return growth is significantly stronger when using future rather than lagged residuals as instruments.} Finally, we note that GMM estimates using only sufficiently lagged residuals as instruments (specification 1 of Tables~\ref{tab: testing HIP noncoll} and \ref{tab: testing HIP coll}) imply $\mu_t$ profiles that are very similar to those shown in Figure~\ref{fig: IV mu}.
%\tcr{(See Appendix Figures~\ref{fig: testing HIP mu noncoll} and \ref{fig: testing HIP mu coll} for the implied $\mu_t$ series.)}

%%%%%%%%%%%%%%%%%%%%%%%%%%%%%%%%%%%%%%%%%%%%%

%\clearpage

\begin{table}
  \caption{Multiple-Equation GMM Estimates: Non-College}
  \label{tab: testing HIP noncoll}
  \small
  \centering
{
\def\sym#1{\ifmmode^{#1}\else\(^{#1}\)\fi}
\begin{tabular}{l*{6}{c}}
  \toprule
  &\multicolumn{6}{c}{Instrument for Each Equation of $\Delta_2 w_{i,t}$:}\\
  &\multicolumn{2}{c}{(1)}               &\multicolumn{2}{c}{(2)}            &\multicolumn{2}{c}{(3)}           \\
  \cmidrule(lr){2-3}  \cmidrule(lr){4-5}  \cmidrule(lr){6-7}
                    &         $w_{i,t-8}$         &         $w_{i,t-12}$         &         $w_{i,t+6}$         &         $w_{i,t+10}$ &         $w_{i,t-8}$         &         $w_{i,t+6}$         \\
  \midrule
  Coefficient on $w_{i,t-2}$ for years\\
\quad 1972--1974          &                   &                   &       0.054       &      -0.013       &                   &                   \\
                    &                   &                   &     (0.057)       &     (0.075)       &                   &                   \\
\addlinespace
\quad 1975--1977          &                   &                   &       0.132       &       0.075       &                   &                   \\
                    &                   &                   &     (0.095)       &     (0.078)       &                   &                   \\
\addlinespace
\quad 1978--1980          &                   &                   &       0.081       &       0.107       &      -0.085       &       0.170       \\
                    &                   &                   &     (0.096)       &     (0.088)       &     (0.054)       &     (0.125)       \\
\addlinespace
\quad 1981--1983          &       0.017       &      -0.057       &       0.267       &       0.300\sym{*}&       0.084       &       0.143       \\
                    &     (0.082)       &     (0.096)       &     (0.137)       &     (0.126)       &     (0.082)       &     (0.085)       \\
\addlinespace
\quad 1984--1986          &      -0.074       &       0.001       &       0.114       &       0.137       &       0.032       &       0.089       \\
                    &     (0.059)       &     (0.065)       &     (0.103)       &     (0.093)       &     (0.109)       &     (0.092)       \\
\addlinespace
\quad 1987--1989          &      -0.199\sym{*}&      -0.161       &       0.050       &       0.026       &      -0.185       &       0.010       \\
                    &     (0.086)       &     (0.135)       &     (0.090)       &     (0.100)       &     (0.117)       &     (0.071)       \\
\addlinespace
\quad 1990--1992          &      -0.069       &      -0.096       &       0.029       &      -0.075       &      -0.151\sym{*}&      -0.045       \\
                    &     (0.059)       &     (0.080)       &     (0.084)       &     (0.079)       &     (0.075)       &     (0.078)       \\
\addlinespace
\quad 1993--1995          &      -0.076       &      -0.125       &       0.139       &       0.084       &      -0.057       &       0.047       \\
                    &     (0.063)       &     (0.085)       &     (0.086)       &     (0.128)       &     (0.062)       &     (0.089)       \\
\addlinespace
\quad 1996--2000          &      -0.079       &      -0.053       &       0.091       &       0.022       &      -0.056       &       0.052       \\
                    &     (0.051)       &     (0.056)       &     (0.062)       &     (0.072)       &     (0.048)       &     (0.051)       \\
\addlinespace
\quad 2002--2006          &      -0.043       &      -0.038       &       0.089       &      -0.052       &      -0.022       &       0.054       \\
                    &     (0.058)       &     (0.052)       &     (0.135)       &     (0.171)       &     (0.065)       &     (0.066)       \\
\addlinespace
\quad 2008--2012          &      -0.049       &      -0.038       &                   &                   &                   &                   \\
                    &     (0.076)       &     (0.079)       &                   &                   &                   &                   \\
\midrule
Observations        &       \multicolumn{2}{c}{5,627}                    &       \multicolumn{2}{c}{6,883}                   &       \multicolumn{2}{c}{5,093}                   \\
Wald \(p\)-value    &       \multicolumn{2}{c}{0.945}                   &       \multicolumn{2}{c}{0.756}                   &       \multicolumn{2}{c}{0.044}                   \\
  \bottomrule
\multicolumn{7}{l}{Notes: $^*$ denotes significance at 0.05 level. }
\end{tabular}
}
\end{table}

\begin{table}
  \caption{Multiple-Equation GMM Estimates: College}
  \label{tab: testing HIP coll}
\small
  \centering
{
\def\sym#1{\ifmmode^{#1}\else\(^{#1}\)\fi}
\begin{tabular}{l*{6}{c}}
  \toprule
  &\multicolumn{6}{c}{Instrument for Each Equation of $\Delta_2 w_{i,t}$:}\\
  &\multicolumn{2}{c}{(1)}               &\multicolumn{2}{c}{(2)}            &\multicolumn{2}{c}{(3)}           \\
  \cmidrule(lr){2-3}  \cmidrule(lr){4-5}  \cmidrule(lr){6-7}
                    &         $w_{i,t-8}$         &         $w_{i,t-12}$         &         $w_{i,t+6}$         &         $w_{i,t+10}$ &         $w_{i,t-8}$         &         $w_{i,t+6}$         \\
  \midrule
  Coefficient on $w_{i,t-2}$ for years\\
\quad 1972--1974          &                   &                   &       0.068       &       0.030       &                   &                   \\
                    &                   &                   &     (0.076)       &     (0.075)       &                   &                   \\
\addlinespace
\quad 1975--1977          &                   &                   &       0.225\sym{*}&       0.065       &                   &                   \\
                    &                   &                   &     (0.091)       &     (0.070)       &                   &                   \\
\addlinespace
\quad 1978--1980          &                   &                   &       0.036       &       0.078       &       0.016       &      -0.004       \\
                    &                   &                   &     (0.077)       &     (0.075)       &     (0.083)       &     (0.069)       \\
\addlinespace
\quad   1981--1983          &      -0.125       &       0.002       &       0.156       &       0.195       &      -0.128\sym{*}&       0.120       \\
                    &     (0.080)       &     (0.096)       &     (0.088)       &     (0.117)       &     (0.061)       &     (0.086)       \\
\addlinespace
\quad 1984--1986          &       0.032       &       0.158       &       0.240\sym{*}&       0.422\sym{*}&       0.018       &       0.174\sym{*}\\
                    &     (0.066)       &     (0.084)       &     (0.071)       &     (0.112)       &     (0.079)       &     (0.074)       \\
\addlinespace
\quad 1987--1989          &      -0.004       &      -0.069       &       0.107\sym{*}&       0.106       &      -0.015       &       0.023       \\
                    &     (0.058)       &     (0.056)       &     (0.046)       &     (0.056)       &     (0.066)       &     (0.067)       \\
\addlinespace
\quad 1990--1992          &      -0.033       &      -0.119\sym{*}&       0.095       &       0.006       &      -0.095       &       0.069       \\
                    &     (0.056)       &     (0.058)       &     (0.061)       &     (0.069)       &     (0.054)       &     (0.055)       \\
\addlinespace
\quad 1993--1995          &      -0.030       &      -0.116       &       0.085       &       0.070       &      -0.071       &       0.069       \\
                    &     (0.053)       &     (0.073)       &     (0.051)       &     (0.060)       &     (0.048)       &     (0.051)       \\
\addlinespace
\quad 1996--2000          &      -0.080\sym{*}&      -0.044       &       0.109\sym{*}&       0.087\sym{*}&      -0.037       &       0.094\sym{*}\\
                    &     (0.035)       &     (0.049)       &     (0.046)       &     (0.040)       &     (0.034)       &     (0.039)       \\
\addlinespace
\quad 2002--2006          &      -0.016       &       0.030       &       0.155       &       0.165       &       0.024       &       0.048       \\
                    &     (0.034)       &     (0.042)       &     (0.084)       &     (0.097)       &     (0.037)       &     (0.040)       \\
\addlinespace
\quad 2008--2012          &      -0.069\sym{*}&      -0.030       &                   &                   &                   &                   \\
                    &     (0.031)       &     (0.036)       &                   &                   &                   &                   \\
\midrule
Observations        &       \multicolumn{2}{c}{7,353}                   &       \multicolumn{2}{c}{9,263}                   &       \multicolumn{2}{c}{7,069}                   \\
Wald \(p\)-value    &       \multicolumn{2}{c}{0.080}                   &       \multicolumn{2}{c}{0.354}                   &       \multicolumn{2}{c}{0.007}                   \\
  \bottomrule
\multicolumn{7}{l}{Notes: $^*$ denotes significance at 0.05 level. }
\end{tabular}
}
\end{table}

%\clearpage
%%%%%%%%%%%%%%%%%%%%%%%%%%%%%%%%%%%%%%%%%%%%%

\begin{figure}%[H]
  \centering
  \subfloat[Cohort Group 3, Non-College]{
    \includegraphics[width=0.45\columnwidth]
    {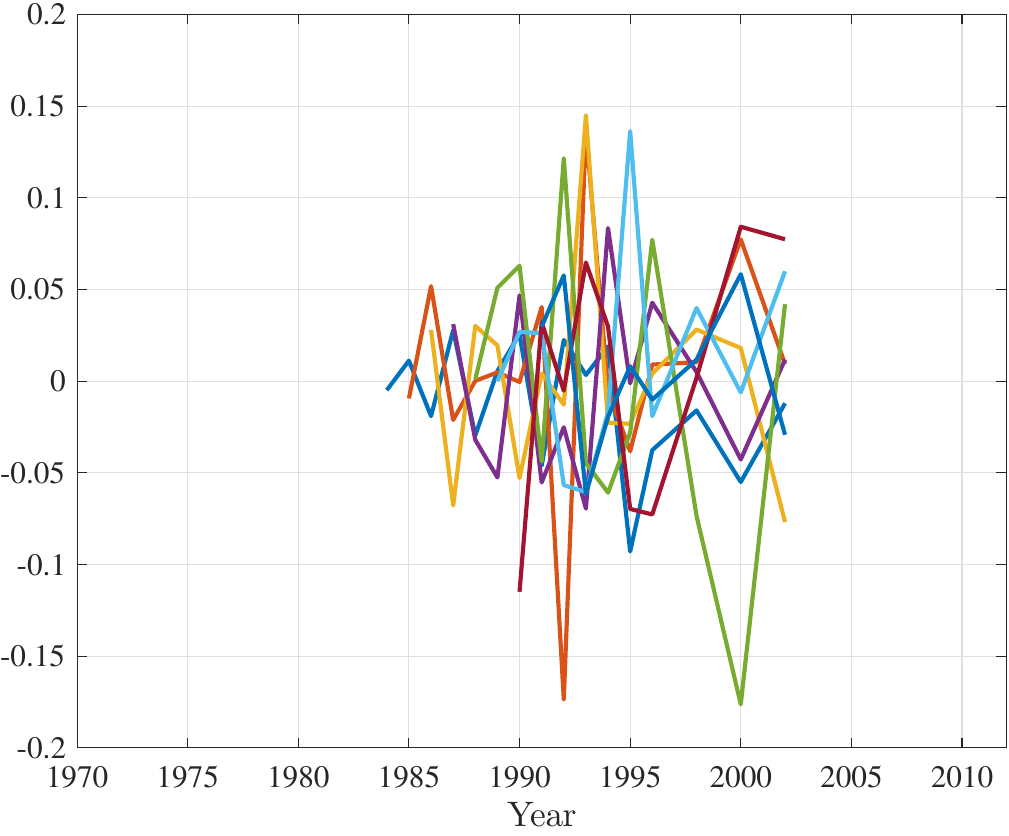}
  }\quad
  \subfloat[Cohort Group 3, College]{
    \includegraphics[width=0.45\columnwidth]
    {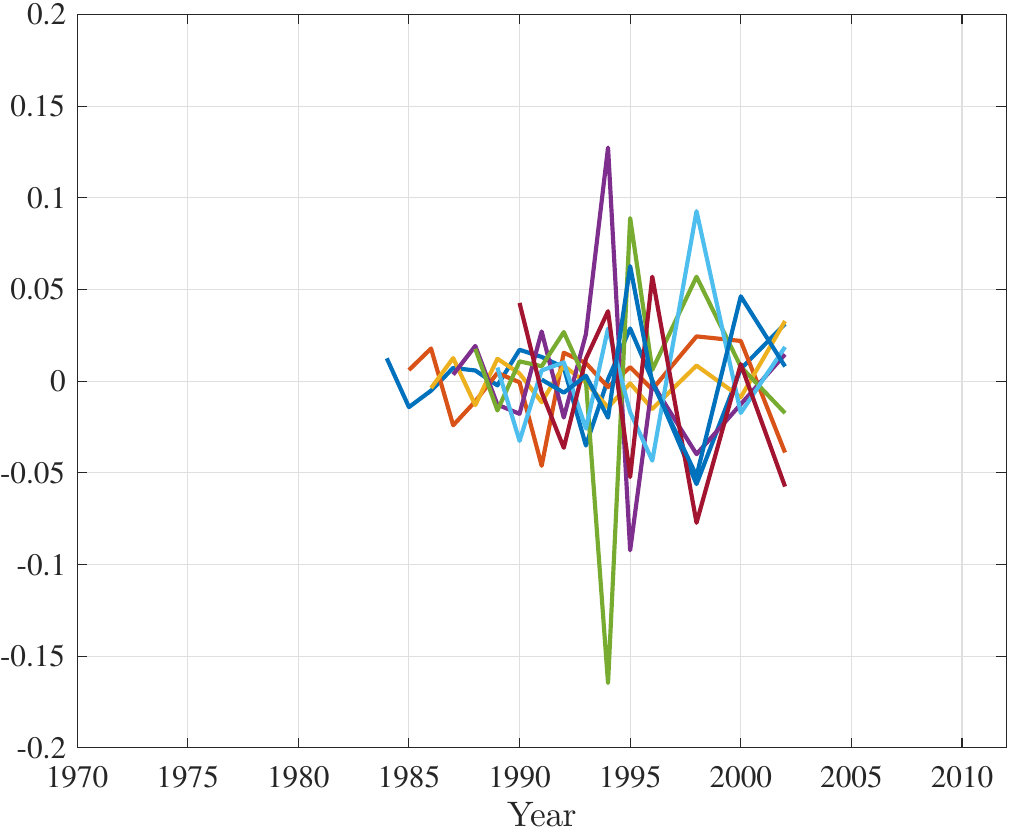}
  }
  \\
  \subfloat[Cohort Group 4, Non-College]{
    \includegraphics[width=0.45\columnwidth]
    {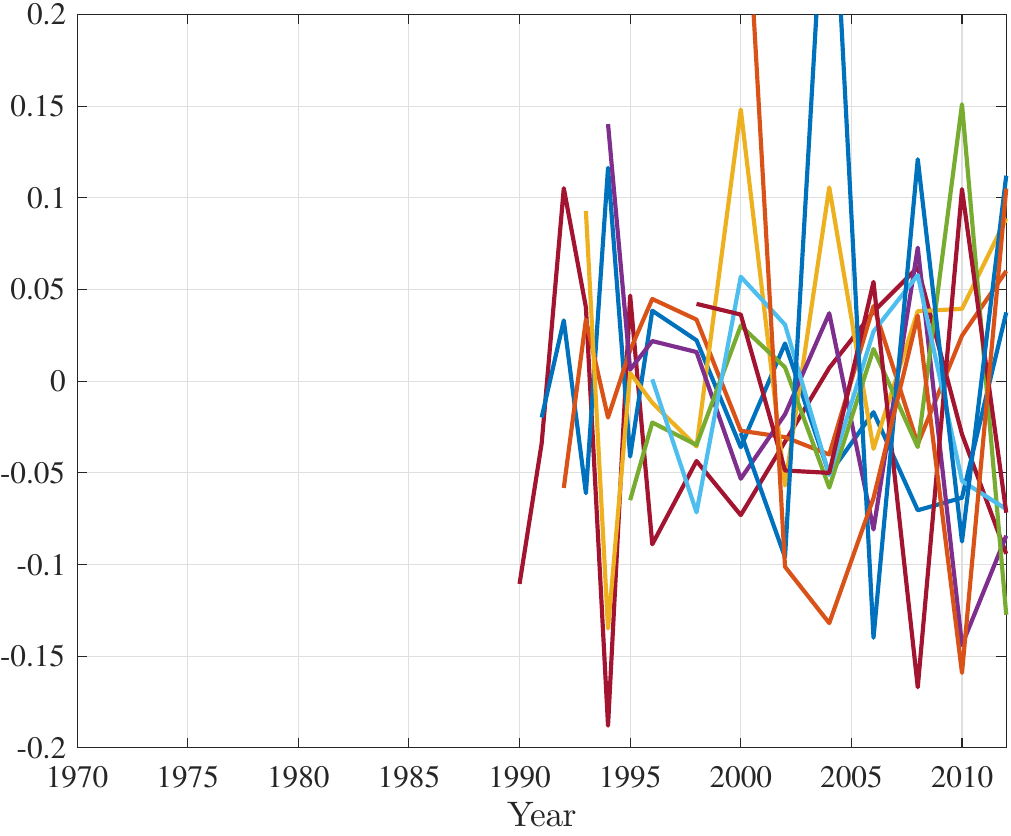}
  }\quad
  \subfloat[Cohort Group 4, College]{
    \includegraphics[width=0.45\columnwidth]
    {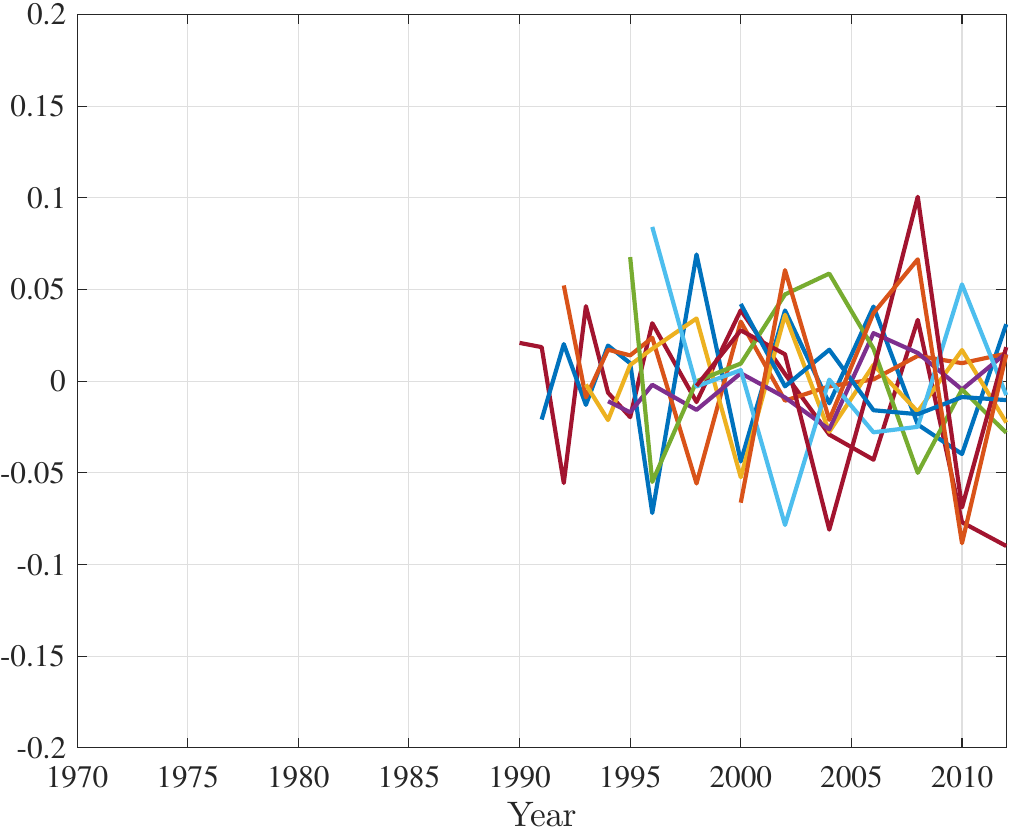}
  }
    \caption{$\cov(\Delta (w_t/\mu_t),\Delta(w_{t'}/\mu_{t'}))$ for Men by Cohort Group}

    \label{fig: cov scaled diff1}

    \caption*{\it Notes: Figure reports covariances for cohort groups $C \in \{3,4\}$  where each line holds $t'$ fixed and varies $t \geq t'+7$.}  %\tcr{Sample restricted to men who had accumulated at least 21 years of experience by year $t'=1992$ ($C=3$) or $2002$ ($C=4$).} }
\end{figure}

\begin{figure}%[H]
  \centering
\includegraphics[width=0.45\columnwidth]{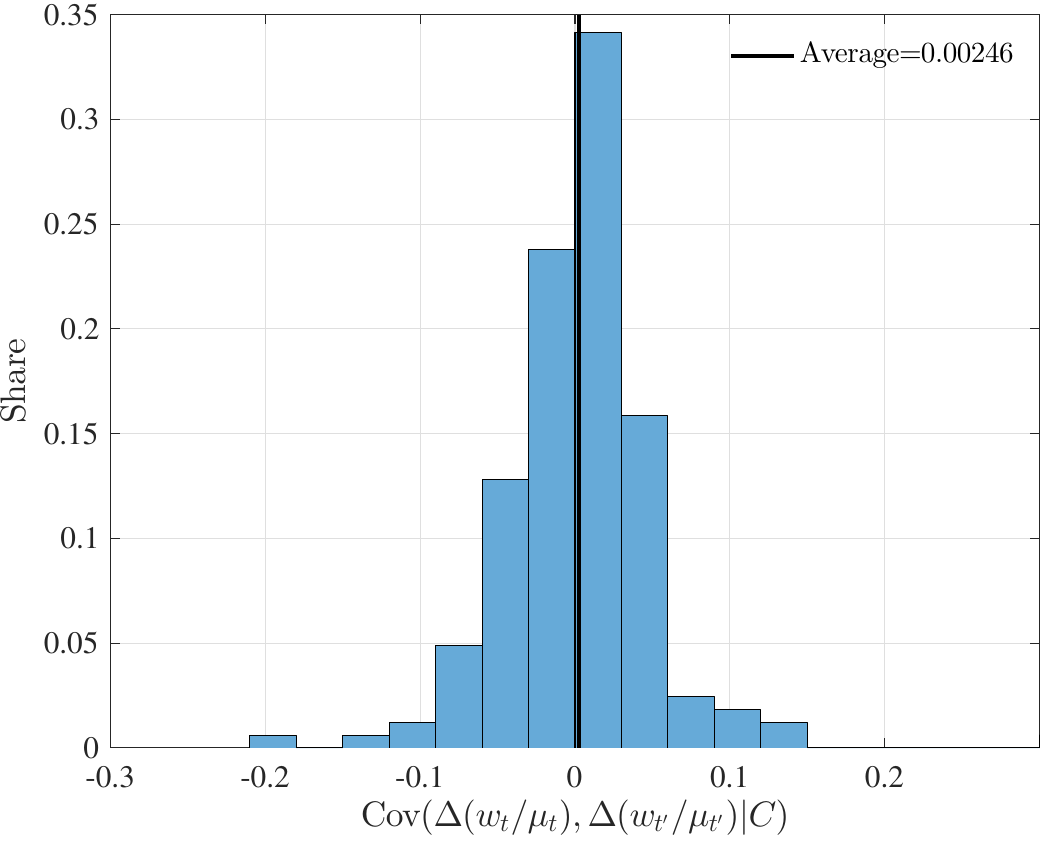}
  \caption{Distribution of $\cov(\Delta (w_t/\mu_t),\Delta(w_{t'}/\mu_{t'}))$  for all $(t,t',C)$ for Low-Experience Men}
    \label{fig: cov scaled diff1 dist}

  \caption*{\it Notes: Figure reports distribution of covariances based on $(t,t')$ satisfying $t \geq t'+7$ for cohort groups $C \in \{3,4\}$ when some individuals in these cohort groups had less than 21 years of experience in year $t$ (i.e., $t\leq 1991$ for $C=3$ or $t\leq 2001$  for $C=4$).}
\end{figure}

As discussed earlier, human capital theory \citep{becker_1964,ben-porath_1967} predicts that optimal skill investment and accumulation become negligible as workers approach the end of their careers.
Assuming no systematic \textit{unobserved} heterogeneity in skill growth among experienced workers (i.e.\ $\lambda_t(c) = 0$ for all workers with at least 21 years of experience), baseline $\mu_t$ estimates (using only experienced workers) reported in Figure~\ref{fig: MD long autocov old mu} can be used to scale log wage residuals to estimate
\begin{equation} \label{eq: cov scaled w growth}
\cov\left(  \Delta\left(\frac{w_{t}}{\mu_{t}}\right), \Delta\left(\frac{w_{t'}}{\mu_{t'}}\right)\big|c\right)
  =\lambda_{t}(c) \lambda_{t'}(c) \var(\delta|c), \quad \text{for $t-t'\geq k+1$,}
\end{equation}
for less-experienced workers.  Systematic heterogeneity in skill growth at younger ages should be reflected in systematically positive covariances in scaled-residual growth. Yet, Figure~\ref{fig: cov scaled diff1} shows that for cohort groups $C \in \{3,4\}$, the covariances in equation~\eqref{eq: cov scaled w growth} fluctuate around zero for all ages.  Figure~\ref{fig: cov scaled diff1 dist} further shows that the distribution of all covariances (for workers with 1--20 years of experience in $t$) is centered around zero. These covariances strongly suggest that systematic skill growth heterogeneity is negligible early in the lifecycle for these cohorts. Related results for $\cov(\Delta (w_t/\mu_t),w_{t'}|c) = \lambda_{t}(c)\mu_{t'}\cov(\delta,\theta_{t'}|c)$ reported in Appendix~\ref{app: test HIP log w growth} further support this conclusion.

Altogether, Tables~\ref{tab: testing HIP noncoll} and \ref{tab: testing HIP coll} and Figures~\ref{fig: cov scaled diff1} and \ref{fig: cov scaled diff1 dist} support condition (i) of Assumption~\ref{assum: gen mu ident}: skill growth is not systematically related to past skill levels throughout the careers of men in our sample.

%\clearpage

%%%%%%%%%%%%%%%%%%%%%%%%%%%%%%%%%%%%%%%%%%%%%%%%%%%%%%%%%%%%%%%%%%%%%%%%%%%%%%%%%%%%%%%%%%%%%%%%%%%%%%

\subsubsection{AR(1) skill dynamics} \label{sec: FE AR(1) skills}

We next consider an alternative model of skill dynamics characterized by a fixed effect, $\psi_i$, and an AR(1) component, $\phi_{i,t}$:
\begin{equation} \label{eq: theta AR(1)}
  \theta_{i,t} = \psi_i+\phi_{i,t},  \qquad \text{where} \quad \phi_{i,t}=\rho_t \phi_{i,t-1}+\nu_{i,t}.  
\end{equation}
For $\rho_t<1$, this specification is consistent with heterogeneous depreciation of skills acquired in the labor market generating mean-reversion to individual-specific baseline skill levels determined by $\psi_i$. Our baseline specification implicitly assumes $\rho_t=1$ for all $t$.

When $\rho_t \neq 1$, the AR(1) skill component violates Assumption~\ref{assum: gen mu ident}(i), since skill growth will be correlated with all past skill levels.  With this more general specification for skills, we assume that all skill components are uncorrelated with non-skill shocks, while maintaining our assumption of limited persistence in non-skill shocks.

\begin{ass}  \label{assum: AR(1) skills ident}
For all cohorts, $c$:
(i) $\cov(\psi,\phi_{t}|c)=0$ for all $t$;
(ii) $\cov(\psi,\eps_{t'}|c)=\cov(\phi_t, \eps_{t'}|c) = 0$ for all $t,t'$;
(iii) $\cov(\phi_{t'},\nu_{t},|c) =\cov(\nu_{t'},\nu_{t},|c) = 0$ for all $t-t'\geq 1$;
(iv) for known $k\geq 1$, $\cov(\eps_t,\eps_{t'}|c)=0$ for all $t-t' \geq k$.
\end{ass}

%Conditions (i) and (ii) are standard for an $AR(1)$ process with an individual fixed effect.  Condition (iii) implies a separation between skill and non-skill wage components, while condition (iv) continues to impose limited persistence on non-skill shocks.

We interpret $\phi_{i,t}$ as skill, regardless of its persistence; however, it is possible to rewrite the problem such that skills are time-invariant and
non-skill shocks include an autoregressive component (along with transitory shocks, $\eps_t$):
\[
  w_{i,t}=\mu_t \psi_i+\tilde{\phi}_{i,t}+\varepsilon_{i,t}, \qquad \text{where} \quad
  \tilde{\phi}_{i,t}=\tilde{\rho}_t\tilde{\phi}_{i,t-1}+\tilde{\nu}_{i,t},
\]
letting $\tilde{\phi}_{i,t}\equiv\mu_t\phi_{i,t}$, $\tilde{\rho}_t\equiv\rho_t \mu_t/\mu_{t-1}$, and $\tilde{\nu}_{i,t}\equiv\mu_t\nu_{i,t}$.
This shows that the distinction between skill vs.\ non-skill persistent shocks is not important from a statistical point of view nor for identification of $\mu_t$.%\footnote{The distinction would be relevant to decomposition of the residual variance into skill vs.\ non-skill components.}

Due to the correlation between skill growth and past skill levels, our IV estimator will generally produce biased estimates for growth rates in skill returns when $\rho_t \neq 1$.\footnote{The IV estimator converges to $\gamma_{t,t'}$ in \eqref{eq: IV HIP}, where
     $\cov(\Delta\theta_t,\theta_{t'})=(\rho_t-1)\cov(\phi_t,\phi_{t'})$.
 The term $\cov(\phi_t,\phi_{t'})$ generally depends on both $t$ and $t'$. For example, when $t'\leq t-k-1$, $\cov(\phi_t,\phi_{t'})=\prod^{t-1}_{j=t'+1}\rho_{j}\var(\phi_{t'})$. The GMM estimates of Tables \ref{tab: testing HIP noncoll} and \ref{tab: testing HIP coll} do not support this dependence of $\gamma_{t,t'}$ on $t'$.
}
%\footnote{For $t' \leq t - k -1$, \tcr{the following holds for each cohort:}
%\[
%\frac{\cov(\Delta w_t,w_{t'})}{\cov(w_{t-1},w_{t'})} =
%\frac{\Delta \mu_{t}}{\mu_{t-1}} + \frac{\mu_{t}}{\mu_{t-1}} \left[\frac{(\rho_t-1)\hat{\rho}_{t'+1,t-1}\var(\phi_{t'})}
%    {\var(\psi) + \hat{\rho}_{t'+1,t-1}\var(\phi_{t'})} \right], \qquad \text{where }  \hat{\rho}_{t,t'} \equiv \prod_{j=t}^{t'} \rho_j.
%\]
% }
However, we show in Appendix~\ref{app: FE AR(1) skills} that if $\var(\psi)>0$ the evolution of both $\rho_t$ and $\mu_t$ over time can still be identified under Assumption~\ref{assum: AR(1) skills ident} and other plausible conditions.

Identification breaks down when $\var(\psi)=0$.  In this knife-edge case, our IV estimator (using past residuals as instruments) identifies $(\rho_t \mu_t - \mu_{t-1})/\mu_{t-1}$, and it is not generally possible to separate growth in skill returns from skill convergence without strong assumptions.
%\footnote{\tcr{For any cohort,} notice that when $\var(\psi)=0$, 
%\[
%\frac{\cov(w_t,w_{t'})}{\cov(w_{t-1},w_{t'})} = \left\{ \begin{array}{ll}
%\frac{\mu_{t}}{\mu_{t-1}} \rho_t & \mbox{if $t' \leq t - k -1$} \vspace{6pt}\\
%\frac{\mu_{t}}{\mu_{t-1}} \left[\rho_t + \frac{\var(\nu_{t})}{\rho_t\var(\phi_{t-1})}\right]  = %\frac{\mu_{t}}{\mu_{t-1}} \left[\frac{\var(\phi_{t})}{\rho_t\var(\phi_{t-1})} \right] & \mbox{if $t' \geq t+k$.}
%\end{array}
%\right.
%\]
%If $\rho_t \leq \var(\phi_t)/\var(\phi_{t-1})$, then $\mu_t/\mu_{t-1}$ is bounded between IV estimates using past residuals as instruments and those using future residuals as instruments.}
This raises concerns that estimated declines in skill returns over the late-1980s and 1990s (see Figure~\ref{fig: MD long autocov old mu}) could instead reflect particularly strong skill convergence (i.e., $\rho_t<1$) over those years.

Our examination of skill growth using test scores in Section~\ref{subsec: AR(1) skills test} suggests that $\rho_t\approx 1$ for experienced workers in the HRS, covering the late-1990s and 2000s.  We now use equally weighted MD estimation to estimate the model with AR(1) skills as defined in equation~\eqref{eq: theta AR(1)} to account for the possibility that $\rho_t<1$ over the longer time period we examine in the PSID.  We begin by assuming that $\rho_t=\rho$ is time-invariant, but also consider the case in which $\rho_t$ follows a cubic polynomial in time.  We estimate $\rho_t$, $\mu_t$ (normalizing $\mu_{1985}=1$),  $\var(\nu_t|c)$, and $\var(\psi|c)$ separately for non-college and college men.  To improve precision and facilitate estimation, we assume that $\var(\psi|c)$ is a cubic polynomial in entry cohort $c$ and that $\var(\nu_t|c)$ is a cubic time trend multiplied by a quadratic experience trend.  Long autocovariances for workers with at least 21 years of experience are targeted.\footnote{Specifically, we target $\widehat{\cov}(w_t,w_{t'}|E_j)$ for all $t-t'\geq 6$ and ten-year experience groups, $E_j$ (21--30 and 31--40 years of experience in year $t$).  There are 729 targeted autocovariances each for non-college and college men.  See Appendix~\ref{app: MD desc} for additional details.}

\begin{figure}[h]
  \centering
    \subfloat[Non-College Men]{
    \includegraphics[width=0.45\columnwidth]{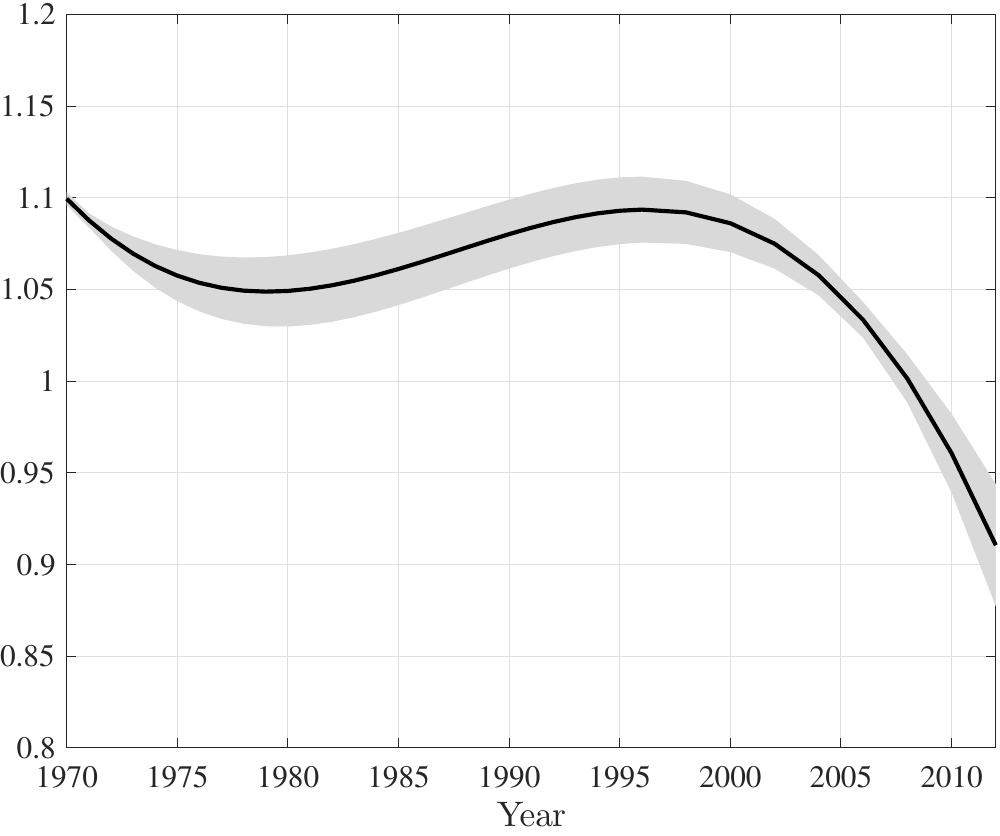}
  }
  \quad
    \subfloat[College Men]{
    \includegraphics[width=0.45\columnwidth]{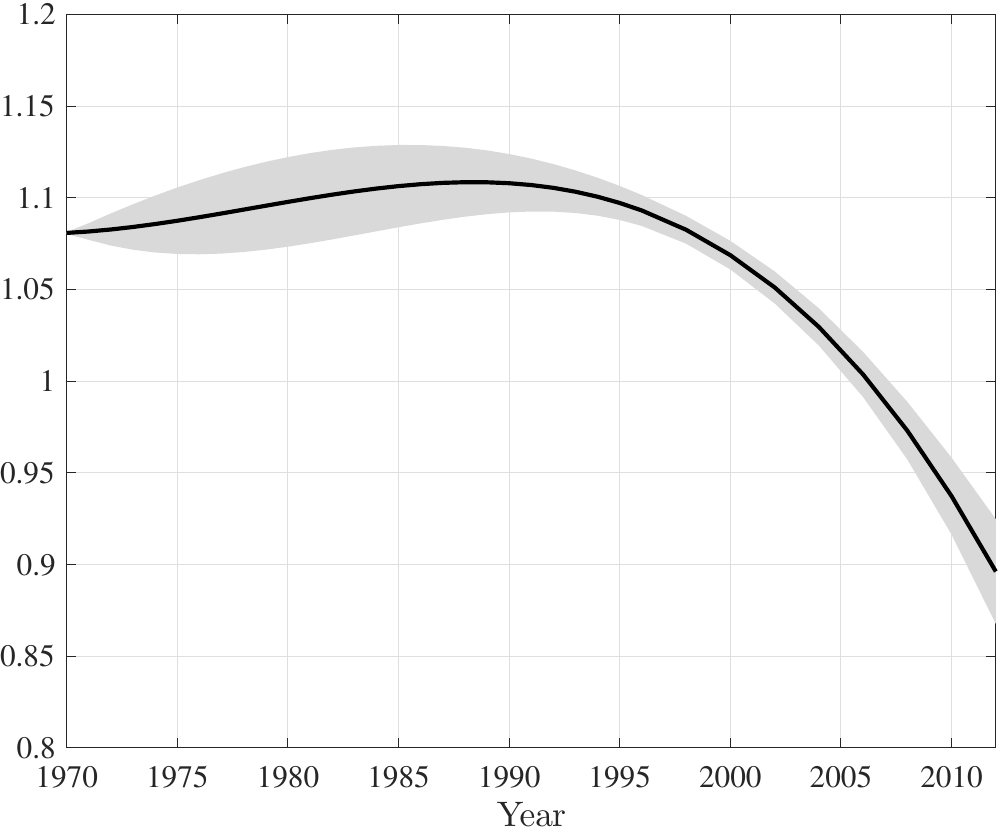}
  }
    \caption{$\rho_t$ implied by MD estimates allowing for time-varying AR(1) skill shocks, 21--40 years of experience}
    \label{fig: rho time-varying AR(1) skill shocks}
  \end{figure}

\begin{figure}[h]
  \centering
    \subfloat[Non-College Men]{
    \includegraphics[width=0.45\columnwidth]{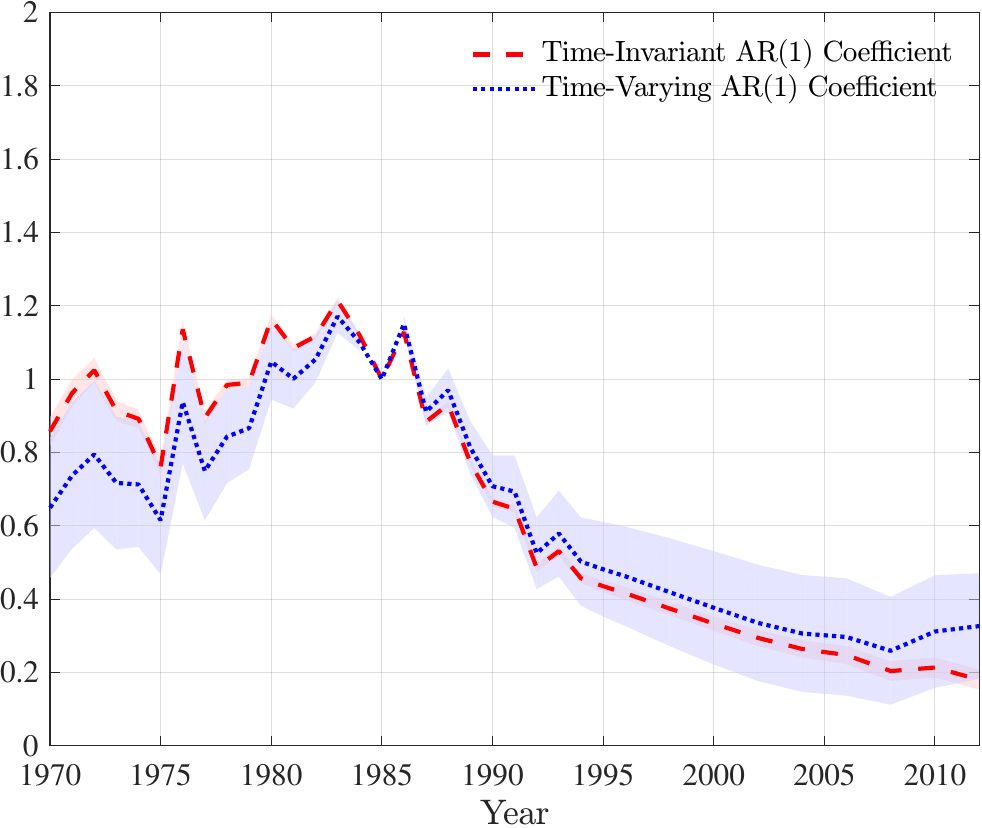}
  }
  \quad
    \subfloat[College Men]{
    \includegraphics[width=0.45\columnwidth]{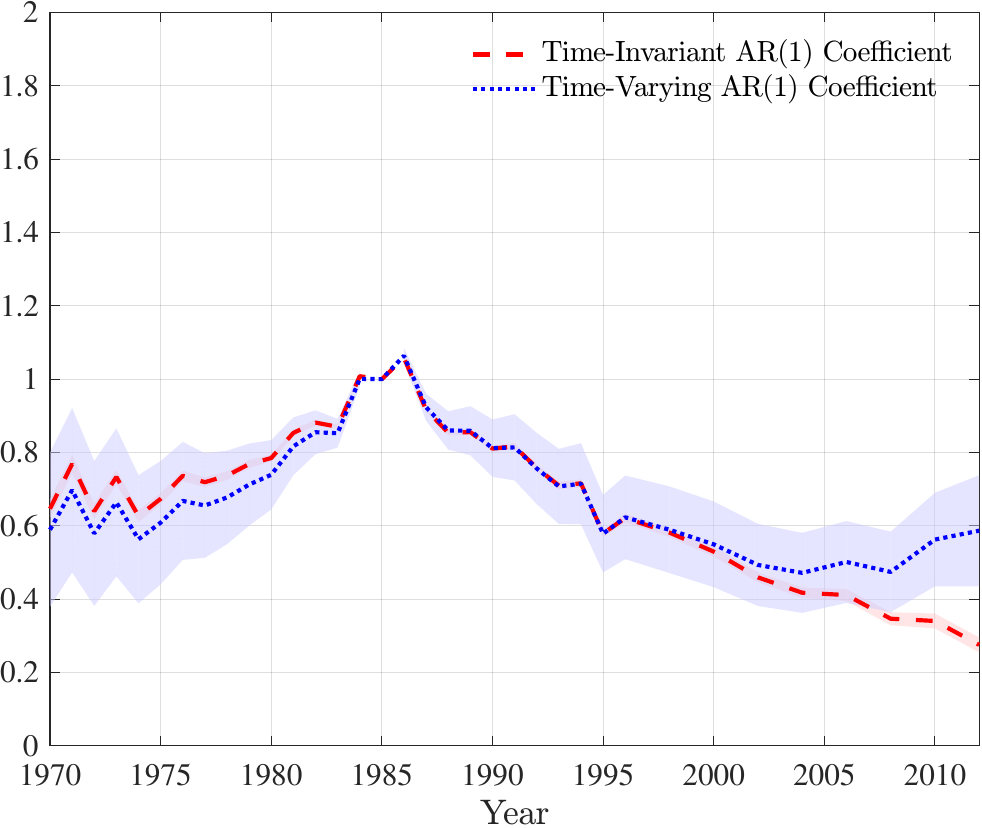}
  }
    \caption{$\mu_t$ implied by MD estimates allowing for time-varying vs. time-invariant AR(1) skill shocks, 21--40 years of experience}
    \label{fig: mu AR(1) skill shocks}
  \end{figure}

Importantly, estimates for $\var(\psi|c)$ are always positive, with estimates significantly greater than zero for cohorts at the heart of our sample (see Appendix Figure~\ref{fig: var(psi) time-varying AR(1) skill shocks}). This means that it is possible to separately identify the process for skill dynamics from the returns to skill over time.
We find that unobserved skills (for experienced men) are not mean-reverting, at least over most of the time period we examine.  When assuming time-invariant $\rho_t=\rho$, its estimated value is 1.071 (0.001) for non-college men and 1.064 (0.001) for college men.  Based on the more general time-varying $\rho_t$ case, Figure~\ref{fig: rho time-varying AR(1) skill shocks} shows a modest increase in $\rho_t$ over the 1980s and early-1990s, falling thereafter.  There is no indication that $\rho_t$ drops over the late-1980s and 1990s, which might explain our sharply falling IV estimated returns to skills over those years. Indeed, Figure~\ref{fig: mu AR(1) skill shocks} shows that the estimated $\mu_t$ series (for both fixed and time-varying $\rho_t$) are very similar to baseline estimates reported in Figure~\ref{fig: MD long autocov old mu}. 
%Appendix Figure~\ref{fig: var(psi) time-varying AR(1) skill shocks} displays estimated $\var(\psi|c)$ for the time-varying $\rho_t$ case.  These estimates are generally consistent with $\var(\psi|c)>0$, with estimates significantly greater than zero over labor-market-entry cohorts at the heart of our sample.  

%%%%%%%%%%%%%%%%%%%%%%%%%%%%%%%%%%%%%%%%%%%%%%%%%%%%%%%%%%%%%%%%%%%%%%%%%%

\subsection{Persistent Non-Skill Shocks} \label{subsec: persistent eps}

Results, thus far, rely on limited persistence in non-skill shocks, $\eps_t$.  Condition (iv) of Assumption~\ref{assum: gen mu ident} is convenient but not critical to our approach using wage dynamics to understand the evolution of skill returns.  In this subsection, we consider the case in which $\eps_t$ contains an autoregressive process.  Specifically, we assume that
\begin{equation} \label{eq: AR(1) eps}
  \eps_{i,t}= \varphi_{i,t} + \teps_{i,t}, \qquad \text{where} \quad \varphi_{i,t} = \rho\varphi_{i,t-1} + \nu_{i,t},
\end{equation}
and $\teps_{i,t}$ has limited persistence.

With this more general process for non-skill shocks, we strengthen conditions (i)--(iii) of Assumption~\ref{assum: gen mu ident} slightly and maintain our baseline assumption on the limited persistence of transitory non-skill shocks.
\begin{ass}  \label{assum: inc dynamics lit} For any cohort $c$,
(i) $\cov(\Delta \theta_t, \theta_{t'}|c)=0$ for all $t-t'\geq 1$;
(ii) $\cov(\theta_t, \varphi_{t'}|c)=\cov(\theta_{t},\nu_{t'} |c)=\cov(\theta_t,\teps_{t'}|c)  = 0$ for all $t,t'$; 
(iii) $\cov(  \varphi_{t'},\nu_t |c)=\cov( \nu_{t'},\nu_t |c)=0$ for all $t-t'\ge 1$, 
$\cov( \varphi_t, \teps_{t'} |c)=\cov( \nu_t, \teps_{t'} |c)=0$ for all $t,t'$;
%$\cov( \nu_t, \teps_{t'} |c)=0$ for all $t,t'$, 
%$\cov( \nu_t, \nu_{t'}|c)=0$ for all $t\neq t'$
and
(iv) for known $k\geq 1$, $\cov(\teps_t,\teps_{t'}|c)=0$ for all $t-t' \geq k$.
\end{ass}

This assumes that skill growth is uncorrelated with past skills and that the evolution of skills is unrelated to the process for non-skill shocks; although, the influence of non-skill shocks on wages never fully disappears.  Appendix~\ref{app: ARMA ident} establishes identification of skill returns, as well as $\rho$ and the variance of initial skills and skill growth, under Assumption~\ref{assum: inc dynamics lit}.\footnote{Indeed, we establish identification for $\varphi_t \sim \text{ARMA}(1,q)$ as sometimes used in the literature on earnings dynamics.}  

\begin{figure}[h]
  \centering
  \subfloat[Non-College]{
    \includegraphics[width=0.45\columnwidth]
    {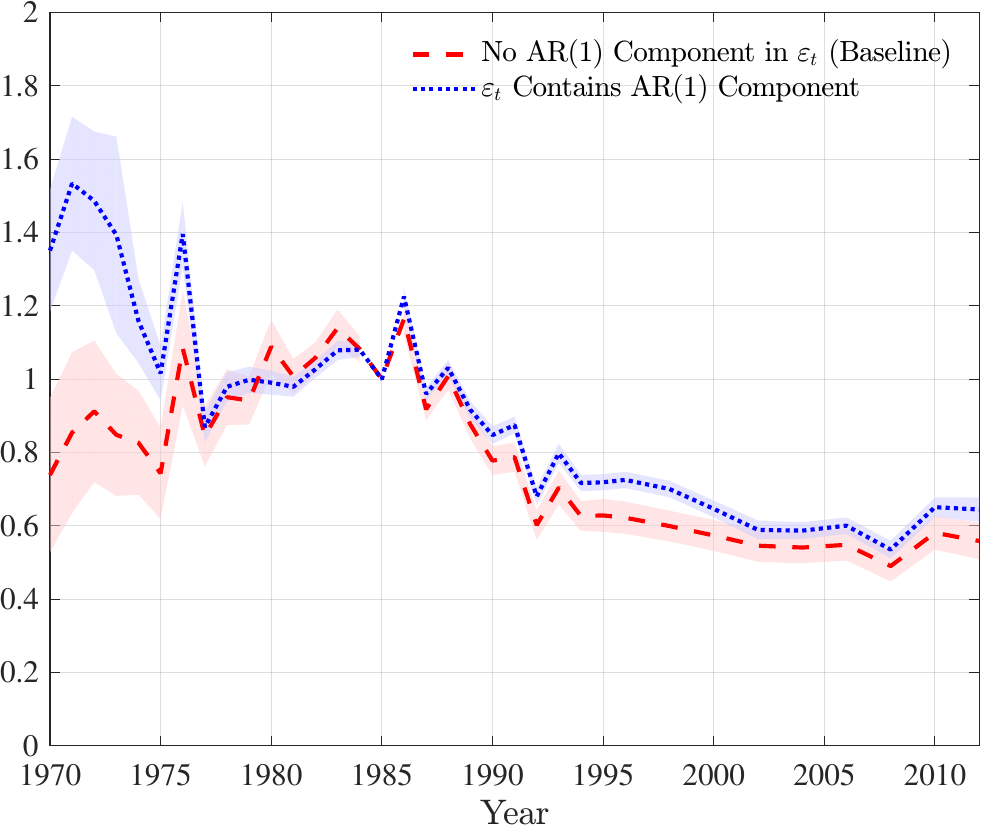}
  }
  \quad
  \subfloat[College]{
    \includegraphics[width=0.45\columnwidth]
    {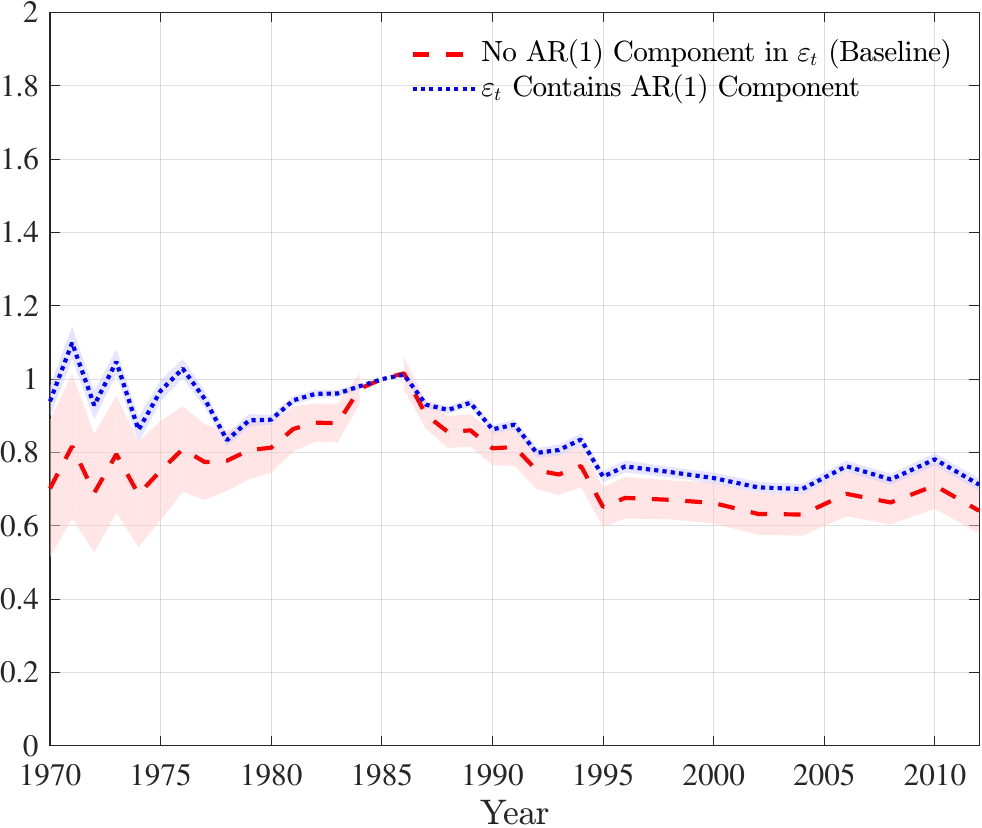}
  }
  \caption{Estimated $\mu_t$ with and without an AR(1) non-skill component}
  \label{fig: AR1 non-skill shocks}
\end{figure}

As in the previous subsection, we turn to equally weighted MD using long residual autocovariances to estimate this very general specification for wage residuals separately for non-college and college men.\footnote{Specifically, we estimate $\mu_t$ (normalizing $\mu_{1985}=1$), $\var(\Delta \theta_t|c)$ (assuming a cubic time trend multiplied by a quadratic experience trend), $\rho$,  $\var(\nu_t|c)$ (assuming a year-specific constant  multiplied by a quadratic experience trend), and $\var(\psi|c)$ (assuming a cubic polynomial in entry cohort $c$).  We target $\widehat{\cov}(w_t,w_{t'}|E_j)$ for all $t-t'\geq 6$ and ten-year experience groups, $E_j$ (1--10,...,  31--40 years of experience in year $t$).  There are 855 targeted autocovariances each for non-college and college men. See Appendix~\ref{app: MD desc} for additional details.}
For $k=6$, Figure~\ref{fig: AR1 non-skill shocks} compares the estimated $\mu_t$  sequence for our baseline specification ($\eps_t=\teps_t$) vs.\ the specification that includes AR(1) non-skill shocks in equation~\eqref{eq: AR(1) eps}.
In both cases, the estimated paths for $\mu_t$ present the familiar pattern of rising returns in the early-1980s, followed by significant declines over the late-1980s and 1990s. Importantly, the evolution of skill returns is largely unaffected by the introduction of an autoregressive non-skill component.\footnote{For both non-college and college men, we estimate $\rho \approx 0.8$.}  %rho=0.80 for non-college; rho=0.78 for college

The literature on earnings dynamics estimates a similar structure for log earnings residuals to that of this subsection; although, this literature has focused primarily on the relative importance of permanent vs.\ transitory shocks while ignoring changes in the returns to unobserved skills.\footnote{See, among others,
\cite{abowd_card_1989,blundell_preston_1998,meghir_pistaferri_2004,blundell_pistaferri_preston_2008,heathcote_perri_violante_2010}. See \cite{macurdy_2007} for a review.}
\cite{haider_2001} and \cite{moffitt_gottschalk_2012} are notable exceptions in that they also estimate the evolution of returns to unobserved skills  using the PSID.\footnote{Other studies exploit different panel data sets on earnings to estimate very similar models to \cite{haider_2001}  and \cite{moffitt_gottschalk_2012}. \cite{debacker_heim_panousi_ramnath_vidangos_2013} use U.S.\ tax return data from 1987 to 2009, while \cite{baker_solon_2003} exploit Canadian tax  return data from 1976 to 1992.}  We show in Appendix~\ref{App: ARMA eps results} that by restricting the autocovariance structure for non-skill shocks and the distributions of skills (over time and across cohorts), these studies estimate upward biased growth in skill returns, $\mu_t$, over time.

%\clearpage

%%%%%%%%%%%%%%%%%%%%%%%%%%%%%%%%%%%%%%%%%%%%%%%%%%%%%%%%%%%%%%%%%%%%%%%%%%%%%%%%%%%%%%%%%%%%%%%%%%%%%%%%%%%%%

\section{Skill Distributions}\label{sec: skill dist}

We now consider identification and estimation of the variance of skills and non-skill shocks over time.  We further decompose the variance of skills into contributions from heterogeneity in initial skills and variation due to idiosyncratic lifecycle skill growth.

To facilitate this analysis, we strengthen conditions (i)--(iii) of Assumption~\ref{assum: gen mu ident} but maintain the limited persistence of non-skill shocks (Assumption~\ref{assum: gen mu ident}(iv)).

\begin{ass}  \label{assum: mu and skill dist ident}
(i) $\cov(\Delta \theta_t, \theta_{t'})=0$ for all $t-t'\geq 1$;
(ii) $\cov(\theta_t, \eps_{t'}) = 0$ for all $t,t'$; and
(iii) for known $k\geq 1$, $\cov(\eps_t,\eps_{t'})=0$ for all $t-t' \geq k$.
\end{ass}

The first two conditions imply that skills follow a random walk and are uncorrelated with non-skill shocks. 
This attributes all transitory wage innovations to the non-skill component.  All three conditions imply that  $\cov(w_t,w_{t'}) = \mu_t\mu_{t'}\var(\theta_{t'})$ for $t-t'\geq k$. As in Proposition~\ref{prop: gen mu ident}, the IV estimator of equation~\eqref{eq: IV estimator} can be used to identify $\mu_{\ul{t}+k},...,\mu_{\bt}$ (with one normalization).  Given skill returns and sufficient panel length, 
%$\bt - \ul{t} \geq 2k$, 
the variance of unobserved skills, $\var(\theta_{t'}) = \cov(w_t,w_{t'})/\mu_t\mu_{t'}$,
can be identified for all but the first and last $k$ periods.  This variance is generally not identified for earlier periods, because it cannot be separated from (unidentified) early skill returns -- only $\mu_t \! \var(\theta_t)$ can be identified for the first $k$ periods.  The unobserved skill variance cannot be identified for later periods, because it is impossible to distinguish between the roles of unobserved skills and transitory non-skill shocks without observing (distant) future wages.

Having identified the variance of unobserved skills over time, it
is straightforward to then identify variation in skill growth, $\var(\Delta \theta_t)=\var(\theta_t)-\var(\theta_{t-1})$,
%for $t=\ul{t}+k+1,...,\bt-k$, 
and the variance of non-skill shocks, $\var(\varepsilon_t) = \var(w_t)-\mu_t^2 \var(\theta_t)$, for all but the first and last $k$ periods.
Proposition~\ref{prop: identification} in Appendix~\ref{app: ident skill distn eps} extends these identification results to the variances of unobserved skills, skill growth innovations, and non-skill transitory shocks when these all vary by cohort.

%%%%%%%%%%%%%%%%%%%%%%%%%%%%%%%%%%%%%%%%%%%%%%%%%%%%%%%%%%%%%%%%%%%%%%%%%%%%%%%%%%%%%%%%%%%%%%%%%%%%%%%%%%%%%%%%%%%%%%

\paragraph{Using future residuals as instruments.}
As discussed in Section~\ref{subsec: returns identification}, future wage residuals are not generally valid instruments in equations~\eqref{eq: IV estimator} or \eqref{eq:  regression} when skills vary over time.   Under Assumption~\ref{assum: mu and skill dist ident}, it is straightforward to show that IV regression using future wage residuals as instruments identifies:
\begin{equation} \label{eq: iv future}
  \frac{\cov(\Delta w_t,w_{t''})}{\cov(w_{t-1},w_{t''})} = \frac{\Delta \mu_t}{\mu_{t-1}} + \frac{\mu_t}{\mu_{t-1}}\frac{\var(\Delta \theta_t)}{\var(\theta_{t-1})} \quad \text{ for } t''\geq t+k.
\end{equation}
Since IV estimates using past residuals consistently estimate $\Delta \mu_t/\mu_{t-1}$, the difference between IV estimates obtained using future vs.\ past residuals as instruments can be used to identify the importance of skill growth innovations (relative to variation in lagged skill levels). 

IV estimates presented in Tables~\ref{tab: testing HIP noncoll} and \ref{tab: testing HIP coll}, as well as Appendix~\ref{app: GMM}, empirically show that using future rather than lagged residuals as instruments nearly always produces higher estimated returns.  Comparing estimates using future vs.\ past residuals as instruments, we show in Appendix~\ref{app: GMM} that the variance of two-year skill growth relative to prior skill levels, $\frac{\var(\Delta \theta_{t-1} + \Delta \theta_{t})}{\var(\theta_{t-2})}$, ranges from 0.16 to 0.29 over our sample period.  Skills are not fixed and unchanging over the lifecycle.

\begin{figure}%[h]
  \centering
  \subfloat[Non-College]{
    \includegraphics[width=0.45\columnwidth]
    {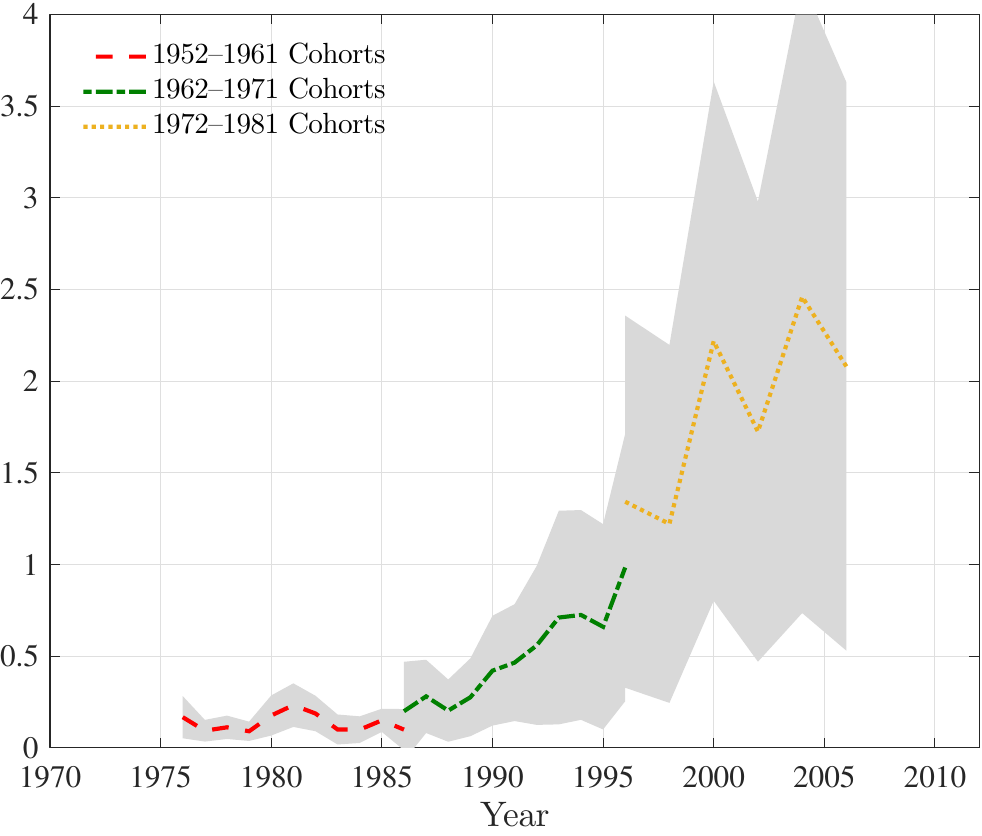}
  }\quad
  \subfloat[College]{
    \includegraphics[width=0.45\columnwidth]
    {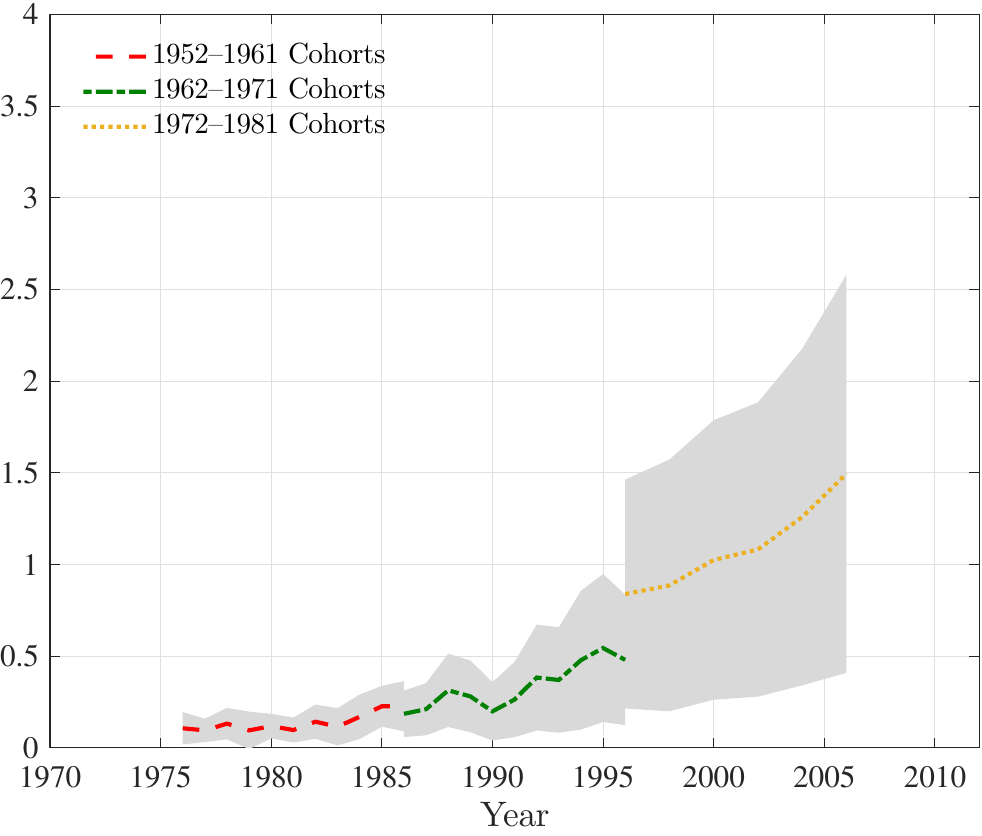}
  }
    \caption{$\var(\theta_{t'}|C)= \Omega_{C,t'}/\mu_{t'}$ implied by MD estimates using long autocovariances, 21--40 years of experience}
    \label{fig: MD long autocov old Omega over mu}
\end{figure}

\paragraph{Evolution of skill variation by cohort.} Figure~\ref{fig: MD long autocov old Omega over mu} reports $\var(\theta_{t'}|C)= \Omega_{C,t'}/\mu_{t'}$ (for older men) obtained from estimates reported in Figures~\ref{fig: MD long autocov old mu} and \ref{fig: MD long autocov old Omega}.  This figure indicates that unobserved skill heterogeneity for the 1952--1961 birth cohorts was largely stable over the late-1970s and early-1980s.  However, beginning in the early-1990s, the variance of unobserved skills grew sharply for both non-college and college men from the 1962--1971 and 1972--1981 birth cohorts.

\begin{figure}[!htbp]
  \centering
  \subfloat[Non-College]{
    \includegraphics[width=0.45\columnwidth]
    {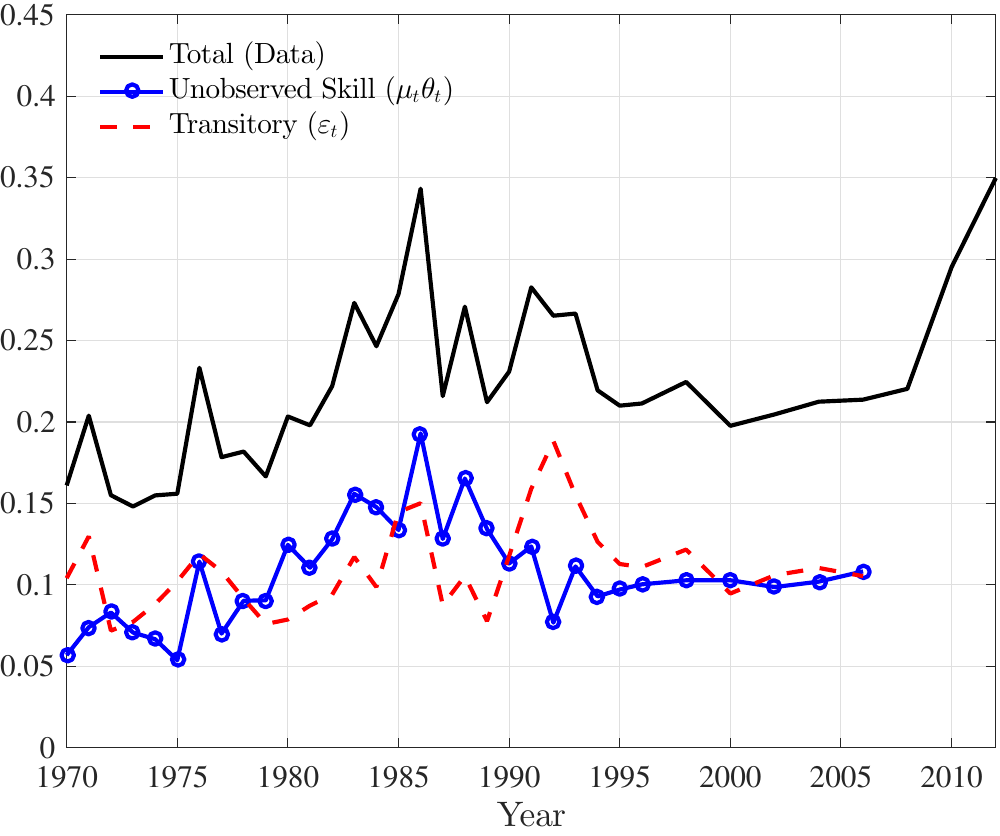}
  }\quad
  \subfloat[College]{
    \includegraphics[width=0.45\columnwidth]
    {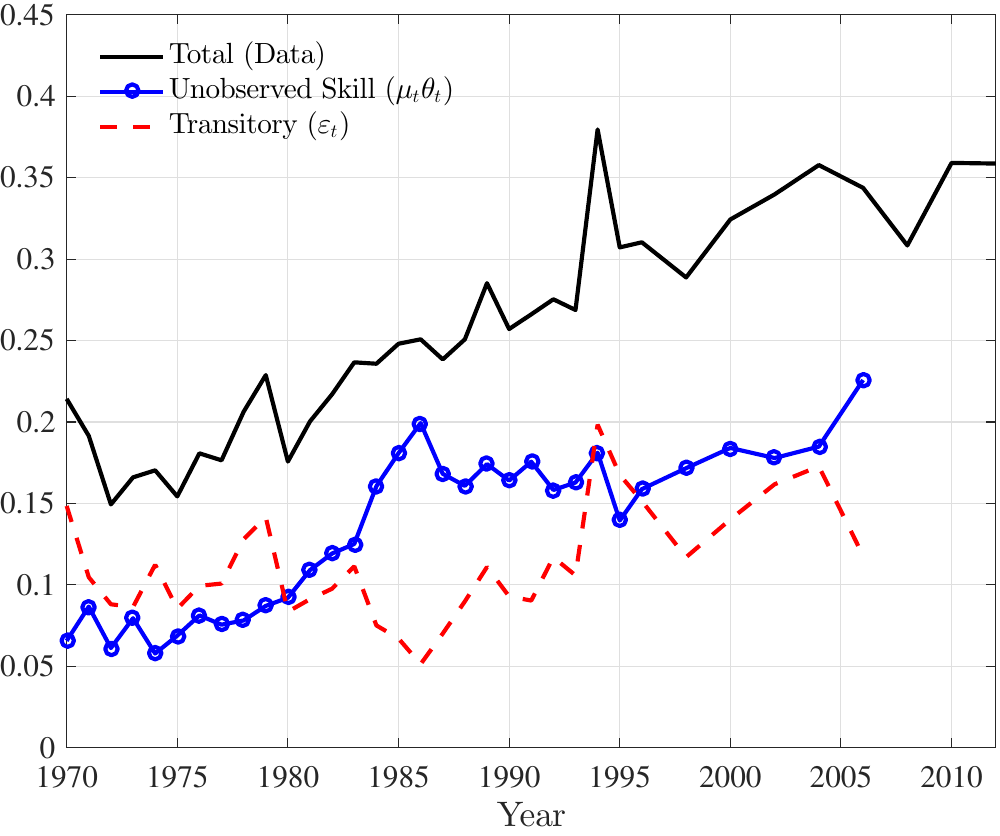}
  }
    \caption{Log wage residual variance decomposition}
    \label{fig: var(w) decomp}
\end{figure}

\paragraph{Decomposing residual variation.} We next explore the extent to which the long-term increase in residual variation reported in Figure~\ref{fig: variance of residuals by sector} is driven by increasing variability of non-skill wage shocks, $\var(\eps_t)$, vs.\ growing dispersion in skills and their returns, $\var(\mu_t\theta_t)$.   Given our interest in understanding these trends for all workers, we focus on our baseline model under Assumption~\ref{assum: mu and skill dist ident}, estimated separately by education (for all ages) using MD estimation.  We have already estimated this model in Section~\ref{subsec: persistent eps}, imposing mild cohort- and time-based smoothness assumptions on the variance of initial skills and skill growth innovations.  Estimated $\mu_t$ profiles are shown as red dashed lines in Figure~\ref{fig: AR1 non-skill shocks}.
Figure~\ref{fig: var(w) decomp} decomposes $\var(w_t)$ into its skill and non-skill components over time.   The main trends in residual inequality are driven by inequality in skills (multiplied by their returns) for both education groups; however, growth in the variance of non-skill shocks contributes to rising residual inequality in the late-1980s and 1990s for college men.

\begin{figure}[!htbp]
  \centering
  \subfloat[Non-College]{
    \includegraphics[width=0.45\columnwidth]
    {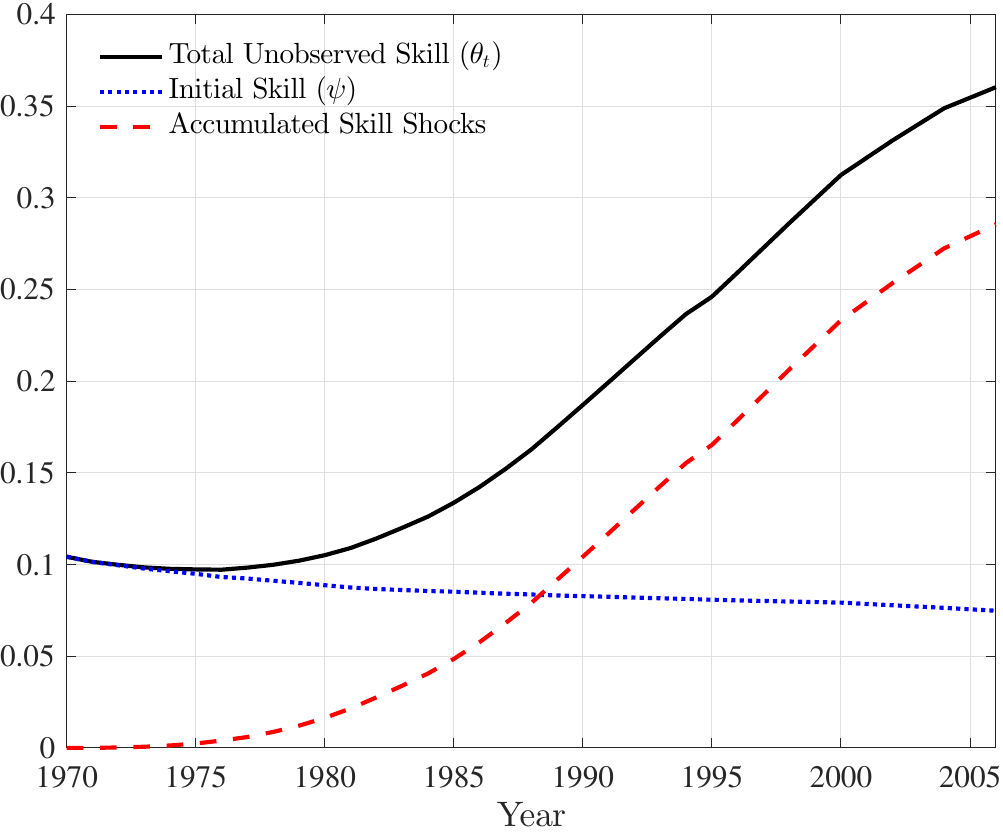}
  }\quad
  \subfloat[College]{
    \includegraphics[width=0.45\columnwidth]
    {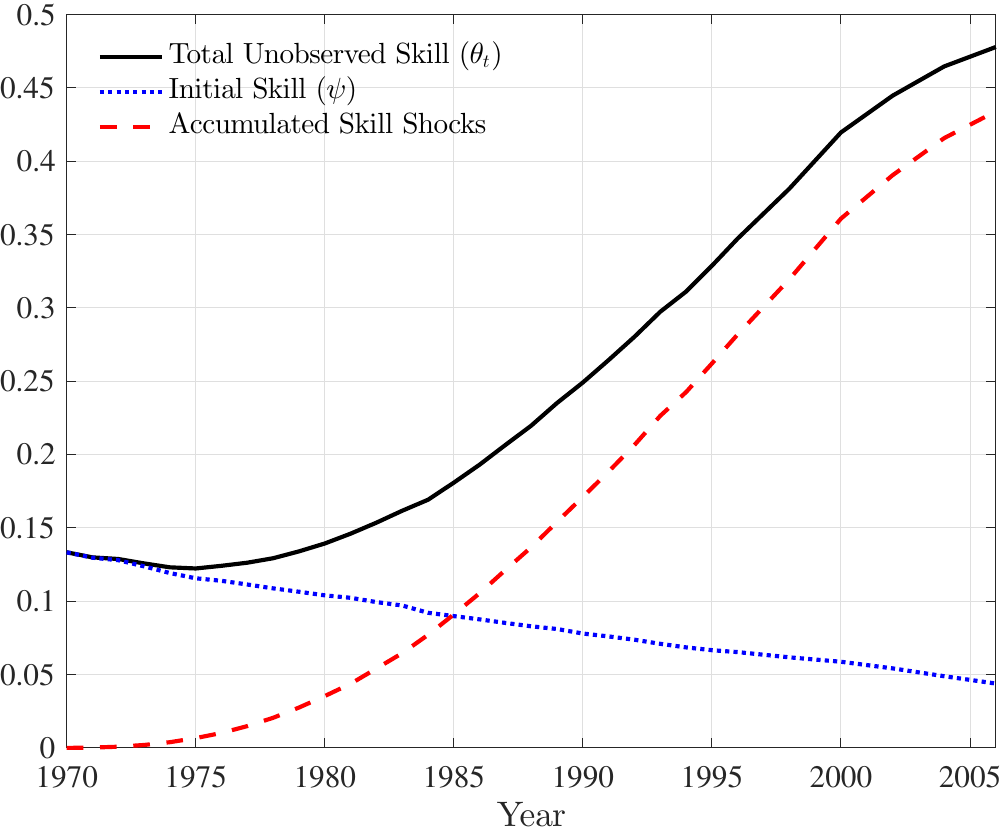}
  }
    \caption{Skill variance decomposition}
    \label{fig: var(theta) decomp}
\end{figure}

\paragraph{Decomposing skill variation.}  We next examine the extent to which changes in the distribution of initial skills vs.\ the distribution of skill growth innovations contribute to the rising skill inequality since the mid-1980s.  Figure~\ref{fig: var(theta) decomp} decomposes the annual variance of skills into the variance of initial skills, $\var(\psi)$, and the variance of skills accumulated since labor market entry, $\var(\theta_t - \psi)$.  This figure reveals modest declines in initial skill inequality due to long-term secular declines in $\var(\psi|c)$ across cohorts.  The large rise in skill inequality is, therefore, driven by a strong increase in the variance of skill growth innovations beginning in the mid-1980s.

\paragraph{Summary.} Altogether, our analysis suggests that rising residual inequality in the late-1970s and early-1980s was driven primarily by increasing returns to unobserved skill for both non-college- and college-educated men. Residual inequality among non-college men declined slightly in the late-1980s, balancing two strong opposing forces: a sharp decline in skill returns, partially offset by a strong increase in skill dispersion.  Meanwhile, college men continued to experience rising residual inequality throughout our sample period.  As with non-college men, a fall in their return to skill was offset by an increase in skill inequality; however, this was accompanied by rising variability of non-skill shocks over the late-1980s and early-1990s.  For both education groups, the strong secular increase in skill dispersion beginning in the 1980s was driven exclusively by increased volatility in skills rather than growing dispersion in skill levels at labor market entry.

%\clearpage

%%%%%%%%%%%%%%%%%%%%%%%%%%%%%%%%%%%%%%%%%%%%%%%%%%%%%%%%%%%%%%%%%%%%%%%%%%%%%%%%%%%%%%%%%%%%

\section{Firms, Occupations, and Multiple Unobserved Skills}  \label{sec: occ multiple skills}

Thus far, we have focused on a specification for log wage residuals that is broadly consistent with traditional assignment models of the labor market, in which each worker is assigned to a different job based on a single-dimensional ranking of worker productivity \citep{tinbergen_1956, sattinger_1993, costinot_vogel_2010}.\footnote{See \cite{lochner_park_shin_2018} for a specification of production technology and skill and job productivity distributions in a traditional assignment model that yields equation~\eqref{eq: resid} as the equilibrium log wage function.}  This section begins by incorporating firm-level differences in pay, estimating the evolution of skill returns using workers who remain on the same job over time.  
We then explore differences in the returns to skill across broad occupation classes and consider the interpretation of our IV estimator when there are multiple skills earning different returns in the market.\footnote{See \cite{sanders_taber_2012} for a survey of the literature on lifecycle wage dynamics in models with multiple skills and occupations.} Throughout this analysis, we continue to account for the fact that skills vary over time for individual workers.

\subsection{Firm-Specific Effects} \label{subsec: firms}

Following \cite{abowd_kramarz_margolis_1999}, a growing literature studies the importance of both worker- and firm-specific wage components.  
\cite{song_price_guvenen_bloom_von_wachter_2018} show that, in the U.S., most of the variation in log earnings residuals is explained by heterogeneity across workers, which also explains most of the rise in residual dispersion. Both \cite{song_price_guvenen_bloom_von_wachter_2018} and \cite{bonhomme_holzheu_lamadon_manresa_mogstad_setzler_2023} further show that the sorting of workers across firms has become more important over time; however, this sorting still explains a relatively small share of residual dispersion.

To explore the implications of heterogeneity in pay across firms for our estimated returns to skill, we incorporate firm fixed effects into our baseline log wage residual specification (assuming a single skill):
\begin{align}
w_{i,t}=\kappa_{j_{i,t}}+\mu_t\theta_{i,t}+\eps_{i,t},\label{eq: AKM}
\end{align}
where $j_{i,t}$ denotes the firm individual $i$ works for in year $t$ and $\kappa_{j}$ represent firm fixed effects.\footnote{A few recent studies estimate time-varying firm-specific effects (i.e., $\kappa_{j,t}$). For example, \cite{card_heining_kline_2013} estimate firm premia separately by subperiods (i.e., rolling-window estimation), while \cite{lachowska_mas_saggio_woodbury_2023} and \cite{engbom_moser_sauermann_2023} allow firm premia to vary freely over time. \cite{song_price_guvenen_bloom_von_wachter_2018} show that the dispersion of estimated firm effects did not change substantially over time in the U.S., providing some support for the common firm fixed effects specification. } 
We invoke the standard ``exogenous job mobility'' assumption of the literature, assuming that the non-skill component $\eps_{i,t}$ satisfies $\E[\eps_{t}|j_{\ul{t}},\ldots,j_{\ol{t}}]=0$ for all $t$.

A version of equation~\eqref{eq: AKM}, abstracting from time-variation in worker skills and their returns (i.e., $\theta_{i,t}=\theta_i$ and $\mu_t=\mu$), is typically estimated using administrative employer-employee matched data, where employer identities are directly observed and a large number of workers transition between firms.\footnote{\cite{bonhomme_lamadon_manresa_2019} include interactions between firm and worker effects in their log wage specification, effectively allowing skill returns to differ across firm types. Their results suggest that these worker--firm interactions explain a small share of overall earnings inequality (in Sweden).}   Since the PSID do not contain firm identifiers, 
we investigate the implications of \emph{unobserved} firm-specific heterogeneity by applying our IV estimator to job stayers, for whom firm fixed effects do not change.  

\begin{figure}[!htbp]
  \centering
    \includegraphics[width=0.45\columnwidth]{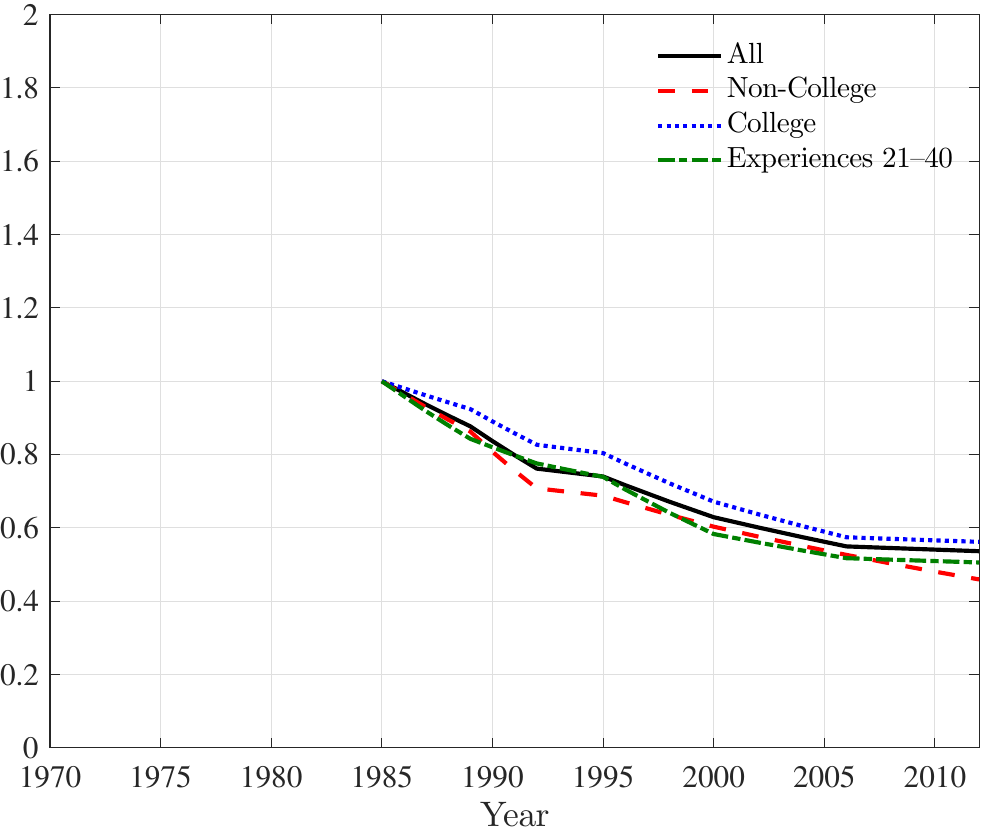}
    \caption{$\mu_t$ Implied by 2SLS for Job Stayers from $t-2$ to $t$}\label{fig: mu job stayer}
\end{figure}

Figure~\ref{fig: mu job stayer} reports the path for skill returns implied by our 2SLS estimator for job stayers (roughly two-thirds of our sample) beginning in 1985.\footnote{It is difficult to identify employer transitions prior to 1985 \citep{brown_light_1992}.  
As discussed in Appendix~\ref{app sec: job stayers}, we identify job stayers based on the most recent start date of the current main job.}
Notably, the evolution of estimated skill returns are quite similar to their counterparts for all workers as reported in Figure~\ref{fig: IV mu}.
Appendix~\ref{app sec: job stayers} shows that IV estimates of skill return growth for stayers are likely to be biased towards zero due to unobserved variation in $\kappa_j$. Using estimated variation in worker- and firm-fixed effects from \cite{song_price_guvenen_bloom_von_wachter_2018}, the appendix argues that this bias is negligible for college men but that the returns to skill for non-college men fell by as much as 50\% more compared to those reported in Figure \ref{fig: mu job stayer}.

\subsection{Occupations}  \label{sec: occ}

Motivated by task-based models of the labor market in which workers are assigned to a limited set of different tasks or jobs \citep{autor_levy_murnane_2003, acemoglu_autor_2011, cortes_2016, acemoglu_restrepo_2022, acemoglu_loebbing_2022}, we next modify our baseline log wage specification to allow for occupation-specific wage functions with residuals given by
\begin{equation}\label{eq: w occ}
    w_{i,t}=\gamma_t^{o_{i,t}}+\mu_t^{o_{i,t}}\theta_{i,t}+\eps_{i,t},
\end{equation}
where $o_{i,t}$ denotes the occupation for worker $i$ in year $t$, and we normalize $\mu_t^o = 1$ and $\gamma^o_t = 0$ for a single occupation-year pair.  Average wages may differ across occupations due to differences in wage functions (i.e., $(\gamma_t^o,\mu_t^o)$) and in the average skill levels of workers in those occupations, with $\E[w_{t}|o_t=o]=\gamma_t^o+\mu_t^o\E[\theta_{t}|o_t=o]$.  For example, management occupations might provide a higher return to skill and be filled by more-skilled workers relative to low-level clerical occupations.
Although assignment and task-based models generally assign all workers of a given skill level to a single task/job, workers with identical skills may choose to work in different occupations due to search and information frictions or heterogeneous preferences for job attributes \citep[e.g.,][]{papageorgiou_2014, taber_vejlin_2020, guvenen_et_al_2020, lise_postel-vinay_2020,  roys_taber_2022, adda_dustmann_2023}.

\cite{cortes_2016} considers a special case of equation~\eqref{eq: w occ}, assuming time-invariant skill returns and skill levels (i.e., $\mu_t^o=\mu^o$ and $\theta_{i,t}=\theta_i$). We further note that equation~\eqref{eq: w occ} generalizes related specifications used in studies of sectoral or firm differences in pay, which typically assume fixed skills over time.\footnote{This specification is broadly consistent with the multi-sector assignment model of \cite{gola_2021}, where wage functions vary with sector (e.g., manufacturing, services) rather than occupation.  
It is also similar to that of \cite{bonhomme_lamadon_manresa_2019}, who allow $\gamma_t^o$ and $\mu_t^o$ to differ across firms instead of occupations, assuming time-invariant worker skills (i.e., $\theta_{i,t}=\theta_i$).  See Section~\ref{subsec: firms} for a discussion of firm-specific pay schedules.  
%Following the canonical \cite{abowd_kramarz_margolis_1999}, the literature on pay differentials across firms typically focuses on estimating differences in the intercept term across \tcr{firms} (i.e.,  $\gamma_t^o = \gamma^o$), assuming no variation in returns to time-invariant individual skills, ($\mu_t^o=\mu$ and $\theta_{i,t}=\theta_i$).  However, several more recent studies estimate time-varying premia. For example, \cite{card_heining_kline_2013} estimate firm premia separately by subperiods (i.e., rolling-window estimation), while \cite{lachowska_mas_saggio_woodbury_2023} and \cite{engbom_moser_sauermann_2023} allow firm premia to vary freely over time. 
}

Occupation-specific skill returns can be identified by strengthening Assumption~\ref{assum: gen mu ident} to condition on recent occupation histories.  
\begin{ass}\label{assum: mu ident occ}
For known $k\geq 1$ and for all $t-t'\geq k+1$: 
(i) $\cov(\Delta \theta_t, \theta_{t'}|o_t,o_{t-1},o_{t'})=0$;
(ii) $\cov(\Delta \theta_t, \eps_{t'}|o_t,o_{t-1},o_{t'}) = 0$;
(iii) $\cov(\eps_t, \theta_{t'}|o_t,o_{t-1},o_{t'}) = \cov(\eps_{t-1}, \theta_{t'}|o_t,o_{t-1},o_{t'}) = 0$; and
(iv) $\cov(\eps_t,\eps_{t'}|o_t,o_{t-1},o_{t'})=\cov(\eps_{t-1},\eps_{t'}|o_t,o_{t-1},o_{t'})=0$.
\end{ass}
Assumption~\ref{assum: mu ident occ} requires that skill dynamics not depend on or influence occupation choices, much as the literature on firm-specific returns assumes that job changes are exogenous \citep{abowd_kramarz_margolis_1999}.  This assumption is likely too strong for young workers simultaneously making early skill investment and career decisions; however, it is more plausible for older workers for whom skill variation is likely to be idiosyncratic and who face weaker incentives to search for a new occupation in response to skill or non-skill wage innovations given their shorter career horizon, greater skill specialization, and stronger (revealed) preferences for current job/occupation amenities \citep{cavounidis_lang_2020}.\footnote{ \cite{gathmann_schoenberg_2010} show that older workers make fewer occupational changes and that those changes entail smaller changes in occupational task content. \cite{gervais_jaimovich_siu_yedid-levi_2016} document declining occupational mobility over the lifecycle.}

As shown in Appendix~\ref{app: ident occ}, $\mu_t^o$ can be identified for all occupations (in all but the first $k$ years) under Assumption~\ref{assum: mu ident occ} using the following IV estimator (for $t-t'\geq k+1$):
\begin{equation} \label{eq: IV occ}
    \frac{\cov(\Delta w_t,w_{t'}|o_t,o_{t-1},o_{t'})}{\cov(w_{t-1},w_{t'}|o_t,o_{t-1},o_{t'})}=\frac{\mu_t^{o_t}-\mu_{t-1}^{o_{t-1}}}{\mu_{t-1}^{o_{t-1}}}.
\end{equation}
Since occupational mobility is low, especially among older workers, we highlight that occupation-specific growth in skill returns can be identified from occupation stayers (from $t-1$ to $t$) alone.\footnote{Assumption~\ref{assum: mu ident occ} implies that estimated return growth for stayers in occupation $o_t=o_{t-1}=o$ should not depend on earlier occupation ($o_{t'}$). Results reported in Appendix~\ref{app: PSID occ} confirm this prediction. }

While occupation-specific growth in skill returns can be identified from occupation stayers (from $t-1$ to $t$), occupation switchers must be incorporated to identify the relative returns to skill across occupations, $\mu_t^o/\mu_t^{o'}$, and the sequence of occupation-specific wage levels, $\gamma_t^o$. In addition to Assumption~\ref{assum: mu ident occ}, identification of $\gamma_t^o$ requires $\E[\Delta \theta_t|o_t,o_{t-1}]=0$ and $\E[\eps_t|o_t,o_{t-1}]=\E[\eps_{t-1}|o_t,o_{t-1}]=0$, which imply
\[
    \E[\Delta w_t|o_t,o_{t-1}]-\left(\frac{\mu_t^{o_t}-\mu_{t-1}^{o_{t-1}}}{\mu_{t-1}^{o_{t-1}}}\right)\E[w_{t-1}|o_t,o_{t-1}]=\gamma_t^{o_t}-\frac{\mu_t^{o_t}}{\mu_{t-1}^{o_{t-1}}}\gamma_{t-1}^{o_{t-1}}.
\]
Given small sample sizes for many  occupation sequences $(o_t,o_{t-1},o_{t'})$, we rely on the stronger assumptions $\E[\Delta \theta_t|o_t,o_{t-1},o_{t'}]=0$ and $\E[\eps_t|o_t,o_{t-1},o_{t'}]=\E[\eps_{t-1}|o_t,o_{t-1},o_{t'}]=0$, which allows us to condition only on $(o_t,o_{t-1})$ in estimation  of $\mu_t^o$ and $\gamma_t^o$.  See Appendix~\ref{app: ident occ}.

\subsubsection{2SLS Estimation of Skill Return Growth for Occupation-Stayers}

We use the PSID to estimate growth in skill returns for occupation-stayers in two broad and exclusive occupation groups (cognitive and routine occupations) considered by \cite{cortes_2016}.\footnote{\cite{cortes_2016} also considers manual occupations, but our sample contains too few observations to obtain precise results for its associated wage parameters.  Appendix~\ref{app: PSID occ} provides details on occupation classifications in the PSID.} 
We also estimate skill returns for those who remain in occupations with high social skill requirements, based on the measure of social skill intensity considered by \cite{deming_2017}.\footnote{We define social occupations as those that fall in the top third of the social skill intensity distribution in the pooled sample of worker-year observations.  See Appendix~\ref{app: PSID occ}.}  As \cite{deming_2017} notes, there is considerable overlap between cognitive occupations and social occupations -- in our sample,  59\% of worker-year observations in cognitive occupations are also in social occupations and  76\% of observations in social occupations are also in cognitive occupations.  

As with equation~\eqref{eq: 2SLS regression} earlier, we use 2SLS (with lagged residuals as instruments) to estimate two-year growth rates in occupation-specific skill returns based on 
\begin{equation} \label{eq: occ 2SLS regression}
\Delta_2 w_{i,t} = \left(\gamma_{t}^{o_{i,t}} - \left[\frac{\mu_t^{o_{i,t}}}{\mu_{t-1}^{o_{i,t-2}}}\right] \gamma_{t-2}^{o_{i,t-2}}\right) +  \left[\frac{\mu_t^{o_{i,t}}-\mu_{t-2}^{o_{i,t-2}} }{\mu_{t-1}^{o_{i,t-2}}}\right] w_{i,t-2} + \underbrace{\left(\varepsilon_{i,t} - \left[\frac{\mu_t^{o_{i,t}}}{\mu_{t-2}^{o_{i,t-2}}}\right] \varepsilon_{i,t-2} + \mu_t^{o_{i,t}}\Delta_2 \theta_{i,t}\right)}_{\equiv \xi_{i,t}}, 
\end{equation}
estimated separately for stayers with $o_{i,t}=o_{i,t-2}=o$ for cognitive, routine, or social occupation groups.\footnote{We use $(w_{t-8},w_{t-9})$ as instruments when both are available (in early survey years); otherwise, we use $(w_{t-8},w_{t-10})$ as instruments.  Our use of two-year differences relies on the natural modification of all assumptions to condition on $(o_t, o_{t-2}, o_{t'})$.}

\begin{figure}%[h]
    \centering
    \subfloat[All Experience Levels]{
    \includegraphics[width=0.45\columnwidth]{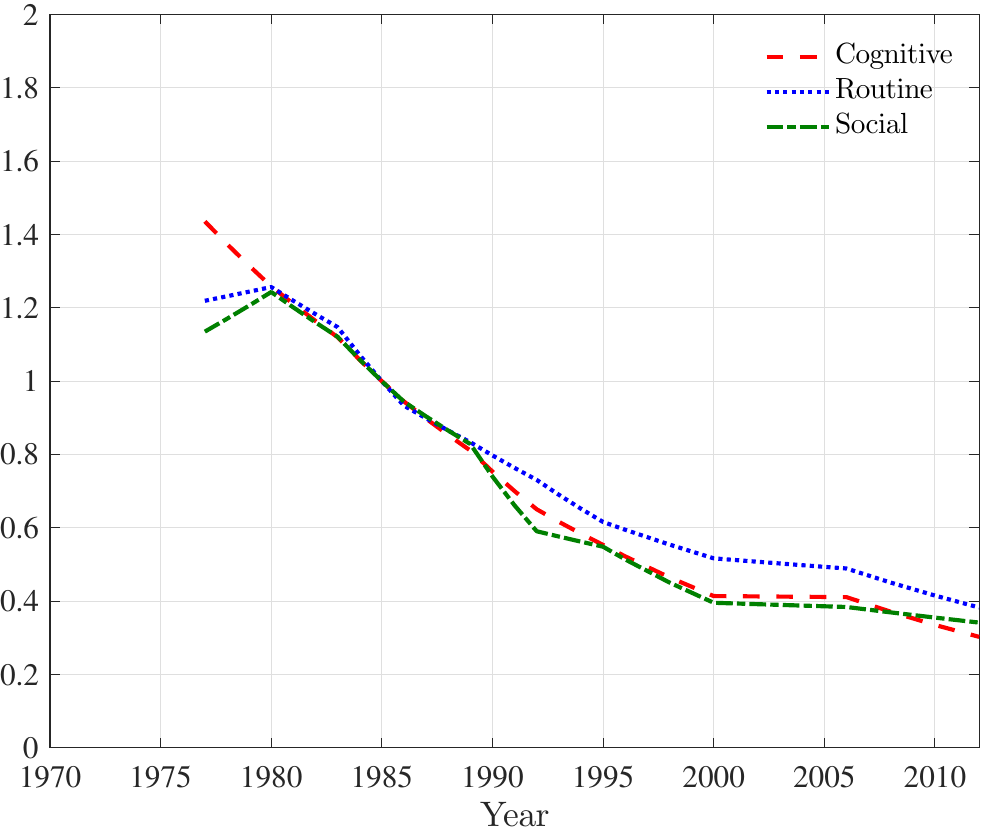}
    }\quad
    \subfloat[21--40 Years of Experiences in Year $t$]{
    \includegraphics[width=0.45\columnwidth]{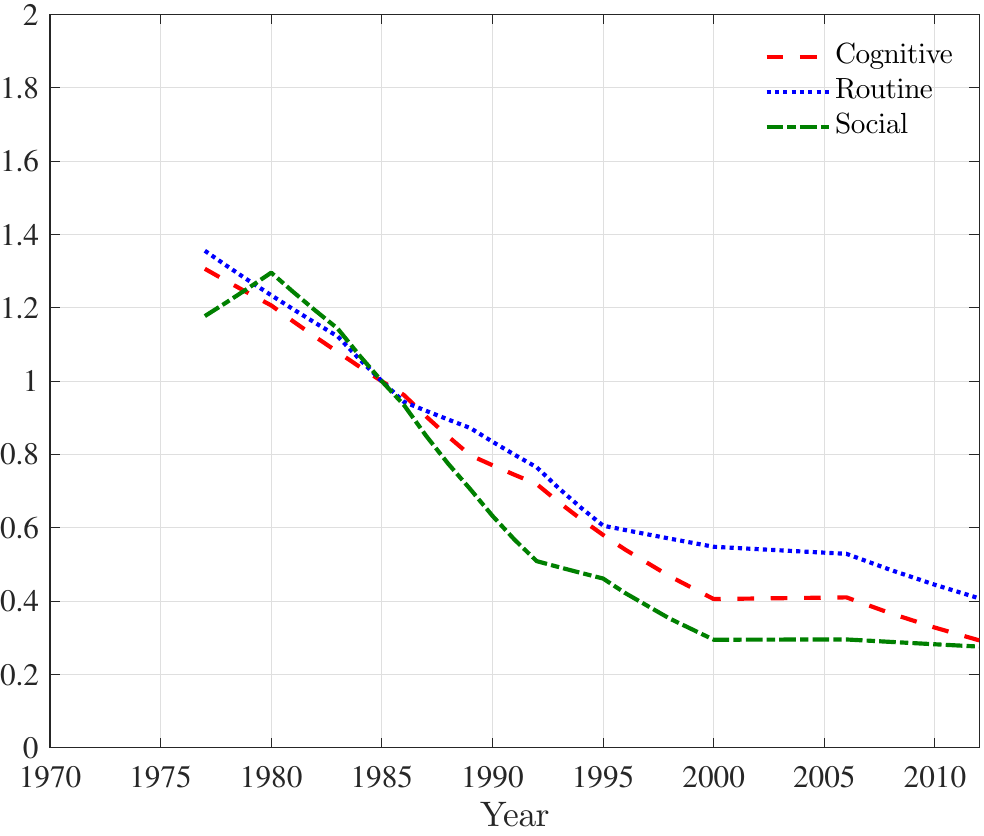}
    }    
\caption{$\mu_t^o / \mu_{1985}^o$ implied by 2SLS estimates for cognitive, routine, and social occupation-stayers between $t-2$ and $t$}
\label{fig: PSID mu occ stayers 2SLS}
\end{figure}

Figure~\ref{fig: PSID mu occ stayers 2SLS} reports implied skill returns (relative to $\mu_{1985}^o$) for all men who remain in cognitive, routine, or social occupations.  Panel~(a) reports estimates based on workers of all experience levels, while panel~(b) reports estimates based on those with 21--40 years of experience.   In both panels, we obtain similar estimated return profiles for job stayers regardless of their occupation type, indicating strong declines in the returns to skill in cognitive, routine, and social occupations. The estimated return profiles also accord well with those estimated earlier for the full sample.

\subsubsection{GMM Estimation of $\gamma_t^o$ and $\mu_t^o$ for Cognitive and Routine Occupations}

In order to estimate differences in the levels of skill returns across occupations and occupation-specific average wage differences, we must also exploit occupational switchers.  We use GMM to simultaneously estimate $\mu_t^o$ and $\gamma_t^o$, now including all occupation stayers and switchers in our sample.
Based on equation~\eqref{eq: occ 2SLS regression}, we exploit the following moments in the PSID:
$\E[\xi_t|o_t,o_{t-2}]=0$ and $\E[w_{t'} \, \xi_t|o_t,o_{t-2}]=0$, where we use lagged residuals $w_{t'}$ from periods $t-8$ and $t-9$ (or $t-10$ in later years) as instruments.
%\[
%    \E\left[\bm z_{t} \left\{\Delta_2 w_{t}-\left(\gamma_t^{o_t}-\frac{\mu_t^{o_t}}{\mu_{t-2}^{o_{t-2}}}\gamma_{t-2}^{o_{t-2}}\right)-\left(\frac{\mu_t^{o_t}}{\mu_{t-2}^{o_{t-2}}}-1\right)w_{t-2}\right\}\Big|o_t,o_{t-2}\right]=\bm 0,\quad \forall (t,o_t,o_{t-2}),
%\]
%where $\bm z_t$ includes $w_{t-8}$ and $w_{t-9}$ (or $w_{t-10}$ in later years).  
See Appendix~\ref{app: PSID occ} for details.

\begin{figure}[!htbp]
  \centering
    \subfloat[$\mu_t^o$]{
      \includegraphics[width=0.45\columnwidth]{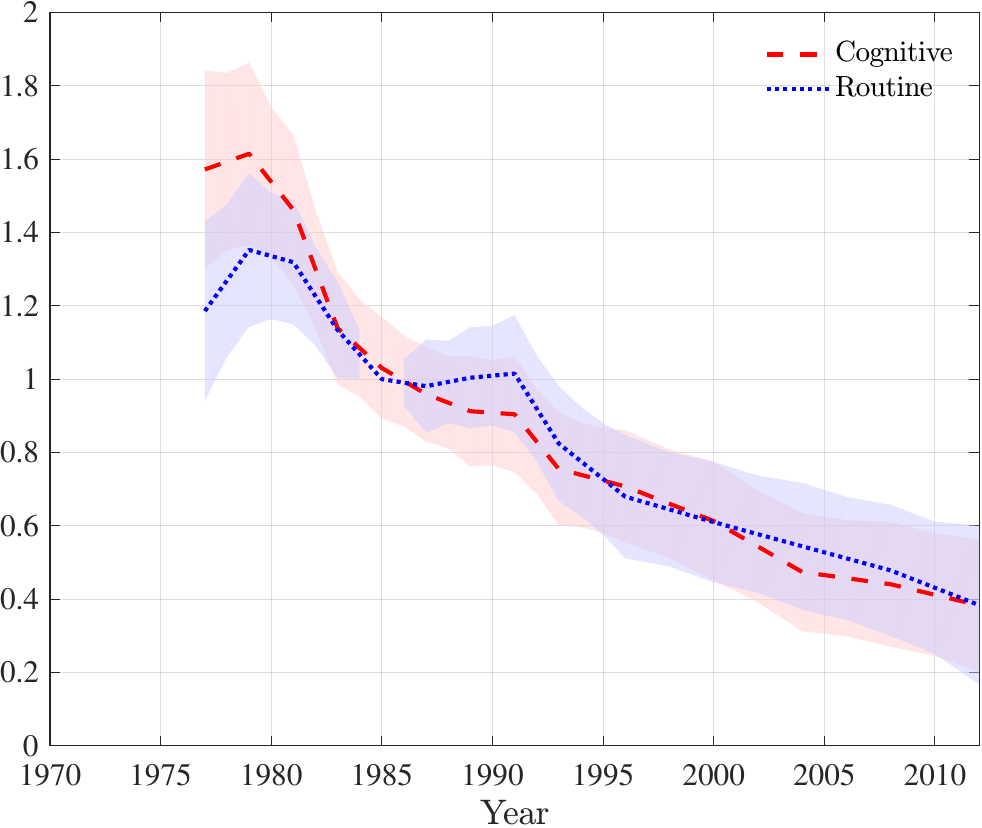}
}\quad
    \subfloat[$\gamma_t^o$]{
      \includegraphics[width=0.45\columnwidth]{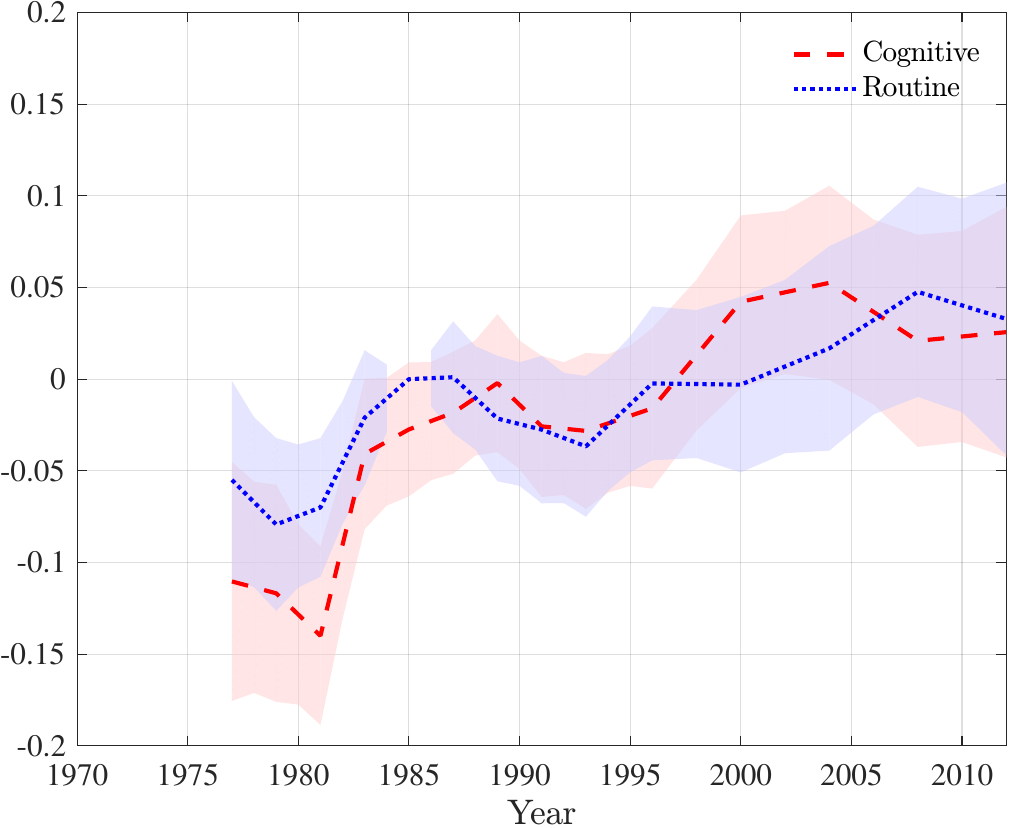}
    }
\caption{GMM estimates of $\mu_t^o$ and $\gamma_t^o$ for cognitive and routine occupations}
\label{fig: GMM occ est mu gamma}
\end{figure}

Given the substantial overlap between cognitive and social occupations (and similar skill return profiles in Figure~\ref{fig: PSID mu occ stayers 2SLS}), we focus this analysis on the two mutually exclusive categories from \cite{cortes_2016}: cognitive and routine occupations.  Here, we normalize $\mu_t^o=1$ and $\gamma_t^o=1$ for routine occupations in 1985; however, no normalizations are needed for cognitive occupations.
Figure~\ref{fig: GMM occ est mu gamma}(a) shows that estimated $\mu_t^o$ series both exhibit substantial declines over time, similar to the 2SLS estimates in Figure~\ref{fig: PSID mu occ stayers 2SLS} and earlier estimates based on the full sample.  We cannot reject that the two skill return series are equal using a standard $J$-test ($p$-value $=0.13$).  Despite sharp drops in the returns to skill in both occupations, Figure~\ref{fig: GMM occ est mu gamma}(b) shows sizeable increases in $\gamma_t^o$ -- nearly 0.20 in cognitive occupations and about 0.12 in routine occupations.   Appendix Figure~\ref{fig: GMM occ est mu gamma OLD} shows similar time patterns for $\mu_t^o$ and $\gamma_t^o$ when using only workers with 21--40 years of experience in year $t$.

\begin{figure}[!htbp]
  \centering
    \subfloat[$\E{[w_t|o_t]}$]{
      \includegraphics[width=0.45\columnwidth]
{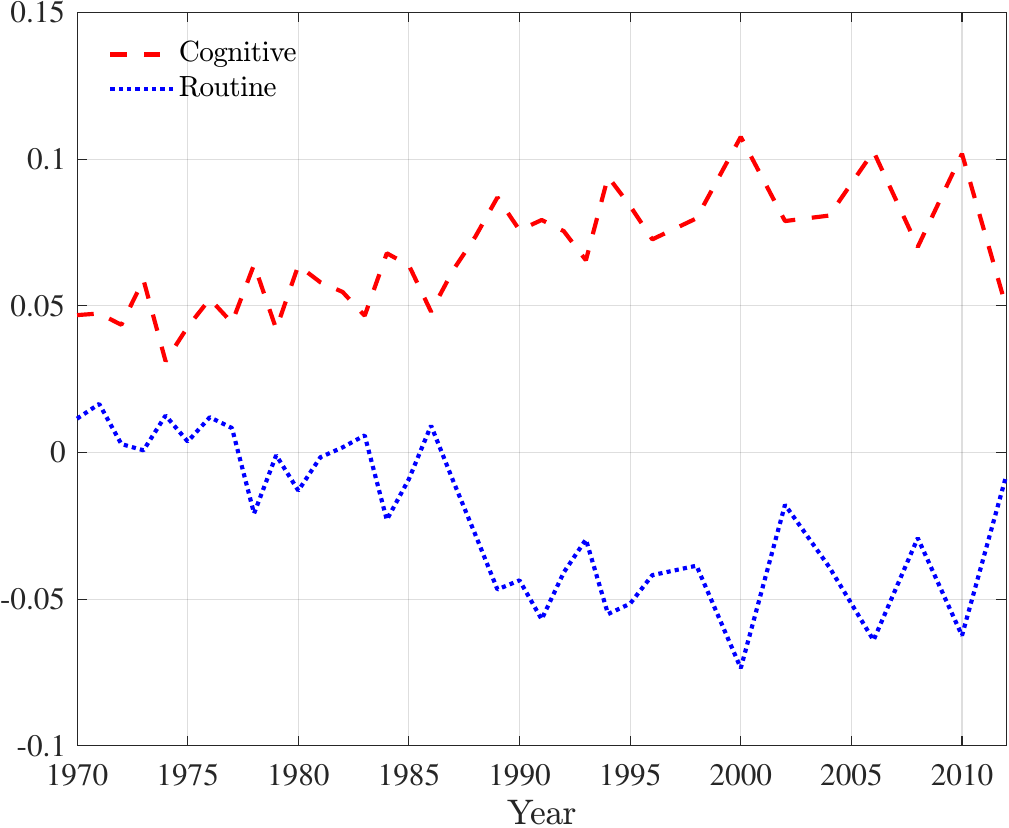}
}\quad
    \subfloat[ $\E{[\theta_t|o_t]}$]{
      \includegraphics[width=0.45\columnwidth]{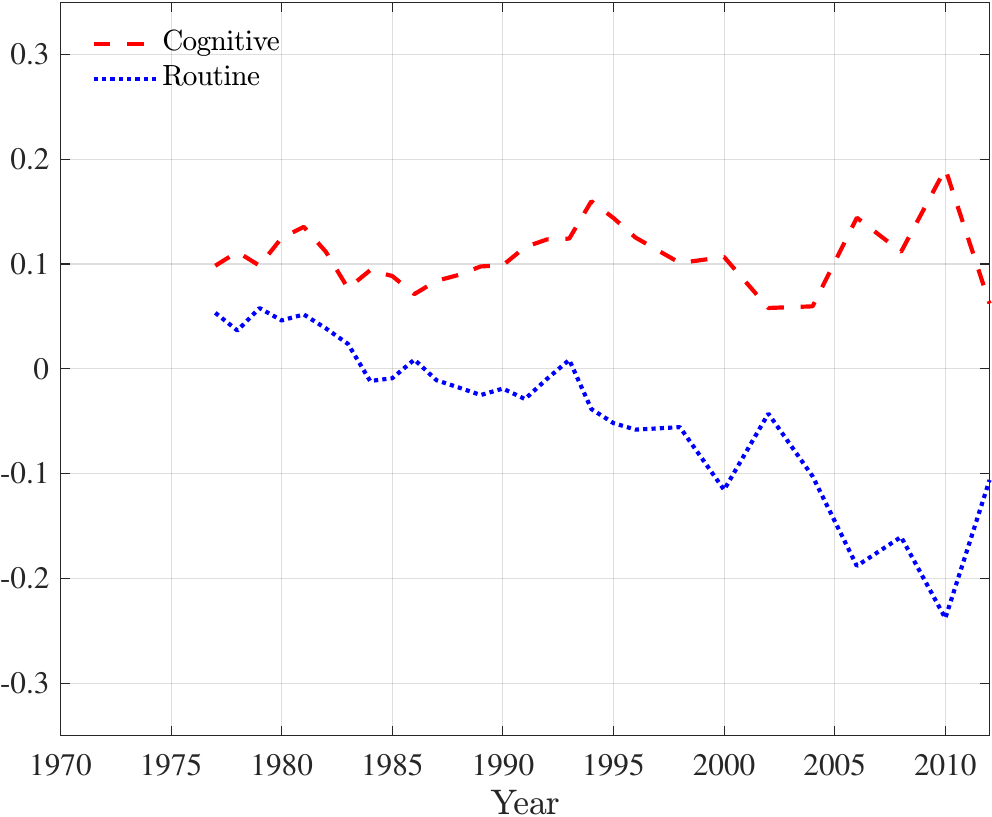}
    }
\caption{Average log wage residual and skill  for cognitive and routine occupations}
\label{fig: GMM occ est avg w theta}
\end{figure}

Using the estimates reported in Figure~\ref{fig: GMM occ est mu gamma} and average log wage residuals by occupation, we can estimate the evolution of average skills by occupation over time from $\E[\theta_t|o_t]=\left(\E[w_t|o_t]-\gamma_t^{o_t}\right)/\mu_t^{o_t}$. Figure~\ref{fig: GMM occ est avg w theta} shows the evolution of average log wage residuals and average skills for cognitive and routine workers.  During our sample period, average log wage residuals rose by about 0.05 for workers in cognitive occupations, while they fell by a similar amount in routine occupations.  Together with estimated $\mu_t^o$ and $\gamma_t^o$, these imply little long-term change in the average unobserved skills of workers in cognitive occupations but roughly 20 log point declines in the average unobserved skills of workers in routine jobs.
%\footnote{\tcr{Our assumptions imply that $\E[\Delta\theta_t |o_t]=0$, so changes in $\E[\theta_t|o_t]$ must be driven by composition changes only.}}  
Notably, these represent skill changes conditional on worker education and experience levels.

Appendix Figure~\ref{fig: GMM occ est constant mu} shows that failing to account for changes in the returns to skill over time (i.e., assuming $\mu_t^o=\mu^o$ for all $t$ as in \cite{cortes_2016}) yields estimates that exhibit little long-term change in $\gamma_t^o$ or average skills, $\E[\theta_t|o_t]$, in both cognitive and routine occupations.    Thus, accounting for the estimated declines in skill returns has important implications for trends in occupation-specific skill levels.

\subsection{Multiple Unobserved Skills} \label{sec: multiple skills}

A growing literature emphasizes the multi-dimensional nature of skills, suggesting that the returns to some types of skills have risen while returns to others have fallen \citep{castex_dechter_2014, deming_2017, edin_et_al_2022}.
Motivated by these  studies, we now consider wage functions that depend on multiple unobserved skills, denoted by $\theta_{i,j,t}$ for $j=1,...,J$:
\begin{equation} \label{eq: w multi skills}
w_{i,t} = \sum\limits_{j=1}^J \mu_{j,t} \theta_{i,j,t} + \eps_{i,t},
\end{equation}
where $\mu_{j,t}$ reflects the market-level value of skill $j$ in year $t$.\footnote{Multi-dimensional assignment and search/matching models of the labor market can give rise to equilibrium log wage functions of the form in equation~\eqref{eq: w multi skills} \citep[e.g.][]{lindenlaub_2017, lise_postel-vinay_2020, lindenlaub_postel-vinay_2023}. These models can also yield more general log wage functions of the entire skill vector, in which case equation~\eqref{eq: w multi skills} can be thought of as a linear approximation. Equation~\eqref{eq: w multi skills} is reminiscent of wage (rather than log wage) functions in \cite{heckman_scheinkman_1987} when worker characteristics can be ``unbundled''.}

We now make use of a multi-dimensional version of Assumption~\ref{assum: mu and skill dist ident} to show that our IV estimator identifies a weighted-average growth rate across all skill returns.  Specifically, we assume that growth in \textit{each} type of skill is uncorrelated with \textit{all} past skill levels.\footnote{A weaker assumption analogous to Assumption~\ref{assum: gen mu ident} (generalized to account for multiple skills) will also ensure that the IV estimator identifies a weighted-average growth in returns.  We impose the stronger conditions based on Assumption~\ref{assum: mu and skill dist ident} here to facilitate interpretation of the weights.}

\begin{ass}  \label{assum: multi skills}
(i) $\cov(\Delta \theta_{j,t},\theta_{j',t'})=0$ for all $j,j'$, and $t-t'\geq 1$;
(ii) $\cov(\theta_{j,t}, \eps_{t'}) = 0$ for all $j,t,t'$; and
(iii) $\cov(\eps_t,\eps_{t'})=0$ for all $t-t' \geq k$.
\end{ass}

This assumption implies:
\begin{equation}\label{eq: multi cov changes}
\cov(\Delta w_t,w_{t'}) = \sum\limits_{j=1}^J \sum\limits_{j'=1}^J \Delta\mu_{j,t} \, \mu_{j',t'}\cov(\theta_{j,t'},\theta_{j',t'}), \quad \text{for $t - t' \geq k+1$},
\end{equation}
where we highlight that the $\cov(\theta_{j,t'},\theta_{j',t'})$ are within-period covariances across skills.
Equation~\eqref{eq: multi cov changes} shows that when all (within-period) correlations between skills are non-negative, a positive (negative) $\cov(\Delta w_t,w_{t'})$ for $t-t' \geq k+1$ implies that total returns $\mu_{j,t}$ are increasing (decreasing) for at least one skill.  Thus, Figures~\ref{fig: PSID autocov} and \ref{fig: PSID autocov by educ} suggest that the returns to at least one skill declined sharply over the late-1980s and 1990s. Consistent with this conclusion,  \cite{castex_dechter_2014} and \cite{deming_2017} estimate strong declines in returns to cognitive skill over this period.

%Given Assumption~\ref{assum: multi skills}, our IV estimator identifies the following:
%\begin{equation}\label{eq: IV multi skill}
%\frac{\cov(\Delta w_t,w_{t'})}{\cov(w_{t-1},w_{t'})} = \frac{\sum\limits_{j=1}^J \Delta\mu_{j,t} \cov(\theta_{j,t'},w_{t'})} {\sum\limits_{j'=1}^J \mu_{j',t-1} \cov(\theta_{j',t'},w_{t'})}, \quad \text{for $t - t' \geq k+1$,}
%\end{equation}
%which implies the following proposition related to skill-specific return growth rates.

As the next result shows, our IV estimator provides a useful summary measure of skill return growth when there are many skills whose returns grow at different rates. For this result, it is useful to define $\ol{\theta}_{i,t} \equiv \sum\limits_{j=1}^J \mu_{j,t}\theta_{i,j,t}$, the total value of a worker's skill vector in period $t$.

\begin{prop} \label{prop: IV multiple skills}
If Assumption~\ref{assum: multi skills} holds, then for all $t - t' \geq k+1$ the IV estimator identifies a weighted-average growth rate across all skill returns:
\begin{equation}\label{eq: avg return growth}
\frac{\cov(\Delta w_t,w_{t'})}{\cov(w_{t-1},w_{t'})} = \sum\limits_{j=1}^J \omega_{j,t',t-1} \left(\frac{\Delta \mu_{j,t}}{\mu_{j,t-1}}\right),
\end{equation}
with weights for $j=1,...,J$ given by $\omega_{j,t',t-1} =  \cov(\theta_{j,t'},\ol{\theta}_{t'})\mu_{j,t-1}/\sum\limits_{j'=1}^J\cov(\theta_{j',t'},\ol{\theta}_{t'})\mu_{j',t-1}$.
%\begin{equation}
%\omega_{j,t',t-1} =  \frac{\cov(\theta_{j,t'},w_{t'})\mu_{j,t-1}}{\sum\limits_{j'=1}^J\cov(\theta_{j',t'},w_{t'})\mu_{j',t-1}}.
%\end{equation}
If $\cov(\theta_{j,t'},\theta_{j',t'})\geq 0, \forall j,j'$, then the weights $\omega_{j,t',t-1} \in [0,1]$ for all $j$.
\end{prop}

When there are multiple skills, our IV estimator identifies the weighted-average growth rate across all skill returns, where the weights, $\omega_{j,t',t-1}$, are larger for skills that are strongly related to wages (in $t'$) and which have a high return (in $t-1$).\footnote{Appendix~\ref{app: ident multiple skills} further shows that the weights are proportional to the extent to which total productivity in period $t'$ predicts the rewards from skill $j$ in period $t-1$.  Proposition~\ref{prop: IV multiple skills2} in Appendix~\ref{app: ident multiple skills} shows that the IV estimator also reflects growth in a weighted-average measure of skill returns.}

The multi-skill problem effectively reduces to the single-skill problem when the relative productivity of different skills is time-invariant: $\mu_{j,t}/\mu_{1,t} = \ol{\mu}_j$ for all $j$ and $t$.
%The multi-skill problem effectively reduces to the single-skill problem when $\alpha_{t}=(\alpha_{1,t},...,\alpha_{J,t})$ is time-invariant, since the evolution of returns to all skills is driven only by variation in $\mu_t$.
%\begin{corol} \label{corr: constant alpha}
%If $\alpha_{t}=\alpha_{t_0}$, then $\frac{\cov(w_t-w_{t_0},w_{t'})}{\cov(w_{t_0},w_{t'})} = \frac{\mu_t-\mu_{t_0}}{\mu_{t_0}}$ for all $t' + k \leq t_0 \leq t-1$.
%\end{corol}
As such, the IV estimator identifies growth rates for all skill returns during periods with fixed relative skill valuations.

\paragraph{Occupations as bundles of skills.}

A simple view of occupations, consistent with multi-dimensional assignment models \citep[e.g.,][]{lindenlaub_2017, lindenlaub_postel-vinay_2023}, is that they  represent different combinations of skill-intensities, $\alpha_{j,t}^o$, leading to different wages by occupation:
\begin{equation} \label{eq: w occ multi skills simple}  
w_{i.t} =  \sum\limits_{j=1}^J \mu_{j,t}\alpha_{j,t}^{o_{i,t}} \theta_{i,j,t} + \eps_{i,t}.
\end{equation}
The returns to skill $j$ in occupation $o$ in year $t$, $\tmu_{j,t}^o = \mu_{j,t}\alpha_{j,t}^o$, depend on the market-level value for that skill, $\mu_{j,t}$, and the occupation-specific skill intensity factor, $\alpha_{j,t}^o$.

Conditioning all covariances in Assumption~\ref{assum: multi skills} on occupation sequence $(o_t=o_{t-1},o_{t'})$, the IV estimator applied to stayers in occupation $o$  (from $t-1$ to $t$) recovers a weighted average of skill-specific return growth, $\Delta \tmu_{j,t}^o/\tmu_{j,t-1}^o = (\Delta \mu_{j,t}\alpha_{j,t} + \mu_{j,t-1}\Delta \alpha_{j,t}^o)/\tmu_{j,t-1}^o$, in occupation $o$, where the returns to skills that are more important for wages in occupation $o$ receive more weight.\footnote{See Appendix~\ref{app: ident multiple skills} for details on all results in this subsection.} Notice that stability of occupation skill intensities (i.e., $\alpha_{j,t}^o=\alpha_j^o$), as assumed by much of the literature \citep[e.g.,][]{autor_dorn_2013, acemoglu_autor_2011, bohm_2020, bohm_von_gaudecker_schran_2024}, would imply that IV estimates using stayers in occupation $o$ identify weighted averages of $\Delta \mu_{j,t}/\mu_{j,t-1}$, where the weights continue to depend on occupation $o$ (e.g.,  IV estimates based on stayers in sales- or communication-based occupations will largely reflect growth in the returns to social skills, while IV estimates based on stayers in manufacturing jobs will primarily reflect growth in returns to manual skills).   Altogether, IV estimators applied to a diverse set of occupations will yield different skill return trends if either (i) relative skill intensities evolve differently across occupations or (ii) returns to various skills evolve differently over time.
The similarity of IV estimated return series across occupation groups reported in Figure~\ref{fig: PSID mu occ stayers 2SLS} are, therefore, consistent with relatively stable occupation skill intensities and similar declines in the returns to a broad range of skills.\footnote{A few recent studies document within-occupation changes in the skill/task content/requirements of jobs \citep{spitz-oener_2006, cavounidis_et_al_2021, cortes_jaimovich_siu_2023}.  Given the inherent challenges of such efforts, some of these documented changes may reflect changes in the skill levels of workers within occupations over time rather than changes in the actual tasks performed by workers.  A separate challenge is that workers may perform different mixes of tasks within the same occupation \citep{autor_handel_2013, spitz-oener_2006}.  In our analysis, any such differences would be interpreted as variation in worker skill bundles. }

\begin{figure}%[h]
    \centering
      \includegraphics[width=0.45\columnwidth]{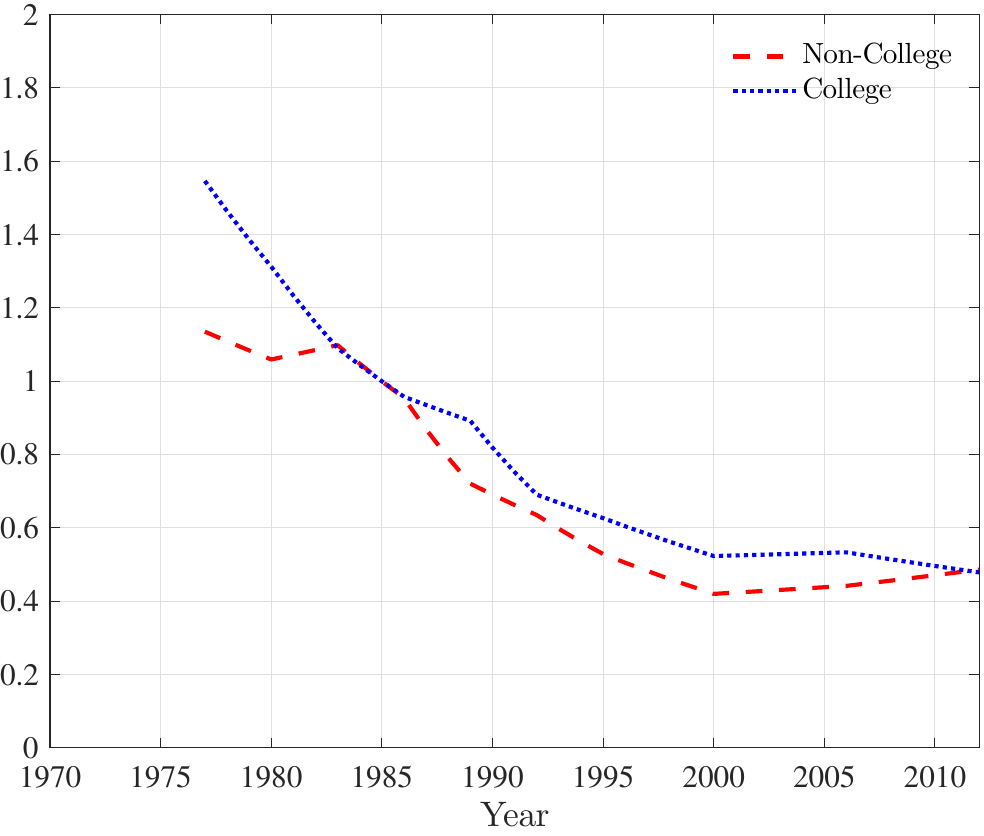}
\caption{$\mu_t$ Implied by 2SLS Estimates for 3-Digit Occupation Stayers Between $t-2$ and $t$}
\label{fig: PSID mu occ stayers}
\end{figure}

Finally, we explore estimated returns for college vs.\ non-college men using a sample of all occupation-stayers (from $t-1$ to $t$), regardless of occupation.  For stable within-occupation skill intensities, this  identifies a weighted average of $\Delta \mu_{j,t}/\mu_{j,t-1}$ where the weights depend on the share of stayers in each occupation $o$.
Figure~\ref{fig: PSID mu occ stayers} displays the implied skill return profiles (by education) for all 3-digit occupation stayers. Both estimated return series are very similar to our baseline estimates for the full sample reported in Figure~\ref{fig: IV mu}.

\subsection{Occupation-Specific Wage Functions with Multiple Skills}

Several studies consider a more substantial role for occupations in multi-skill models of the labor market \citep[see, e.g.,][]{gathmann_schoenberg_2010, yamaguchi_2012,  yamaguchi_2018, bohm_2020, guvenen_et_al_2020, roys_taber_2022, bohm_von_gaudecker_schran_2024}.\footnote{The canonical Roy model \citep{roy_1951} is a special case in which there are an equal number of occupations and skills with each skill rewarded only in its ``own'' sector.  See \cite{heckman_sedlacek_1985}, \cite{keane_wolpin_1997}, and \cite{kambourov_manovskii_2009} for important empirical applications of this framework to occupational choice and wages.
%See \cite{heckman_sedlacek_1985} for an early empirical analysis using this framework. \cite{keane_wolpin_1997} provide an influential early empirical analysis of a dynamic model of skill accumulation with multiple skills and occupations within the \cite{roy_1951} tradition. \cite{kambourov_manovskii_2009} incorporate search frictions in a framework with occupation-specific skills. 
}
Motivated by this literature, we interpret our IV estimator when wages depend on multiple unobserved skills that are rewarded differently across occupations: 
\begin{equation} \label{eq: w occ multi skills}
w_{i,t} = \gamma_t^{o_{i,t}} + \sum\limits_{j=1}^J \mu_{t}^{o_{i,t}}\alpha_{j,t}^{o_{i,t}} \theta_{i,j,t} + \eps_{i,t}.
\end{equation}
In this case, occupation- and skill-specific returns, $\tmu_{j,t}^o = \mu_{t}^o\alpha_{j,t}^o$, arise when occupations reward a worker's total productivity differently, where total productivity depends on the intensity of each skill used in that occupation.\footnote{Equation~\eqref{eq: w occ multi skills} is analogous to wage functions in the skill-weights model of \cite{lazear_2009}; although, $w_t$ reflects \textit{log} wage (residuals) here rather than wages as in \cite{lazear_2009}.} 
We focus on estimating growth in occupation-specific returns to skills, $\mu_t^o$; however, wage levels may also differ across occupations and over time, $\gamma_t^o$, due to, for example, compensating differences or market frictions.  Estimating changes in $\gamma_t^o$ is challenging with multiple skills, so we leave that to future work. 

Conditioning covariances in Assumption~\ref{assum: multi skills} on occupation sequence $(o_t=o_{t-1},o_{t'})$, it is straightforward to show that for occupation-stayers, our IV estimator identifies:
\begin{equation} \label{eq: IV occ multi skills}
    \frac{\cov(\Delta w_t,w_{t'}|o_t=o_{t-1}=o,o_{t'}=o')}{\cov(w_{t-1},w_{t'}|o_t=o_{t-1}=o,o_{t'}=o')}=\frac{\Delta \mu_t^{o}}{\mu_{t-1}^{o} }+
    \frac{\mu_t^{o}}{\mu_{t-1}^{o} } \sum\limits_{j=1}^J \tilde{\omega}_{j,t',t-1}^{o,o'} \left(\frac{\Delta \alpha_{j,t}^{o}}{\alpha_{j,t-1}^{o}}\right), 
\end{equation}
where the weights on skill intensity growth, $\tilde{\omega}_{j,t',t-1}^{o,o'}$, sum to one and are non-negative if all skills are non-negatively correlated conditional on occupations $(o,o')$.  See Appendix~\ref{app: ident occ multiple skills} for details.

If skill intensities do not vary over time within occupations (i.e., $\alpha_{j,t}^o=\alpha_j^o$), then our IV estimator for stayers identifies occupation-specific skill return growth, $\Delta \mu_t^o/\mu_{t-1}^o$, as in Section~\ref{sec: occ}.\footnote{Appendix~\ref{app: ident occ multiple skills} discusses our IV estimator applied to the sample of stayers in occupation $o_t=o_{t-1}=o$, regardless of past occupation $o_{t'}$, as well as for all stayers in any occupation $o_t=o_{t-1}$.}   More generally, the IV estimator for stayers in occupation $o$ also reflects any growth in skill intensities within that occupation. %\footnote{As noted earlier, \cite{spitz-oener_2006, cavounidis_et_al_2021}, and \cite{cortes_jaimovich_siu_2023} suggest that the task content of occupations has evolved over time.  \cite{yamaguchi_2018} and \cite{roys_taber_2022} directly link $\alpha_{j,t}^o$ to occupational task importance/requirements as recorded in the Dictionary of Occupational Titles (DOT) and O*NET, respectively.  Given infrequent updates of these data sources, both studies restrict the extent of time variation in $\alpha_{j,t}^o$.}   
If  log wage residuals are characterized by equation~\eqref{eq: w occ multi skills}, the results in Figure~\ref{fig: PSID mu occ stayers 2SLS} are consistent with similar growth in all skill intensities within manual, routine, and social occupations, coupled with similar declines in the returns to these skills within these occupations.

%%%%%%%%%%%%%%%%%%%%%%%%%%%%%%%%%%%%%%%%%%%%%%%%%%%%%%%%%%%%%%%%%%%%%%%%%%%%%%%%%%%%%%%%%%%%%%%%%%%%%%%%%%%%%%%
\section{Returns Estimated from Administrative Earnings Data} \label{sec: GSF}

Previous studies have documented different trends in income volatility when using administrative data rather than the PSID \citep[see, e.g.,][]{sabelhaus_song_2010, debacker_heim_panousi_ramnath_vidangos_2013}.\footnote{See \cite{moffitt_et_al_2022} for a useful effort to reconcile disparate findings across data sources.}
We show in this section that estimated patterns for skill returns are similar to those already presented when using earnings records from IRS W-2 Forms (maintained by the Social Security Administration, SSA) linked with survey data from the SIPP.  These data include the full SSA history of wage and salary measures for all linked respondents from 1951 to 2011.

Our analysis begins with a sample of US-born white men ages 16--64 who could be linked to any of nine SIPP panels (i.e., panels from 1984--2008).  We work with log wage residuals constructed as with the PSID and restrict observations to years when individuals were no longer enrolled in school. We focus mainly on results using Detailed Earnings Records (DER), which are uncapped and available from 1978 onward; however, we also take advantage of Summary Earnings Records (SER) available since 1951, which report earnings capped at the FICA taxable maximum.
See Appendix~\ref{app: GSF} for a detailed discussion of these data and our sample.  We highlight Appendix Figures~\ref{fig: GSF pred w by educ} and \ref{fig: GSF autocov by educ}, which show very similar patterns to Figures \ref{fig: resid quart pred} and \ref{fig: PSID autocov by educ} regarding convergence in predicted wage residuals given base-year residual quartiles and sharp declines in residual autocovariances $\cov(w_t,w_b)$ over years $t\geq b+6$ for fixed base year $b$.  Together, these indicate declines in the return to skills over the late-1980s and 1990s, consistent with our PSID results.

\begin{figure}[h]
    \centering
    \subfloat[Non-College]{
      \includegraphics[width=0.45\columnwidth]{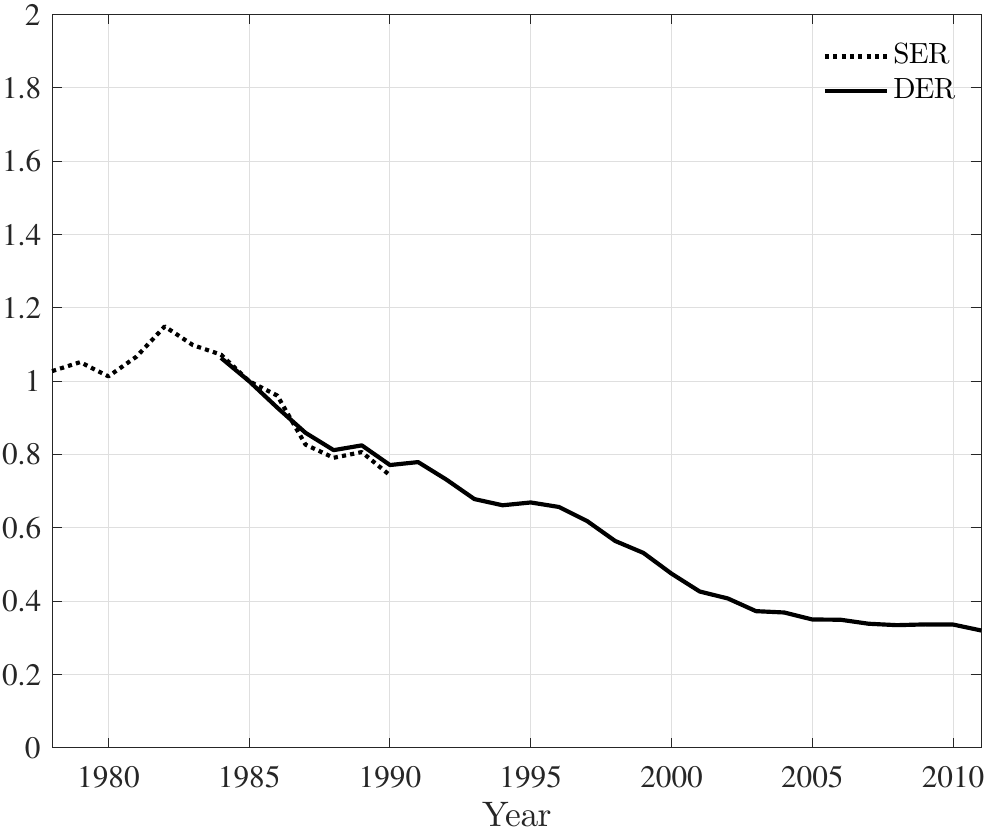}
}\quad
    \subfloat[College]{
      \includegraphics[width=0.45\columnwidth]{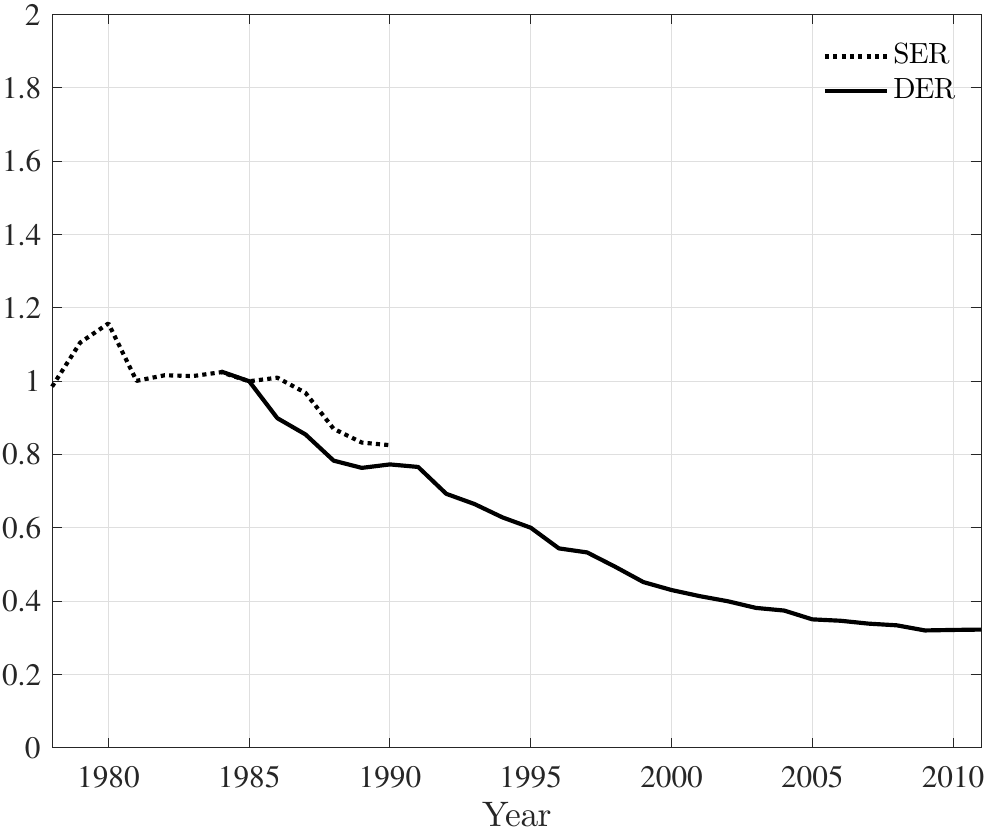}
}
\caption{$\mu_t$ Implied by IV Estimates (instrument: $w_{t-7}$), Experience 32--40 in $t$ (SIPP/W-2)}
    \label{fig: IV mu by educ GSF}
\end{figure}

We use our IV estimator in equation~\eqref{eq: IV estimator} to estimate growth rates for skill returns using $w_{t-7}$ as an instrument (consistent with $k=6$).  Since sample sizes are much larger than in the PSID, we limit our sample to men with 32--40 years of experience to focus on years when wage growth is generally negligible, yet before most men begin retiring.\footnote{Preliminary results were similar for broader experience ranges like those used in the PSID. }  Figure~\ref{fig: IV mu by educ GSF} reports the implied 1984--2011 time series for $\mu_t$ (normalizing $\mu_{1985}=1$) when using only DER-based residual earnings. To identify $\mu_t$ over earlier years, we combine DER- and SER-based residuals, using the latter only as lagged instruments.  Both sets of estimates are very similar to analogous PSID-based estimates reported in Figure~\ref{fig: IV mu}.\footnote{See Appendix Tables \ref{tab: IV SER GSF} and \ref{tab: IV DER GSF} for the estimates, standard errors, and sample sizes when using the SER and DER earnings residuals as instruments.}

\paragraph{Occupational stayers.}  We next explore growth in skill returns for occupation-stayers as in Section~\ref{sec: occ multiple skills}.  Here, we must limit our sample to those observed in one of the SIPP panels during years $t-1$ and $t$, since only the survey data contains occupation information.  We estimate skill return growth for (i) those remaining in the same occupation, (ii) those remaining in a cognitive occupation, and (iii) those remaining in a routine occupation during years $t-1$ and $t$.\footnote{Occupations are based on 24 categories created by the Census Bureau.  As with the PSID, we consider routine and cognitive occupation groupings.  See Appendix~\ref{app: GSF} for details.}  Given the timing of SIPP panels and sample sizes, we estimate annual skill return growth rates for two separate periods: 1991--1999 and 2002--2011.  Table~\ref{tab: IV stayers GSF} reports these IV results using $w_{t-7}$ as instruments, again focusing on men with 32--40 years of experience.  The first two columns suggest that skill returns fell by about 2\% per year over the 1990s and 2000s, consistent with earlier estimates.  The remaining columns suggest that skill returns were fairly stable for cognitive occupations but declined by 3.7--4.9\% per year for routine occupations.  While we cannot reject equality of skill return growth rates (within periods) across the two occupation groups,\footnote{The difference in return growth across occupation groups is 0.036 (SE=0.031) for 1991--1999 and 0.046 (SE=0.024) for 2002--2011.} stronger declines in routine occupations could be driven by routine-biased technical change  \citep{autor_dorn_2013}.
%\tcg{I think most people will relate RBTC with a stronger decline in $\gamma_{t}^{routine}$ relative to $\gamma_t^{cognitive}$ (rather than differential changes in $\mu_t^o$). Maybe we can cite \cite{costinot_vogel_2010} and \cite{acemoglu_loebbing_2022}? \cite{costinot_vogel_2010} show that ``extreme-biased technical change,'' a simultaneous increase in the demand for tasks with low- and high-skill intensities  (relative to middle), leads to wage polarization, a decrease in skill returns  among low-skill workers (routine?) and an increase in return to skill among high-skill workers (cognitive?). \cite{acemoglu_loebbing_2022} show that such a fall in the relative demand for tasks with middle-skill intensities could occur due to automation. I guess we can say that these papers belong to the RBTC literature?}

\begin{table}[!htbp]
  \centering\small
  \caption{2SLS estimates of $\Delta\mu_t/\mu_{t-1}$ for occupational stayers with Experience 32--40 in $t$  (SIPP/W-2)}
\label{tab: IV stayers GSF}
\begin{tabular}{ccccccc}
\toprule
& \multicolumn{2}{c}{Same occupation} & \multicolumn{2}{c}{Cognitive occupations} &  \multicolumn{2}{c}{Routine occupations}    \\
\cmidrule(lr){2-3}  \cmidrule(lr){4-5}  \cmidrule(lr){6-7}
& 1991--1999 & 2002--2011 & 1991--1999 & 2002--2011 & 1991--1999
  & 2002--2011 \\
  \midrule
$\Delta\mu_t/\mu_{t-1}$ & -0.021 & -0.017 & -0.013 & 0.009 & -0.049$^*$ & -0.037$^*$ \\
  & (0.013) & (0.011) & (0.021) & (0.016) & (0.023) & (0.018)\\
Observations & 8,400 & 11,000 & 2,900 & 4,400 & 5,200 & 6,100 \\
  \bottomrule
\multicolumn{7}{l}{Notes: Reports coefficient estimates from 2SLS regression of $\Delta w_t$ on $w_{t-1}$ using $w_{t-7}$ as an}\\
\multicolumn{7}{l}{ instrument. $^*$ denotes significance at 0.05 level. The number of observations is rounded to the}\\
\multicolumn{7}{l}{ nearest 100 due to confidentiality requirements.}
\end{tabular}

\end{table}

%\clearpage

%%%%%%%%%%%%%%%%%%%%%%%%%%%%%%%%%%%%%%%%%%%%%%%%%%%%%%%%%%%%%%%%%%%%%%%%%%%%%%%%%%%%%%%%%%%%%%%%%%%%%%%%%%%%%%%%%%%%%%

\section{Conclusions} \label{sec: conclusion}

Economists have devoted considerable effort to understand the underlying causes of rising residual wage inequality over the past few decades. Most studies have relied on repeated cross-sectional data on wages with a few recent studies incorporating additional measures of worker skills or job tasks.  While these efforts have yielded important insights and motivated robust theoretical literatures, they typically assume that distributions of skills or early-career skill growth have remained stable across cohorts born decades apart.

This paper takes a very different approach, demonstrating that traditional panel data sets can be used to separately identify changes in the returns to unobserved skill from changes in the distributions of unobserved skill and in the distribution of transitory non-skill shocks.  Based on transparent identifying assumptions, we show that a simple IV strategy (that exploits panel date on log wage residuals) can be used to estimate the returns to unobserved skill over time.  We test and cannot reject key assumptions, further showing that our main conclusions are robust to relaxing most assumptions.  Once skill returns have been identified, it is straightforward to identify and estimate the evolution of skill (and skill growth) distributions as well as distributions of transitory non-skill shocks.  Importantly, none of this requires measures of the tasks workers perform nor direct measures of worker skill levels; although, future work could incorporate such measures (when available) within our framework to relax various assumptions, improve the precision of estimates, and/or identify the full complement of task- or skill-specific returns.

Using survey data on wages from the PSID and administrative earnings records from W2 forms, we show that skill returns for American men were fairly stable or increasing in the 1970s and early-1980s, but then fell sharply over the late-1980s and 1990s (especially among non-college men) before stabilizing again. The decline in returns was offset by a strong increase in the variance of unobserved skills beginning in the early-1980s, driven by rising skill volatility (rather than changes in the dispersion of skills at labor market entry).  We also estimate a moderate increase in the variance of transitory non-skill wage innovations during the late-1980s and 1990s for college-educated men, contributing to growth in their residual inequality over that period.   These conclusions stand in stark contrast to prevailing views, which attribute rising residual inequality primarily to rising skill returns, despite recent evidence by \cite{castex_dechter_2014} suggesting that the returns to cognitive skill fell by half between the late-1980s and 2010 (consistent with our estimated declines in skill returns).

Given growing interest in the importance of tasks, occupations, and the multiplicity of skills for recent trends in wage inequality, we extend our analysis to account for heterogeneous pricing of multiple unobserved skills across occupations.
Our analysis of PSID data indicates that skill returns fell similarly for men working in routine, cognitive, and social occupations.  This finding is consistent with similar changes (or stability) in the skill-intensities of these occupation types, coupled with similar declines in the returns to  heterogeneous skills used within those occupations.
We find that the substantial decline in log wage residuals among workers in routine relative to cognitive occupations can be attributed to (i)~weaker growth in wages paid to similarly skilled workers in routine relative to cognitive occupations,  and (ii) substantial (unobserved) skill declines among workers in routine relative to cognitive jobs.
%\tcr{Additionally, we find that log wage residuals declined substantially for workers in routine relative to cognitive jobs over the past few decades, because (i) the wages paid to similarly skilled workers diverged across these occupations and (ii) average unobserved skills of workers in routine jobs deteriorated considerably while there was little change in the average skills of workers in cognitive occupations.}  
Our estimates based on administrative W2 earnings records suggest that the returns to skill may have fallen more for workers in routine relative to cognitive occupations; however, we cannot reject that the declines in returns were equal (as estimated in the PSID).  Whether skill and wage levels (as suggested by the PSID) or the returns to skill (as suggested by W2 records) fell more strongly within routine occupations, our findings are broadly consistent with some form of routine-biased technical change \citep{autor_dorn_2013}.

All of our conclusions derive directly from the time patterns for long-autocovariances for log wage residuals,  notably the sharp declines over the late-1980s and 1990s (see Figures~\ref{fig: PSID autocov} and \ref{fig: PSID autocov by educ}).  These drops are broad-based, evident for young and old, non-college and college workers.  They are equally pronounced for firm- and occupation-stayers, suggesting that they are not simply explained by shifts in firm/occupation structure or by changes in firm/occupational switching \citep[see, e.g.,][]{kambourov_manovskii_2008}.  
%\tcr{[Argue why it is probably not due to increasing firm-pay inequality and job mobility.]} \tcg{(Not sure. Maybe we can argue that increasing firm-pay inequality would look like increasing $\mu_{t}$ under sorting? Perhaps we could say that occupation-stayers are likely to stay in the same firm, but it might be a stretch.)}
Viewed through the lens of the canonical wage function for unobserved skills introduced by \cite{juhn_murphy_pierce_1993}, it is difficult to reconcile these robust trends with rising returns to unobserved skill, as is so often done. While we do not attempt to explain why unobserved skill returns fell over a period when returns to education rose,\footnote{Results in \cite{juhn_murphy_pierce_1993} raised an alternative challenge: why did the returns to unobserved skills (apparently) rise while the returns to education fell in the late-1970s.} we hope that our findings spur new thinking on this issue.

Equally important, our results suggest that more attention  be devoted to understanding the dramatic increase in unobserved skill volatility. This may simply reflect a different type of technological change -- one characterized by the frequent introduction of new tasks that displace others \citep[e.g.,][]{andolfatto_smith_2001, acemoglu_restrepo_2018}. Defining workers' skill levels by the most productive task(s) they can perform, this type of technological change would generate growing volatility in skills over the lifecycle (or economic turbulence as in \cite{ljungqvist_sargent_1998}), which could, in turn, reduce skill returns \citep[see, e.g.,][]{lochner_park_shin_2018}.  Growing knowledge/task specialization in the workforce 
%\tcr{(References?)} \tcg{(I'm not sure this is true. I thought technical progress would flatten hierarchies and enable workers to perform a wider range of tasks. Maybe we can interpret increasing complexity of tasks as evidence of specialization?  Empirically, I guess greater specialization over time would imply greater earnings losses after a job loss, which seems consistent with the findings of Schmieder, von Wachter, and Heining (2023) from Germany.  Related to economic turbulence?)} 
is likely to  further exacerbate these forces.  An alternative explanation may be that more able workers are simply more capable of learning and adapting to new tasks \citep{nelson_phelps_1966}, which would imply greater variation in lifecycle wage growth during periods of rapid innovation.\footnote{See Section 3.2 of \cite{hornstein_krusell_violante_2005} for a survey of theory and evidence on this view of technological change and skills.}  
Finally, if firms possess imperfect information about workers' skills, our estimated ``skill distributions'' would instead reflect the distributions of beliefs about worker skills. Thus, our estimates may also reflect changes in firms' abilities to identify (and reward) workers' skill levels over their careers \citep[e.g., see][]{lemieux_macleod_parent_2009, jovanovic_2014}.

\clearpage

\baselineskip=16pt

\bibliography{LPS}

\newpage
\renewcommand{\thepage}{A\arabic{page}}
\setcounter{page}{1}

\appendix

%\section*{Online Appendix for ``The Evolution of Unobserved Skill Returns in the U.S.: A New Approach Using Panel Data"}

\begin{center}
    \Large \bfseries Appendix
\end{center}

%%%% Re-Number Tables and figures for the Appendix %%%%%%%%
\renewcommand{\thetable}{\thesection-\arabic{table}}
\setcounter{table}{0}

\renewcommand{\thefigure}{\thesection-\arabic{figure}}
\setcounter{figure}{0}

\section{Prior Assumptions in the Literature} \label{app: prior assum}

\subsection{\cite{juhn_murphy_pierce_1993}} \label{app: JMP}

Let $c$ reflect the year an individual enters the labor market and $e=t-c$ labor market experience.  Then,
\begin{align*}
\var(w_{t}|c) &= \mu_t^2 \var(\theta_t|c) + \var(\eps_t|c) %\label{eq: var decomp 0}
%        &= \mu_t^2 \left[\var(\theta_{t-1}|c) + \var(\Delta\theta_t|c)\right] + \var(\eps_t|c) \label{eq: var decomp 1} 
        = \mu_t^2 \left[\var(\theta_{c}|c) + \sum\limits_{\tau=c+1}^t \var(\Delta\theta_{\tau}|c)\right] + \var(\eps_t|c).  %\label{eq: var decomp 2}
\end{align*}
where the second equality holds by assuming $\cov(\Delta\theta_t,\theta_{t'}|c)=0$ for $t\geq t'+1$.

The period $t$ to $t+\ell$ time difference for this variance for a given cohort can be written as follows:
\begin{align*} %\label{eq: Delta_c}
\Delta_c(t,\ell) & \equiv  \var(w_{t+\ell}|c) - \var(w_{t}|c)  \\
% & =  \left[\mu_{t+\ell}^2 - \mu_t^2\right] \var(\theta_t|c) + \mu_{t+\ell}^2\sum\limits_{\tau=t+1}^{t+\ell} \var(\Delta\theta_\tau|c) + \left[\var(\eps_{t+\ell}|c) - \var(\eps_t|c)\right] \\
  & =  \left(\mu_{t+\ell}^2 - \mu_t^2\right) \left[\var(\theta_c|c) + \sum\limits_{\tau=c+1}^{t} \var(\Delta\theta_\tau|c) \right] + \mu_{t+\ell}^2\sum\limits_{\tau=t+1}^{t+\ell} \var(\Delta\theta_\tau|c) + \left[\var(\eps_{t+\ell}|c) - \var(\eps_t|c)\right].
\end{align*}

Next, consider the time difference for the residual variance following an experience group over time  (assuming $c+\ell<t$):
%
%\small
%
\begin{align*} %\label{eq: Delta_x} 
\Delta_e(t,\ell)  \equiv&  \var(w_{t+\ell}|c+\ell) - \var(w_{t}|c)  \\
% & =  \left[\mu_{t+\ell}^2 - \mu_t^2\right] \var(\theta_t|c) + \mu_{t+\ell}^2\left[\var(\theta_{t+\ell}|c+\ell) - \var(\theta_t|c)\right]+ \left[\var(\eps_{t+\ell}|c+\ell) - \var(\eps_t|c)\right] \\
   =&  \left(\mu_{t+\ell}^2 - \mu_t^2\right) \left[\var(\theta_{c}|c) + \sum\limits_{\tau=c+1}^t \var(\Delta\theta_{\tau}|c)\right]\\
   &+ \mu_{t+\ell}^2 \left[\var(\theta_{c+\ell}|c+\ell) + \! \sum\limits_{\tau=c+\ell+1}^{t+\ell} \var(\Delta\theta_\tau|c+\ell) - \var(\theta_c|c) - \! \sum\limits_{\tau=c+1}^{t} \var(\Delta\theta_\tau|c) \right] \\
   & + \left[\var(\eps_{t+\ell}|c+\ell) - \var(\eps_t|c)\right].
\end{align*}

\normalsize

To simplify the comparison between cohort- and experienced-based growth in residual inequality, assume that shocks only depend on time and not cohort/experience: $\var(\eps_t|c)=\var(\eps_t)$ %\sigma_{\eps_t}^2$ 
and $\var(\Delta\theta_t|c) = \var(\Delta\theta_t)$ %\sigma_{\Delta\theta_t}^2 
 for all $c,t$.
In this case, we have (for $c+\ell<t$):
\begin{align}
\Delta_c(t,\ell)  = & \left(\mu_{t+\ell}^2 - \mu_t^2\right) \left[\var(\theta_c|c) + \sum\limits_{\tau=c+1}^{t} \var(\Delta\theta_\tau) \right] + \mu_{t+\ell}^2\sum\limits_{\tau=t+1}^{t+\ell} \var(\Delta\theta_\tau) + \left[\var(\eps_{t+\ell}) - \var(\eps_t)\right], \label{eq: Delta_c simple} 
%\Delta_e(t,k)  =&\left[\mu_{t+\ell}^2 - \mu_t^2\right] \left[\var(\theta_c|c) + \sum\limits_{\tau=c+1}^{t} \var(\Delta\theta_\tau) \right] + \mu_{t+\ell}^2 \left[\var(\theta_{c+\ell}|c+\ell) - \var(\theta_c|c) + \sum\limits_{\tau=t+1}^{t+\ell} \var(\Delta\theta_\tau) - \sum\limits_{\tau=c+1}^{c+\ell} \var(\Delta\theta_\tau) \right] \notag \\
% & + \left[\sigma_{\eps_{t+\ell}}^2 - \sigma_{\eps_t}^2\right] \notag \\
%    = & \Delta_c(t,k) + \mu_{t+\ell}^2 \left[\var(\theta_{c+\ell}|c+\ell) - \var(\theta_c|c) \right]  - \mu_{t+\ell}^2\sum\limits_{\tau=c+1}^{c+\ell} \sigma^2_{\Delta\theta_\tau} \notag \\
%   = & \Delta_c(t,k) + \mu_{t+\ell}^2 \left[\var(\theta_{c+\ell}|c+\ell) - \var(\theta_{c+\ell}|c) \right] \label{eq: Delta_x Delta_c}
\end{align}
and
%\small
\begin{align*}
&    \Delta_e(t,\ell)-\Delta_c(t,\ell)\\
    =&\mu_{t+\ell}^2 \left[\var(\theta_{c+\ell}|c+\ell) - \var(\theta_c|c) + \! \sum\limits_{\tau=c+\ell+1}^{t+\ell} \var(\Delta\theta_\tau) - \! \sum\limits_{\tau=c+1}^{t} \var(\Delta\theta_\tau)-\sum\limits_{\tau=t+1}^{t+\ell} \var(\Delta\theta_\tau) \right]\\
    =&\mu_{t+\ell}^2 \left[\var(\theta_{c+\ell}|c+\ell) - \var(\theta_c|c) -\sum\limits_{\tau=c+1}^{c+\ell} \var(\Delta\theta_\tau) \right]   \\
    =&\mu_{t+\ell}^2 \left[\var(\theta_{c+\ell}|c+\ell) - \var(\theta_{c+\ell}|c)\right]        
\end{align*}

As discussed in \cite{juhn_murphy_pierce_1993}, equation \eqref{eq: Delta_c simple} shows that the change in variances over time for a given cohort incorporates both time effects and experience effects.  The experience effects are reflected in the accumulation of permanent skill shocks from $t+1$ to $t+\ell$ (second term), while the time effects reflect changes in skill returns (first term) \textit{and} in non-skill transitory shocks (third term).  The evolution of variances over time for a given experience group includes the same three effects plus a fourth reflecting the difference the variance of skill levels between the cohorts as of the later time period. This is important, since it suggests that similar time patterns for residual variances obtained by following cohorts or experience groupings (i.e.,\ $\Delta_e(t,\ell)\approx \Delta_c(t,\ell)$) implies that there is little variation across cohorts in early skill levels (i.e.,\ $\var(\theta_{c+\ell}|c+\ell) \approx \var(\theta_{c+\ell}|c)$).  This would be the case if the variance of initial skill levels were identical across cohorts ($\var(\theta_{c+\ell}|c+\ell) =\var(\theta_{c}|c)$) and if there were no early skill shocks over the first $\ell+1$ years of working careers.  Alternatively, growth in the variance of initial skills across cohorts could offset growth in the variance of skills accumulated via labor market experience.

In the absence of initial cohort differences and early career skill shocks, changes in the variance of residuals should be the same over time whether we follow cohorts or experience groups: $\Delta_e(t,\ell)= \Delta_c(t,\ell)$.  Put another way, we should observe similar growth over time in the variance when following cohorts or experience groups regardless of whether that growth is due to an increase in skill returns (i.e.,\ the first term in equation \eqref{eq: Delta_c simple}), the existence of skill growth shocks (i.e.,\ the second term in equation \eqref{eq: Delta_c simple}), or growth in transitory non-skill shocks (i.e.,\ the third term in equation \eqref{eq: Delta_c simple}).  Thus, comparing growth in the variance of residuals for given cohorts or experience groups (as in \cite{juhn_murphy_pierce_1993}) is not directly informative about changes in the returns to skill unless there are no skill shocks and the variance of non-skill shocks is time-invariant. Stated differently, $\Delta_e(t,\ell)= \Delta_c(t,\ell)>0$ is consistent with growth in skill returns, permanent skill shocks, or growth in the variance of non-skill shocks.

Finally, if cohorts are initially identical (i.e.,\ $\var(\theta_c|c)$ does not depend on $c$) and shocks depend only on time, then the variance of residuals will grow less (or decrease more) over time when following an experience group than when following a cohort if skill growth shocks are important early in the lifecycle (i.e.,\ $\var(\Delta\theta_\tau)>0$ for $\tau=c+1,...,c+\ell$).

\iffalse
\color{blue}

\subsubsection{Simple Case}
Write the skill level as the sum of initial skill at the labor market entry and accumulated skill changes:
\begin{align*}
  \theta_{i,t}=\psi_i+\sum_{j=0}^{t-c-1}\Delta\theta_{i,t-j}.
\end{align*}
Suppose that the skill variance can be written as the sum of cohort and experience effects:
\begin{align*}
\var(\theta_t|c)=\underbrace{\var(\psi|c)}_{\text{cohort}}+\underbrace{\chi(t-c)\var(\Delta\theta)}_{\text{experience}}.
\end{align*}
Then,
\begin{align*}
\var(w_t|c)=\underbrace{\mu_t^2}_{\text{time}}\times\big[\underbrace{\var(\psi|c)}_{\text{cohort}}+\underbrace{\chi(t-c)\var(\Delta\theta)}_{\text{experience}}\big]+\var(\eps_t|c),
\end{align*}
Assuming $\var(\eps_t|c)$ does not depend on time, experience, or cohort (or depends only on time or cohort), we have
\begin{align*}
  &  \big[\underbrace{\var(w_{t+\ell}|c+\ell)-\var(w_{t}|c+\ell)}_{\text{change within cohort}}\big]-  \big[\underbrace{\var(w_{t+\ell}|c+\ell)-\var(w_{t}|c)}_{\text{change within experience}}\big]\\
=&\mu_{t}^2\times\Big\{\underbrace{\big[\chi(t-c)-\chi(t-c-k)\big]\var(\Delta\theta)}_{\text{change across experience}}-\underbrace{[\var(\psi|c+\ell)-\var(\psi|c)]}_{\text{change across cohort}}\Big\}.
\end{align*}
The fact that this equals zero does not imply time-invariant distributions of skill or of the non-skill component, even in this simple case.

\color{black}
\fi

\subsection{\cite{lemieux_2006}} \label{app: Lemieux}

We use data from the Health and Retirement Study (HRS), described in Appendix \ref{app: HRS}, to test \cite{lemieux_2006}'s assumption. We first residualize the word recall scores by regressing them on indicators of race, education, experience, and birth year. Next, we regress the squared residuals of test scores on the indicators of race, education, experience, and calendar year, and jointly test whether the estimated coefficients on year indicators are identical (or jointly equal to zero excluding the base year 1996).

Table \ref{tab: lemieux test} reports the $p$-values of the Wald tests conducted on the full sample, and college- and non-college subsamples. Since all $p$-values are smaller than 0.05, we reject the hypothesis that the variance of unobserved skill stays constant over time at the 5\% significance level. 
%Figure \ref{fig: lemieux test} plots estimated coefficients on year indicators along with 95\% confidence intervals, showing that  many of the estimated coefficients on year indicators are individually significant.

\begin{table}[h]
  \centering
\caption{Wald test $p$-values}\label{tab: lemieux test}
\begin{tabular}{cccc}

\toprule
Variables & All Men  & Non-College & College \\
  \midrule
Year &   0.0000 & 0.0003 & 0.0000 \\
\bottomrule
\end{tabular}
\end{table}

\if0
\begin{figure}[h]
    \centering
    \subfloat[All Men]{
      \includegraphics[width=0.45\columnwidth]
      {./figures/HRS_Lemieux/tscore_year}    
    }\\
    \subfloat[Non-College]{
      \includegraphics[width=0.45\columnwidth]{./figures/HRS_Lemieux/tscore_year_sector0}
}\quad
    \subfloat[College]{
      \includegraphics[width=0.45\columnwidth]{./figures/HRS_Lemieux/tscore_year_sector1}
}
\caption{Estimated coefficients (solid lines) on year dummies %from Regression of $T_{i,t}^2$ on race, education, experience, and
  %year dummies,
 with 95\% confidence intervals (dashed lines) (base year: 1996)}\label{fig: lemieux test}
\end{figure}
\fi

\subsection{\cite{castex_dechter_2014}} \label{app: CD}

We consider regression log wage residuals in period $t+\ell$ on lagged test score residuals, $\tilde{T}_{j,t}$, where we explicitly allow for different test measurements denoted by $j$. This yields
\[
 \hat{\beta}_{j,t,t+\ell}\parrow  \frac{\cov(w_{t+\ell},\tilde{T}_{j,t})}{\var(\tilde{T}_{j,t})}
  = \frac{\mu_{t+\ell}}{\tau_j} \underbrace{\left[1+\frac{ \cov(\theta_{t+\ell}-\theta_{t},\theta_{t})}{\var(\theta_{t})} \right]}_{\text{Skill Dynamics } (SD_{t,t+\ell})} \underbrace{\left[\frac{\tau_j^2\var(\theta_{t})}{\tau_j^2\var(\theta_{t})+\var(\eta_{j,t})} \right]}_{\text{Test Reliability Ratio } (R_{j,t})},
\]
The ratio of these estimators (using the same test measurement $j$) for two different cohorts observed in years $t$ and $t'$, respectively, yields the following:
\[
 \frac{\hat{\beta}_{j,t,t+\ell}}{\hat{\beta}_{j,t',t'+\ell}}
 \parrow   \frac{\mu_{t+\ell}}{\mu_{t'+\ell}} \left[\frac{SD_{t,t+\ell}}{SD_{t',t'+\ell}} \right] \left[\frac{R_{j,t}}{R_{j,t'}} \right]
\]
Growth in skill returns is biased when skill dynamics or the reliability of measurements vary across cohorts.  If test measurement error is time-invariant, i.e., $\var(\eta_{j,t})=\sigma^2_j$, then the reliability ratios will differ if and only if the variance of skills differs across the cohorts (i.e., $\var(\theta_t) \neq \var(\theta_{t'})$).

\paragraph{Using different test measurements across cohorts.}
Notice that the ratio of estimators for two different cohorts in $t$ and $t'$ using different measurements $j$ and $j'$ yields the following:
\[
 \frac{\hat{\beta}_{j,t,t+\ell}}{\hat{\beta}_{j',t',t'+\ell}}
 \parrow   \frac{\mu_{t+\ell}}{\mu_{t'+\ell}} \left[\frac{SD_{t,t+\ell}}{SD_{t',t'+\ell}} \right] \left[\frac{R_{j,t}}{R_{j,t'}} \right] \left[\frac{\tau_{j'}}{\tau_j}\right] \left[\frac{R_{j,t'}}{R_{j',t'}} \right],
\]
where additional bias arises due to differences in the test score measurement quality as determined by $\tau_j/\tau_{j'}$ and the reliability ratio, $R_{j,t'}/R_{j',t'}$.

\paragraph{Re-scaling different test measurements across cohorts.}
\cite{deming_2017} scales measurements by their standard deviations, $\sigma_{\tilde{T}_j} = \sqrt{\tau_j^2\var(\theta_{t})+\var(\eta_{j,t})}$, before regressing log wage residuals on test score residuals. Denote these regression coefficients as
\begin{eqnarray*}
 \tilde{\beta}_{j,t,t+\ell} &\parrow &  \frac{\cov(w_{t+\ell},\tilde{T}_{j,t}/\sigma_{\tilde{T}_j})}{\var(\tilde{T}_{j,t}/\sigma_{\tilde{T}_j})} \\
  &=& \frac{\mu_{t+\ell}}{\tau_j} \ SD_{t,t+\ell} R_{j,t} \sigma_{\tilde{T}_j} \\
  &=& \frac{\mu_{t+\ell}}{\tau_j} \left[\frac{ \cov(\theta_{t+\ell},\theta_{t})}{\var(\theta_{t})} \right] \left[\frac{\tau_j^2\var(\theta_{t})}{\tau_j^2\var(\theta_{t})+\var(\eta_{j,t})} \right]  \sqrt{\tau_j^2\var(\theta_{t})+\var(\eta_{j,t})} \\
    &=& \frac{\mu_{t+\ell}}{\tau_j} \left[\frac{ \cov(\theta_{t+\ell},\theta_{t})}{\var(\theta_{t})} \right] \left[\frac{\tau_j^2\var(\theta_{t})}{\sqrt{\tau_j^2\var(\theta_{t})+\var(\eta_{j,t})} } \right]   \\
    &=&  \mu_{t+\ell}\, \tau_j \left[\frac{ \cov(\theta_{t+\ell},\theta_{t})}{\sqrt{\tau_j^2\var(\theta_{t})+\var(\eta_{j,t})} } \right].   \\
\end{eqnarray*}
Clearly, this re-scaling of measurements will not help eliminate any biases for $\mu_{t+\ell}/\mu_{t'+\ell}$  when taking the ratio $\tilde{\beta}_{j,t,t+\ell}/\tilde{\beta}_{j',t',t'+\ell}$.

\paragraph{No measurement error.}
In the absence of measurement error in test scores, we have
\[
\hat{\beta}_{j,t,t+\ell} \parrow    \frac{\mu_{t+\ell}}{\tau_j} \ SD_{t,t+\ell}
\qquad \text{and} \qquad
 \tilde{\beta}_{j,t,t+\ell} \parrow  \mu_{t+\ell} \ SD_{t,t+\ell} \, \sqrt{\var(\theta_t)},
\]
which implies
\[
\frac{\hat{\beta}_{j,t,t+\ell}}{\hat{\beta}_{j',t',t'+\ell}} \parrow \frac{\mu_{t+\ell}}{\mu_{t'+\ell}} \left[\frac{\tau_{j'}}{\tau_j}\right]\left[\frac{SD_{t,t+\ell}}{SD_{t',t'+\ell}} \right]
\qquad \text{and} \qquad
 \frac{\tilde{\beta}_{j,t,t+\ell}}{\tilde{\beta}_{j',t',t'+\ell}} \parrow \frac{\mu_{t+\ell}}{\mu_{t'+\ell}}\left[\frac{SD_{t,t+\ell}}{SD_{t',t'+\ell}} \right] \left[\frac{\sqrt{\var(\theta_t)}}{\sqrt{\var(\theta_{t'})}} \right].
\]
This highlights that re-scaling does not help in addressing the challenge that without the same measure over time, it is impossible to sort out changes in skill variation across cohorts from the ``strength" of skill measurements used for the different cohorts.

%%%%%%%%%%%%%%%%%%%%%%%%%%%%%%%%%%%%%%%%%%%%%%%%%%%%%%%%%%%%%%%%%%%%%%%%%%%%%%%%%%
%%%% Re-Number Tables and figures for the Appendix %%%%%%%%
\renewcommand{\thetable}{\thesection-\arabic{table}}
\setcounter{table}{0}

\renewcommand{\thefigure}{\thesection-\arabic{figure}}
\setcounter{figure}{0}

%%%%%%%%%%%%%%%%%%%%%%%%%%%%%%%%%%%%%%%%%%%%%%%%%%%%%%%%%%%%%%%%%%%%%%%%%%%%%%%%%%
\section{Early Occupational Experiences in NLSY} \label{app: NLSY occ}

We use the data provided by \cite{castex_dechter_2014} to calculate the fraction of years each individual in the NLSY79 and NLSY97 worked in different occupations over ages 17--26.  We restrict the sample to male respondents who took the Armed Forces Vocational Aptitude Battery (ASVAB) tests between ages 16 and 17.5, so the skill measurements are comparable.  The AFQT test is based on four ASVAB subtests: arithemtic reasoning, mathematics knowledge, word knowledge, and paragraph comprehension.  Occupations are coded based on the current (or most recent) job at the time of each interview.

\begin{figure}[h]
  \centering
  \includegraphics{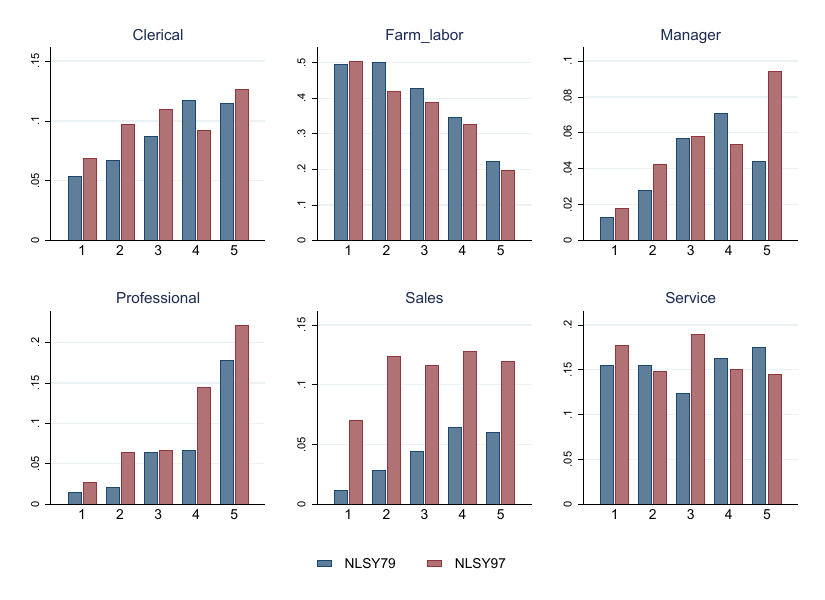}
\caption{Average fraction of years (over ages 17--26) spent working in occupations by AFQT quintile}
    \label{fig: occ exp afqt nlsy}
\end{figure}

Table~\ref{tab: occ exp nlsy} reports the average fraction of years individuals reported working in 6 different occupation categories over ages 17--26.  Figure~\ref{fig: occ exp afqt nlsy} reports the fraction of years in different occupations separately by AFQT quintile.

\begin{table}[h]
  \centering
\caption{Average fraction of years (over ages 17--26) spent working in occupations}
\label{tab: occ exp nlsy}
\begin{tabular}{lcccccc}
\toprule
NLSY Cohort &  Clerical &  Farm labor &  Manager  &  Professional &  Sales & Service  \\
  \midrule
NLSY79 &   0.088    & 0.401  &   0.043    & 0.068  &   0.042  &    0.154 \\
NLSY97 & 0.099  &    0.369  &   0.053   &   0.105 &     0.112   &   0.162 \\
\bottomrule
\multicolumn{7}{l}{Notes: Sample sizes are 1,200 in NLSY79 and 1,007 in NLSY97.}
\end{tabular}
\end{table}

\clearpage
%%%%%%%%%%%%%%%%%%%%%%%%%%%%%%%%%%%%%%%%%%%%%%%%%%%%%%%%%%%%%%%%%%%%%%%%%%%%%%%%%%
%%%% Re-Number Tables and figures for the Appendix %%%%%%%%
\renewcommand{\thetable}{\thesection-\arabic{table}}
\setcounter{table}{0}

\renewcommand{\thefigure}{\thesection-\arabic{figure}}
\setcounter{figure}{0}

%%%%%%%%%%%%%%%%%%%%%%%%%%%%%%%%%%%%%%%%%%%%%%%%%%%%%%%%%%%%%%%%%%%%%%%%%%%%%%%%%%

\section{Identification Results}

\subsection{Identifying skill returns and distributions of skill and non-skill shocks by cohort} \label{app: ident skill distn eps}

In addition to establishing identification for $\mu_t$ over time, the following proposition establishes identification of the variances of unobserved skills, skill growth innovations, and non-skill transitory shocks when these are all allowed to vary by cohort.
To achieve the results, we refine Assumption \ref{assum: mu and skill dist ident} to ensure its applicability to each cohort, $c$. For the completeness of the arguments, we present the revised condition.

\begin{assbis}{assum: mu and skill dist ident}\label{assum: mu and skill dist ident 2}
Let $c$ be a cohort denoting the year of labor market entry. 
For all cohorts, $c$: (i) $\cov(\Delta \theta_t, \theta_{t'}|c)=0$ for all $t-t'\geq 1$;
(ii) $\cov(\theta_t, \eps_{t'}|c) = 0$ for all $t,t'$; and
(iii) for known $k\geq 1$, $\cov(\eps_t,\eps_{t'}|c)=0$ for all $t-t' \geq k$.
\end{assbis}

\begin{prop}\label{prop: identification}
Suppose $\overline{t}-\ul{t} \geq  2k$ for some $k\ge 1$, and Assumption~\ref{assum: mu and skill dist ident 2} holds. Then,
(i) $\mu_t$ is identified for all $t\geq \ul{t} + k$ up to a normalization $\mu_{t^*}=1$ for some period $t^*\geq \ul{t} + k$,
(ii) $\var(\theta_t|c)$ and $\var(\varepsilon_t|c)$ are identified for all $(c,t)$ such that $\ul{t}+k\leq t\leq \ol{t}-k$ and cohort $c$ is observed both in period $t$ and some later period
$t'\geq t+k$, and
(iii) $\var(\Delta \theta_t|c)$ is identified for all $(c,t)$ such that $\var(\theta_t|c)$ and $\var(\theta_{t-1}|c)$ are identified.
\end{prop}

\begin{proof}  
%\tcr{[This ``proof'' is mostly copied and pasted from the text, so it must be modified to incorporate cohorts!]}
%\tcb{[YS notes: This proof should be cleaned up. Different wordings. Maybe more formal proof.]}
\textbf{(i) Identification of $\mu_t$.} Without loss of generality, let ${t^*}=\ul{t} + k$ and $t'=\ul{t}$ so that $t^*-t'\ge k$ and $\mu_{t^*}=\mu_{k+\ul{t}}=1$. We first proceed with the following derivation: 
\begin{align}
    \cov(w_{t^*},w_{t'}|c) & = \mu_{t^*}\mu_{t'} \cov(\theta_{t^*},\theta_{t'}|c) & \text{[Assum \ref{assum: mu and skill dist ident 2} (ii)--(iii)]} \nonumber \\
                         & = \mu_{t^*}\mu_{t'} \cov(\theta_{t'}+\Delta \theta_{t'+1} + \cdots +\Delta \theta_{t^*}, \theta_{t'}|c) \nonumber \\
                         & = \mu_{t^*}\mu_{t'} \var(\theta_{t'}|c) & \text{[Assum \ref{assum: mu and skill dist ident 2} (i)].} \label{eq:cov_w's}
\end{align}
Using the IV estimation formula and the normalization $\mu_{t^*}=1$, we identify $\mu_{t^*+1}$:
\begin{align*}
    \frac{\cov(w_{t^*+1}, w_{t'}|c)}{\cov(w_{t^*},w_{t'}|c)} 
    & = \frac{\mu_{t^*+1} \mu_{t'} \var(\theta_{t'}|c)} {\mu_{t^*} \mu_{t'} \var(\theta_{t'}|c)} \\
    & = \frac{\mu_{t^*+1} } {\mu_{t^*} } \\
    & = \mu_{t^*+1}.
\end{align*}
We identify $\mu_{\ul{t} + k},\ldots,\mu_{\overline{t}}$ by applying the above arguments recursively. 
\newline
\textbf{(ii) Identification of $\var(\theta_t|c)$ and $\var(\varepsilon_t|c)$.} For any $(t,t')$ such that $t'-t \ge k$, we can rearrange equation \eqref{eq:cov_w's} and get the following expression: 
\begin{align*}
    \var(\theta_{t}|c) = \frac{\cov(w_{t'},w_{t}|c)}{\mu_{t'}\mu_{t}}.
\end{align*}
Since $\mu_t$ is now known for $t\ge \ul{t} + k$, we can identify $\var(\theta_t|c)$ for $t= \ul{t} + k,\ldots, \overline{t}-k$ by varying $(t,t') \in \{(\ul{t} + k, \ul{t} + 2k),\ldots, ( \overline{t}-k, \overline{t}) \}$. For the same time periods, $\var(\varepsilon_t|c)$ is identified using:
\begin{align*}
   \var(\varepsilon_t|c) = \var(w_t|c) - \mu_t^2 \var(\theta_t|c). 
\end{align*}
\textbf{(iii) Identification of $\var(\Delta \theta_t|c)$.} Assumption~\ref{assum: mu and skill dist ident 2}(i) implies that 
\begin{align*}
    \var(\Delta \theta_t|c) = \var(\theta_t|c) - \var(\theta_{t-1}|c). 
\end{align*}Therefore, $\var(\Delta \theta_t|c)$ is identified for $t=\ul{t} + k+1,\ldots,\overline{t}-k$ since $\var(\theta_t|c)$ is already determined in the previous step. 

\begin{comment}
Condition (i) of Assumption~\ref{assum: mu and skill dist ident} implies that $\var(\theta_t) = \var(\theta_{t-1}) + \var(\Delta \theta_t)$.  Together with conditions (ii) and (iii), it implies that  $\cov(w_t,w_{t'}) = \mu_t\mu_{t'}\var(\theta_{t'})$ for $t-t'\geq k$. As in Proposition~\ref{prop: gen mu ident}, the IV estimator of equation~\eqref{eq: IV estimator} can be used to identify $\mu_{k+1},...,\mu_T$ (with one normalization).  Moreover, for $T \geq 2k+1$:
\[
\var(\theta_{t'}) = \frac{\cov(w_t,w_{t'})}{\mu_t\mu_{t'}}
\]
is identified for all $t'=k+1,...,T-k$.  The variance of unobserved skills can be identified for all but the first and last $k$ periods.  Intuitively, this variance is not identified for earlier periods (without additional assumptions), because it cannot be separated from skill returns --- only $\mu_t^2 \! \var(\theta_t)$ can be identified for the first $k$ periods.  The unobserved skill variance cannot be identified for later periods, because it is impossible to distinguish between the roles of unobserved skills and transitory non-skill shocks without observing (sufficiently distant) future wages.

Having identified the variance of unobserved skills over time, it
is straightforward to then identify variation in skill growth from $\var(\Delta \theta_t)=\var(\theta_t)-\var(\theta_{t-1})$ for $t=k+2,...,T-k$ and the variance of non-skill shocks, $\var(\varepsilon_t) = \var(w_t)-\mu_t^2 \var(\theta_t)$ for $t=k+1,...,T-k$.
\end{comment}

\end{proof}

\subsection{Identifying early skill returns, $\mu_t$} \label{app: identify early mu}

The bias for $\Delta \mu_t/\mu_{t-1}$ when using future residuals as instruments (equation~\eqref{eq: iv future}) presents the key identification challenge for $\mu_t$ in early sample periods.
Here, we show that the identification results can be extended to earlier years by utilizing additional cohort information. 
\begin{propbis}{prop: identification}\label{prop: identification 2}
    Suppose that Assumption~\ref{assum: mu and skill dist ident 2} holds for two cohorts $c$ and $\tilde{c}$. Furthermore, the following two conditions hold:  (a) $\var(\theta_{t-1}|c)\neq \var(\theta_{t-1}|\tilde{c})$ and (b) $\var(\Delta\theta_t|c)=\var(\Delta\theta_t|\tilde{c})$. 
    Then,
    (i) %$\mu_t$ is identified for all $\ul{t} \le t\le \overline{t} -k$ up to a normalization $\mu_{t^*}=1$ for some period $ t^*\geq \ul{t} + k$
    $\mu_t$ is identified for all $t$ up to a normalization for some period $t^*$,
    (ii) $\var(\theta_t|c)$ and $\var(\varepsilon_t|c)$ are identified for all $(c,t)$ such that %cohort $c$ is observed and $\ul{t} \le t \le \overline{t} -k$ 
    $t\leq\ol{t}-k$ and cohort $c$ is observed both in period $t$ and some later period $t'\geq t+k$, and
    (iii) $\var(\Delta \theta_t|c)$ is identified for %observed $c$ and $ \ul{t} + 1 \le t \le \overline{t}-k$ 
    for all (c,t) such that $\var(\theta_t|c)$ and $\var(\theta_{t-1}|c)$ are identified. 
\end{propbis}
\begin{proof} 
%For any $t$ and some $k\ge1$, pick $t'$ such that $t'\ge t+k$. Then, we have.
Assumption \ref{assum: mu and skill dist ident 2} and conditions (a) and (b) imply that, for $t'\geq t+k$,
    \begin{align} \label{eq: early mu ident}
\frac{\cov(w_{t}, w_{t'}|c)-\cov(w_{t},w_{t'}|\tilde{c})}{\cov(w_{t-1}, w_{t'}|c)-\cov(w_{t-1},w_{t'}|\tilde{c})}
 =\frac{\mu_t\mu_{t'}\big[\var(\theta_{t-1}|c)-\var(\theta_{t-1}|\tilde{c})\big]}{\mu_{t-1}\mu_{t'}\big[\var(\theta_{t-1}|c)-\var(\theta_{t-1}|\tilde{c})\big]} 
 =\frac{\mu_{t}}{\mu_{t-1}},
\end{align}
which identifies $\mu_t/\mu_{t-1}$ for all $t\leq \ol{t}-k$.
We combine the identification result of $\mu_t$ in Proposition \ref{prop: identification} with this, and the recursive arguments establish the desired results. 
The remaining identification results follow directly from the previous results in  Proposition \ref{prop: identification}.
\end{proof}

%if Assumption~\ref{assum: mu and skill dist ident} holds and two cohorts exist, $c$ and $\tilde{c}$, such that (a) $\var(\theta_{t-1}|c)\neq \var(\theta_{t-1}|\tilde{c})$ and (b) $\var(\Delta\theta_t|c)=\var(\Delta\theta_t|\tilde{c})$, then $\mu_t/\mu_{t-1}$ can be identified for early years by differencing out biases across cohorts:

\begin{comment}
This ensures identification for all $\mu_t$, satisfying $t \leq T-k$.  Coupled with identification of all $\mu_t$ for $t\geq k+1$ (given normalization for some $t$) from Proposition~\ref{prop: gen mu ident}, if $T\geq 2k+1$, then we can  identify $\var(\theta_t|c)$ and $\var(\eps_t|c)$ for all $t \leq T-k$, and $\var(\Delta\theta_t|c)$ for all $2 \leq t \leq T-k$ (for cohorts observed in the relevant periods--see Proposition~\ref{prop: identification}).
\end{comment}
Condition (a) on cohorts is likely to hold quite generally.  For example, differences in the variance of initial skill levels would
contribute to different variances later in life.  Even if initial skill distributions were identical across cohorts,
the older cohort is likely to have accumulated more skill growth innovations over its longer career. Condition (b) holds when the skill growth variance depends only on time (and not experience) or
when there is a non-monotonic experience trend in the variance of skill changes. For example, young workers may experience
greater variation in skill growth than middle age workers due to differences in training or learning opportunities, while older workers
may have a greater variance in skill changes due to differences in health shocks or skill obsolescence.  Indeed,
\cite{baker_solon_2003} and \cite{blundell_graber_mogstad_2015} estimate a U-shaped age profile
for the variance of earnings shocks.  %In what follows, we presume that $\mu_t$ can be identified for all $\ul{t} \le t \le \bt - k$.

\subsection{Identification when skills have a permanent and AR(1) component} \label{app: FE AR(1) skills}

Here, we consider the case of Section~\ref{sec: FE AR(1) skills} in which skills are characterized by a permanent component $\psi_i$ and persistent component $\phi_{i,t}$ that follows an AR(1) process:
\begin{align*}
  \theta_{i,t}&= \psi_i+\phi_{i,t},\\
  \phi_{i,t}&= \rho_t \phi_{i,t-1}+\nu_{i,t},
\end{align*}
where we exclude the possibility of $\rho_t=1$ because Assumption \ref{assum: gen mu ident}(i) holds in that case.

For $t' \leq t - k -1$, Assumption~\ref{assum: AR(1) skills ident} implies the following:
\[
\frac{\cov(\Delta w_t,w_{t'})}{\cov(w_{t-1},w_{t'})} =
\frac{\Delta \mu_{t}}{\mu_{t-1}} + \frac{\mu_{t}}{\mu_{t-1}} \left[\frac{(\rho_t-1)\hat{\rho}_{t'+1,t-1}\var(\phi_{t'})}
    {\var(\psi) + \hat{\rho}_{t'+1,t-1}\var(\phi_{t'})} \right],
\]
where $ \hat{\rho}_{t,t'} \equiv \prod_{j=t}^{t'} \rho_j$. Clearly, the IV estimator is not consistent when $\rho_t\neq 1$. For example, when $\var(\psi)=0$, the term in brackets simplifies to $\rho_t-1$, which implies that the IV estimator converges to $\rho_t\mu_t/\mu_{t-1} -1$. However, identification of skill returns over time for the case $\rho_t\neq 1$ is still feasible  as long as $\var(\psi)>0$.

%Clearly, for $\var(\psi)=0$, the term in brackets simplifies to $\rho_t-1$, which implies that the IV estimator converges to $\rho_t\mu_t/\mu_{t-1} -1$.  Thus, when $\rho_t\neq 1$, our simple IV estimator will be inconsistent; however, identification of skill returns over time is still feasible as long as $\var(\psi)>0$.

%We show identification when Assumption~\ref{assum: AR(1) skills ident} holds. for each cohort $c$ and $\tilde{c}$.
%\tcb{[Minor changes. Assumption 3 should hold conditional on cohort, which is slightly stronger than the unconditional version.]}
%Let $c_i$ be the year when individual $i$ starts working and his wage is observed. We assume $\phi_{i,t}=0$ for all $t< c_i$ (zero for all pre-observation periods). Although $\phi_{i,c_i}$ cannot be distinguished from $\psi_{i}$ when $\rho_t=1$ for all $t$, we exclude this possibility for now (because Assumption \ref{assum: gen mu ident}(i) holds in that case) and do not impose the condition $\phi_{i,c_i}=0$.

To show identification, it is convenient to re-write the log wage equation as follows:
\begin{align*}
  w_{i,t}&=\mu_t \psi_i+\tilde{\phi}_{i,t}+\varepsilon_{i,t},\\
  \tilde{\phi}_{i,t}&=\tilde{\rho}_t\tilde{\phi}_{i,t-1}+\tilde{\nu}_{i,t},
\end{align*}
where $\tilde{\phi}_{i,t}\equiv\mu_t\phi_{i,t}$, $\tilde{\rho}_t\equiv\rho_t \mu_t/\mu_{t-1}$, and $\tilde{\nu}_{i,t}\equiv\mu_t\nu_{i,t}$.

Notice that Assumption \ref{assum: AR(1) skills ident} and the AR(1) process modeled in Equation~\eqref{eq: theta AR(1)} imply the following orthogonality conditions in terms of the transformed variables:
\begin{assbis}{assum: AR(1) skills ident}\label{assum: identification_of_rho}
For all cohorts, c: 
(i) $\cov(\psi,\tphi_{t}|c)=0$ for all $t$; (ii) $\cov(\psi,\eps_{t'}|c)=\cov(\tphi_{t},\eps_{t'}|c)=0$ for all $t,t'$;
(iii) $\cov(\tphi_{t'}, \tnu_{t}|c)=\cov( \tnu_{t'},\tnu_t|c)=0$ for all $t-t'\ge 1$;
(iv) for known $k\ge 1$, $\cov(\eps_t,\eps_{t'}|c)=0$ for all $t-t' \ge k$.
\end{assbis}

\paragraph{Identification of $\tilde{\rho}_t$.}
Under Assumption \ref{assum: identification_of_rho}, we can construct the following moment condition: for $t'\leq t-k-1$,
\begin{align}\label{eq: quasi-differenced cov}
  \cov(w_{t'},w_{t}|c)-\tilde{\rho}_t\cov(w_{t'},w_{t-1}|c)
=\mu_{t'} (\mu_{t}-\tilde{\rho}_t\mu_{t-1})\var(\psi|c).
%=\mu_{t'} \mu_{t}(1-\tcr{\rho_t})\var(\psi|c).
\end{align}
Suppose that there exist two cohorts $c$ and $\tilde{c}$ such that $\var(\psi|c)>0$ and $\var(\psi|\tilde{c})>0$. Taking the ratio of \eqref{eq: quasi-differenced cov} for cohort $c$ relative to $\tilde{c}$ yields
\[
  \frac{\cov(w_{t'},w_{t}|c)-\tilde{\rho}_t\cov(w_{t'},w_{t-1}|c)  }{  \cov(w_{t'},w_{t}|\tilde{c})-\tilde{\rho}_t\cov(w_{t'},w_{t-1}|\tilde{c})}=\frac{\var(\psi|c)}{\var(\psi|\tilde{c})}.
\]
Similarly, for $t''\leq t-k-1$,
\[
  \frac{\cov(w_{t''},w_{t}|c)-\tilde{\rho}_t\cov(w_{t''},w_{t-1}|c)  }{  \cov(w_{t''},w_{t}|\tilde{c})-\tilde{\rho}_t\cov(w_{t''},w_{t-1}|\tilde{c})}=\frac{\var(\psi|c)}{\var(\psi|\tilde{c})}
\]
Combining these two equations yields
\begin{align*}
  \frac{\cov(w_{t'},w_{t}|c)-\tilde{\rho}_t\cov(w_{t'},w_{t-1}|c)  }{
  \cov(w_{t'},w_{t}|\tilde{c})-\tilde{\rho}_t\cov(w_{t'},w_{t-1}|\tilde{c})}=
  \frac{\cov(w_{t''},w_{t}|c)-\tilde{\rho}_t\cov(w_{t''},w_{t-1}|c)  }{
  \cov(w_{t''},w_{t}|\tilde{c})-\tilde{\rho}_t\cov(w_{t''},w_{t-1}|\tilde{c})},
\end{align*}
which becomes
\begin{align}
A \trho_t^2 + B \trho_t + C = 0,\label{eq: quadratic}
\end{align}
where
\begin{align*}
  A=&\cov(w_{t'},w_{t-1}|c)\cov(w_{t''},w_{t-1}|\tilde{c})-\cov(w_{t''},w_{t-1}|c)\cov(w_{t'},w_{t-1}|\tilde{c}),\\
  B=&\cov(w_{t''},w_{t}|{c}) \cov(w_{t'},w_{t-1}|\tilde{c})+\cov(w_{t''},w_{t-1}|{c}) \cov(w_{t'},w_{t}|\tilde{c})\\
    &-\cov(w_{t'},w_{t}|c)\cov(w_{t''},w_{t-1}|\tilde{c})-\cov(w_{t'},w_{t-1}|c) \cov(w_{t''},w_{t}|\tilde{c}),\\
  C=&\cov(w_{t'},w_{t}|c) \cov(w_{t''},w_{t}|\tilde{c})-\cov(w_{t''},w_{t}|c)\cov(w_{t'},w_{t}|\tilde{c}).
\end{align*}
We can investigate some cases that equation \eqref{eq: quadratic}  has a unique solution. First, if $A=0$ and $B\neq 0$, then the unique solution is 
\begin{align*}
    \trho_t = - \frac{C}{B}.
\end{align*}
Second, if $A\neq0$ and $B^2-4AC=0$, then the unique solutions becomes
\begin{align*}
    \trho_t = - \frac{B}{2A}.
\end{align*}
Notice that there exist other set of sufficient conditions, especially by constructing the additional moment conditions using different cohorts or applying instruments from different time periods.

\paragraph{Identification of $\mu_t$.}
From Equation \eqref{eq: quasi-differenced cov}, we have, for $t\leq t'-k-1$,
\begin{align}  \frac{\cov(w_{t},w_{t'}|c)-\tilde{\rho}_{t'}\cov(w_{t},w_{t'-1}|c)}{\cov(w_{t-1},w_{t'}|c)-\tilde{\rho}_{t'}\cov(w_{t-1},w_{t'-1}|c)}
=\frac{\mu_{t}}{\mu_{t-1}}.\label{eq: quasi-differenced cov ratio 1}
\end{align}
Because $\tilde{\rho}_t$ is identified for all $t\geq \ul{t} + k+1$, $\mu_{t}/\mu_{t-1}$ is identified for all $t\leq \ol{t}-k-1$.

Equation \eqref{eq: quasi-differenced cov} also implies that, for $t'\leq t-k-2$,
\begin{align}  \frac{\cov(w_{t'},w_{t}|c)-\tilde{\rho}_t\cov(w_{t'},w_{t-1}|c)}{\cov(w_{t'},w_{t-1}|c)-\tilde{\rho}_{t-1}\cov(w_{t'},w_{t-2}|c)}
=\frac{\mu_t-\trho_t\mu_{t-1}}{\mu_{t-1}-\trho_{t-1}\mu_{t-2}}=\frac{\mu_{t-1}}{\mu_{t-2}}\frac{\left(\frac{\mu_t}{\mu_{t-1}}-\trho_t\right)}{\left(\frac{\mu_{t-1}}{\mu_{t-2}}-\trho_{t-1}\right)}.\label{eq: quasi-differenced cov ratio 2}
\end{align}
Because $\tilde{\rho}_t$ is identified for all $t\geq \ul{t} + k+1$ and $\mu_{t}/\mu_{t-1}$ is identified for all $t\leq \ol{t}-k-1$ based on Equation \eqref{eq: quasi-differenced cov ratio 1}, $\mu_{t}/\mu_{t-1}$ for $t\geq \ol{t}-k$ is also identified from Equation \eqref{eq: quasi-differenced cov ratio 2} as long as $\ol{t}-k-1\geq \ul{t}+k+1$. Therefore,  $\mu_t$ is identified for all $t$ (up to a normalization $\mu_{t^*}=1$) if $\ol{t}-\ul{t}\geq 2(k+1)$.
%\ys{[YS: changed from $\ol{t} \geq 2k+4$.]}
%\tcr{[Is it strange that we can identify all $\mu_t$ here while we claim not to identify early $\mu_t$ in baseline case? YS: This result is based on stronger assumptions (conditional on cohort and the unique solution requirements in the previous page) and the identification arguments are different from the simple IV approach of the baseline model. So, it looks okay to me.]} \tcg{I think we need the condition $\rho_t\neq 1$ for later years because the RHS of \eqref{eq: quasi-differenced cov} is zero when $\rho_t=1$. [YS: Yes, $\rho_t=1$ goes back to the random walk case,  which is identified by the simple IV estimator. I added $\rho_t \neq 1$ when we define the model. DELETE green comments if everything looks okay.]}

\paragraph{Identification of $\rho_t$.}
$\rho_t=\trho_t\mu_{t-1}/\mu_t$ is identified for $t \ge \ul{t}+ k+1$ because $\tilde{\rho}_{t}$ is identified for $t \ge \ul{t}+ k+1$ and $\mu_{t}$ is identified for all $t$.

\paragraph{Identification of $\var(\psi|c)$.}
Equation \eqref{eq: quasi-differenced cov} implies $\var(\psi|c)$ is identified for all cohorts observed in $t'$, $t-1$, and $t$ such that $t'\leq t-k-1$.

\paragraph{Identification of $\var(\phi_t|c)$.}
For $t'\geq t+k$, 
\begin{align*}
  \cov(w_t,w_{t'}|c)=\mu_t\mu_{t'}\left[\var(\psi|c)+\left(\sum_{j=t+1}^{t'} \rho_j\right)   \var(\phi_t|c)\right].
\end{align*}
Since $\rho_t$ is identified for $t \ge \ul{t}+ k+1$, $\var(\phi_t|c)$ is identified for $t\geq \ul{t}+k$.

\medskip

Finally, we note that it is straightforward to identify the covariance structure for $\eps_t$ from ``close'' autocovariances given identification of everything else.

\subsection{Identification with $\varepsilon_{i,t} \sim \text{ARMA}(1,q)$ } \label{app: ARMA ident}

We demonstrate identification for the model in Sections~\ref{subsec: returns identification} generalized so that the transitory component $\eps_{i,t}$ include an ARMA(1,$q$) process $\varphi_{i,t}$:
%$\varepsilon_{i,t}=\xi_{i,t}$ for $t=c_i+1$ and, for $t>c_i+1$,
\begin{align}
  \varepsilon_{i,t} & = \varphi_{i,t} + \teps_{i,t} \\
  \varphi_{i,t} & = \rho_t \varphi_{i,t-1} + \sum_{j=0}^{\min\{q,t-c_i-1\}} \beta_j \nu_{i,t-j},
\end{align}
where $\beta_0=1$. Similar to Assumption \ref{assum: inc dynamics lit}, we assume that $\nu_{i,t}$ is an \emph{i.i.d.}~innovation term, not correlated with any other variables including $\nu_{i,t'}$ for $t'\neq t$ and $\varphi_{i,t'}$ for $t-t'\ge 1$. Without loss of generality, we assume that $k\ge q$. Otherwise, we can redefine $k'=\max\{k,q\}$ and use $k'$ instead of $k$.

%We assume conditions (i) and (ii) of Assumption~\ref{assum: mu and skill dist ident} are met for each cohort, so skill growth is uncorrelated with past skill levels and skills are uncorrelated with non-skill shocks.

\paragraph{Identification of $\rho_t$.}
%\ys{YS: Identification of $\rho_t$ is re-written.}

For %$(c,t,t')$ such that $c<t \leq t'-k-1$,
$t'\geq t+k+1$, we have
%\ys{Changed from $c< t \leq t'+k$? It should be $t'-t \ge k+1$.}
\begin{align*}
\cov(\varepsilon_{t}, \varepsilon_{t'}-\rho_{t'}\varepsilon_{t'-1}|c)=\cov\left(\varphi_{t}+\tilde{\eps}_t, \sum_{j=0}^{\min\{q,t-c-1\}}\beta_j \nu_{t'-j}  + (\teps_{t'} -\rho_{t'} \teps_{t'-1})  \Big|c\right)=0, 
\end{align*}
and therefore
\begin{align}
  \cov(w_t,w_{t'}|c)-\rho_{t'}\cov(w_t,w_{t'-1}|c)
%=\cov(w_t,w_{t'}-\rho_{t'} w_{t'-1}|c)
% =&\cov(\mu_t\theta_t,\mu_{t'}\theta_{t'}-\rho
%    \mu_{t'-1}\theta_{t'-1}|c)\\
=\mu_t (\mu_{t'}-\rho_{t'}\mu_{t'-1})\var(\theta_t|c).\label{eq: quasi-differenced cov 2}
\end{align}
Taking the ratio of \eqref{eq: quasi-differenced cov 2} for cohort $c$ relative to $\tilde{c}$ yields
\[
  \frac{\cov(w_t,w_{t'}|c)-\rho_{t'}\cov(w_t,w_{t'-1}|c)  }{  \cov(w_t,w_{t'}|\tilde{c})-\rho_{t'}\cov(w_t,w_{t'-1}|\tilde{c})}=\frac{\var(\theta_t|c)}{\var(\theta_t|\tilde{c})}.
\]
Similarly, for $t'' \geq t+k+1$,
%\ys{Changed from $t \leq t''+k$.}
\[
  \frac{\cov(w_t,w_{t''}|c)-\rho_{t''}\cov(w_t,w_{t''-1}|c)  }{  \cov(w_t,w_{t''}|\tilde{c})-\rho_{t''}\cov(w_t,w_{t''-1}|\tilde{c})}=\frac{\var(\theta_t|c)}{\var(\theta_t|\tilde{c})}.
\]
Combining these two equations yields
\begin{align}
  \frac{\cov(w_t,w_{t'}|c)-\rho_{t'}\cov(w_t,w_{t'-1}|c)  }{
  \cov(w_t,w_{t'}|\tilde{c})-\rho_{t'}\cov(w_t,w_{t'-1}|\tilde{c})}=
  \frac{\cov(w_t,w_{t''}|c)-\rho_{t''}\cov(w_t,w_{t''-1}|c)  }{
  \cov(w_t,w_{t''}|\tilde{c})-\rho_{t''}\cov(w_t,w_{t''-1}|\tilde{c})}.\label{eq:
  rho equation}
\end{align}
Equation \eqref{eq: rho equation} can be written as
\begin{align}
  A_1 \rho_{t'} \rho_{t''} + B_1\rho_{t'} + C_1 \rho_{t''} + D_1 =0,\label{eq: bilinear1}
\end{align}
where
\begin{align*}
  A_1=&\cov(w_t,w_{t'-1}|c)\cov(w_t,w_{t''-1}|\tilde{c})-\cov(w_t,w_{t''-1}|c)\cov(w_t,w_{t'-1}|\tilde{c}),\\
  B_1=&\cov(w_t,w_{t'-1}|\tilde{c}) \cov(w_t,w_{t''}|c)-\cov(w_t,w_{t'-1}|c) \cov(w_t,w_{t''}|\tilde{c}),\\
  C_1=&\cov(w_t,w_{t'}|\tilde{c}) \cov(w_t,w_{t''-1}|c)-\cov(w_t,w_{t'}|c)\cov(w_t,w_{t''-1}|\tilde{c}),\\
  D_1=&\cov(w_t,w_{t'}|c) \cov(w_t,w_{t''}|\tilde{c})-\cov(w_t,w_{t'}|\tilde{c})\cov(w_t,w_{t''}|c).
\end{align*}
By changing $t$, $c$, or $\tilde{c}$, we can also construct an equation
\begin{align}
  A_2 \rho_{t'} \rho_{t''} + B_2 \rho_{t'} + C_2 \rho_{t''} + D_2 =0,\label{eq: bilinear2}    
\end{align}
where $A_2, B_2, C_2,$ and $D_2$ are defined in a similar way.  

We investigate some cases that this system of equations \eqref{eq: bilinear1}--\eqref{eq: bilinear2} has a unique solution. When $A_1=A_2=0$, it becomes a system of linear equations. If $B_1C_2-B_2C_1\neq 0$, it has a unique solution
\begin{align*}
\begin{pmatrix}
\rho_{t'} \\
\rho_{t''}
\end{pmatrix}
= - 
\begin{pmatrix}
B_{1} & C_{1} \\
B_{2} & C_{2}
\end{pmatrix}^{-1}
\begin{pmatrix}
D_1 \\
D_2
\end{pmatrix}.
\end{align*}
When $A_1\neq 0$ and $A_2 \neq 0$, it becomes a set of rectangular hyperbolas. We first rearrange the equations:
\begin{align*}
    \left(\rho_{t'} + \frac{C_1}{A_1}\right) \left(\rho_{t''} + \frac{B_1}{A_1}\right) = \frac{B_1C_1-A_1D_1}{A_1^2}\\
    \left(\rho_{t'} + \frac{C_2}{A_2}\right) \left(\rho_{t''} + \frac{B_2}{A_2}\right) = \frac{B_2C_2-A_2D_2}{A_2^2}.
\end{align*}
If the constants on the right hand side have different signs and the graphs share only one asymptote, it always has a unique solution. Therefore, we conclude that $(\rho_{t'},\rho_{t''})$ are jointly identified when (i) $(B_1C_1-A_1D_1)(B_2C_2-A_2D_2)<0$; and (ii-1) $C_1/A_1 = C_2/A_2$ and $(B_1/A_1)\neq (B_2/A_2)$ or (ii-2) $C_1/A_1 \neq C_2/A_2$ and $(B_1/A_1) = (B_2/A_2)$. Once we identify $\rho_{t'}$ for some $t'$, we can recover $\rho_{t^{\dagger}}$ recursively by constructing equation \eqref{eq: bilinear1} with $t^{\dagger}$ instead of $t''$ or with different cohort if necessary. Then, we can identify $\rho_{t}$ for all $t\geq  \ul{t} + k+1$. Notice that these conditions are sufficient for identification. Additional equations generated by varying $t,c,$ or $\tilde{c}$ can provide a different set of sufficient conditions for identification.  
\begin{comment}
If $A=0$ or $B^2-4AC=0$ holds for some $(c,\tilde{c},t,t',t'')$ such that $c<t$, $\tilde{c}<t$, $t'-t\geq k$, and $t''-t\geq k$, then $\rho$ is identified from the unique solution to \eqref{eq: quadratic}. Otherwise, there may be two real solutions (if $B^2-4AC>0$).

When $A\neq 0$ and $B^2-4AC>0$ holds for other available $(c,\tilde{c},t,t',t'')$ such that $c<t$, $\tilde{c}<t$, $t'-t\geq k$, and $t''-t\geq k$, the autocorrelation parameter $\rho$ can be uniquely identified by combining at least two distinct quadratic equations \eqref{eq: quadratic}. To see this, re-write \eqref{eq: quadratic} for two different sets of $(c,\tilde{c},t,t',t'')$:
\begin{align}
  \rho^2+\frac{B_1}{A_1}\rho+\frac{C_1}{A_1}=&0, \label{eq: quadratic 1}\\
  \rho^2+\frac{B_2}{A_2}\rho+\frac{C_2}{A_2}=&0. \label{eq: quadratic 2}
\end{align}
If $A_1\neq 0$, $B_1^2-4A_1 C_1>0$, $A_2\neq 0$, and $B_2^2-4A_2 C_2>0$, then $\rho$ is identified from the common solution to \eqref{eq: quadratic 1} and \eqref{eq: quadratic 2}:
\[
  \rho=\left(\frac{B_1}{A_1}-\frac{B_2}{A_2}\right)^{-1}\left(\frac{C_2}{A_2}-\frac{C_1}{A_1}\right)=\frac{A_1 C_2-A_2 C_1}{A_2 B_1-A_1 B_2}
\]
as long as $B_1/A_1\neq B_2/A_2$.  
\end{comment}

\paragraph{Identification of $\mu_t$.}
For $t'\geq t+k+1$, suppose that there exists $(c,\tilde{c})$ such that
$\var(\theta_{t-1}|c)\neq\var(\theta_{t-1}|\tilde{c})$ and
$\var(\Delta\theta_t|c)=\var(\Delta\theta_t|\tilde{c})$. Then, from Equation \eqref{eq: quasi-differenced cov 2}, we have
\begin{align}
&\frac{  \big[\cov(w_t,w_{t'}|c)-\rho_{t'}\cov(w_t,w_{t'-1}|c)\big]-\big[\cov(w_t,w_{t'}|\tilde{c})-\rho_{t'}\cov(w_t,w_{t'-1}|\tilde{c})\big]}{  \big[\cov(w_{t-1},w_{t'}|c)-\rho_{t'}\cov(w_{t-1},w_{t'-1}|c)\big]-\big[\cov(w_{t-1},w_{t'}|\tilde{c})-\rho_{t'}\cov(w_{t-1},w_{t'-1}|\tilde{c})\big]
} \nonumber \\
=&\frac{\mu_t
   (\mu_{t'}-\rho_{t'}\mu_{t'-1})\big[\var(\theta_{t-1}|c)-\var(\theta_{t-1}|\tilde{c})\big]}{\mu_{t-1}
   (\mu_{t'}-\rho_{t'}\mu_{t'-1})\big[\var(\theta_{t-1}|c)-\var(\theta_{t-1}|\tilde{c})\big]} \nonumber \\
=&\frac{\mu_t}{\mu_{t-1}}. \label{eq: mu iden 1 for ARMA}
\end{align}
Since $\rho_t$ is identified for all $t\geq \ul{t}+k+1$, $\mu_t/\mu_{t-1}$ is identified for all $t\leq \ol{t}-k-1$.

Equation \eqref{eq: quasi-differenced cov 2} also implies that, for $t' \leq t-k-2$, 
\begin{align}
 \frac{ \cov(w_{t'},w_{t}|c)-\rho_t\cov(w_{t'},w_{t-1}|c)}{\cov(w_{t'},w_{t-1}|c)-\rho_{t-1}\cov(w_{t'},w_{t-2}|c)}
%=\frac{\mu_{t'} (\mu_{t}-\rho_t\mu_{t-1})\var(\theta_{t'}|c)}{\mu_{t'}
%  (\mu_{t-1}-\rho_{t-1}\mu_{t-2})\var(\theta_{t'}|c)}
=\frac{\mu_t-\rho_{t}\mu_{t-1}}{\mu_{t-1}-\rho_{t-1}\mu_{t-2}}=\frac{\mu_{t-1}}{\mu_{t-2}}\frac{\left(\frac{\mu_t}{\mu_{t-1}}-\rho_t\right)}{\left(\frac{\mu_{t-1}}{\mu_{t-2}}-\rho_{t-1}\right)} \label{eq: mu iden 2 for ARMA}
\end{align}
Because $\rho_t$ is identified for all $t\geq \ul{t} + k+1$ and $\mu_{t}/\mu_{t-1}$ is identified for all $t\leq \ol{t}-k-1$ based on Equation \eqref{eq: mu iden 1 for ARMA}, $\mu_{t}/\mu_{t-1}$ for $t\geq \ol{t}-k$ is also identified from Equation \eqref{eq: mu iden 2 for ARMA} as long as $\ol{t}-k-1\geq \ul{t}+k+1$. Therefore,  $\mu_t$ is identified for all $t$ (up to a normalization $\mu_{t^*}=1$) if $\ol{t}-\ul{t}\geq 2(k+1)$.

%Using Equations \eqref{eq: mu iden 1 for ARMA}--\eqref{eq: mu iden 2 for ARMA}, we can identify $\mu_t$ for all $t \ge \ul{t} + (k+1)$. Notice that we need the variance conditions for $\theta_t$ only for one time period given the normalization. 

\paragraph{Identification of $\var(\theta_t|c)$.}
For $t'\geq t+k+1$, Equation \eqref{eq: quasi-differenced cov 2}
 implies
 \[
\var(\theta_t|c)= \frac{ \cov(w_t,w_{t'}|c)-\rho_{t'}\cov(w_t,w_{t'-1}|c)}{\mu_t (\mu_{t'}-\rho_{t'}\mu_{t'-1})}.
\]
Because $\rho_t$ is identified for all $t\geq \ul{t} + k+1$ and $\mu_t$ is identified for all $t$, $\var(\theta_t|c)$ is identified for all $t\leq \ol{t}-k-1$. %$\ul{t} + (k+1) \le t\leq \ol{t}-(k+1)$.

\medskip

Finally, we note that it is straightforward to identify $\{\beta_j \}$ and $\var(\nu_t|c)$ from ``close'' autocovariances given identification of everything else.

\begin{comment}
\color{red}
Should we drop the following identification of non-skill shock process (beyond $\rho$)?

\paragraph{Identification of $\beta_j$}
First, note that the ARMA(1,$q$) process can be written as an
MA($t-c+1$) process:
\[
\varepsilon_{i,t}=  \sum_{j=0}^{t-c_i-1}\tilde{\beta}_j\xi_{i,t-j},
\]
where $\tilde{\beta}_j=1$ for $j=0$,
$\tilde{\beta}_j=\rho\tilde{\beta}_{j-1}+\beta_j$ for $1\leq j\leq q$, and
$\tilde{\beta}_j=\rho\tilde{\beta}_{j-1}$ for $j>q$.

Then, for $t=c+1$ and $j\geq 0$,
\begin{align*}
  \frac{\cov(\varepsilon_t,\varepsilon_{t+j}|c)}{\var(\varepsilon_t|c)}=\tilde{\beta}_j  \frac{ \var(\xi_{t}|c)}{\var(\xi_t|c)}=\tilde{\beta}_j.
\end{align*}
Therefore, $\tilde{\beta}_j$'s are identified from cohort-specific
autocovariances of $\varepsilon_t$, which can be obtained from
\[
  \cov(\varepsilon_t,\varepsilon_{t+j}|c)=\cov(w_t,w_{t+j}|c)-\mu_t\mu_{t+j}\var(\theta_t|c).
\]
Given $\rho$ and $\tilde{\beta}_j$, $\beta_j$ for $1\leq j\leq q$ is identified from
$ \beta_j=\tilde{\beta}_j-\rho\tilde{\beta}_{j-1}$.

\paragraph{Identification of $\var(\xi_t|c)$.}
For the initial period $t=c+1$ for cohort $c$,
$\var(\xi_t|c)=\var(\varepsilon_t|c)$. For $c+1<t\leq\ol{t}-k$,
\[
  \var(\xi_t|c)=\var(\varepsilon_{t}|c)-\sum_{j=1}^{\min\{q,t-c-1\}}\tilde{\beta}_j^2\var(\xi_{t-j}|c).
\]

\end{comment}

\subsection{Identification with Heterogeneous Skill Growth Rates} \label{app: HIP ident}

We now demonstrate identification for the model in Section~\ref{subsect: HIP} with systematic heterogeneity in lifecycle skill growth.

Letting $\psi_i$ reflect the initial skill for an individual entering the labor market, the skill growth process \eqref{eq: HIP skill growth} implies that the level of unobserved skill for individual $i$ from cohort $c_i$ in year $t$ can be written as
%\begin{equation} \label{app eq: HIP skills}
%  \theta_{i,t}=\psi_i + \sum_{j=0}^{t-c_i-1}\big[ \lambda_{t-j}(c_i) \delta_i +\nu_{i,t-j}\big].
%\end{equation}
\begin{equation} \label{eq: HIP skills}
  \theta_{i,t}
=\psi_i + \Lambda_t(c_i) \delta_i + \sum_{j=0}^{t-c_i-1} \nu_{i,t-j},
\end{equation}
where $\Lambda_t(c)\equiv \sum^{t-c-1}_{j=0}\lambda_{t-j}(c)$ reflects the accumulated influence of skill growth heterogeneity.

To facilitate an identification analysis, assume that idiosyncratic skill growth shocks $\nu_{i,t}$ are serially uncorrelated and uncorrelated with initial skills and systematic skill growth; however, we make no assumptions about the correlation between heterogeneous skill growth rates and initial skill levels. Consistent with the literature estimating HIP models, we strengthen conditions (ii) and (iii) of Assumption~\ref{assum: gen mu ident} to assume that non-skill shocks are uncorrelated with all skill-related components. Formally, we assume the following, explicitly conditioning on cohorts.
\begin{ass}  \label{assum: HIP skills}
For all cohorts, $c$:
(i) $\cov(\psi,\nu_{t}|c)=\cov(\delta,\nu_{t}|c)=0$ for all $t$;
(ii) $\cov(\psi,\eps_{t'}|c)=\cov(\delta,\eps_{t'}|c)=\cov(\nu_t, \eps_{t'}|c) = 0$ for all $t,t'$;
(iii) for known $k\geq 1$, $\cov(\eps_t,\eps_{t'}|c)=0$ for all $t-t' \geq k$. 
\end{ass}

Assumption~\ref{assum: HIP skills} implies that the covariance between skills in periods $t$ and $t'<t$ can be written as
\begin{align*}
\cov(\theta_{t'},\theta_{t}|c)  %&=  \var(\theta_{t'}|c) + [\Lambda_t(c)- \Lambda_{t'}(c)]\cov(\theta_{t'},\delta|c)  \\
  =&  \var(\psi|c) + \Lambda_{t'}(c)\Lambda_t(c) \var(\delta|c) + [\Lambda_{t'}(c) + \Lambda_t(c)]\cov(\psi,\delta|c) + \sum_{j=0}^{t'-c-1} \var(\nu_{t-j}|c).
\end{align*}

In addition to Assumption \ref{assum: HIP skills}, we assume that there exists $\ol{e}$ such that $\lambda_t(c)=0$ for $e=t-c\geq \ol{e}$.

\paragraph{Identification of $\mu_t$.}
 $\mu_t$ can be identified based on experienced workers with $\lambda_t(c)=0$ for which Propositions \ref{prop: identification}
 and \ref{prop: identification 2} can be applied. First, $\mu_t/\mu_{t-1}$ for $t\geq \ul{t} + (k+1)$ is identified if
there exists some cohort $c$ such that (i) the cohort has experience $e=t-c \geq \ol{e}$ in year $t$ and (ii) the
cohort is observed in years $t'\leq t-k-1$, $t-1$, and $t$.  Moreover, $\mu_t/\mu_{t-1}$ for $t\leq k+1$ is identified if
there exist two cohorts $c$ and $\tilde{c}$ such that (i) both cohorts
have experience of at least $\ol{e}$ in year $t$,
(ii) both cohorts are observed in years $t-1$, $t$, and some year
$t'\geq t+k$, and (iii)
$\var(\theta_{t-1}|c)\neq \var(\theta_{t-1}|\tilde{c})$ and
$\var(\nu_t|c)=\var(\nu_t|\tilde{c})$.

% Suppose that we have already identified $\mu_t$ (from old workers
% $e\geq \ol{e}$), so consider cohorts with $e=t-c<\ol{e}$.
\paragraph{Identification of $\lambda_t(c)$.}
By dividing the residual by $\mu_t$, we get
\[
  \frac{w_{i,t}}{\mu_t}= \theta_{i,t}+\frac{\varepsilon_{i,t}}{\mu_t},
\]
and its first difference is
\[
  \Delta\left(\frac{w_{i,t}}{\mu_t}\right)
=\Delta\theta_{i,t}+\Delta\left(\frac{\varepsilon_{i,t}}{\mu_t}\right)
=\lambda_t(c_i)\delta_i+\nu_{i,t}+\Delta\left(\frac{\varepsilon_{i,t}}{\mu_t}\right).
\]

For $|t'-t|\geq k+1$, 
 $\cov(\Delta\varepsilon_t,\Delta\varepsilon_{t'}|c)=0$ and 
\begin{align*}
  \cov\left(  \Delta\left(\frac{w_{t}}{\mu_t}\right),
  \Delta\left(\frac{w_{t'}}{\mu_{t'}}\right)\Big|c\right)=&\cov(\Delta
  \theta_t,\Delta \theta_{t'}|c)=\lambda_t(c) \lambda_{t'}(c)\var(\delta|c).
\end{align*}
Therefore, we can identify changes in $\lambda_t(c)$:
\[
  \frac{\lambda_{t}(c)}{\lambda_{t-1}(c)}=\frac{\cov\left(  \Delta\left(\frac{w_{t}}{\mu_t}\right),
  \Delta\left(\frac{w_{t'}}{\mu_{t'}}\right)|c\right)}{\cov\left(  \Delta\left(\frac{w_{t-1}}{\mu_{t-1}}\right),
  \Delta\left(\frac{w_{t'}}{\mu_{t'}}\right)|c\right)}, \qquad \forall (t,t') \text{ such that }t-t'\geq k+2 \text{ or } t'-t\geq k+1.
\]
Normalizing $\lambda_{t^*(c)}(c)=1$ for some $t^*(c)$, all $\lambda_t(c)$'s can be identified for which the covariances in (scaled) residual wage changes are observed.

\paragraph{Identification of $\var(\delta|c)$.}
Once $\lambda_t(c)$'s have been identified, $\var(\delta|c)$ is identified from
\begin{align*}
\var(\delta|c)= \frac{ \cov\left(  \Delta\left(\frac{w_{t}}{\mu_{t}}\right),
  \Delta\left(\frac{w_{t'}}{\mu_{t'}}\right)|c\right)}{\lambda_{t}(c) \lambda_{t'}(c)}.
\end{align*}

\paragraph{Identification of $\cov(\psi,\delta|c)$.}
For $t'-t\geq k+1$,
$\cov(\varepsilon_t,\Delta\varepsilon_{t'}|c)=0$ and 
\begin{align*}
  \cov\left(  \frac{w_{t}}{\mu_t},
  \Delta\left(\frac{w_{t'}}{\mu_{t'}}\right)\Big|c\right)
=\cov(\theta_t,\Delta \theta_{t'}|c)
=\lambda_{t'}(c)\cov(\theta_t,\delta|c),
% =&\lambda_{t'}(c)\cov\left(\psi+\sum_{j=0}^{t-c-1}\tau_{t-j}(c)\delta,\delta\Big|c\right)\\
% =&\lambda_{t'}(c)\left\{\cov(\psi,\delta|c) +\var(\delta|c) \sum_{j=0}^{t-c-1}\lambda_{t-j}(c) \right\}.
\end{align*}
where
\begin{equation}
\cov(\theta_{t},\delta|c)=
\cov(\psi,\delta|c) +\var(\delta|c)
  \sum_{j=0}^{t-c-1}\lambda_{t-j}(c).\label{eq: cov of theta and delta}
\end{equation}
Therefore,
\[
\cov(\psi,\delta|c)=
\frac{\cov\left(  \frac{w_{t}}{\mu_t},
  \Delta\left(\frac{w_{t'}}{\mu_{t'}}\right)|c\right)}{\lambda_{t'}(c)}  -\var(\delta|c) \sum_{j=0}^{t-c-1}\lambda_{t-j}(c) .
\]

\paragraph{Identification of $\var(\theta_t|c)$.}
For $t'-t\geq k$, write
\[
\theta_{i,t'}% =\theta_{i,t}+\sum_{\tau=t+1}^{t'}\big[\lambda_{\tau}(c)\delta_i
  % +\tilde{\nu}_{i,\tau}\big]
% =\theta_{i,t}+\sum_{j=1}^{t'-t}\big[\lambda_{t+j}(c)\delta_i
%   +\tilde{\nu}_{i,t+j}\big]
=\theta_{i,t}+\sum_{j=0}^{t'-t-1}\big[\lambda_{t'-j}(c_i)\delta_i  +\nu_{i,t'-j}\big].
\]
Then,
\[
  \cov\left(  \frac{w_{t}}{\mu_{t}},
  \frac{w_{t'}}{\mu_{t'}}\Big|c\right)
=\cov(  \theta_{t}, \theta_{t'}|c)
=\var(\theta_t|c)+\cov(\theta_{t},\delta|c)\sum_{j=0}^{t'-t-1}\lambda_{t'-j}(c). % \\
% =&\var(\theta_t|c)+
% \left\{\cov(\psi,\delta|c) +\var(\delta|c) \sum_{j=0}^{t-c-1}\lambda_{t-j}(c) \right\}\left\{\sum_{j=0}^{t'-t-1}\tau_{t'-j}(c)\right\}
\]
Therefore,
\[
\var(\theta_t|c)=
  \cov\left(  \frac{w_{t}}{\mu_{t}},
  \frac{w_{t'}}{\mu_{t'}}\Big|c\right) -
\cov(\theta_t,\delta|c)\sum_{j=0}^{t'-t-1}\lambda_{t'-j}(c).
\]

\paragraph{Identification of $\var(\nu_t|c)$.}

Note that
\[
  \var(\theta_{t}|c)=\var(\theta_{t-1}|c)+\var(\delta|c)\lambda_{t}(c)^2+2\cov(\theta_{t-1},\delta|c)\lambda_{t}(c)+\var(\nu_{t}|c),
\]
from which $\var(\nu_{t}|c)$ is identified once all the other components have been identified.

\medskip

Finally, we note that it is straightforward to identify $\{\beta_j \}$ and $\var(\xi_t|c)$ from ``close'' autocovariances given identification of everything else.

\if0
\paragraph{Identification of $\eps_t \sim MA(q)$ process.}

Having already identified skill returns $\mu_t$, as well as cohort- and time-specific
variances of skills, $\var(\theta_t|c) = \cov(w_t,w_{t'}|c)/\mu_t\mu_{t'}$, for most cohorts and years.
Next, $\cov(\varepsilon_t,\varepsilon_{t'}|c)=\cov(w_t,w_{t'}|c)-\mu_t\mu_{t'}\var(\theta_t|c)$
for $t'\geq t$ is identified for $(c,t)$ such that $\var(\theta_t|c)$
is identified. Then, $\beta_j$ is identified from some cohort $c$ and period $t=c+1$ as follows:
\begin{equation*}
\frac{\cov(\varepsilon_{t},\varepsilon_{t+j}|c)}{\var(\varepsilon_t|c)}=
\beta_j\frac{ \var(\xi_t|c)}{\var(\xi_t|c)}=\beta_j.
\end{equation*}
Given $\beta_j$'s, we can recover $\var(\xi_t|c)$ each period by
following cohorts over time.\footnote{Consider a cohort $c$. For the initial
period $t=c+1$, $\var(\xi_t|c)=\var(\varepsilon_t|c)$. For
$t>c+1$,
\[
\var(\xi_{t}|c)=\var(\varepsilon_{t}|c)-\sum^{\min\{q,t-c-1\}}_{j=1}\beta_j^2
\var(\xi_{t-j}|c).
\]
}

\fi

\subsection{Identification with Occupations} \label{app: ident occ}

\paragraph{Identification of $\mu_t^o$.}
With Assumption~\ref{assum: mu ident occ}(iii)--(iv), the long autocovariance for log wage residuals  for $t-t'\geq k+1$ can be written as follows:
\begin{align}
\cov(w_t,w_{t'}|o_t,o_{t-1},o_{t'})
=&\mu_t^{o_t}\big[\mu_{t'}^{o_{t'}}\cov(\theta_t,\theta_{t'}|o_t,o_{t-1},o_{t'})+\cov(\theta_t,\eps_{t'}|o_t,o_{t-1},o_{t'})\big],\label{eq: long cov occ t}\\
\cov(w_{t-1},w_{t'}|o_t,o_{t-1},o_{t'})
=&\mu_{t-1}^{o_{t-1}}\big[\mu_{t'}^{o_{t'}}\cov(\theta_{t-1},\theta_{t'}|o_t,o_{t-1},o_{t'})+\cov(\theta_{t-1},\eps_{t'}|o_t,o_{t-1},o_{t'})\big].\label{eq: long cov occ t-1}
\end{align}
Moreover, Assumption~\ref{assum: mu ident occ}(i)--(ii) imply $\cov(\theta_t,\theta_{t'}|o_t,o_{t-1},o_{t'})=\cov(\theta_{t-1},\theta_{t'}|o_t,o_{t-1},o_{t'})$ and $\cov(\theta_t,\eps_{t'}|o_t,o_{t-1},o_{t'})=\cov(\theta_{t-1},\eps_{t'}|o_t,o_{t-1},o_{t'})$, so equations \eqref{eq: long cov occ t} and \eqref{eq: long cov occ t-1} imply equation \eqref{eq: IV occ}.

\paragraph{IV Estimator without Conditioning on $o_{t'}$.}
Next, we show that $\mu_t^o$ is identified based on covariances conditioned only on $(o_t,o_{t-1})$ when we assume $\E[\Delta\theta_t|o_t,o_{t-1},o_{t'}]=\E[\eps_t|o_t,o_{t-1},o_{t'}]=\E[\eps_{t-1}|o_t,o_{t-1},o_{t'}]=0$ in addition to Assumption \ref{assum: mu ident occ}. Consider the long autocovariance \eqref{eq: long cov occ t} that is not conditioned on $o_{t'}$:
\begin{align}
    \cov(w_t,w_{t'}|o_t,o_{t-1})
    =&\E\big[\cov(w_t,w_{t'}|o_t,o_{t-1},o_{t'})|o_{t},o_{t-1}\big]\nonumber\\
    &+\cov\big(\E[w_t|o_t,o_{t-1},o_{t'}],\E[w_{t'}|o_t,o_{t-1},o_{t'}]|o_t,o_{t-1}\big).\label{eq: long cov occ t avg}
\end{align}
The second term in equation \eqref{eq: long cov occ t avg} is
\begin{align*}
&\cov\big(\E[w_t|o_t,o_{t-1},o_{t'}],\E[w_{t'}|o_t,o_{t-1},o_{t'}]|o_t,o_{t-1}\big)\\
=&\cov\big(\gamma_t^{o_t}+\mu_t^{o_t}\E[\theta_t|o_t,o_{t-1},o_{t'}],\E[w_{t'}|o_t,o_{t-1},o_{t'}]|o_t,o_{t-1}\big)\\
=&\mu_t^{o_t}\cov\big(\E[\theta_t|o_t,o_{t-1},o_{t'}],\E[w_{t'}|o_t,o_{t-1},o_{t'}]|o_t,o_{t-1}\big).
\end{align*}
where we used the additional assumption $\E[\eps_t|o_t,o_{t-1},o_{t'}]=0$.

Thus, the long autocovariances \eqref{eq: long cov occ t} and \eqref{eq: long cov occ t-1} that are not conditioned on $o_{t'}$ are given by
\begin{align*}
\cov(w_t,w_{t'}|o_t,o_{t-1})
=&    \mu_t^{o_t}\Xi^{o_t,o_{t-1}}_{t},\\
\cov(w_{t-1},w_{t'}|o_t,o_{t-1})
=&\mu_{t-1}^{o_{t-1}}\Xi^{o_t,o_{t-1}}_{t-1},
\end{align*}
where
\begin{align*}
\Xi^{o_t,o_{t-1}}_{t}
\equiv&\E\big[\mu_{t'}^{o_{t'}}\cov(\theta_t,\theta_{t'}|o_t,o_{t-1},o_{t'})+\cov(\theta_t,\eps_{t'}|o_t,o_{t-1},o_{t'})|o_t,o_{t-1}\big]\\
&+\cov\big(\E[\theta_t|o_t,o_{t-1},o_{t'}],\E[w_{t'}|o_t,o_{t-1},o_{t'}]|o_t,o_{t-1}\big).
\end{align*}
Assumption~\ref{assum: mu ident occ}(i)--(ii) and $\E[\Delta\theta_t|o_t,o_{t-1},o_{t'}]=0$ imply $\Xi^{o_t,o_{t-1}}_{t}=\Xi^{o_t,o_{t-1}}_{t-1}$, so
\begin{align*}
 \frac{\cov(\Delta w_t,w_{t'}|o_t,o_{t-1})}{\cov(w_{t-1},w_{t'}|o_t,o_{t-1})}=\frac{\mu_t^{o_t}-\mu_{t-1}^{o_{t-1}}}{\mu_{t-1}^{o_{t-1}}}.
\end{align*}
The IV estimator for stayers in an occupation ($o_t=o_{t-1}=o$) identifies growth in returns to skill in that occupation. Moreover, the IV estimator for occupational switchers ($o_{t}\neq o_{t-1}$) identifies the differences in the level of returns to skill across occupations, given a normalization $\mu_{t^*}^{o^*}=1$ for some $(t^*,o^*)$.

\paragraph{Identification of $\gamma_t^o$.}
Given $\mu_t^o$, we show that $\gamma_t^o$ is identified under the  assumptions $\E[\Delta\theta_t|o_t,o_{t-1}]=\E[\eps_t|o_t,o_{t-1}]=\E[\eps_{t-1}|o_t,o_{t-1}]=0$.

Since $\mu_t^o$ is identified, we can use it to scale the average log wage residuals as follows:
\begin{align*}
   \frac{ \E\left[w_t|o_t,o_{t-1}\right]}{\mu^{o_t}_t}=\frac{\gamma_t^{o_t}}{\mu^{o_t}_t}+\E[\theta_t|o_t,o_{t-1}].
\end{align*}

Using $\E[\Delta\theta_t|o_t,o_{t-1}]=0$, the average growth of scaled log wage residual is
\begin{align*}
   \frac{ \E\left[w_t|o_t,o_{t-1}\right]}{\mu^{o_t}_t}-\frac{ \E\left[w_{t-1}|o_t,o_{t-1}\right]}{\mu^{o_{t-1}}_{t-1}}=\frac{\gamma_t^{o_t}}{\mu^{o_t}_t}-\frac{\gamma_{t-1}^{o_{t-1}}}{\mu^{o_{t-1}}_{t-1}}.
\end{align*}
Therefore, with a normalization $\gamma_{t^*}^{o^*}=0$ for some $(t^*,o^*)$, $\gamma_t^{o^*}$ for $t\neq t^*$ is identified based on  stayers in occupation $o^*$. On the other hand, $\gamma_t^{o}$ for $o\neq o^*$ is identified from occupation switchers.

\subsection{Identification with Multiple Skills} \label{app: ident multiple skills}
 
Recall that we define $\ol{\theta}_{i,t} \equiv \sum_j \mu_{j,t}\theta_{i,j,t}$.
Given Assumption~\ref{assum: multi skills}, our IV estimator identifies the following:
\begin{equation*}%\label{eq: IV multi skill}
\frac{\cov(\Delta w_t,w_{t'})}{\cov(w_{t-1},w_{t'})} = \frac{\sum\limits_{j=1}^J \Delta\mu_{j,t} \cov(\theta_{j,t'},\ol{\theta}_{t'})} {\sum\limits_{j'=1}^J \mu_{j',t-1} \cov(\theta_{j',t'},\ol{\theta}_{t'})}, \quad \text{for $t - t' \geq k+1$,}
\end{equation*}
which implies Proposition~\ref{prop: IV multiple skills}.

\begin{proof}[Proof of Proposition~\ref{prop: IV multiple skills}]

We consider skill return growth from period $t_0$ to $t$, where the text assumes $t_0=t-1$.  More generally, we require $t'+k \leq t_0 \leq t-1$. Empirically, we use $t_0=t-2$ given the biennial nature of the PSID in later years. 

Assumption~\ref{assum: multi skills} implies the following:
\begin{eqnarray}
\frac{\cov(w_t - w_{t_0},w_{t'})}{\cov(w_{t_0},w_{t'})} &=& \frac{\sum\limits_{j=1}^J (\mu_{j,t}-\mu_{j,t_0})  \cov(\theta_{j,t'},\ol{\theta}_{t'})} {\sum\limits_{j=1}^J \mu_{j,t_0} \cov(\theta_{j,t'},\ol{\theta}_{t'})} \label{eq: IV multi skill} \\
&=& \frac{\sum\limits_{j=1}^J \left(\frac{\mu_{j,t}-\mu_{j,t_0}}{\mu_{j,t_0}}\right) \mu_{j,t_0}  \cov(\theta_{j,t'},\ol{\theta}_{t'})} {\sum\limits_{j=1}^J \mu_{j,t_0} \cov(\theta_{j,t'},\ol{\theta}_{t'})} \nonumber \\
&=& \sum\limits_{j=1}^J \omega_{j,t',t_0} \left(\frac{\mu_{j,t}-\mu_{j,t_0}}{\tmu_{j,t_0}}\right) \nonumber
\end{eqnarray}
where the weights,
\[
\omega_{j,t',t_0} \equiv  \frac{\cov(\theta_{j,t'},\ol{\theta}_{t'})\mu_{j,t_0}}{\sum\limits_{j'=1}^J \cov(\theta_{j',t'},\ol{\theta}_{t'})\mu_{j',t_0}}.
\]
Notice that $\cov(\theta_{j,t'},\theta_{j',t'})\geq 0, \forall j,j'$, implies that $\cov(\theta_{j,t'},\ol{\theta}_{t'}) \geq 0, \forall j$.
Since $\mu_{j,t}\geq 0, \forall j,t$, the weights $\omega_{j,t',t_0}\geq 0$ whenever $\cov(\theta_{j,t'},\theta_{j',t'})\geq 0, \forall j,j'$. Since the weights sum to one, non-negativity of the weights further implies that none exceeds one.

\end{proof}

We can re-write the weights in terms of linear projections:
\[
\omega_{j,t',t_0}  
= \frac{L(\theta_{j,t'}|\ol{\theta}_{t'})\mu_{j,t_0}}{\sum\limits_{j'=1}^J L(\theta_{j',t'}|\ol{\theta}_{t'})\tmu_{j',t_0}} 
= \frac{L(\theta_{j,t_0}|\ol{\theta}_{t'})\mu_{j,t_0}}{\sum\limits_{j'=1}^J L(\theta_{j',t_0}|\ol{\theta}_{t'})\tmu_{j',t_0}} 
= \frac{L(\mu_{j,t_0}\theta_{j,t_0}|\ol{\theta}_{t'})}{\sum\limits_{j'=1}^J L(\mu_{j',t_0}\theta_{j',t_0}|\ol{\theta}_{t'})} 
\]
where $L(a|b)=\cov(a,b)b/\var(b)$ is the linear projection of $a$ onto $b$.   The second equality follows from condition (i) of Assumption~\ref{assum: multi skills}, which implies that $L(\theta_{j,t}|\ol{\theta}_{t'})=L(\theta_{j,t'}|\ol{\theta}_{t'})$ for all $ t \geq t'$.
Thus, the weight on growth in returns to skill $j$ depends on the (linearly) predicted rewards from skill $j$ in period $t_0$, $\mu_{j,t_0}\theta_{j,t_0}$, given total worker productivity in period $t'$, $\ol{\theta}_{t'}$.

\begin{prop} \label{prop: IV multiple skills2}
If Assumption~\ref{assum: multi skills} holds, then for all $t-t' \geq k+1$, the IV estimator identifies growth in the weighted-average return to skills,  $m_{t,t'} = \sum\limits_{j=1}^J \varphi_{j,t'}\mu_{j,t}$:
\[
\frac{\cov(w_t-w_{t_0},w_{t'})}{\cov(w_{t_0},w_{t'})} = \frac{m_{t,t'} - m_{t_0,t'}}{m_{t_0,t'} }
\]
with weights
\begin{equation} \label{eq: varphi}
\varphi_{j,t'} \equiv \frac{\cov(\theta_{j,t'},\ol{\theta}_{t'})}{\sum\limits_{j'=1}^J\cov(\theta_{j',t'},\ol{\theta}_{t'})},\quad \text{for $j=1,...,J$.}
\end{equation}
If $\cov(\theta_{j,t'},\theta_{j',t'})\geq 0, \forall j,j'$, then the weights $\varphi_{j,t'} \in [0,1], \forall j$.
\end{prop}

\begin{proof}[Proof of Proposition~\ref{prop: IV multiple skills2}]
Using the definitions of $m_{t,t'}$ and $\varphi_{j,t'}$, growth in the weighted-average return to skills can be written as
\begin{align*}
\frac{m_{t,t'} - m_{t_0,t'}}{m_{t_0,t'} }=    \frac{\sum\limits_{j=1}^J \varphi_{j,t'}(\mu_{j,t}-\mu_{j,t_0})}{\sum\limits_{j=1}^J \varphi_{j,t'}\mu_{j,t_0}} 
=\frac{\sum\limits_{j=1}^J \cov(\theta_{j,t'},\ol{\theta}_{t'})(\mu_{j,t}-\mu_{j,t_0})  } {\sum\limits_{j'=1}^J \cov(\theta_{j',t'},\ol{\theta}_{t'})\mu_{j',t_0} }=\frac{\cov(w_t - w_{t_0},w_{t'})}{\cov(w_{t_0},w_{t'})},
\end{align*}
where the last equality reflects equation \eqref{eq: IV multi skill}.

Notice that $\cov(\theta_{j,t'},\theta_{j',t'})\geq 0, \forall j,j'$, implies that $\cov(\theta_{j,t'},\ol{\theta}_{t'}) \geq 0, \forall j$.
So the weights $\varphi_{j,t'}\geq 0$. Since the weights sum to one, non-negativity of the weights further implies that none exceeds one.

\end{proof}

Following the argument above, we can also write these weights in terms of linear projections:
\[
\varphi_{j,t'} \equiv \frac{L(\theta_{j,t'}|\ol{\theta}_{t'})}{\sum\limits_{j'=1}^J L(\theta_{j',t'}|\ol{\theta}_{t'})}.
\]
Condition (i) of Assumption~\ref{assum: multi skills} implies that $L(\theta_{j,t}|\ol{\theta}_{t'})=L(\theta_{j,t'}|\ol{\theta}_{t'}), \forall t \geq t'$, so the weights are proportional to the predicted level of skill $j$ in periods $t\geq t'$ conditional on total worker productivity in period $t'$, $\ol{\theta}_{t'}$.

\subsubsection{Occupations as Bundles of Skills}

We now consider the case in which log wage residuals are given by equation~\eqref{eq: w occ multi skills simple}, where we define $\tilde{\mu}_{j,t}^o \equiv \mu_{j,t}\alpha_{j,t}^o$ and $\ol{\theta}_{i,t}^{o_{i,t}} \equiv \sum_{j=1}^J \tilde{\mu}_{j,t}^{o_{i,t}}\theta_{i,j,t}$. 

Focusing on occupation stayers, we make the following assumption to accommodate multiple skills and occupations.
\begin{ass}  \label{assum: occ multi skills}
 (i) $\cov(\Delta \theta_{j,t},\theta_{j',t'}|o_t=o_{t-1},o_{t'})=0$ for all $j,j'$, and $t-t'\geq 1$; for known $k\geq 1$ and for all $t-t'\geq k+1$:
(ii) $\cov(\theta_{j,t}, \eps_{t'}|o_t=o_{t-1},o_{t'})=\cov(\theta_{j,t-1}, \eps_{t'}|o_t=o_{t-1},o_{t'}) = 0$ for all $j$; 
(iii) $\cov(\eps_{t},\theta_{j,t'}|o_t=o_{t-1},o_{t'})=\cov(\eps_{t-1},\theta_{j,t'} |o_t=o_{t-1},o_{t'}) = 0$ for all $j$; and
(iv) $\cov(\eps_t,\eps_{t'}|o_t=o_{t-1},o_{t'})=\cov(\eps_{t-1},\eps_{t'}|o_t=o_{t-1},o_{t'})=0$.
\end{ass}

\paragraph{IV Estimator Conditional on $o_{t}=o_{t-1}=o$ and $o_{t'}=o'$.}
 Given Assumption~\ref{assum: occ multi skills}, our IV estimator conditional on $o_t=o_{t-1}=o$ and $o_{t'}=o'$ identifies the following for $t-t'\geq k+1$:\footnote{To simplify notation, we let $\cov(x,y|o,o')$ represent $\cov(x,y|o_t=o_{t-1}=o, o_{t'}=o')$, $\cov(x,y|o)$ represent $\cov(x,y|o_t=o_{t-1}=o)$, and $\E[x|o]$ represent $\E[x|o_t=o_{t-1}=o]$.}
\begin{align*}
\frac{\cov( \Delta w_t,w_{t'}|o,o')}{\cov(w_{t-1},w_{t'}|o,o')} 
=& \frac{\sum\limits_{j=1}^J \Delta\tilde{\mu}_{j,t}^o \cov(\theta_{j,t'},\ol{\theta}_{t'}^{o'}|o,o')} {\sum\limits_{j'=1}^J \tilde{\mu}_{j',t-1}^o \cov(\theta_{j',t'},\ol{\theta}_{t'}^{o'}|o,o')}\\
=& \frac{\sum\limits_{j=1}^J \left(\frac{\Delta\tilde{\mu}_{j,t}^o}{\tilde{\mu}_{j,t-1}^o}\right)\tilde{\mu}_{j,t-1}^o \cov(\theta_{j,t'},\ol{\theta}_{t'}^{o'}|o,o')} {\sum\limits_{j'=1}^J \tilde{\mu}_{j',t-1}^o \cov(\theta_{j',t'},\ol{\theta}_{t'}^{o'}|o,o')}\\
=&\sum^J_{j=1}\upsilon_{j,t',t-1}^{o,o'}\left(\frac{\Delta\tilde{\mu}_{j,t}^o}{\tilde{\mu}_{j,t-1}^o}\right)\\
=&\sum^J_{j=1}\upsilon_{j,t',t-1}^{o,o'}\left(\frac{\Delta\mu_{j,t}}{\mu_{j,t-1}}+\frac{\mu_{j,t}}{\mu_{j,t-1}}\frac{\Delta\alpha_{j,t}^o}{\alpha_{j,t-1}^o}\right),
\end{align*}
where
\begin{align*}
\upsilon^{o,o'}_{j,t',t-1}\equiv \frac{\tilde{\mu}_{j,t-1}^o \cov(\theta_{j,t'},\ol{\theta}_{t'}^{o'}|o,o')} {\sum\limits_{j'=1}^J \tilde{\mu}_{j',t-1}^o \cov(\theta_{j',t'},\ol{\theta}_{t'}^{o'}|o,o')}    .
\end{align*}
Therefore, if $\cov(\theta_{j,t'},\ol{\theta}_{t'}^{o'}|o,o')\geq 0$ for all $j$ and $(o,o')$, the IV estimator for all occupational stayers reflects weighted average of the growth rate of skill-specific returns in occupation $o$. Notice that $\cov(\theta_{j,t'},\theta_{j',t'}|o,o')\geq 0, \forall j,j'$, implies that $\cov(\theta_{j,t'},\ol{\theta}_{t'}^{o'}|o,o') \geq 0, \forall j$.

\paragraph{IV Estimator for Stayers in Occupation $o$.}
Next, we show the IV estimator formula based on covariances conditioned only on $o_t=o_{t-1}=o$. 
Consider the long residual autocovariance that is not conditioned on $o_{t'}$:
\begin{align}
    \cov(w_t,w_{t'}|o)
    =\E\big[\cov(w_t,w_{t'}|o,o_{t'})|o\big]
    +\cov\big(\E[w_t|o,o_{t'}],\E[w_{t'}|o,o_{t'}]|o\big).\label{eq: long cov occ multi t avg}
\end{align}
With Assumption~\ref{assum: occ multi skills}, the first term in equation~\eqref{eq: long cov occ multi t avg} is
\begin{align*}
\E\big[\cov(w_t,w_{t'}|o,o_{t'})|o\big]
=&\sum\limits_{j=1}^J\tilde{\mu}_{j,t}^o \E\big[\cov(\theta_{j,t'},\ol{\theta}_{t'}^{o_{t'}}|o,o_{t'})\big|o\big].
\end{align*}
With additional assumptions $\E[\theta_{j,t}-\theta_{j,t'}|o_t,o_{t-1},o_{t'}]=0$ for all $j$ and $\E[\eps_t|o_t,o_{t-1},o_{t'}]=\E[\eps_{t'}|o_t,o_{t-1},o_{t'}]=0$, the second term in equation \eqref{eq: long cov occ multi t avg} is 
\begin{align*}
\cov\big(\E[w_t|o,o_{t'}],\E[w_{t'}|o,o_{t'}]|o\big)
=&\sum_{j=1}^J\tilde{\mu}_{j,t}^o\cov\big(\E[\theta_{j,t'}|o,o_{t'}],\E[\ol{\theta}_{t'}^{o_{t'}}|o,o_{t'}]|o\big).
\end{align*}
Therefore,
\begin{align*}
    \cov(w_t,w_{t'}|o)
    =&\sum\limits_{j=1}^J \tilde{\mu}_{j,t}^o\Big\{\E\big[\cov(\theta_{j,t'},\ol{\theta}_{t'}^{o_{t'}}|o,o_{t'})\big|o\big]+\cov\big(\E[\theta_{j,t'}|o,o_{t'}],\E[\ol{\theta}_{t'}^{o_{t'}}|o,o_{t'}]|o\big)\Big\}\\
    =&\sum\limits_{j=1}^J \tilde{\mu}_{j,t}^o\cov(\theta_{j,t'},\ol{\theta}_{t'}^{o_{t'}}|o).
\end{align*}

Altogether, Assumption~\ref{assum: occ multi skills}, $\E[\theta_{j,t}-\theta_{j,t'}|o_t,o_{t-1},o_{t'}]=\E[\theta_{j,t-1}-\theta_{j,t'}|o_t,o_{t-1},o_{t'}]=0$ for all $j$, and $\E[\eps_t|o_t,o_{t-1},o_{t'}]=\E[\eps_{t-1}|o_t,o_{t-1},o_{t'}]=\E[\eps_{t'}|o_t,o_{t-1},o_{t'}]=0$ imply that the IV estimator conditional on $o_t=o_{t-1}=o$ is
\begin{align*}
\frac{\cov( \Delta w_t,w_{t'}|o)}{\cov(w_{t-1},w_{t'}|o)} 
= \frac{\sum\limits_{j=1}^J \Delta\tilde{\mu}_{j,t}^o\cov(\theta_{j,t'},\ol{\theta}_{t'}^{o_{t'}}|o)} {\sum\limits_{j'=1}^J \tilde{\mu}_{j',t-1}^o\cov(\theta_{j',t'},\ol{\theta}_{t'}^{o_{t'}}|o)}
=\sum^J_{j=1}\tilde{\upsilon}_{j,t',t-1}^{o}\left(\frac{\Delta\tilde{\mu}_{j,t}^o}{\tilde{\mu}_{j,t-1}^o}\right)
=\sum^J_{j=1}\tilde{\upsilon}_{j,t',t-1}^{o}\left(\frac{\Delta\mu_{j,t}}{\mu_{j,t-1}} + \frac{\mu_{j,t}}{\mu_{j,t-1}}\frac{\Delta\alpha_{j,t}^o}{\alpha_{j,t-1}^o}\right),
\end{align*}
where
\begin{align*}
\tilde{\upsilon}^{o}_{j,t',t-1}\equiv \frac{\tilde{\mu}_{j,t-1}^o \cov(\theta_{j,t'},\ol{\theta}_{t'}^{o_{t'}}|o)} {\sum\limits_{j'=1}^J \tilde{\mu}_{j',t-1}^o \cov(\theta_{j',t'},\ol{\theta}_{t'}^{o_{t'}}|o)}    .
\end{align*}
If $\cov(\theta_{j,t'},\ol{\theta}_{t'}^{o_{t'}}|o)\geq 0$ for all $j$ and $o$, the IV estimator for stayers in occupation $o$ reflects a weighted average of the growth rate of skill-specific returns.

\paragraph{IV Estimator for All Occupation Stayers.}

Finally, consider the IV estimator for all occupation stayers (i.e., $o_t=o_{t-1}$), regardless of occupation. The log wage residual autocovariance for occupational stayers is
\begin{align*}
    \cov(w_t,w_{t'}|o_t=o_{t-1})=&\E\big[\cov(w_t,w_{t'}|o_t,o_{t-1})\big|o_t=o_{t-1}\big]\\
    &+\cov\big(\E[w_t|o_t,o_{t-1}],\E[w_{t'}|o_t,o_{t-1}]\big|o_t=o_{t-1}\big).
\end{align*}
The first term is
\begin{align*}
\E\big[\cov(w_t,w_{t'}|o_t,o_{t-1})\big|o_t=o_{t-1}\big]=\sum\limits_{j=1}^J \mu_{j,t}\E\big[\alpha_{j,t}^{o_t}\cov(\theta_{j,t'},\ol{\theta}_{t'}^{o_{t'}}|o_t,o_{t-1}) |o_t=o_{t-1}\big].
\end{align*}
Assuming $\E[\theta_{j,t}-\theta_{j,t'}|o_t,o_{t-1}]=0$ for all $j$ and $\E[\eps_t|o_t,o_{t-1}]=\E[\eps_{t'}|o_t,o_{t-1}]=0$, the second term is
\begin{align*}
&\cov\big(\E[w_t|o_t,o_{t-1}],\E[w_{t'}|o_t,o_{t-1}]\big|o_t=o_{t-1}\big)\\
=&\E\Big[\E[w_t|o_t,o_{t-1}]\big(\E[w_{t'}|o_t,o_{t-1}]-\E[w_{t'}|o_t=o_{t-1}]\big)\big|o_t=o_{t-1}\Big]\\
=&\sum_{j=1}^J\mu_{j,t}\E\left[\alpha_{j,t}^{o_t}\E\left[\theta_{j,t'}|o_t,o_{t-1}\right]\big(\E[\ol{\theta}_{t'}^{o_{t'}}|o_t,o_{t-1}]-\E[\ol{\theta}_{t'}^{o_{t'}}|o_t=o_{t-1}]\big)\big|o_t=o_{t-1}\right]\\
=&\sum_{j=1}^J\mu_{j,t}\cov\left(\alpha_{j,t}^{o_t}\E\left[\theta_{j,t'}|o_t,o_{t-1}\right],\E[\ol{\theta}_{t'}^{o_{t'}}|o_t,o_{t-1}]\big|o_t=o_{t-1}\right).
\end{align*}

Altogether,
\begin{align*}
\cov( w_t,w_{t'}|o_t=o_{t-1})
=&\sum_{j=1}^J\mu_{j,t}\Big\{\E[\alpha_{j,t}^{o_t}\cov(\theta_{j,t'},\ol{\theta}_{t'}^{o_{t'}}|o_t,o_{t-1})|o_t=o_{t-1}]\\
&+\cov\big(\alpha_{j,t}^{o_t}\E[\theta_{j,t'}|o_t,o_{t-1}],\E[\ol{\theta}_{t'}^{o_{t'}}|o_t,o_{t-1}]\big|o_t=o_{t-1}\big)\Big\}\\
=&\sum_{j=1}^J\mu_{j,t}\cov(\alpha_{j,t}^{o_t}\theta_{j,t'},\ol{\theta}_{t'}^{o_{t'}}|o_t=o_{t-1}).
\end{align*}

Therefore, Assumption~\ref{assum: occ multi skills}, $\E[\theta_{j,t}-\theta_{j,t'}|o_t,o_{t-1}]=\E[\theta_{j,t-1}-\theta_{j,t'}|o_t,o_{t-1}]=0$ for all $j$, and $\E[\eps_t|o_t,o_{t-1}]=\E[\eps_{t-1}|o_t,o_{t-1}]=\E[\eps_{t'}|o_t,o_{t-1}]=0$ imply that the IV estimator for all stayers is
\begin{align*}
\frac{\cov(\Delta w_t,w_{t'}|o_t=o_{t-1})}{\cov( w_{t-1},w_{t'}|o_t=o_{t-1})}
=&\frac{\sum_{j=1}^J\mu_{j,t}\cov(\alpha_{j,t}^{o_t}\theta_{j,t'},\ol{\theta}_{t'}^{o_{t'}}|o_t=o_{t-1})}{\sum_{j'=1}^J\mu_{j',t-1}\cov(\alpha_{j',t-1}^{o_{t-1}}\theta_{j',t'},\ol{\theta}_{t'}^{o_{t'}}|o_t=o_{t-1})}-1\\
=&\sum_{j=1}^J\hat{\upsilon}_{j,t',t-1}\left(\frac{\mu_{j,t}}{\mu_{j,t-1}}\frac{\cov(\alpha_{j,t}^{o_t}\theta_{j,t'},\ol{\theta}_{t'}^{o_{t'}}|o_t=o_{t-1})}{\cov(\alpha_{j,t-1}^{o_{t-1}}\theta_{j,t'},\ol{\theta}_{t'}^{o_{t'}}|o_t=o_{t-1})}-1\right)\\
=&\sum_{j=1}^J\hat{\upsilon}_{j,t',t-1}\left(\frac{\Delta\mu_{j,t}}{\mu_{j,t-1}}+\frac{\mu_{j,t}}{\mu_{j,t-1}}\frac{\cov(\Delta\alpha_{j,t}^{o_t}\theta_{j,t'},\ol{\theta}_{t'}^{o_{t'}}|o_t=o_{t-1})}{\cov(\alpha_{j,t-1}^{o_{t-1}}\theta_{j,t'},\ol{\theta}_{t'}^{o_{t'}}|o_t=o_{t-1})}\right),
\end{align*}
where
\begin{align*}
    \hat{\upsilon}_{j,t',t-1}\equiv\frac{\mu_{j,t-1}\cov(\alpha_{j,t-1}^{o_{t-1}}\theta_{j,t'},\ol{\theta}_{t'}^{o_{t'}}|o_t=o_{t-1})}{\sum_{j'=1}^J\mu_{j',t-1}\cov(\alpha_{j',t-1}^{o_{t-1}}\theta_{j',t'},\ol{\theta}_{t'}^{o_{t'}}|o_t=o_{t-1})}.
\end{align*}
The weights are positive if and only if
\begin{align*}
\cov(\alpha_{j,t-1}^{o_{t-1}}\theta_{j,t'},\ol{\theta}_{t'}^{o_{t'}}|o_t=o_{t-1})
=&\E\big[\alpha_{j,t-1}^{o_{t-1}}\cov(\theta_{j,t'},\ol{\theta}_{t'}^{o_{t'}}|o_t,o_{t-1})|o_t=o_{t-1}\big]\\
&+\cov\big(\alpha_{j,t-1}^{o_{t-1}}\E[\theta_{j,t'}|o_t,o_{t-1}],\E[\ol{\theta}_{t'}^{o_{t'}}|o_t,o_{t-1}]|o_t=o_{t-1}\big)\geq 0.
\end{align*}
The first term is positive if $\cov(\theta_{j,t'},\ol{\theta}_{t'}^{o_{t'}}|o_t=o_{t-1}=o)\geq 0$ for all $(j,o)$, which is the condition for positive weights among stayers in occupation $o$ as shown above. 
The second term is zero when $\E[\ol{\theta}_{t'}^{o_{t'}}|o_t=o_{t-1}=o]$ does not vary with $o$.

\subsection{Identification with Occupation-Specific Wage Functions and Multiple Skills} \label{app: ident occ multiple skills}

We now consider the case in which log wage residuals are given by equation~\eqref{eq: w occ multi skills}, where we now define $\tilde{\mu}_{j,t}^o \equiv \mu_t^o\alpha_{j,t}^o$ and $\ol{\theta}_{i,t}^{o_{i,t}} \equiv \sum_{j=1}^J \tilde{\mu}_{j,t}^{o_{i,t}}\theta_{i,j,t}$. 

\paragraph{Identification of $\mu_t^o$.}
With Assumption \ref{assum: occ multi skills}, the long autocovariance of log wage residuals for stayers in occupation $o_t=o_{t-1}=o$ and $o_{t'}=o'$ can be written as follows for $t-t'\geq k+1$:
\begin{align}
\cov(w_t,w_{t'}|o,o')=&\mu_t^o\sum_{j=1}^J\alpha_{j,t}^o\cov(\theta_{j,t'},\ol{\theta}_{t'}^{o'}|o,o'),\label{eq: long cov occ multi t}\\
\cov(w_{t-1},w_{t'}|o,o')=&\mu_{t-1}^o\sum_{j=1}^J\alpha_{j,t-1}^o\cov(\theta_{j,t'},\ol{\theta}_{t'}^{o'}|o,o').\label{eq: long cov occ multi t-1}
\end{align}
Together, equations~\eqref{eq: long cov occ multi t} and \eqref{eq: long cov occ multi t-1} imply
\begin{align*}
    \frac{\cov(\Delta w_t,w_{t'}|o,o')}{\cov(w_{t-1},w_{t'}|o,o')}
    =\frac{\mu_t^o}{\mu_{t-1}^o}\frac{\sum_{j=1}^J\alpha_{j,t}^o\cov(\theta_{j,t'},\ol{\theta}_{t'}^{o'}|o,o')}{\sum_{j'=1}^J\alpha_{j',t-1}^o\cov(\theta_{j',t'},\ol{\theta}_{t'}^{o'}|o,o')}-1
    =\frac{\Delta\mu_t^o}{\mu_{t-1}^o}+\frac{\mu_t^o}{\mu_{t-1}^o}\left(\sum_{j=1}^J\tilde{\omega}^{o,o'}_{j,t',t-1}\frac{\Delta\alpha_{j,t}^o}{\alpha_{j,t-1}^o}\right),
\end{align*}
where
\begin{align*}
\tilde{\omega}_{j,t',t-1}^{o,o'}\equiv\frac{\alpha_{j,t-1}^o\cov(\theta_{j,t'},\ol{\theta}_{t'}^{o'}|o,o')}{\sum_{j'=1}^J\alpha_{j',t-1}^o\cov(\theta_{j',t'},\ol{\theta}_{t'}^{o'}|o,o')}.    
\end{align*}
Occupation-specific returns, $\mu_t^o$, are identified based on stayers in occupation $o$ if $\alpha_{j,t}^o$ does not change over time. Furthermore, $\cov(\theta_{j,t'},\theta_{j',t'}|o,o')\geq 0, \forall j,j'$, implies that $\cov(\theta_{j,t'},\ol{\theta}_{t'}^{o'}|o,o') \geq 0, \forall j$, in which case the weights $\tilde{\omega}_{j,t',t-1}^{o,o'}$ are non-negative.

\paragraph{IV Estimator for Stayers in Occupation $o$.}
Next, we show that $\mu_t^o$ is identified based on covariances conditioned only on $o_t=o_{t-1}=o$ when we assume $\E[\theta_{j,t}-\theta_{j,t'}|o_t,o_{t-1},o_{t'}]=\E[\theta_{j,t-1}-\theta_{j,t'}|o_t,o_{t-1},o_{t'}]=0$ for all $j$ and $\E[\eps_t|o_t,o_{t-1},o_{t'}]=\E[\eps_{t-1}|o_t,o_{t-1},o_{t'}]=\E[\eps_{t'}|o_t,o_{t-1},o_{t'}]=0$ in addition to Assumption \ref{assum: occ multi skills}.

Consider  the long autocovariance that is not conditioned on $o_{t'}$, i.e., equation \eqref{eq: long cov occ multi t avg}.
Given Assumption~\ref{assum: occ multi skills} and equation~\eqref{eq: long cov occ multi t}, the first term in equation~\eqref{eq: long cov occ multi t avg} is
\begin{align*}
\E\big[\cov(w_t,w_{t'}|o,o_{t'})|o\big]
=&\mu_t^o\E\left[\sum_{j=1}^J\alpha_{j,t}^o\cov(\theta_{j,t'},\ol{\theta}_{t'}^{o_{t'}}|o,o_{t'})\big|o\right].
%=&\mu_t^o\E\left[\mu_{t'}^{o_{t'}}\sum_{j=1}^J\sum_{j'=1}^J\alpha_{j}^o\alpha_{j'}^{o_{t'}}\cov(\theta_{j,t-1},\theta_{j',t'}|o,o_{t'})\big|o\right].
\end{align*}
The second term in equation \eqref{eq: long cov occ multi t avg} is
\begin{align*}
\cov\big(\E[w_t|o,o_{t'}],\E[w_{t'}|o,o_{t'}]|o\big)
=&\cov\left(\gamma_t^{o}+\mu_t^{o}\sum_{j=1}^J\alpha_{j,t}^o\E[\theta_{j,t}|o,o_{t'}],\E[\ol{\theta}_{t'}^{o_{t'}}|o,o_{t'}]\Big|o\right)\\
=&\mu_t^{o}\sum_{j=1}^J\alpha_{j,t}^o\cov\big(\E[\theta_{j,t'}|o,o_{t'}],\E[\ol{\theta}_{t'}^{o_{t'}}|o,o_{t'}]|o\big),
%=&\mu_t^{o}\sum_{j=1}^J\alpha_j^o\cov\big(\E[\theta_{j,t-1}|o,o_{t'}],\E[w_{t'}|o,o_{t'}]|o\big).
\end{align*}
using additional assumptions $\E[\theta_{j,t}-\theta_{j,t'}|o_t,o_{t-1},o_{t'}]=0$ and $\E[\eps_t|o_t,o_{t-1},o_{t'}]=\E[\eps_{t'}|o_t,o_{t-1},o_{t'}]=0$.

Therefore, the long autocovariances in equations \eqref{eq: long cov occ multi t} and \eqref{eq: long cov occ multi t-1} that are not conditioned on $o_{t'}$ are given by
\begin{align*}
\cov(w_t,w_{t'}|o)=&\mu_t^o\sum_{j=1}^J\alpha_{j,t}^o\cov(\theta_{j,t'},\ol{\theta}_{t'}^{o_{t'}}|o),\\
\cov(w_{t-1},w_{t'}|o)=&\mu_{t-1}^o\sum_{j=1}^J\alpha_{j,t-1}^o\cov(\theta_{j,t'},\ol{\theta}_{t'}^{o_{t'}}|o),
\end{align*}
where we use the law of total covariance:
\begin{align*}
\cov(\theta_{j,t'},\ol{\theta}_{t'}^{o_{t'}}|o)=\E\left[\cov(\theta_{j,t'},\ol{\theta}_{t'}^{o_{t'}}|o,o_{t'})\big|o\right]+\cov\big(\E[\theta_{j,t'}|o,o_{t'}],\E[\ol{\theta}_{t'}^{o_{t'}}|o,o_{t'}]|o\big).
\end{align*}
Combining the two equations gives
\begin{align*}
    \frac{\cov(\Delta w_t,w_{t'}|o)}{\cov(w_{t-1},w_{t'}|o)}
    %=\frac{\mu_t^o}{\mu_{t-1}^o}\frac{\sum_{j=1}^J\alpha_{j,t}^o\cov(\theta_{j,t'},\ol{\theta}_{t'}^{o_{t'}}|o)}{\sum_{j'=1}^J\alpha_{j',t-1}^o\cov(\theta_{j',t'},\ol{\theta}_{t'}^{o_{t'}}|o)}-1
    =\frac{\Delta\mu_t^o}{\mu_{t-1}^o}+\frac{\mu_t^o}{\mu_{t-1}^o}\left(\sum_{j=1}^J\iota^{o}_{j,t',t-1}\frac{\Delta\alpha_{j,t}^o}{\alpha_{j,t-1}^o}\right),
\end{align*}
where
\begin{align*}
\iota_{j,t',t-1}^{o}\equiv\frac{\alpha_{j,t-1}^o\cov(\theta_{j,t'},\ol{\theta}_{t'}^{o_{t'}}|o)}{\sum_{j'=1}^J\alpha_{j',t-1}^o\cov(\theta_{j',t'},\ol{\theta}_{t'}^{o_{t'}}|o)}.    
\end{align*}
If $\cov(\theta_{j,t'},\ol{\theta}_{t'}^{o_{t'}}|o)\geq 0$ for all $j$, then these weights are non-negative.

\paragraph{IV Estimator for All Occupation Stayers.}
Finally, consider the IV estimator for all occupational stayers (i.e., $o_t=o_{t-1}$), regardless of occupation. 
We show that if $\gamma_t^o=0$, then this estimator identifies a weighted average of occupation-specific skill returns.

The log wage residual autocovariance for occupational stayers is
\begin{align*}
    \cov(w_t,w_{t'}|o_t=o_{t-1})=&\E\big[\cov(w_t,w_{t'}|o_t,o_{t-1})\big|o_t=o_{t-1}\big]\\
    &+\cov\big(\E[w_t|o_t,o_{t-1}],\E[w_{t'}|o_t,o_{t-1}]\big|o_t=o_{t-1}\big).
\end{align*}

Under Assumption \ref{assum: occ multi skills}, the first term is
\begin{align*}
\E\big[\cov(w_t,w_{t'}|o_t,o_{t-1})\big|o_t=o_{t-1}\big]=\E\left[\mu_t^{o_t}\sum_{j=1}^J\alpha_{j,t}^{o_t}\cov(\theta_{j,t'},\ol{\theta}_{t'}^{o_{t'}}\Big|o_t,o_{t-1})|o_t=o_{t-1}\right].
\end{align*}

If $\gamma_t^o=0$ and $\E[\theta_{j,t}-\theta_{j,t'}|o_t,o_{t-1},o_{t'}]=0$ for all $j$, the second term is
\begin{align*}
&\cov\big(\E[w_t|o_t,o_{t-1}],\E[w_{t'}|o_t,o_{t-1}]\big|o_t=o_{t-1}\big)\\
=&\E\Big[\E[w_t|o_t,o_{t-1}]\big(\E[w_{t'}|o_t,o_{t-1}]-\E[w_{t'}|o_t=o_{t-1}]\big)\big|o_t=o_{t-1}\Big]\\
=&\E\left[\mu_t^{o_t}\sum_{j=1}^J\alpha_{j,t}^{o_t}\E\left[\theta_{j,t'}|o_t,o_{t-1}\right]\Big(\E[\ol{\theta}_{t'}^{o_{t'}}|o_t,o_{t-1}]-\E[\ol{\theta}_{t'}^{o_{t'}}|o_t=o_{t-1}]\Big)\Big|o_t=o_{t-1}\right]\\
\end{align*}

Therefore, the long autocovariances for occupational stayers are
\begin{align*}
\cov( w_t,w_{t'}|o_t=o_{t-1})
=&\sum_o\Pr(o_t=o_{t-1}=o|o_t=o_{t-1})\mu_t^o\Psi^o_{t',t},\\
\cov( w_{t-1},w_{t'}|o_t=o_{t-1})
=&\sum_o\Pr(o_t=o_{t-1}=o|o_t=o_{t-1})\mu_{t-1}^o\Psi^o_{t',t-1},
\end{align*}
where $\Pr(o_t=o_{t-1}=o|o_t=o_{t-1})$ denotes the share of stayers in occupation $o$ among all occupation stayers and
\begin{align*}
\Psi^o_{t',t}\equiv&\sum_{j=1}^J\alpha_{j,t}^o\left\{\cov(\theta_{j,t'},\ol{\theta}_{t'}^{o_{t'}}|o_t=o_{t-1}=o)
+\E[\theta_{j,t'}|o_t=o_{t-1}=o]\big(\E[\ol{\theta}_{t'}^{o_{t'}}|o_t=o_{t-1}=o]-\E[\ol{\theta}_{t'}^{o_{t'}}|o_t=o_{t-1}]\big)\right\} .
\end{align*}
The term $\cov(\theta_{j,t'},\ol{\theta}_{t'}^{o_{t'}}|o_t=o_{t-1}=o)$ is positive if $\cov(\theta_{j,t'},\theta_{j',t'}|o_t=o_{t-1}=o)\geq 0$ for all $j'$, while the
second term is zero when $\E[\ol{\theta}_{t'}^{o_{t'}}|o_t=o_{t-1}=o]$ does not vary with $o$.

The two long residual autocovariances imply
\begin{align*}
    \frac{\cov(\Delta w_t,w_{t'}|o_t=o_{t-1})}{\cov(w_{t-1},w_{t'}|o_t=o_{t-1})}
    =\sum_o \zeta_{t',t-1}^o\left(\frac{\Delta\mu_t^o}{\mu_{t-1}^o}+\frac{\mu_t^o}{\mu_{t-1}^o}\frac{\Psi^o_{t',t}-\Psi^o_{t',t-1}}{\Psi^o_{t',t-1}}\right),   
\end{align*}
where
\begin{align*}
\zeta_{t',t-1}^o\equiv\frac{\Pr(o_t=o_{t-1}=o|o_t=o_{t-1})\Psi_{t',t-1}^o\mu_{t-1}^o}{\sum_{o'}\Pr(o_t=o_{t-1}=o'|o_t=o_{t-1})\Psi_{t',t-1}^{o'}\mu_{t-1}^{o'}}.
\end{align*}
The weights are non-negative if and only if $\Psi_{t',t-1}^o\geq 0$ for all $o$.

If $\alpha_{j,t}^o=\alpha_{j,t-1}^o$, then $\Psi^o_{t',t}=\Psi^o_{t',t-1}$ and the IV estimator identifies a weighted average of $\Delta \mu_t^o/\mu_{t-1}^o$ across occupations.

\section{MD Estimation and Standard Errors} \label{app: MD desc}

\subsection{MD Estimation}

For a given parameter vector $\bm \Lambda$, we can compute theoretical counterparts
for $\cov(w_t,w_{t'}|s,c)$, where $s\in\{\text{Non-college, College}\}$ indicates non-college and college status,  implied by any specific model and compare them with the sample covariances. Since some
cohort (or, equivalently, experience $e=t-c$) cells have few
observations when calculating residual covariances, we generally partition
the cohort set into ten-year cohort groups (e.g., $C_1$, $C_2$, $C_3$, and $C_4$ corresponding to cohorts born 1942--1951, 1952--1961, 1962--1971, and 1972--1981, respectively) or the experience set into 10-year experience groups
$E_1$, $E_2$, $E_3$, and $E_4$, corresponding to
1--10, 11--20, 21--30, and 31--40 years, respectively, aggregating within these cohort or experience groups.

In the case of cohort grouping in Section~\ref{sec: MD long autocov}, the minimum distance estimator $\hat{\bm \Lambda}$ solves
  \[
    \min_{\bm \Lambda} \sum_{(s,j,t,t')\in\Gamma} \Big\{
    \widehat{\cov}(w_t,w_{t'}|s,
    C_j)-\cov(w_t,w_{t'}|s,C_j,\bm \Lambda)\Big\}^2,
  \]
where $\Gamma$ is described in Table~\ref{tab: cohort groups}; $\widehat{\cov}(w_{t},w_{t'}|s,C_j)$ is the sample covariance for residuals in years $t$ and $t'$ conditional on education group $s$ and cohort group $C_j$;
and $\cov(w_t,w_{t'}|s,C_j,\bm \Lambda)$ is the corresponding theoretical covariance given parameter vector $\bm\Lambda$.

In the case of experience grouping in Section~\ref{sec: FE AR(1) skills}, the minimum distance estimator $\hat{\bm \Lambda}$ solves
  \[
    \min_{\bm \Lambda} \sum_{(s,j,t,t')\in\Gamma} \Big\{
    \widehat{\cov}(w_t,w_{t'}|s,
    E_j)-\cov(w_t,w_{t'}|s,E_j,\bm \Lambda)\Big\}^2,
  \]
where $\Gamma=\{s,j,t,t'|1970\leq t'\leq t\leq 2012, t-t' \geq 6, s\in\{\text{Non-college, College}\}, j\in\{3,4\}\}$; $\widehat{\cov}(w_{t},w_{t'}|s,E_j)$  is the sample covariance for residuals in years $t$ and $t'$ conditional on education group $s$ and experience group $E_j$;
and $\cov(w_t,w_{t'}|s,E_j,\bm \Lambda)$ is the corresponding theoretical covariance given parameter vector $\bm\Lambda$.  In Sections~\ref{subsec: persistent eps} and \ref{sec: skill dist}, we also include covariance moments for less-experienced workers, $E_1$ and $E_2$, covering the same time periods.

\subsection{Standard Errors}

Consider the case of experience-based moments, and
let $m=1,2,\ldots,M$ be the index of all moments. Let $d_{i,m}$ be the
indicator of whether individual $i$ contributes to the $m^{th}$ moment
$\cov(w_t,w_{t'}|s,E_j)$. That is, both $w_{i,t}$ and $w_{i,t'}$ are
non-missing and $s_{i,t}=s_{i,t'}=s$ and $e_{i,t}\in E_j$. Also let
$p_m(\bm\Lambda)=\cov(w_t,w_{t'}|s,E_j,\bm \Lambda)$.
Then, we can write
  \begin{align*}
    h_{m}(\bm z_i,\bm \Lambda)=d_{i,m}\big[w_{i,t}w_{i,t'}- p_m(\bm \Lambda)\big],
  \end{align*}
  where $\bm z_i$ includes $w_{i,t}$ $d_{i,m}$ for all $t$ and $m$ for
  individual $i$.  Let
  $\bm h(\bm z,\bm \Lambda)=[h_1(\bm z,\bm \Lambda)\, h_2(\bm z,\bm \Lambda)\,\ldots\, h_M(\bm
  z,\bm \Lambda)]^{\top}$.
  Then the following moment condition holds for the true parameter
  $\bm \Lambda_0$:
  \begin{align*}
    \E[\bm h(\bm z,\bm \Lambda_0)]=\bm 0.
  \end{align*}
  The minimum distance estimator $\hat{\bm \Lambda}$ is
  equivalent to the GMM estimator that solves
  \begin{align*}
    \min_{\bm \Lambda} \left[\frac{1}{N}\sum^N_{i=1}\bm h(\bm z_i,\bm \Lambda)\right]^{\top} \bm W \left[\frac{1}{N}\sum^N_{i=1}\bm h(\bm z_i,\bm \Lambda)\right],
  \end{align*}
  where
  $\bm W=\text{diag}(\frac{N^2}{N_1^2}, \frac{N^2}{N_2^2},\ldots, \frac{N^2}{N_M^2})$ and
  $N_m=\sum^N_{i=1}d_{i,m}$.

  The GMM estimator $\hat{\bm \Lambda}$ is asymptotically normal with
  a variance matrix
  \begin{align*}
  \bm V=(\bm H^{\top}\bm W \bm H)^{-1}(\bm H^{\top} \bm W \Omega \bm W \bm
  H)(\bm H^{\top} \bm W \bm H)^{-1},
  \end{align*}
  % \begin{align*}
  %   \text{Avar}(\hat{\bm\Lambda})=(\bm H' \bm W \bm H)^{-1} (\bm H'
  %   \bm W \Omega \bm W \bm H)(\bm H' \bm W \bm H)^{-1}
  % \end{align*}
  where $\bm H$ is the Jacobian of the vector of moments,
  $\E[\partial \bm h(\bm z,\bm\Lambda_0)/\partial\bm\Lambda^{\top}]$, and
  $\bm \Omega=\E[\bm h(\bm z,\bm\Lambda_0) \bm h(\bm
  z,\bm\Lambda_0)^{\top}]$.
  Both expectations are replaced by sample averages and evaluated at
  the estimated parameter:
    \begin{align*}
      \hat{\bm H}=&\frac{1}{N}\sum^N_{i=1}\frac{\partial \bm h(\bm
      z_i,\hat{\bm\Lambda})}{\partial \bm\Lambda^{\top}}={\bm W}^{-\frac{1}{2}}\frac{\partial
       \bm p(\hat{\bm \Lambda})}{\partial \bm\Lambda^{\top}},\\
      \hat{\bm\Omega}=&\frac{1}{N}\sum^N_{i=1}\bm h(\bm
      z_i,\hat{\bm\Lambda}) \bm h(\bm
      z_i,\hat{\bm\Lambda})^{\top},
    \end{align*}
    where
    ${\bm W}^{-\frac{1}{2}}=\text{diag}(\frac{N_1}{N}, \frac{N_2}{N},\ldots,
    \frac{N_M}{N})$.

\clearpage

%%%%%%%%%%%%%%%%%%%%%%%%%%%%%%%%%%%%%%%%%%%%%%%%%%%%%%%%%%%%%%%%%%%%%%%%%%%%%%%%%%
%%%% Re-Number Tables and figures for the Appendix %%%%%%%%
\renewcommand{\thetable}{\thesection-\arabic{table}}
\setcounter{table}{0}

\renewcommand{\thefigure}{\thesection-\arabic{figure}}
\setcounter{figure}{0}

%%%%%%%%%%%%%%%%%%%%%%%%%%%%%%%%%%%%%%%%%%%%%%%%%%%%%%%%%%%%%%%%%%%%%%%%%%%%%%%%%%

\section{PSID Data Details and Additional Results} \label{app: PSID}

\subsection{Data Description} \label{app: PSID data description}

The PSID is a longitudinal survey of a representative sample of
individuals and families in the U.S.\ beginning in 1968.  The survey
was conducted annually through 1997 and biennially since. We use data
collected from 1971 through 2013. Since earnings
were collected for the year prior to each survey, our analysis
studies hourly wages from 1970 to 2012.

Our sample is restricted to male heads of households from the core
(SRC) sample and excludes those from any PSID oversamples
(SEO, Latino) as well as those with zero individual weights.\footnote{The
earnings questions we use are asked only of household heads. We
also restrict our sample to those who were heads of household and
not students during the survey year of the observation of interest
as well as two years earlier.  Our sampling scheme is very similar
to that of \cite{moffitt_gottschalk_2012}.}  We use earnings (total
wage and salary earnings, excluding farm and business income) from
any year these men were ages 16--64, had potential experience of 1--40 years,
had positive wage and salary income, had positive hours worked, and
were not enrolled as a student.

Our sample is composed of 92\% whites, 6\% blacks, and 1\% hispanics
with an average age of 39 years old. We create seven education
categories based on current years of completed schooling: 1-5 years,
6-8 years, 9-11 years, 12 years, 13-15 years, 16 years, and 17 or more
years. College workers are defined as those with more than 12 years of
schooling. In our sample, 13\% of respondents finished less than 12
years of schooling, 35\% had exactly 12 years of completed schooling,
21\% completed some college (13-15 years), 21\% completed college (16
years), and 10\% had more than 16 years of schooling.

The wage measure we use divides annual earnings by annual hours worked, trimming
the top and bottom 1\% of all wages within year and college/non-college status by ten-year experience
cells. The resulting sample contains 3,766 men and 44,547 person-year
observations.

\begin{figure}[h]
  \centering
  \includegraphics[width=0.45\columnwidth]{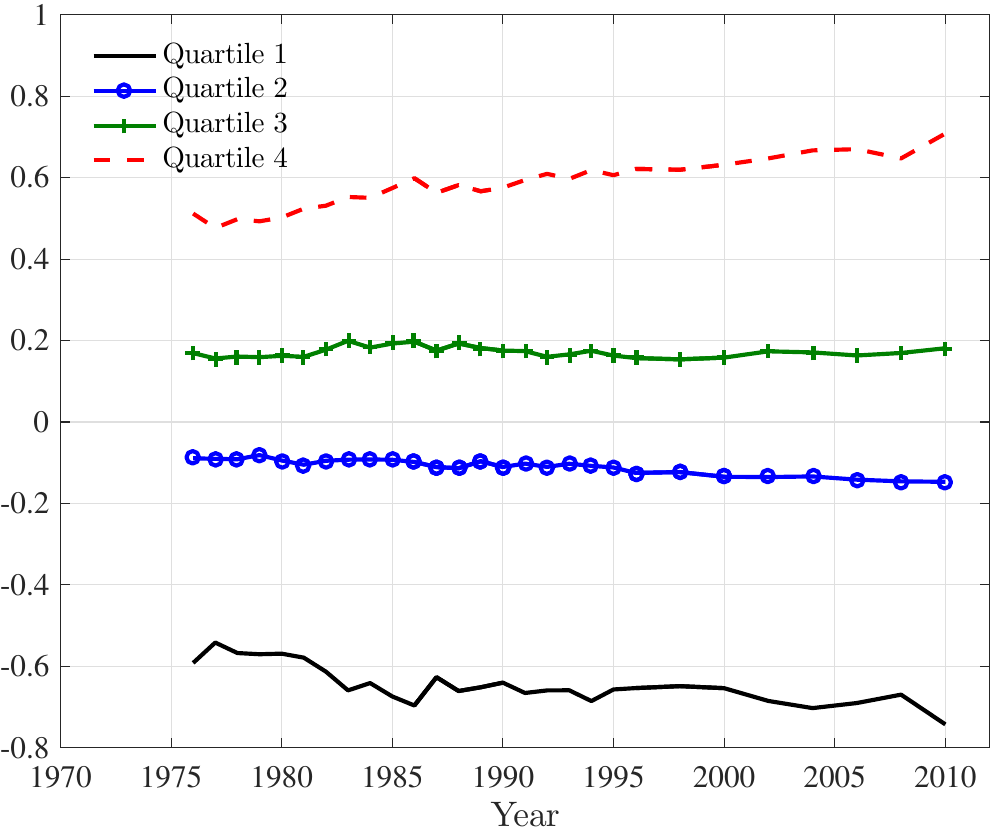}
\caption{Average Log Wage Residuals by Quartile}
    \label{fig: resid quart actual}
\end{figure}

Figure~\ref{fig: resid quart actual} shows the widening of the residual distribution over time, reporting average log wage residuals within each quartile.  Consistent with Figure~\ref{fig: variance of hourly earnings}, the distribution widened most during the early 1980s and then again after 2000.

\begin{figure}[h]
  \centering
  \includegraphics[width=0.45\columnwidth]{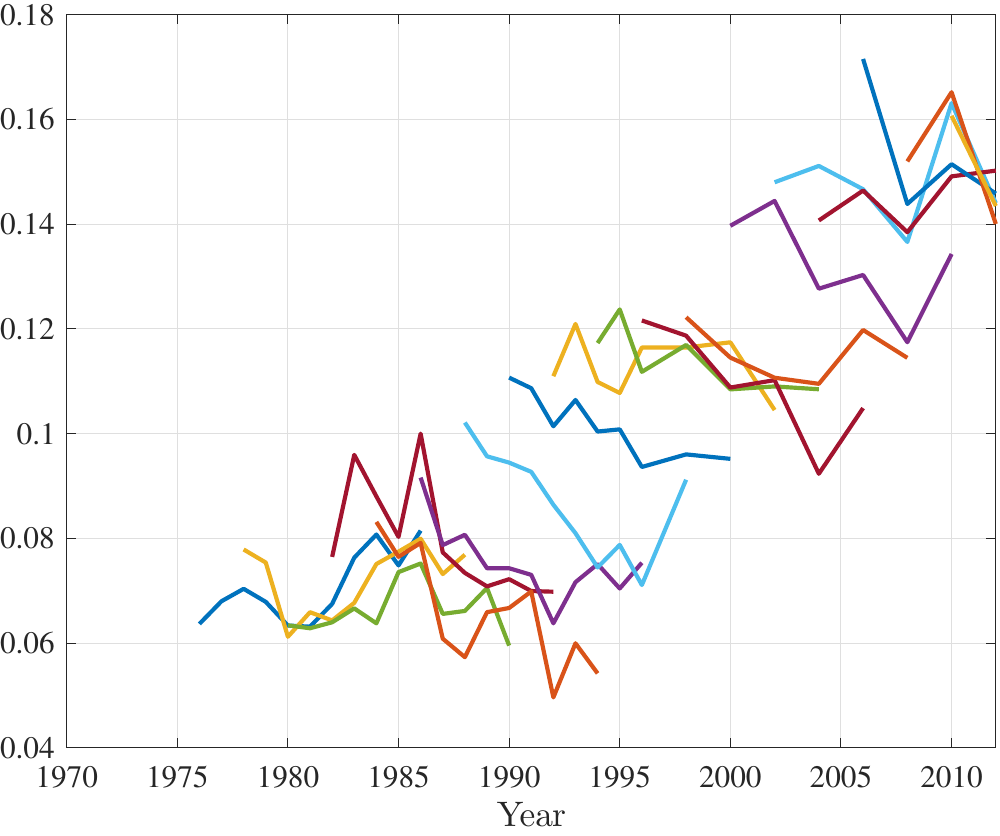}
\caption{Log Wage Residual Autocovariances (`Balanced' Sample)}
    \label{fig: PSID autocov no attrition}
\end{figure}

To examine whether attrition affects the residual autocovariances reported in Figure~\ref{fig: PSID autocov}, Figure~\ref{fig: PSID autocov no attrition} shows the autocovariances, $\cov(w_b,w_t)$ for $6 \leq t-b \leq 16$, where the samples for each line (representing different base years, $b$) are restricted to those individuals observed in the base year as well as at least one of the last two years used for that line (i.e.\ $t-b=15$ or $16$ in early years or $t-b=14$ or $16$ in later years with biannual surveys).  Comparing Figures~\ref{fig: PSID autocov} and \ref{fig: PSID autocov no attrition}, the autocovariance patterns are quite similar, indicating little effect of sample attrition (due to non-response or retirement) on the key moments used in our analysis.

\begin{figure}[h]
  \centering
  \includegraphics[width=0.45\columnwidth]{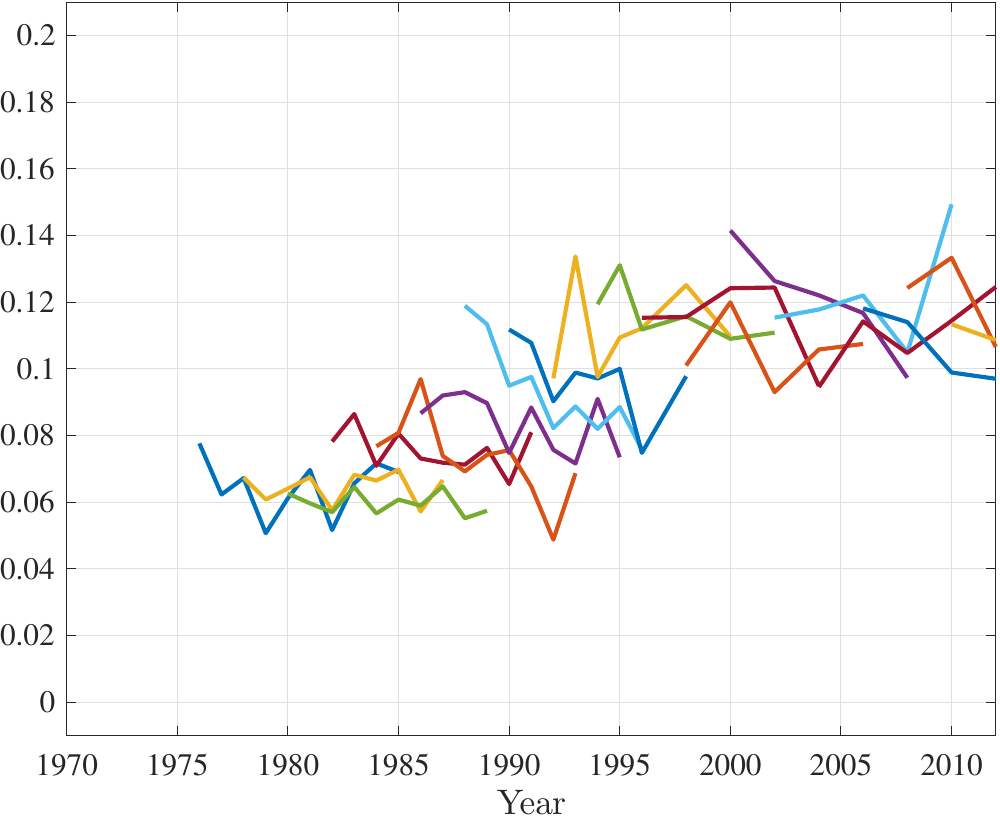}
\caption{Autocovariances for Log Wage Residuals (1--14 Years of Experience)}
    \label{fig: PSID autocov expr 1-14}
\end{figure}

Figure~\ref{fig: PSID autocov expr 1-14} shows the residual autocovariances for individuals with 1--14 years of experience in the base years.  Regardless of the base year, the autocovariances are typically declining from late 1980s through the 1990s as in Figures~\ref{fig: PSID autocov}(a) (full sample) and~\ref{fig: PSID autocov}(b) (men with 15--30 years experience) in the text.  The lines also shift upwards over time, consistent with rising skill variances.

\clearpage

%%%%%%%%%%%%%%%%%%%%%%%%%%%%%%%%%%%%%%%%%%%%%%%%%%%%%%%%%%%%%%%%%%%%%%%%%%%%%%%%%%%%%%%%%%%%

\subsection{2SLS Estimates of Skill Returns by Education (PSID)} \label{app: 2SLS by educ}

\begin{table}[h]
\def\sym#1{\ifmmode^{#1}\else\(^{#1}\)\fi}
\caption{2SLS estimates of $\Delta_2\mu_{t}/\mu_{t-2}$ for men with 21--40 years experience, 1979--1995}\label{tab: iv1 old} \center
\center \small
\begin{tabular}{l*{6}{c}}
\hline
                    & 1979--1980 & {1981--1983}& {1984--1986}  & {1987--1989}    & {1990--1992}   &  {1993--1995}\\
 \hline
\multicolumn{7}{c}{\underline{A. All men }} \vspace{2pt}\\
$\Delta_2\mu_t/\mu_{t-2}$ &   -0.052   &   -0.088$^{*}$  &   -0.031      &   -0.100$^{*}$  &  -0.036     &      -0.104$^{*}$   \\
                        &    (0.050)         &    (0.043)       &    (0.050)    &    (0.046)      &    (0.044)  &    (0.045)        \\
\addlinespace
Observations        & {928}         & {1,323}         & {1,244}         & {1,211}         & {1,244}         & {1,300}               \\
1st stage \(F\)-Statistic&     117.23         &     132.19         &      66.26     &     130.53     &     132.83     &     201.62   \\
\\
\multicolumn{7}{c}{\underline{B. Non-college men}} \vspace{2pt}\\
$\Delta_2\mu_t/\mu_{t-2}$&      -0.108         &     0.009         &     -0.019         &      -0.101         &     -0.051         &      -0.105         \\
                    &    (0.061)         &    (0.062)         &    (0.072)         &    (0.070)         &    (0.066)         &    (0.065)         \\
\addlinespace
  Observations        &{552}         &{777}         &{678}         &{609}         &{555}         &{542}         \\
1st stage \(F\)-Statistic&      66.06         &      59.12
                                             &      24.04         &   55.22   &    65.32 &  72.32  \\
  \\
\multicolumn{7}{c}{\underline{C. College men}} \vspace{2pt}\\
$\Delta_2\mu_t/\mu_{t-2}$&     -0.031         &      -0.166\sym{**} &    -0.003         &     -0.088         &     -0.024         &      -0.104        \\
                    &    (0.068)         &    (0.053)         &    (0.074)         &    (0.060)         &    (0.059)         &    (0.060)         \\
\addlinespace
  Observations        &{314}         &{491}         &{509}         &{524}         &{594}         &{758}        \\
1st stage \(F\)-Statistic&      73.87         &      90.56         &      99.30         &      71.46         &      66.14         &     142.24      \\

\hline
\multicolumn{7}{l}{Notes: Estimates from 2SLS regression of $w_{i,t}-w_{i,t-2}$ on $w_{i,t-2}$ using instruments $(w_{i,t-8}, w_{i,t-9})$.  }\\
\multicolumn{7}{l}{Experience restrictions based on year $t$. $^*$ denotes significance at 0.05 level. }
\end{tabular}
\end{table}
%\clearpage

\begin{table}
\def\sym#1{\ifmmode^{#1}\else\(^{#1}\)\fi}
\caption{2SLS estimates of $\Delta_2\mu_t/\mu_{t-2}$ for men with 21--40 years experience, 1996--2012}\label{tab: iv2 old}  \center \small
\begin{tabular}{l*{4}{c}}
\hline
                    &\multicolumn{1}{c}{1996--2000}&\multicolumn{1}{c}{2002--2006}&\multicolumn{1}{c}{2008--2012}\\
\hline
\multicolumn{4}{c}{\underline{A. All men }} \vspace{2pt}\\
$\Delta_2\mu_t/\mu_{t-2}$ &   -0.084$^{*}$ &     -0.040         &     -0.058         \\
                                &    (0.030)         &    (0.032)         &    (0.031)         \\
\addlinespace
Observations               & {1,427}             & {1,591}                 & {1,493}         \\
1st stage \(F\)-Statistic&    295.75         &     281.91         &     267.83     \\
\\
\multicolumn{4}{c}{\underline{B. Non-college men}} \vspace{2pt}\\

$\Delta_2\mu_t/\mu_{t-2}$&      -0.073         &     -0.064         &      0.011         \\
&    (0.053)         &    (0.046)         &    (0.082)         \\
\addlinespace
  Observations             &{589}         &{624}         &{481}         \\
1st stage \(F\)-Statistic&  96.00         &     126.69         &     114.93         \\
\\
\multicolumn{4}{c}{\underline{C. College men}} \vspace{2pt}\\
$\Delta_2\mu_t/\mu_{t-2}$&    -0.094\sym{**} &     -0.040         &     -0.074\sym{*}  \\
                    &   (0.036)         &    (0.042)         &    (0.032)         \\
\addlinespace
  Observations          &{834}         &{960}         &{866}         \\
1st stage \(F\)-Statistic&       212.60         &     169.90         &     163.07         \\
\hline
\multicolumn{4}{l}{Notes: Estimates from 2SLS regression of $w_{i,t}-w_{i,t-2}$ on}\\
\multicolumn{4}{l}{ $w_{i,t-2}$ using instruments $(w_{t-8},w_{t-9})$ for 1996--2000 and}\\
\multicolumn{4}{l}{ $(w_{t-8},w_{t-10})$ for 2002--2006 and 2008--2012. Experience } \\
\multicolumn{4}{l}{restrictions based on year $t$. $^*$ denotes significance at 0.05} \\
\multicolumn{4}{l}{level. }
\end{tabular}
\end{table}

\clearpage

%%%%%%%%%%%%%%%%%%%%%%%%%%%%%%%%%%%%%%%%%%%%%%%%%%%%%%%%%%%%%%%%%%%%%%%%%%%%%%%%%%%%%%%%%%%%

\subsection{GMM Estimates of Skill Returns, Over-Identification Tests, and Variance of Skill Growth} \label{app: GMM}

In this appendix, we report GMM estimates for the returns to skill using the same model and moments (i.e.\ lagged residuals serve as instruments) as with our 2SLS approach in Section~\ref{sec: IV} along with $J$-statistics to test for overidentification. We also report analogous GMM estimates that use both past and future wage residuals as instruments, reporting $J$-statistics to test the validity of the latter.  Finally, we combine estimates using past vs.\ future residuals as instruments to estimate the variance of skill growth relative to lagged skill levels.

To begin, rewrite the two-period wage growth equation~\eqref{eq: 2SLS regression} as follows:
\begin{align}
\Delta_2 w_{i,t}=&\left(\frac{\Delta_2 \mu_t}{\mu_{t-2}}\right)
                    w_{i,t-2}+u_{i,t}, \label{eq: diff regression}
\end{align}
where $u_{i,t}\equiv\varepsilon_{i,t}-\frac{\mu_t}{\mu_{t-2}}\varepsilon_{i,t-2}+\mu_t\Delta_2\theta_{i,t}$.

Serially uncorrelated skill shocks implies the following moment condition:
\begin{align}
  \E[w_{t'}u_t]=0, \quad \text{ for } t'\leq t-2-k. \label{eq: valid lags}
\end{align}
Under the stronger assumption that $\var(\Delta\theta_{t})=0, \forall t$, the
following additional moment condition holds:
\begin{align}
  \E[w_{t''}u_t]=0, \quad \text{ for } t''\geq t+k. \label{eq: valid leads}
\end{align}
Equation~\eqref{eq: valid leads} will not hold when $ \var(\Delta_2\theta_{t})> 0$, and the IV estimate using future residuals as instruments is
asymptotically biased with probability limit
\begin{align*}
  \frac{\cov(\Delta_2 w_{t},w_{t'})}{\cov(w_{t-2},w_{t'})}=\frac{\Delta_2 \mu_t}{\mu_{t-2}}+\frac{\mu_t}{\mu_{t-2}}
 \frac{\var(\Delta_2\theta_t)}{\var(\theta_{t-2})}>\frac{\Delta_2\mu_t}{\mu_{t-2}},
  \quad \text{ for } t'\geq t+k.
\end{align*}

The difference between estimates using future and past residuals as instruments
identifies the magnitude of the skill shock variance relative to the skill variance: for $t'\leq t-2-k$ and $t''\geq t+k$,
\begin{equation}\label{eq: relative magnitude of skill shock}
\frac{\var(\Delta_2\theta_t)}{\var(\theta_{t-2})} = \left[\frac{\cov(\Delta_2 w_{t},w_{t''})}{\cov( w_{t-2},w_{t''})}-\frac{\cov(\Delta_2 w_{t},w_{t'})}{\cov(w_{t-2},w_{t'})}
  \right]\left[1+\frac{\cov(\Delta_2 w_{t},w_{t'})}{\cov(w_{t-2},w_{t'})}\right]^{-1}.
\end{equation}

\subsubsection{Overidentification Tests}

We begin by testing the moments in equation~\eqref{eq: valid leads} using Hansen's $J$-test,
assuming $k=6$ and using the two nearest valid instruments. This amounts to
using $w_{i,t-8}$ and $w_{i,t-9}$ (or $w_{i,t-10}$) for equation~\eqref{eq:
  valid lags} and the first two available out of $w_{i,t+6},w_{i,t+7},w_{i,t+8},w_{i,t+9}$ for \eqref{eq: valid leads}.

Table \ref{tab: balanced} reports the two-step optimal GMM estimates
(allowing for heteroskedasticity and serial correlation within
individual) for the coefficient on $w_{i,t-2}$ along with Hansen's
$J$-statistics when estimating the wage growth equation \eqref{eq: diff regression}.  Panel~A
reports estimates when moments from both equations~\eqref{eq: valid lags}
and \eqref{eq: valid leads} are used (i.e., lags and leads), while Panel~B reports estimates
when only the moment condition from equation~\eqref{eq: valid lags} is used (i.e., lags only).
The sample is restricted to be the same in both panels.\footnote{Because use of both leads and lags requires observations that are as many as 19 years apart, this restriction reduces the sample size substantially relative to that used in our baseline 2SLS analysis (see Tables~\ref{tab: iv1} and \ref{tab: iv2}). Panel~A of Table~\ref{tab: nonbalanced} below reports GMM estimates when
this sample selection is not imposed.  Those results are directly comparable and quite similar to those in Tables~\ref{tab: iv1} and \ref{tab: iv2}.}

Comparing the $J$-statistics in Panels~A and B in Table \ref{tab: balanced}, we can test the validity of using leads as instruments (i.e.\ moments in equation~\eqref{eq: valid leads}). Since the differences are greater than 5.991 (critical value for $\chi^2_{2}$ at significance level 0.05) except for 1979--1980 and 2002--2004, we reject the `leads' moments in equation~\eqref{eq: valid leads} at 5\% significance level for 1981--2000.  (See Panel~C for $p$-values of these tests.) Moreover, all $J$-statistics in Panel~B are smaller than 3.841 (critical value for $\chi^2_{1}$ at significance level 0.05), implying that we cannot reject the lags as instruments (i.e.\ moments in equation~\eqref{eq: valid lags}) at the 5\% level.  Altogether, these results suggest that the lagged residuals are valid instruments, while the leads are not (in most years).

Finally, note that the estimates using both leads and lags as instruments are always greater than their counterparts using only lags. This reflects the positive bias induced from using leads when there are idiosyncratic skill growth shocks.

\begin{table}[H]
\caption{GMM Estimates of Skill Return Growth using Leads and Lags as Instruments (Balanced Samples)}
\label{tab: balanced}
\footnotesize
\centering
\def\sym#1{\ifmmode^{#1}\else\(^{#1}\)\fi}
\begin{tabular}{l*{8}{c}}
\toprule
                    &\multicolumn{1}{c}{1979--80}&\multicolumn{1}{c}{1981--83}&\multicolumn{1}{c}{1984--86}&\multicolumn{1}{c}{1987--89}&\multicolumn{1}{c}{1990--92}&\multicolumn{1}{c}{1993--95}&\multicolumn{1}{c}{1996--2000}&\multicolumn{1}{c}{2002--04}\\
\midrule
\multicolumn{9}{l}{A. 2 Nearest Valid Lags and 2 Nearest (Potentially Valid) Leads as Instruments} \\
\ \ Coeff. on $w_{i,t-2}$&  -0.019  &  0.088\sym{*}  & 0.053  &  0.007    &  -0.030   &    0.026  &     0.008      &      0.022     \\
                    &    (0.053)  &    (0.044)    &    (0.046)   &    (0.034) &    (0.038)    &   (0.035)  &    (0.0235)   &    (0.035)  \\

\ \ Observations        &\multicolumn{1}{c}{818}         &\multicolumn{1}{c}{1,251}         &\multicolumn{1}{c}{1,325}         &\multicolumn{1}{c}{1,356}         &\multicolumn{1}{c}{1,313}         &\multicolumn{1}{c}{1,311}         &\multicolumn{1}{c}{1,375}         &\multicolumn{1}{c}{777}         \\
\ \ \(J\)-Statistic     &       4.400         &      10.392         &      11.743         &       9.579         &       9.461         &       6.991         &       8.922         &       1.646         \\

\addlinespace
\multicolumn{9}{l}{B. 2 Nearest Valid Lags as Instruments} \\

\ \ Coeff. on $w_{i,t-2}$&     -0.070  &    -0.010   &     -0.065   &     -0.057   &    -0.103\sym{*}  &  -0.025   &  -0.041    &  -0.003 \\
                    &    (0.056)   &    (0.053)  &    (0.055) &    (0.040)   &    (0.046)   &    (0.039)  &  (0.029)    &    (0.0389)   \\

\ \ Observations        &\multicolumn{1}{c}{818}         &\multicolumn{1}{c}{1,251}         &\multicolumn{1}{c}{1,325}         &\multicolumn{1}{c}{1,356}         &\multicolumn{1}{c}{1,313}         &\multicolumn{1}{c}{1,311}         &\multicolumn{1}{c}{1,375}         &\multicolumn{1}{c}{777}         \\

\ \ \(J\)-Statistic     &       0.009         &       0.187         &       0.632         &       0.869         &       0.064         &       0.238         &       0.107         &       0.016         \\

\addlinespace
\multicolumn{9}{l}{C. $p$-Values for $J$-Tests of the Validity of Leads as Instruments} \\
\ \ Leads& 0.111 & 0.006 & 0.004 & 0.013 & 0.009 & 0.034 & 0.012 & 0.443 \\
\ \ Lags &0.924 & 0.665 & 0.427 & 0.351 & 0.800 & 0.626 & 0.744 & 0.899 \\

\bottomrule
\multicolumn{9}{l}{Notes: GMM estimates for a regression of $(w_{i,t}-w_{i,t-2})$ on $w_{i,t-2}$.  Panel A uses as instruments the 2 nearest}\\
\multicolumn{9}{l}{available lags from $(w_{t-8},w_{t-9},w_{t-10})$ and 2 nearest available leads from $(w_{t+6},...,w_{t+9})$. Panel B uses only} \\
\multicolumn{9}{l}{the 2 lags as instruments. Panel~C reports $p$-values based on a comparison of $J$-statistics from Panels A and B.}\\
\multicolumn{9}{l}{\sym{*} denotes significance at 0.05 level.  }
\end{tabular}
\end{table}

\subsubsection{Inferring Relative Magnitude of Skill Shocks} \label{app: iv var of skill growth}

Table \ref{tab: nonbalanced} reports GMM estimates using only lags or leads as instruments where all available observations are used (i.e.,\ samples are not restricted to be the same across specifications).  Panel~A reports estimates when only the moments in equation~\eqref{eq: valid lags} are used (i.e.,\ 2 nearest valid lags). These results are analogous to the 2SLS estimates in Tables~\ref{tab: iv1} and \ref{tab: iv2}, using the same samples.  Comparing estimates across the tables, we see that they are quite similar. Panel~B reports GMM estimates when only the moments in equation~\eqref{eq: valid leads} are used (i.e., 2 nearest potentially valid leads), also based on all available observations. Finally, we compare the estimates in Panels~A and B using equation~\eqref{eq: relative magnitude of skill shock} to estimate the relative importance of skill growth shocks.  These estimates are reported in Panel~C.  The variance of (two-year) skill growth relative to the variance of prior skill levels ranges from 0.16 to 0.29 over our entire sample period.

\begin{table}[H]
\caption{GMM Estimates of Skill Return Growth using Leads vs.\ Lags as Instruments and Relative Skill Shock Variance (Unbalanced Samples)}
\label{tab: nonbalanced}
\footnotesize
\centering
\def\sym#1{\ifmmode^{#1}\else\(^{#1}\)\fi}
\begin{tabular}{l*{8}{c}}
\toprule
                    &\multicolumn{1}{c}{1979--80}&\multicolumn{1}{c}{1981--83}&\multicolumn{1}{c}{1984--86}&\multicolumn{1}{c}{1987--89}&\multicolumn{1}{c}{1990--92}&\multicolumn{1}{c}{1993--95}&\multicolumn{1}{c}{1996--2000}&\multicolumn{1}{c}{2002--04}\\
\midrule
\multicolumn{9}{l}{A. 2 Nearest Valid Lags as Instruments} \\
\ \ \ Coeff. on $w_{i,t-2}$& -0.033  &   -0.045  &  -0.044  &  -0.084\sym{*}  &  -0.083\sym{*} &  -0.067    & -0.076\sym{*} &  -0.090\sym{*} \\
                    &    (0.045)    &    (0.038)    &  (0.038)  &    (0.033)  &  (0.035) &  (0.035)    &  (0.025)    &    (0.035)       \\
\ \ \ Observations        &\multicolumn{1}{c}{1,349}         &\multicolumn{1}{c}{2,077}         &\multicolumn{1}{c}{2,188}         &\multicolumn{1}{c}{2,245}         &\multicolumn{1}{c}{2,189}         &\multicolumn{1}{c}{2,095}         &\multicolumn{1}{c}{2,122}         &\multicolumn{1}{c}{1,377}         \\

\addlinespace
\multicolumn{9}{l}{B. 2 Nearest (Potentially Valid) Leads as Instruments} \\

\ \ \ Coeff. on $w_{i,t-2}$&  0.165\sym{*} &  0.229\sym{*}& 0.193\sym{*}&  0.099\sym{*}  & 0.067   &  0.087\sym{*}  & 0.073\sym{*} & 0.115\sym{*} \\
                    &    (0.059)   &    (0.053)  &  (0.047) &  (0.042)  & (0.043)  &    (0.038)   &  (0.028)   &    (0.039)   \\

\ \ \ Observations        &\multicolumn{1}{c}{1,500}         &\multicolumn{1}{c}{2,229}         &\multicolumn{1}{c}{2,159}         &\multicolumn{1}{c}{2,100}         &\multicolumn{1}{c}{2,042}         &\multicolumn{1}{c}{1,994}         &\multicolumn{1}{c}{2,178}         &\multicolumn{1}{c}{1,249}         \\
\addlinespace
\multicolumn{9}{l}{C. Estimated Shock Variances Relative to Skill Variances} \\
\ \ \ $\var(\Delta_2\theta_t)/\var(\theta_{t-2})$ &.204 &0.287& 0.248& 0.200 &0.163& 0.166& 0.161& 0.225\\

\bottomrule

\multicolumn{9}{l}{Notes: GMM estimates for a regression of $(w_{i,t}-w_{i,t-2})$ on $w_{i,t-2}$.  Panel A uses 2 nearest available lags as}\\
\multicolumn{9}{l}{instruments from  $(w_{t-8},w_{t-9},w_{t-10})$.  Panel B uses 2 nearest available leads as instruments from $(w_{t+6},...,w_{t+9})$.}\\
\multicolumn{9}{l}{  Panel C reports estimates of skill growth shock variance relative to skill variance based on equation \eqref{eq: relative magnitude of skill shock}.}\\
\multicolumn{9}{l}{\sym{*} denotes significance at 0.05 level.}
\end{tabular}
\end{table}

\subsection{Testing HIP based on growth in log wage residuals} \label{app: test HIP log w growth}

This appendix shows results for $\cov(\Delta (w_t/\mu_t),w_{t'})$ in PSID.

\begin{figure}[h]
  \centering
  \subfloat[Cohort Group 3, Non-College]{
    \includegraphics[width=0.45\columnwidth]
    {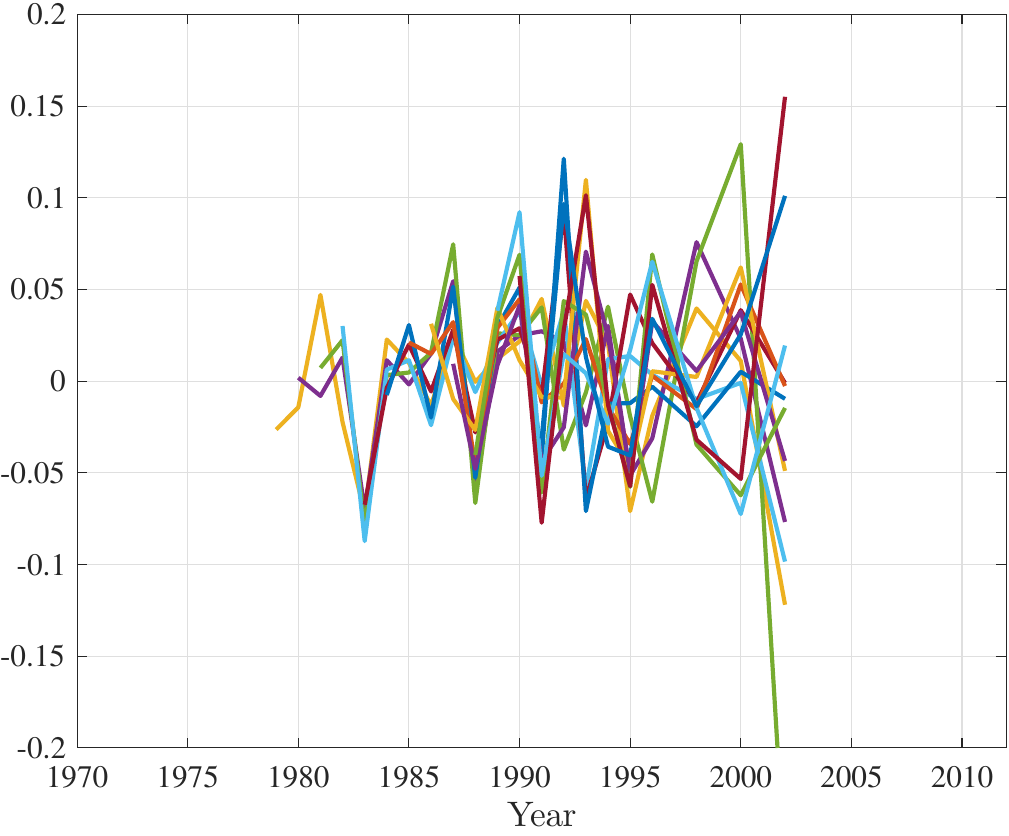}
  }\quad
  \subfloat[Cohort Group 3, College]{
    \includegraphics[width=0.45\columnwidth]
    {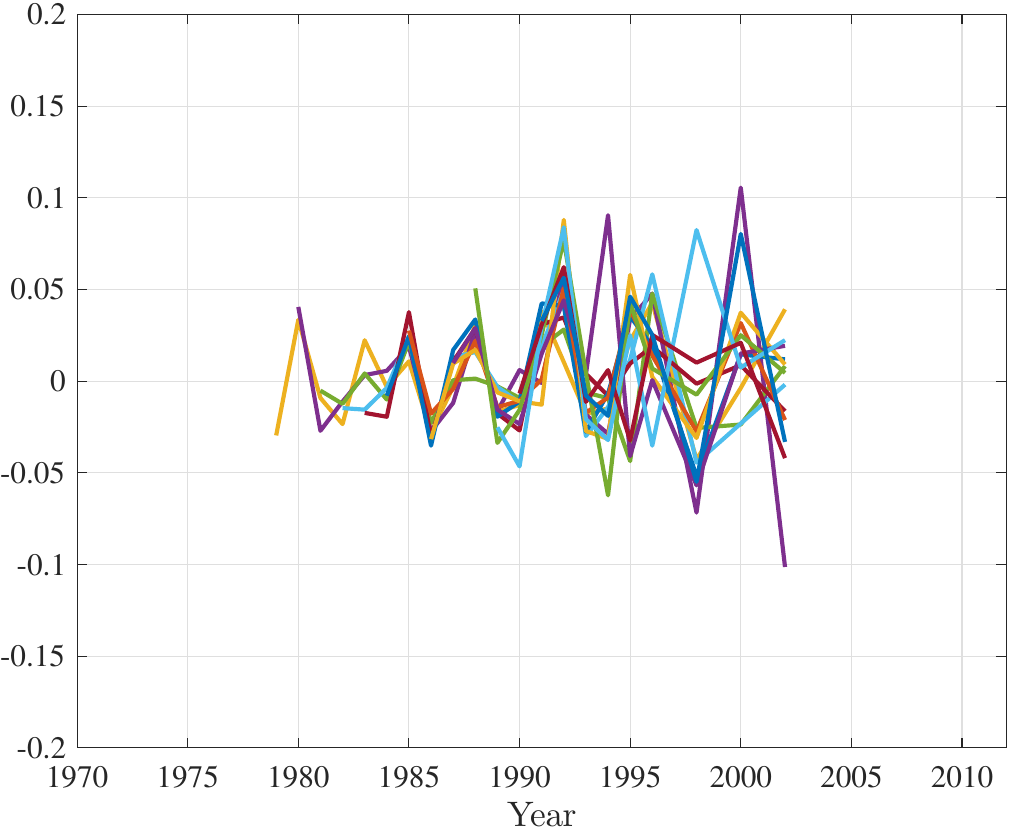}
  }
  \\
  \subfloat[Cohort Group 4, Non-College]{
    \includegraphics[width=0.45\columnwidth]
    {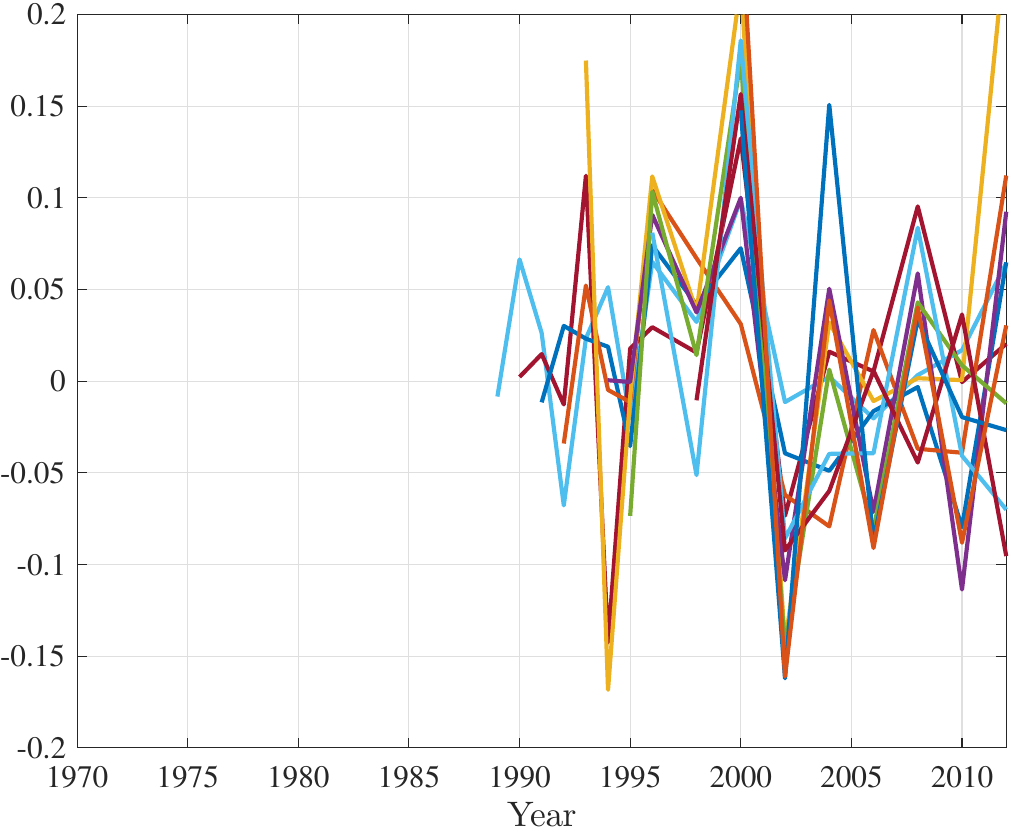}
  }\quad
  \subfloat[Cohort Group 4, College]{
    \includegraphics[width=0.45\columnwidth]
    {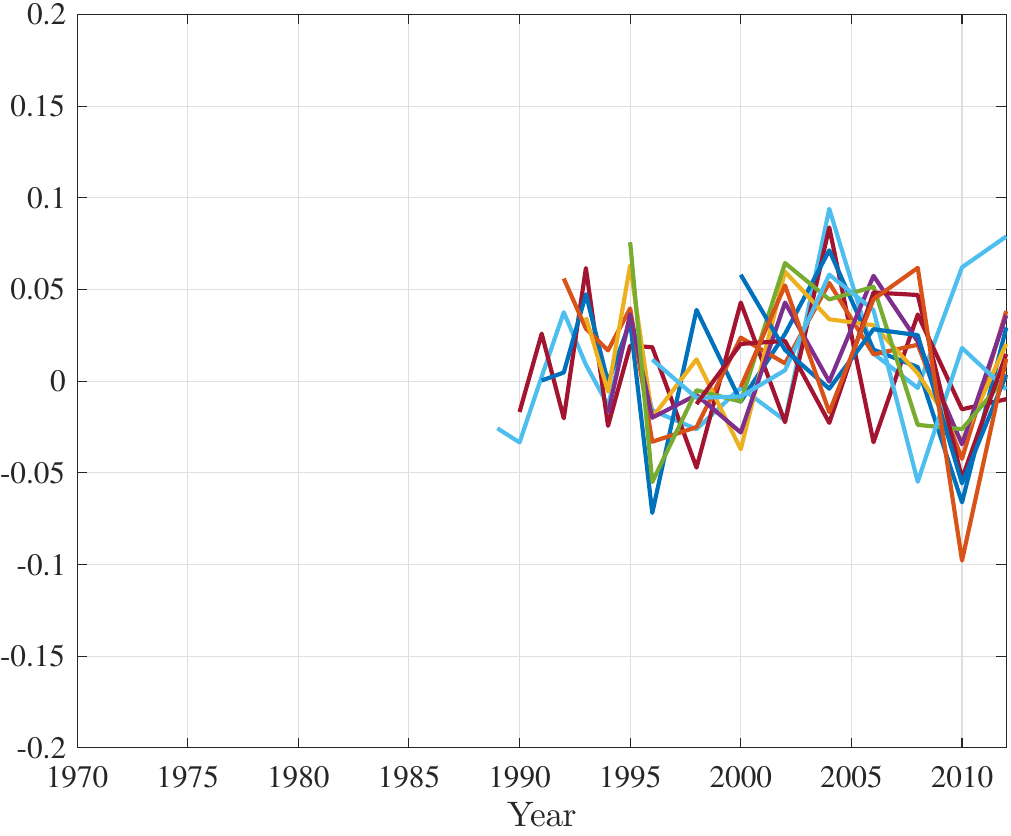}
  }
    \caption{$\cov(\Delta (w_t/\mu_t),w_{t'})$ for each $t,t'$ by Cohort Group}

    \label{fig: cov scaled diff2}
\end{figure}

\begin{figure}%[h]
  \centering
\includegraphics[width=0.45\columnwidth]{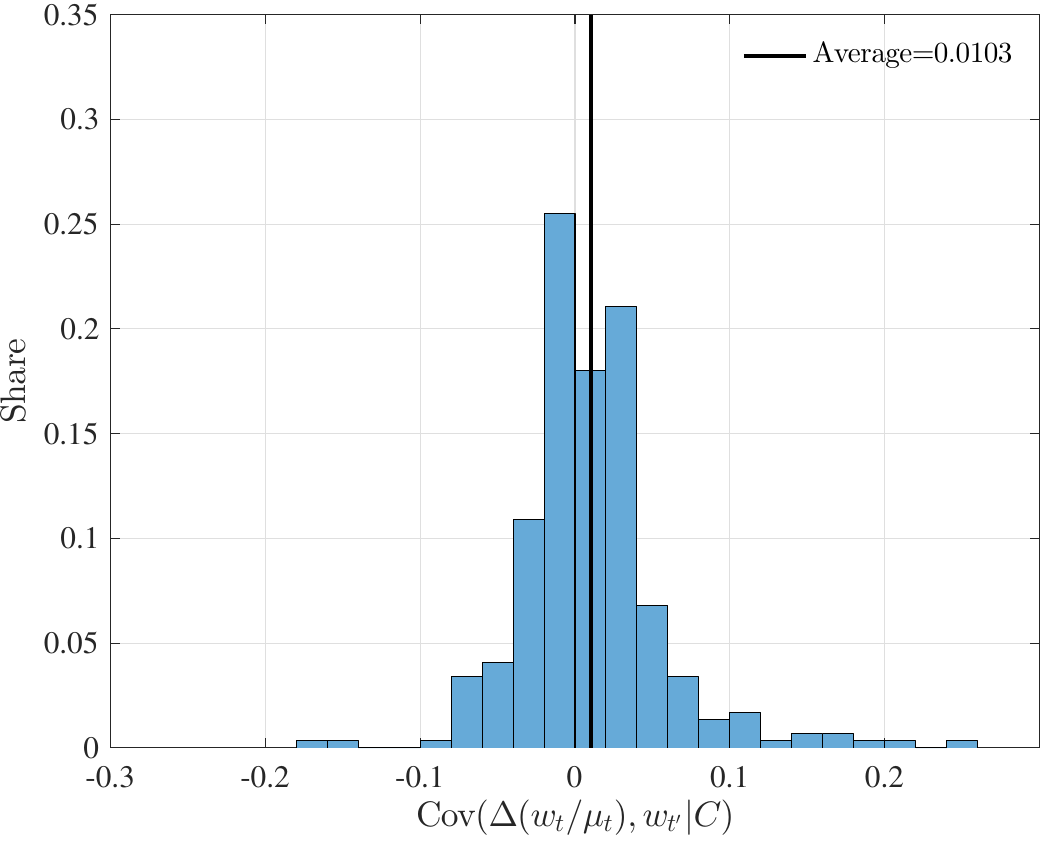}
  \caption{Distribution of $\cov(\Delta (w_t/\mu_t),w_{t'})$  for all $(t,t',C)$ for for Low-Experience Men}
    \label{fig: cov scaled diff2 dist}
\end{figure}

\clearpage

%%%%%%%%%%%%%%%%%%%%%%%%%%%%%%%%%%%%%%%%%%%%%%%%%%%%%%%%%%%%%%%%%%%%%%%%%%%%%%%%%%%%%

\subsection{Additional estimates for model with AR(1) skill dynamics}

\begin{figure}[h]
  \centering
    \subfloat[Non-College Men]{
    \includegraphics[width=0.45\columnwidth]{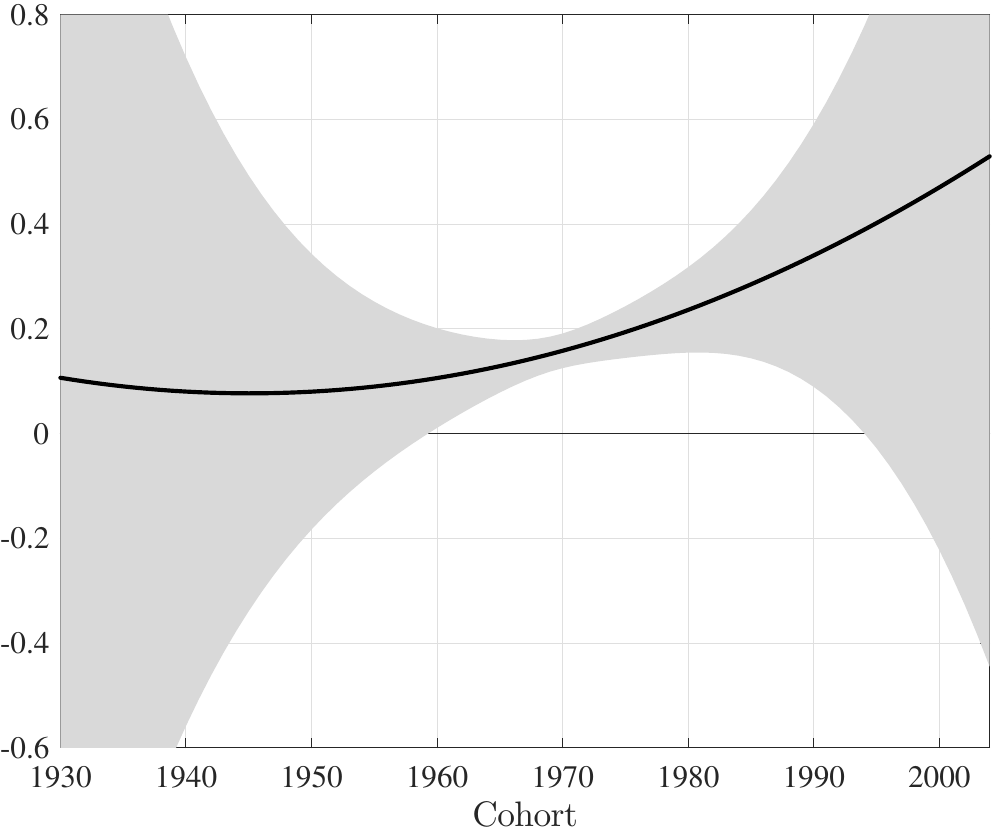}
  }
  \quad
    \subfloat[College Men]{
    \includegraphics[width=0.45\columnwidth]{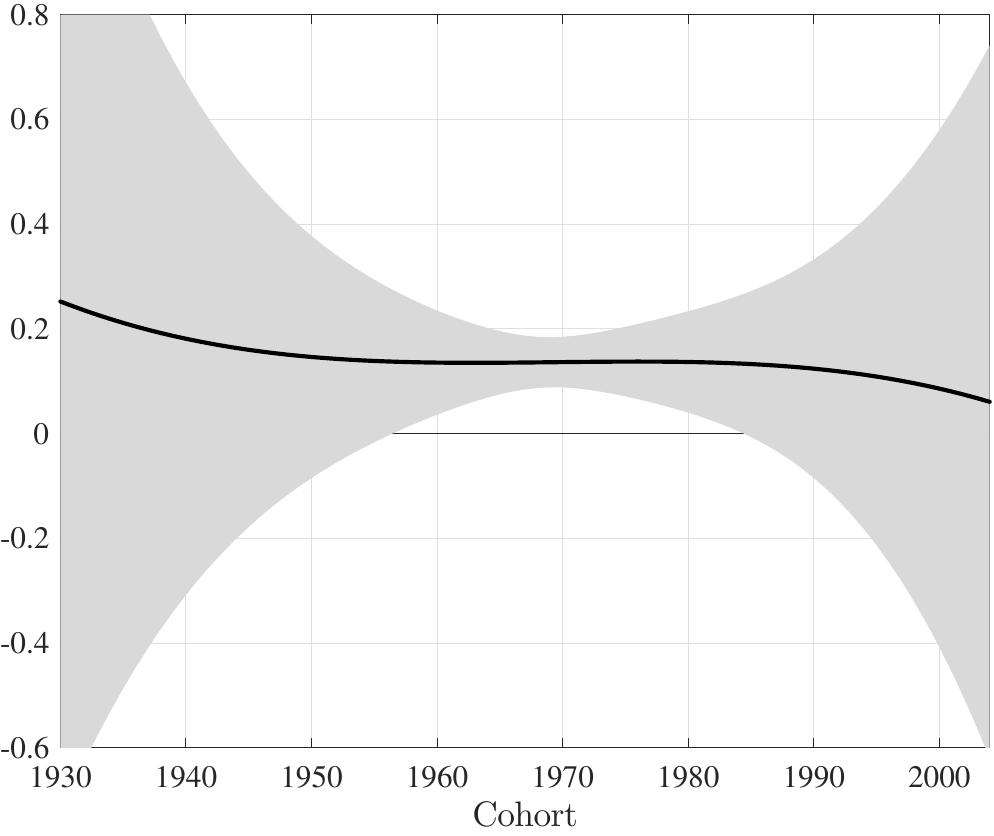}
  }
    \caption{$\var(\psi|c)$ implied by MD estimates allowing for time-varying AR(1) skill shocks, 21--40 years of experience}
    \label{fig: var(psi) time-varying AR(1) skill shocks}
  \end{figure}

Figure~\ref{fig: var(psi) time-varying AR(1) skill shocks} reports estimated $\var(\psi|c)$ when allowing for time-varying AR(1) skill shocks as discussed in Section~\ref{sec: FE AR(1) skills}.  See the text for additional details on the specification.

%%%%%%%%%%%%%%%%%%%%%%%%%%%%%%%%%%%%%%%%%%%%%%%%%%%%%%%%%%%%%%%%%%%%%%%%%%%%%%%%%%%%%

\subsection{Comparison with specifications used in literature on earnings dynamics} \label{App: ARMA eps results}

As noted in the text, the literature on earnings dynamics estimates similar models; although, it rarely considers the returns to unobserved skills.  \cite{haider_2001} and \cite{moffitt_gottschalk_2012} are notable exceptions.  Since their estimates suggest qualitatively different patterns for the returns to skill over time, this appendix explores whether those can be explained by seemingly modest differences in specifications.

\begin{figure}[h]
  \centering
  \subfloat[Non-College]{
    \includegraphics[width=0.45\columnwidth]
    {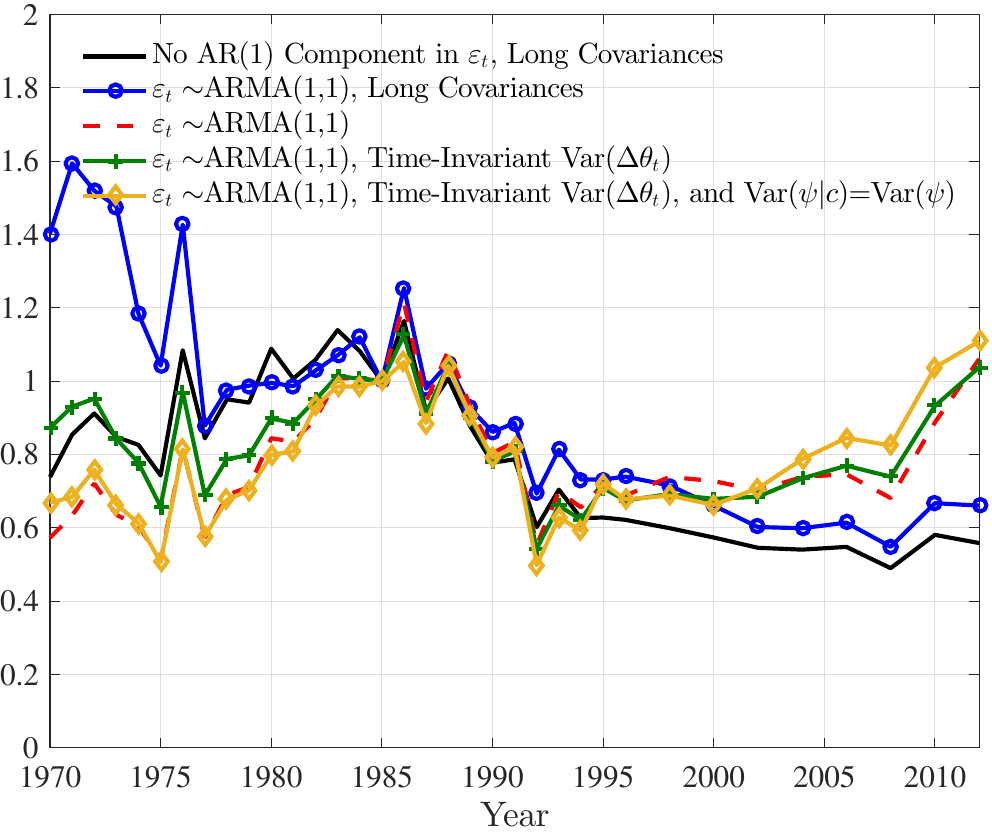}
  }
  \quad
  \subfloat[College]{
    \includegraphics[width=0.45\columnwidth]
    {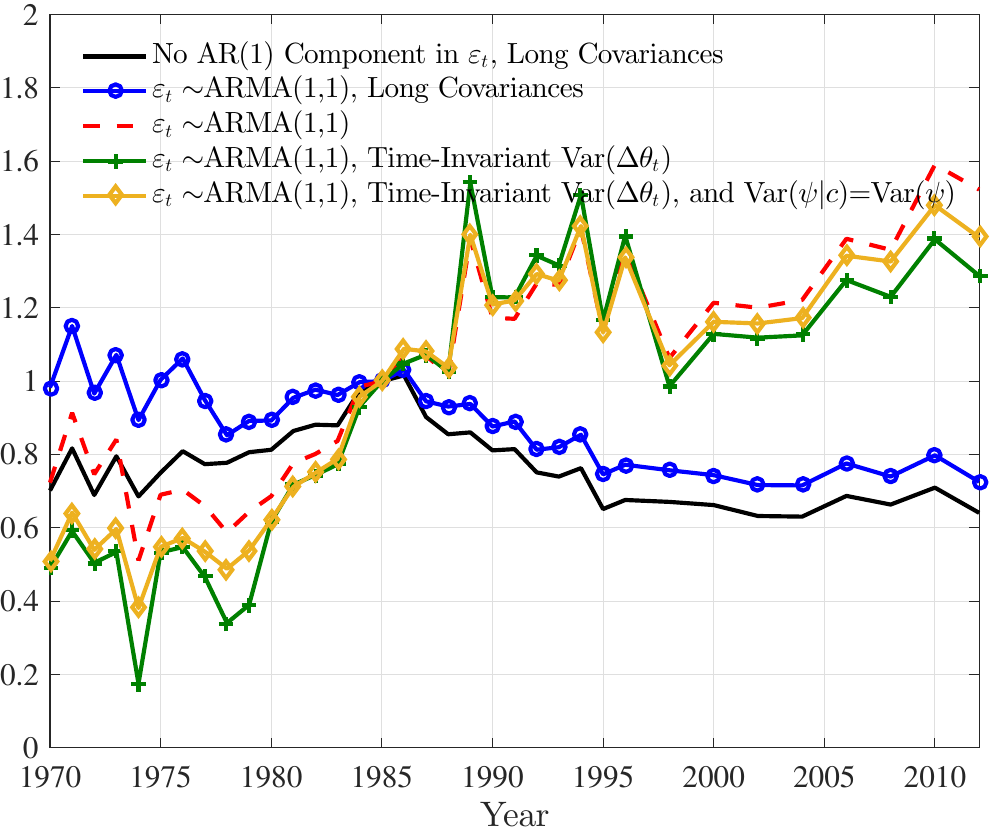}
  }
  \caption{Estimated $\mu_t$ under Different Restrictions: Non-College and College}
  \label{app fig: compare MG by educ}
\end{figure}

As with most of the earnings dynamics literature, \cite{haider_2001} and \cite{moffitt_gottschalk_2012} assume $\eps_t \sim \text{ARMA}(1,1)$, which is very similar, though not identical, to our specification in equation~\eqref{eq: AR(1) eps} with $k=2$.  We prefer the latter given our desire to maintain a completely flexible (time-varying) autocorrelation structure for transitory shocks, $\teps_t$.  Regardless, the two specifications yield very similar estimated $\mu_t$ series when using the same long autocovariances (i.e., $\cov(w_t,w_{t'})$ for $|t-t'|\geq 6$) in MD estimation, as can be seen from comparing the estimated returns given by the blue lines with circles in Figure~\ref{app fig: compare MG by educ} and the returns reported in Figure~\ref{fig: AR1 non-skill shocks}.  The former assumes $\eps_t \sim \text{ARMA}(1,1)$, while the latter assumes $\eps_t$ contains an AR(1) component, $\varphi_t$, and transitory component (with $k=6$), $\teps_t$, as described by equation~\eqref{eq: AR(1) eps}.\footnote{Formally, we specify ARMA(1,1) non-skill shocks as $\eps_{i,t}=\nu_{i,t}$ for $t=c_i+1$ and $\eps_{i,t}= \rho\eps_{i,t-1}+ \nu_{i,t} + \beta_1 \nu_{i,t-1}$for $t>c_i+1$, with $\cov(\nu_t,\nu_{t'})=0$ for all $t\neq t'$.}  Figure~\ref{app fig: compare MG by educ} also reproduces estimates from our baseline specification, which assumes that $\eps_t$ only contains a transitory component (with $k=6$).  Altogether, these estimates indicate that accounting for persistent non-skill shocks has little effect on estimated returns to skills over time.

The red dashed lines in Figure~\ref{app fig: compare MG by educ} show that estimating the $\eps_t \sim \text{ARMA}(1,1)$ specification using all autocovariances, as in \cite{haider_2001} and \cite{moffitt_gottschalk_2012} and the earnings dynamics literature more generally, yields quite different $\mu_t$ patterns for college men.  This suggests that accounting for a flexible short-term autocorrelation structure -- accommodated by using only long autocovariances -- has important implications for skill returns among college men.  (Failing to allow for time variation in the short-term autocorrelation structure for $\eps_t$ is primarily responsible for the different patterns.)
The remaining two estimated $\mu_t$ series in Figure~\ref{app fig: compare MG by educ} impose additional assumptions about skills made by \cite{haider_2001} and \cite{moffitt_gottschalk_2012}: time-invariance of non-skill growth innovations and cohort-invariance of initial skill variation.  Neither of these assumptions has major implications for our estimated $\mu_t$ series for non-college or college men.

\begin{figure}[h]
  \centering

    \includegraphics[width=0.45\columnwidth]
    {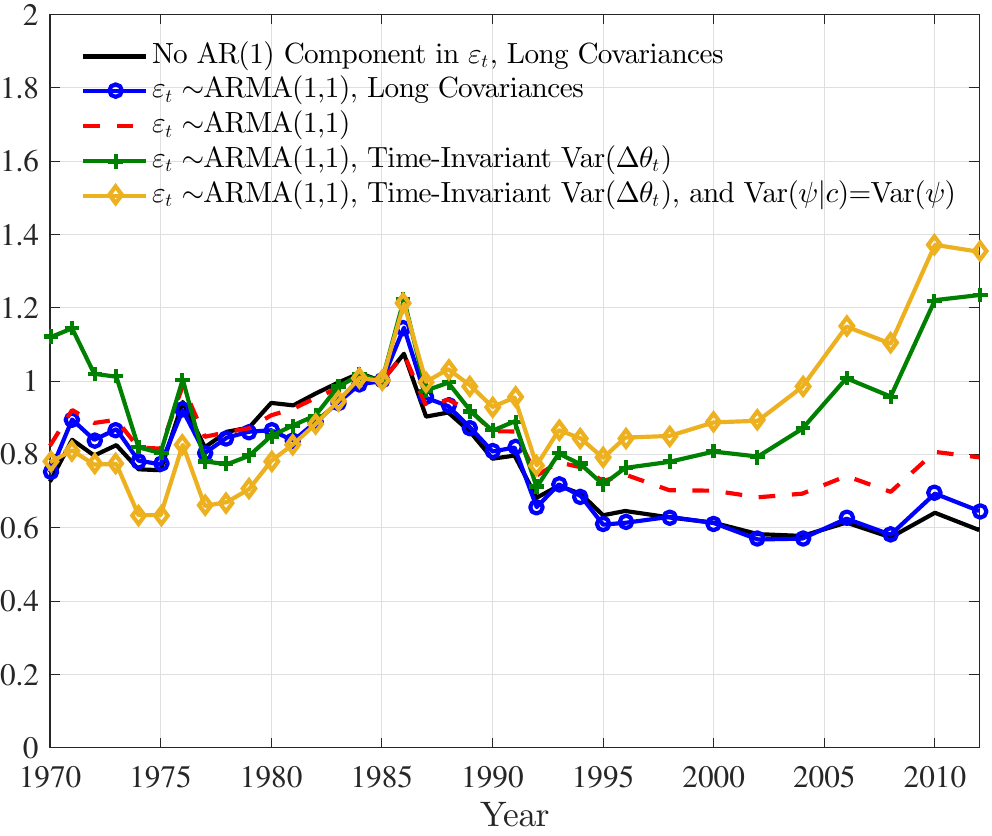}
  \caption{Estimated $\mu_t$ under Different Restrictions: All Men}
  \label{app fig: compare MG all}
\end{figure}

Since \cite{haider_2001} and \cite{moffitt_gottschalk_2012} estimate their models for all men (rather than by education), we reproduce our analysis pooling all non-college and college men. The estimated $\mu_t$ series is reported in Figure~\ref{app fig: compare MG all}.  In this case, we find that their assumptions imposing stability of initial skill distributions and of the distribution of skill growth innovations produce upward biased estimates in skill return growth beginning in the mid-1990s.  Restrictions on the short-term autocorrelation structure for non-skill shocks have relatively modest effects on their estimated $\mu_t$ series, similar to patterns observed for the non-college sample.

\clearpage
%%%%%%%%%%%%%%%%%%%%%%%%%%%%%%%%%%%%%%%%%%%%%%%%%%%%%%%%%%%%%%%%%%%%%%%%%%%%%%%%%%

\subsection{Estimation with Job Stayers in PSID} \label{app sec: job stayers}

This appendix considers log wage residuals that include firm fixed effects as in equation~\eqref{eq: AKM}. Since the PSID do not contain firm identifiers, we investigate the implications of \emph{unobserved} firm-specific heterogeneity for our IV estimator, focusing on job stayers for whom firm fixed effects do not change.  
In addition to the standard ``exogenous job mobility'' assumption of the literature (i.e., $\E[\eps_{t}|j_{\ul{t}},\ldots,j_{\ol{t}}]=0, \forall t$), we assume the following.

\begin{ass}  \label{assum: gen mu ident job stayer}
For known $k\geq 1$ and for all $t-t' \geq k+1$:
(i) $\cov(\Delta \theta_t, \theta_{t'}|j_t=j_{t-1})=0$;
(ii) $\cov(\Delta \theta_t, \eps_{t'}|j_t=j_{t-1}) = 0$;
(iii) $\cov(\eps_t, \theta_{t'}|j_t=j_{t-1})=\cov(\eps_{t-1}, \theta_{t'}|j_t=j_{t-1}) = 0$; 
(iv) $\cov(\eps_t,\eps_{t'}|j_t=j_{t-1})=\cov(\eps_{t-1},\eps_{t'}|j_t=j_{t-1})=0$; and (v) $\cov(\Delta\theta_t, \kappa_{j_{t'}}|j_t=j_{t-1})=0$.
\end{ass}
Conditions (i)--(iv) strengthen Assumption \ref{assum: gen mu ident} to condition on individuals remaining in the same job between periods $t-1$ and $t$, while condition~(v) further assumes that recent skill changes for job stayers are orthogonal to the identity of previous employers sufficiently long ago.

Under Assumption \ref{assum: gen mu ident job stayer}, the IV estimator for job stayers, for $t-t'\geq k+1$, is given by:
\begin{eqnarray}
  \frac{\cov(\Delta
  w_t,w_{t'}|j_t=j_{t-1})}{\cov(w_{t-1},w_{t'}|j_t=j_{t-1})}
  &=&\frac{\Delta\mu_t\cov(\theta_{t-1},\kappa_{j_{t'}}+\mu_{t'}\theta_{t'}|j_t=j_{t-1})}{\cov(\kappa_{j_{t-1}},\kappa_{j_{t'}}+\mu_{t'}\theta_{t'}|j_t=j_{t-1})+\cov(\mu_{t-1}\theta_{t-1},
  \kappa_{j_{t'}}+\mu_{t'}\theta_{t'}|j_t=j_{t-1})} \nonumber \\
&=&\frac{\Delta\mu_t}{\mu_{t-1}}\times\left(1+\frac{\cov(\kappa_{j_{t-1}},\kappa_{j_{t'}}+\mu_{t'}\theta_{t'}|j_t=j_{t-1})}{\cov(\mu_{t-1}\theta_{t-1},
   \kappa_{j_{t'}}+\mu_{t'}\theta_{t'}|j_t=j_{t-1})}  \right)^{-1}. \label{eq: IV for job stayers}
\end{eqnarray}
If $\cov(\kappa_{j_{t-1}},\kappa_{j_{t'}}+\mu_{t'}\theta_{t'}|j_t=j_{t-1})> 0$ and
$\cov(\mu_{t-1}\theta_{t-1}, \kappa_{j_{t'}}+\mu_{t'}\theta_{t'}|j_t=j_{t-1})>
0$, then the IV estimator is biased towards zero due to unobserved variation in $\kappa_{j_t}$. With exogenous job mobility, these conditions are equivalent to assuming that both $\kappa_{j_{t-1}}$ and $\theta_{t-1}$ are positively correlated with lagged residuals $w_{t'}$ for job stayers.
Importantly, equation~\eqref{eq: IV for job stayers} suggests that our IV estimates (reported in Figure~\ref{fig: mu job stayer}) likely under-estimate the actual decline in skill returns over time.

\subsubsection{Measuring Job Transitions in the PSID}

\begin{figure}[!htbp]
  \centering
    \includegraphics[width=0.45\columnwidth]
    {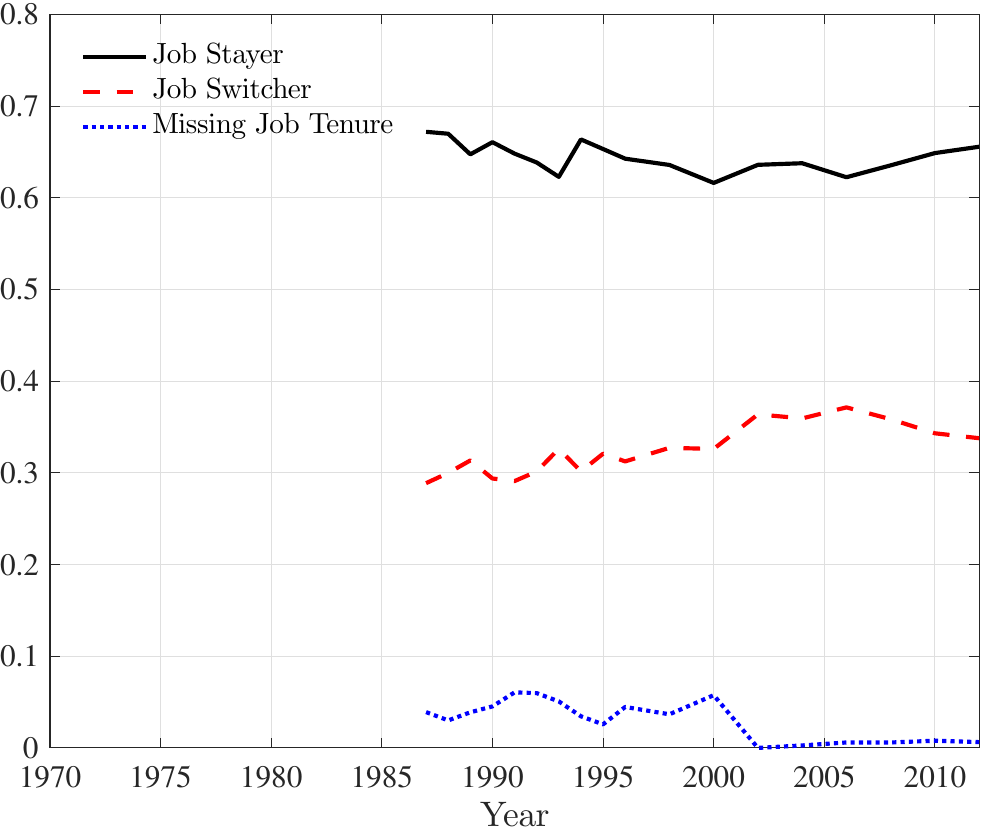}
    \caption{Job Transition Rates from $t-2$ to $t$ (PSID)}\label{app fig: job share}
  \end{figure}

We measure two-year job changes in the PSID based on the most recent start year of the current main job, which has been available since the 1988 wave.\footnote{Before 1988, respondents only report the total years of experience at the current main job, which may not be continuous.} Specifically, we define job stayers between earnings years $t-2$ and $t$ as those whose job in survey year $t+1$ (interview usually takes place between March and May) started before $t-2$. Figure~\ref{app fig: job share} shows the shares of job stayers, job switchers, and missing job start years for individuals with non-missing log wage residuals in years $t$ and $t-2$. The share of job stayers is substantial (over 60\%) and remains relatively stable over time, with a modest decline until 2000 followed by a rebound thereafter.
The drop in the share with missing job tenure between 2000 and 2002 is likely due to a modest change in survey wording about start date(s) on the current job (asking about all start/stop dates if multiple spells rather than only the most recent start date).

\subsubsection{Accounting for the Bias in Estimated Skill Returns}

As noted above, our IV estimates likely under-estimate the decline in skill returns over time.
We next show that it is possible to exploit estimates from studies using worker--firm matched data to correct for the bias associated with our IV estimator.

If  $\cov(\Delta\theta_{t''},\theta_{t'}|j_t=j_{t-1}) =\cov(\Delta\theta_{t''},\kappa_{j_{t'}}|j_t=j_{t-1})=0$ and
$\cov(\kappa_{j_{t''}}-\kappa_{j_{t''-1}},\theta_{t'}|j_t=j_{t-1}) = \cov(\kappa_{j_{t''}}-\kappa_{j_{t''-1}},\kappa_{j_{t'}}|j_t=j_{t-1})=0$ for $t''\geq t'+1$, 
then the intertemporal covariances for job stayers in the bias term in equation~\eqref{eq: IV for job stayers} can be written as within-period covariances, leading to
\begin{equation}
  \frac{\cov(\Delta
  w_t,w_{t'}|j_t=j_{t-1})}{\cov(w_{t-1},w_{t'}|j_t=j_{t-1})}
=\frac{\Delta\mu_t}{\mu_{t-1}}\times\Bigg(1+\frac{\mu_{t'}}{\mu_{t-1}}\underbrace{\frac{\var(\kappa_{j_{t'}}|j_t=j_{t-1}) + \cov(\kappa_{j_{t'}},\mu_{t'}\theta_{t'}|j_t=j_{t-1})}{\var(\mu_{t'}\theta_{t'}|j_t=j_{t-1}) + \cov(
  \kappa_{j_{t'}},\mu_{t'}\theta_{t'}|j_t=j_{t-1})}}_{\equiv \Upsilon_{t'}}
   \Bigg)^{-1}.\label{eq: simplified IV for job stayers}
\end{equation}
With our IV estimator for stayers and estimates of $\Upsilon_{t'}$, equation~\eqref{eq: simplified IV for job stayers} can be used to identify $\mu_t$ over time.  Unfortunately, the PSID do not allow us to identify the within-period covariances that make up $\Upsilon_{t'}$.  We, therefore, use estimates of the variance and covariance terms that compose $\Upsilon_{t'}$ from \cite{song_price_guvenen_bloom_von_wachter_2018}, which are based on earnings records and employer identifiers from IRS W-2 forms. Since their estimates are based on the sample of all workers, not only job stayers, this approach further assumes that $\Upsilon_{t'}$  is the same for job stayers and switchers.  The variance/covariance components from \cite{song_price_guvenen_bloom_von_wachter_2018} are presented in Table~\ref{tab: Upsilon}, which reports separate estimates over ``rolling windows'' of 7 years. When $\Upsilon_{t'}\neq 0$,  $\mu_{t'}$ is needed for earlier years to identify $\mu_t$ for later years. Since $\mu_{t'}$ prior to 1985 cannot be identified from our IV estimates, we normalize $\mu_{t'}=1$ for early years.  This is generally consistent with weak time trends for estimated returns prior to 1985 when using the full sample.

\begin{table}[htbp]
  \centering
  \caption{Within-Period Covariances (from Table III of
    \cite{song_price_guvenen_bloom_von_wachter_2018}) and Implied $\Upsilon_t$}\label{tab: Upsilon}
\begin{tabular}{cccccc}
\toprule
&1980--1986&	1987--1993&	1994--2000&	2001--2007&	2007--2013\\
\midrule
$\var(\kappa_{j_t})$ & 0.084 & 0.075 & 0.067 & 0.075 & 0.081 \\
$\cov(\kappa_{j_t},\mu_t\theta_t)$ & 0.017 & 0.029 & 0.038 & 0.047 & 0.054 \\
  $\var(\mu_t\theta_t)$ & 0.330 & 0.375 & 0.422 & 0.452 & 0.476 \\
  \midrule
Implied $\Upsilon_t$ & 0.290 & 0.257 & 0.228 & 0.244 & 0.255 \\
\bottomrule
\end{tabular}
\end{table}

Figure \ref{app fig: mu job stayer} reports estimates for $\mu_t$ that correct for bias associated with firm fixed effects using the values of $\Upsilon_{t'}$ reported in Table \ref{tab: Upsilon}.  These are based on (uncorrected) IV estimates for job stayers like those reported in Figure~\ref{fig: mu job stayer}; however, these estimates only use a single lag, $w_{t-8}$, as an instrument (rather than two lags, $w_{t-8}$ and $w_{t-10}$) to align with equation~\eqref{eq: simplified IV for job stayers}.\footnote{Note that estimates in Figure~\ref{app fig: mu job stayer} are obtained from two-year growth rates using data from every other year starting in 1986 (instead of all available years grouped in 2 or 3 years). We construct $\mu_t$ based on equation~\eqref{eq: simplified IV for job stayers} modified for two-year growth rates, the IV estimates, and $\Upsilon_{t'}$ from Table \ref{tab: Upsilon}, normalizing $\mu_{t'}=1$ for $t'\leq 1986$.}   Compared to (uncorrected) estimates reported in Figure~\ref{app fig: mu job stayer}, the (corrected) $\mu_t$ series displayed in Figure~\ref{app fig: mu job stayer} implies very little bias for college men but notably stronger bias for non-college men.

\begin{figure}[!htbp]
  \centering
    \includegraphics[width=0.45\columnwidth]{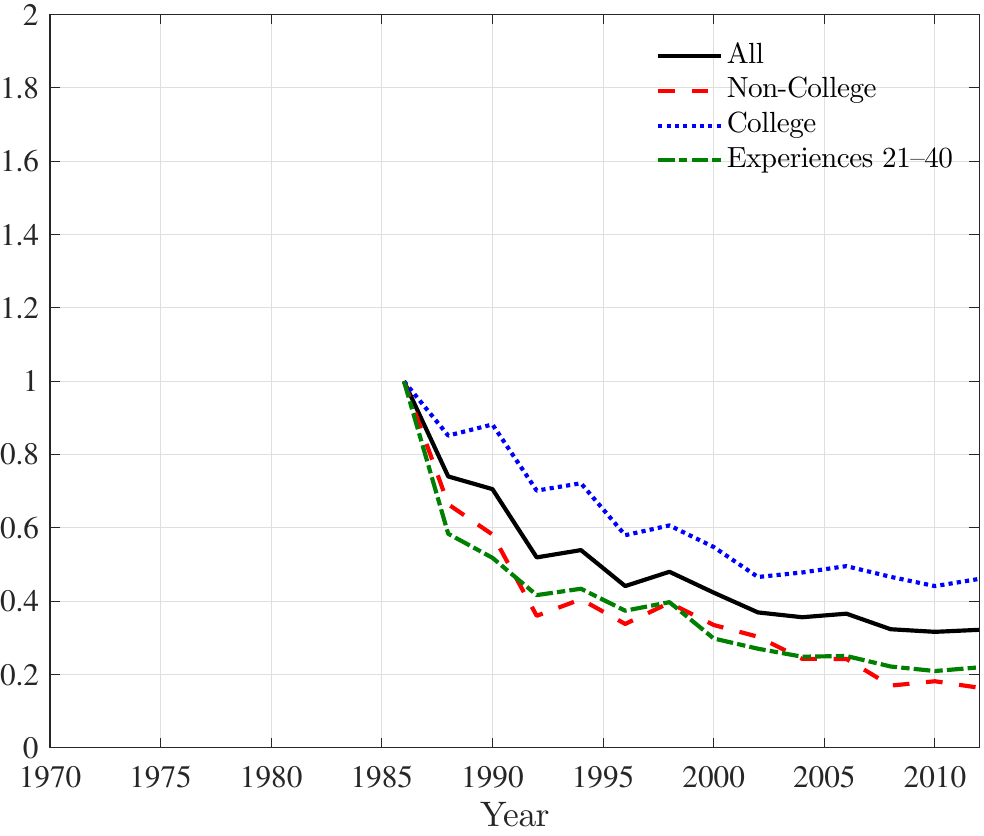}
    \caption{$\mu_t$ Corrected for Bias due to Firm Fixed Effects}\label{app fig: mu job stayer}
  \end{figure}

\if0
\begin{figure}[!htbp]
  \centering
  \subfloat[Assuming $\Upsilon_{t'}=0$]{
    \includegraphics[width=0.45\columnwidth]
    {./figures/PSID_job_stayer/mu_job_stayer}
  }\quad
  \subfloat[
  Using $\Upsilon_{t'}$ Values from Table \ref{tab: Upsilon}]{
    \includegraphics[width=0.45\columnwidth]
    {./figures/PSID_job_stayer/mu_job_stayer2}
  }  
    \caption{$\mu_t$ Implied by 2SLS for Job Stayers from $t-2$ to $t$}\label{app fig: mu job stayer}
  \end{figure}
\fi

\clearpage

%%%%%%%%%%%%%%%%%%%%%%%%%%%%%%%%%%%%%%%%%%%%%%%%%%%%%%%%%%%%%%%%%%%%%%%%%%%%%%%%%%

\subsection{Estimation with Multiple Occupations in PSID} \label{app: PSID occ}

In creating occupation codes for our sample period, we combine retrospective (1968--1980) and original (1981--2013) measures, which creates a break in occupational mobility trends (in 1981) due to lower measurement error in the retrospective measures \citep{kambourov_manovskii_2008}.  The 3-digit codes are based on the 1970 Census classification prior to 2002, while they are based on the 2000 Census classification afterwards. Therefore, we do not measure occupation changes between years 2000 and 2002.\footnote{Since we pool observations across several years (assuming constant growth of skill returns within each pooled sample) for 2SLS estimation, the change in skill return between 2000 and 2002 reflects an extrapolation from adjacent years.} We create 1- and 2-digit codes from the first and first two digits of the 3-digit codes, respectively. 

We  use the 3-digit codes to create 3 broad and exclusive occupation categories (cognitive, routine, and manual) considered by \cite{cortes_2016}.  Given small sample sizes for manual occupations in the PSID, our analysis focuses on cognitive and routine occupations.

We also estimate skill returns for those who stay in occupations with high social skill requirements, as measured by data from the Occupational Information Network (O*NET). The O*NET is a survey of U.S. workers that asks about the abilities, skills, knowledge, and work activities required in an occupation.  Following \cite{deming_2017}, we measure an occupation's social skill intensity as the average of the four items in the 1998 O*NET module on ``social skills'' (coordination, negotiation,  persuasion, and social perceptiveness). The social skill intensity measures are then assigned to individuals in the PSID sample based on their current 3-digit occupation in each year. We define social occupations as occupations that fall in the top third of the social skill intensity distribution in the pooled sample of worker-year observations. As noted by \cite{deming_2017}, cognitive occupations are also very likely to be social occupations. Among worker-year observations in cognitive occupations, around 59\% are in social occupations. Conversely, around 76\% of observations in social occupations are also in cognitive occupations.

\subsubsection{GMM Estimation using Occupation Stayers and Switchers} \label{app: GMM occ}

We estimate occupation-specific $\gamma_t^o$ and $\mu_t^o$ for routine and cognitive occupations (normalizing $\gamma_{1985}^{o}=0$ and $\mu_{1985}^{o}=1$ for routine occupations) using optimal two-step GMM.  Because we use the PSID data, we use the following moments based on equation~\eqref{eq: occ 2SLS regression}:
\[
    \E\left[\bm z_{t} \left\{\Delta_2 w_{t}-\left(\gamma_t^{o_t}-\frac{\mu_t^{o_t}}{\mu_{t-2}^{o_{t-2}}}\gamma_{t-2}^{o_{t-2}}\right)-\left(\frac{\mu_t^{o_t}}{\mu_{t-2}^{o_{t-2}}}-1\right)w_{t-2}\right\}\Big|o_t,o_{t-2}\right]=\bm 0,\quad \forall (t,o_t,o_{t-2}),
\]
where $\bm z_{i,t}=(1, w_{i,t-8},w_{i,t-9})^\top$ (or $(1, w_{i,t-8},w_{i,t-10})^\top$ in later sample years).
We use linear splines for $\gamma_t^o$ and $\mu_t^o$, each with 14 knots in $t$, restricting to moment conditions with at least 20 observations (9 switcher moments are dropped).
Altogether, there are 54 parameters with 303 moment conditions. 

We also estimate the model imposing equal skill returns, $\mu_t^{\text{routine}}=\mu_t^{\text{cognitive}} = \mu_t$ for all $t$.  The estimated $\mu_t$ and $\gamma_t^o$ series are nearly identical to their counterparts reported in Figure~\ref{fig: GMM occ est mu gamma}, while the $J$-statistic comparing the unrestricted and restricted criterion functions equals 20.08 and is distributed $\chi^2_{14}$ under the null hypothesis of equal skill returns.  Thus, we cannot reject the null of identical returns at conventional levels ($p$-value $= 0.128$).

\if0
    \begin{itemize}
    \item The number of parameters reduces to 41 when imposing identical
    growth rate of $\mu_t^o$ for both occupations ($J$: 292.97). The
    difference in $J$-statistic (18.10) follows $\chi^2_{13}$ under
    the null hypothesis of identical growth. % with 95\%
    % critical value 22.362
    The $p$-value is 0.154, so we cannot reject the null
    hypothesis at any conventional levels.
    \item The number of parameters reduces to 40 when imposing identical
     $\mu_t^o$ for both occupations ($J$: 294.948). The
    difference in $J$-statistic (20.08) follows $\chi^2_{14}$ under
    the null hypothesis of identical $\mu_t^o$.
    The $p$-value is 0.128, so we cannot reject the null
    hypothesis at any conventional levels.
    \item When using experienced workers only (21+ in $t$), the number
    of moments reduces to 261.
    \end{itemize}
\fi

\begin{figure}[!htbp]
  \centering
    \subfloat[$\gamma_t^o$]{
      \includegraphics[width=0.45\columnwidth]{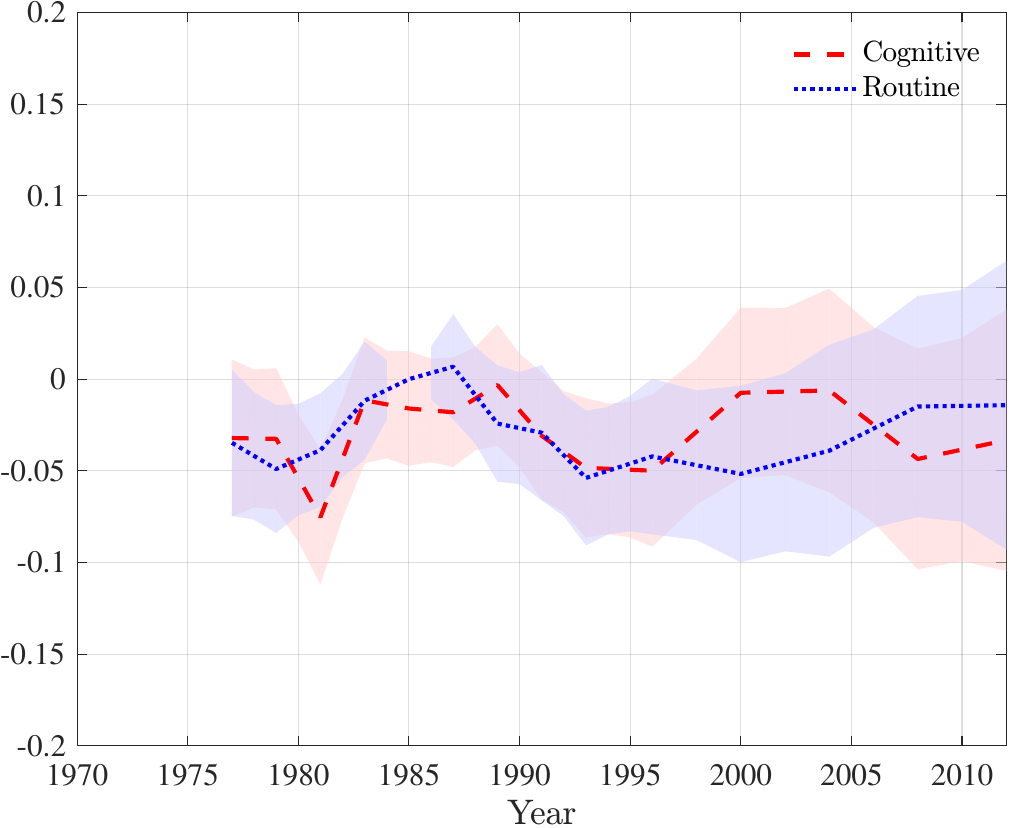}
}\quad
    \subfloat[ $\E{[\theta_t|o_t]}$]{
      \includegraphics[width=0.45\columnwidth]{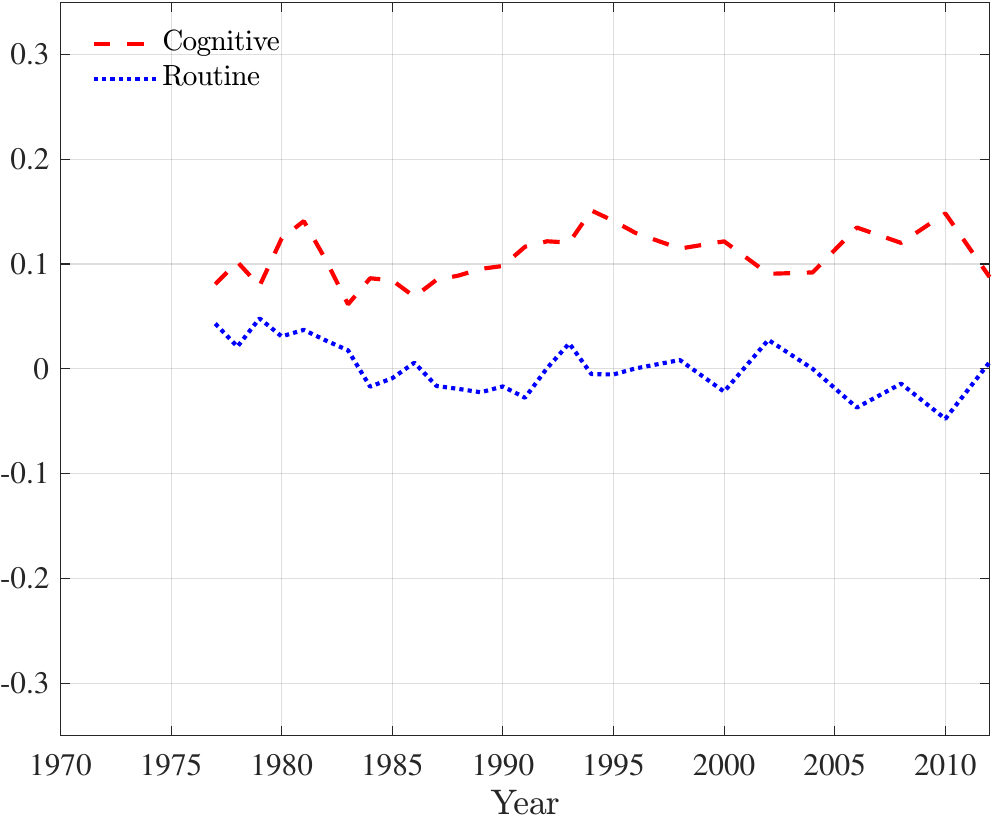}
    }
\caption{GMM estimates imposing time-invariant $\mu_t^o$}
\label{fig: GMM occ est constant mu}
\caption*{\footnotesize \it Notes: Imposing $\mu^{\text{routine}}=1$, $\mu^{\text{cognitive}}$ is estimated to be 0.946 (SE$=$0.026).}
\end{figure}

For comparison with \cite{cortes_2016}, Figure~\ref{fig: GMM occ est constant mu} reports estimates of $\gamma_t^o$ and $\E[\theta_t|o_t]$ when imposing time-invariant $\mu_t^o=\mu^o$.  Estimated time profiles for $\gamma_t^o$ and $\E[\theta_t|o_t]$ are notably flatter than their counterparts allowing for variation in skill returns (see Figures~\ref{fig: GMM occ est mu gamma}(b) and \ref{fig: GMM occ est avg w theta}(b)).

\begin{figure}[!htbp]
  \centering
    \subfloat[$\mu_t^o$]{
      \includegraphics[width=0.45\columnwidth]{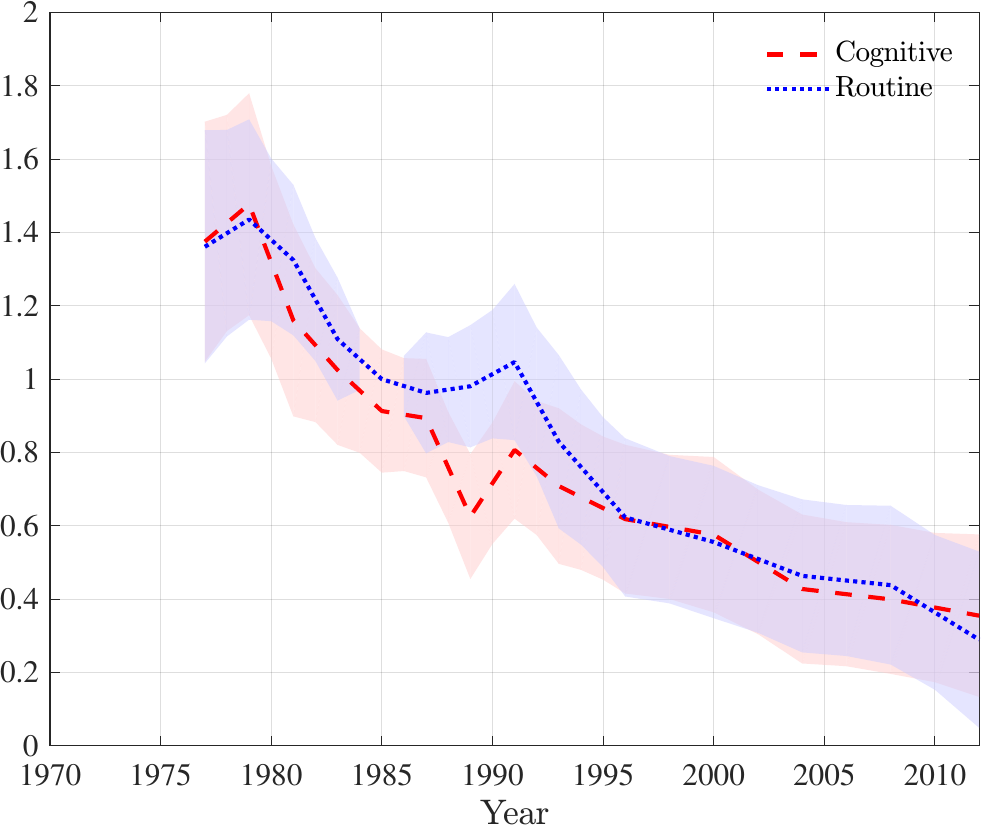}
}\quad
    \subfloat[$\gamma_t^o$]{
      \includegraphics[width=0.45\columnwidth]{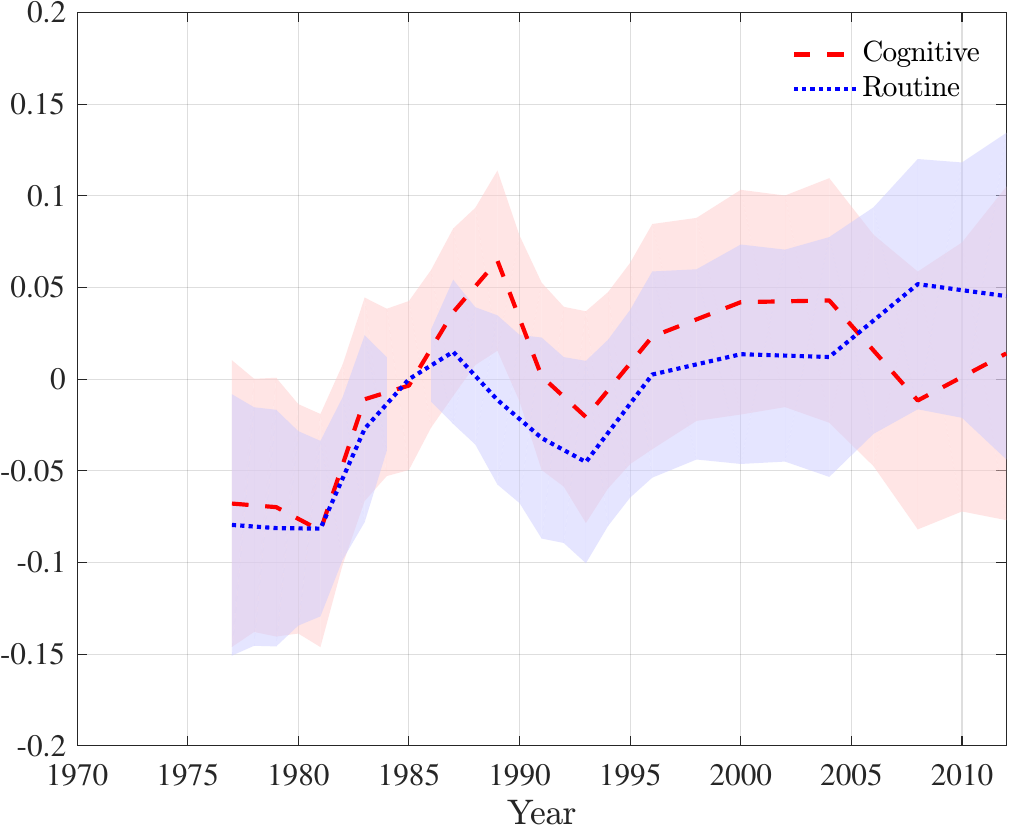}
    }
\caption{GMM estimates of $\mu_t^o$ and $\gamma_t^o$, 21--40 years of experience in $t$}
\label{fig: GMM occ est mu gamma OLD}
\end{figure}

\begin{figure}[!htbp]
  \centering
    \subfloat[$\E{[w_t|o_t]}$]{
      \includegraphics[width=0.45\columnwidth]{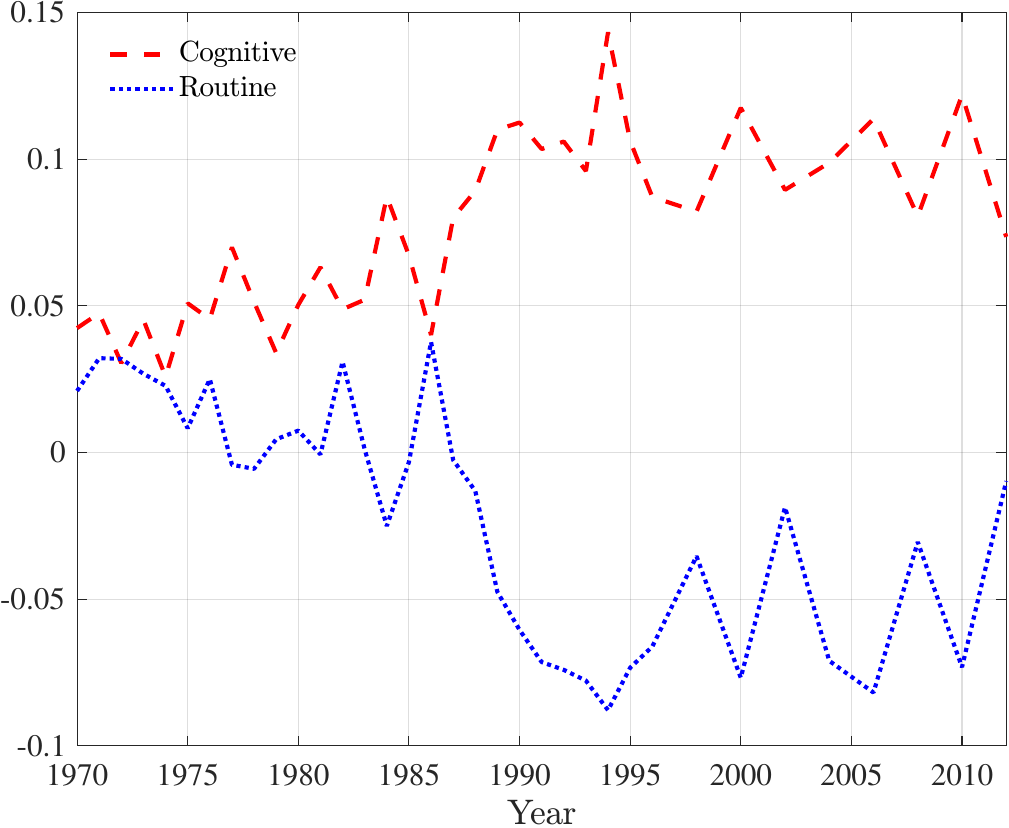}
}\quad
    \subfloat[ $\E{[\theta_t|o_t]}$]{
      \includegraphics[width=0.45\columnwidth]{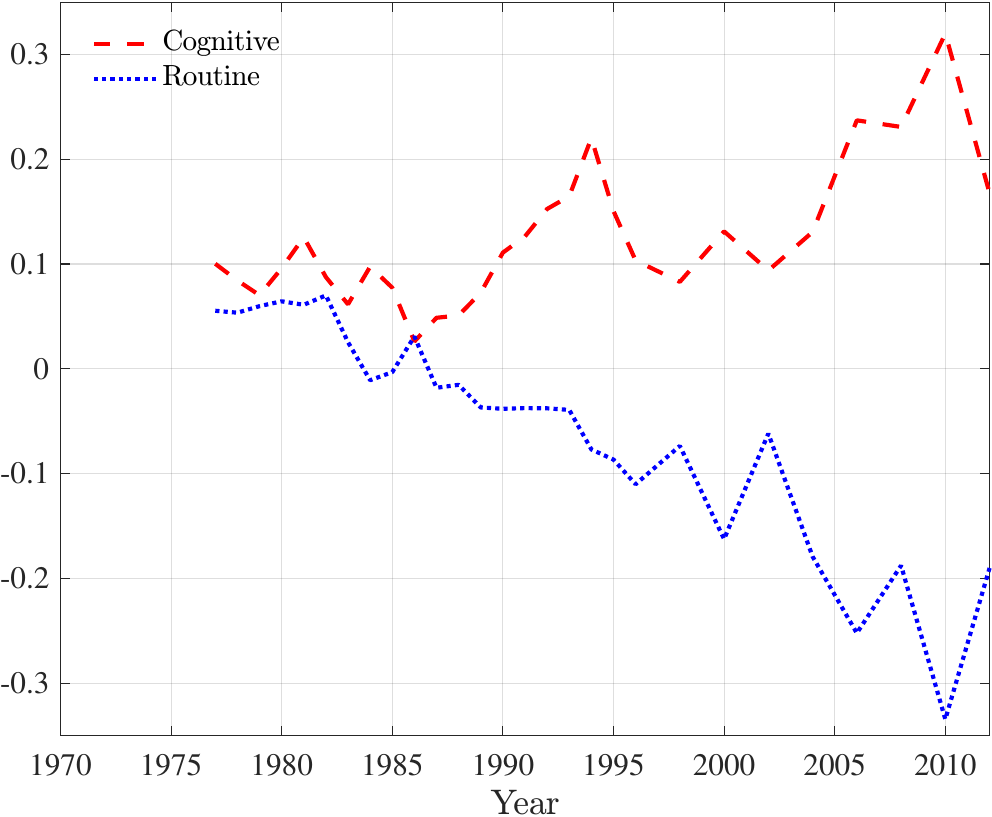}
    }
\caption{Average log wage residual and skill by occupation, 21--40 years of experience in $t$}
\label{fig: GMM occ est avg w theta OLD}
\end{figure}

Finally, we use the same estimation strategy, restricting the sample to men who had 21--40 years of experience in year $t$.  Due to fewer observations, this reduces the number of moments used in estimation to 261.  These results are reported in Figures~\ref{fig: GMM occ est mu gamma OLD} and \ref{fig: GMM occ est avg w theta OLD}.

\subsubsection{2SLS Estimated Returns for Occupational Stayers}  \label{app: 2SLS occ}

Tables~\ref{tab: 2SLS stayers 1} and \ref{tab: 2SLS stayers 2} report 2SLS estimates (and standard errors) for skill return growth underlying Figures~\ref{fig: PSID mu occ stayers 2SLS} and \ref{fig: PSID mu occ stayers} in the main text.  First-stage $F$-statistics for the instruments are also reported.

\begin{figure}[h]
    \centering
    \subfloat[Constant Occupation between $t-2$ and $t$]{
      \includegraphics[width=0.42\columnwidth]{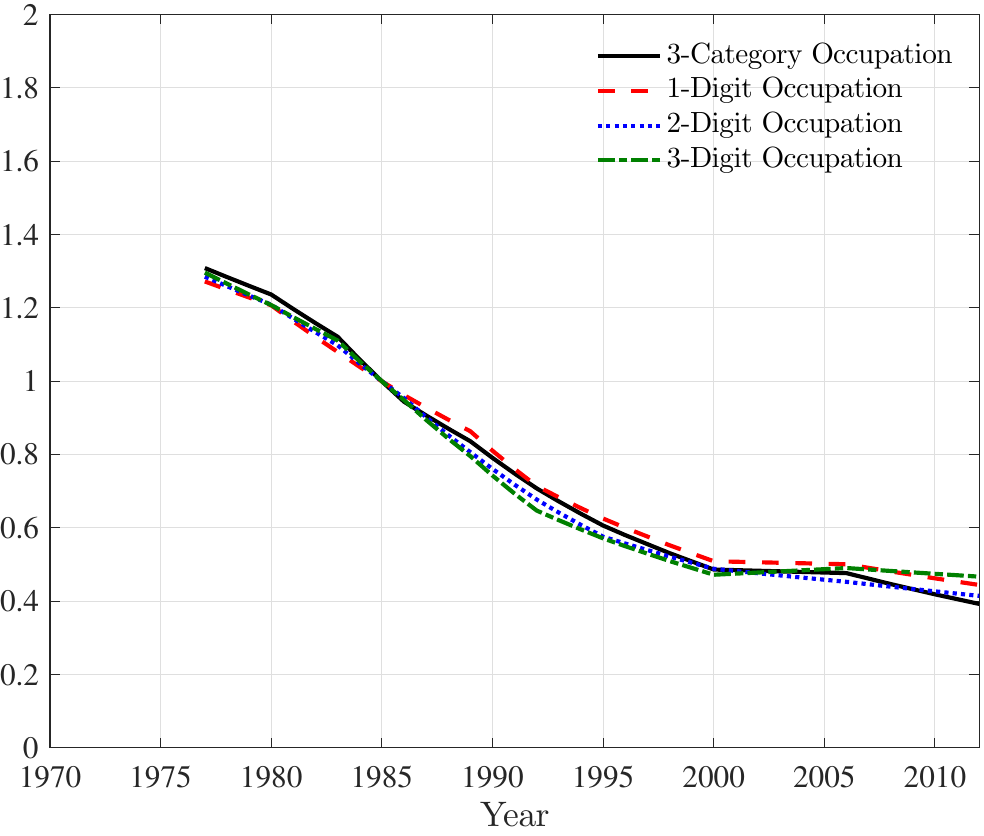}
    }\quad
    \subfloat[Constant Occupation between $t-2$ and $t$: \\ 21--40 Years of Experience]{
      \includegraphics[width=0.42\columnwidth]{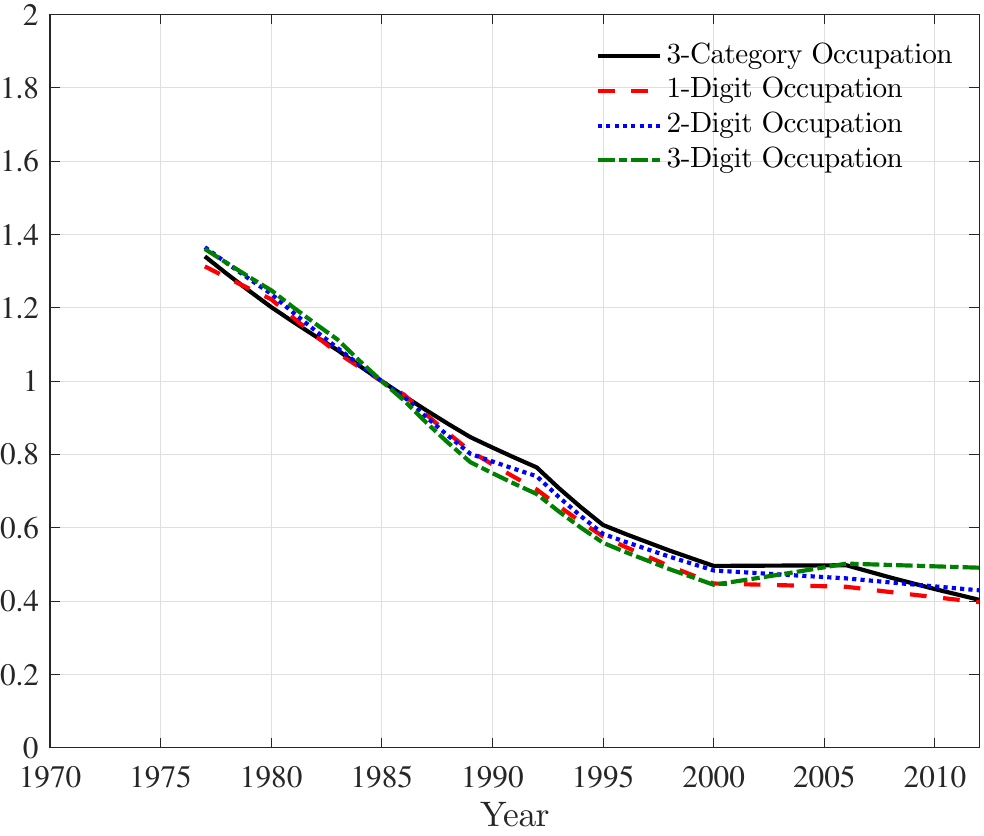}
    }\\
    \subfloat[Constant Occupation between $t-2$ and $t$: Non-College]{
      \includegraphics[width=0.42\columnwidth]{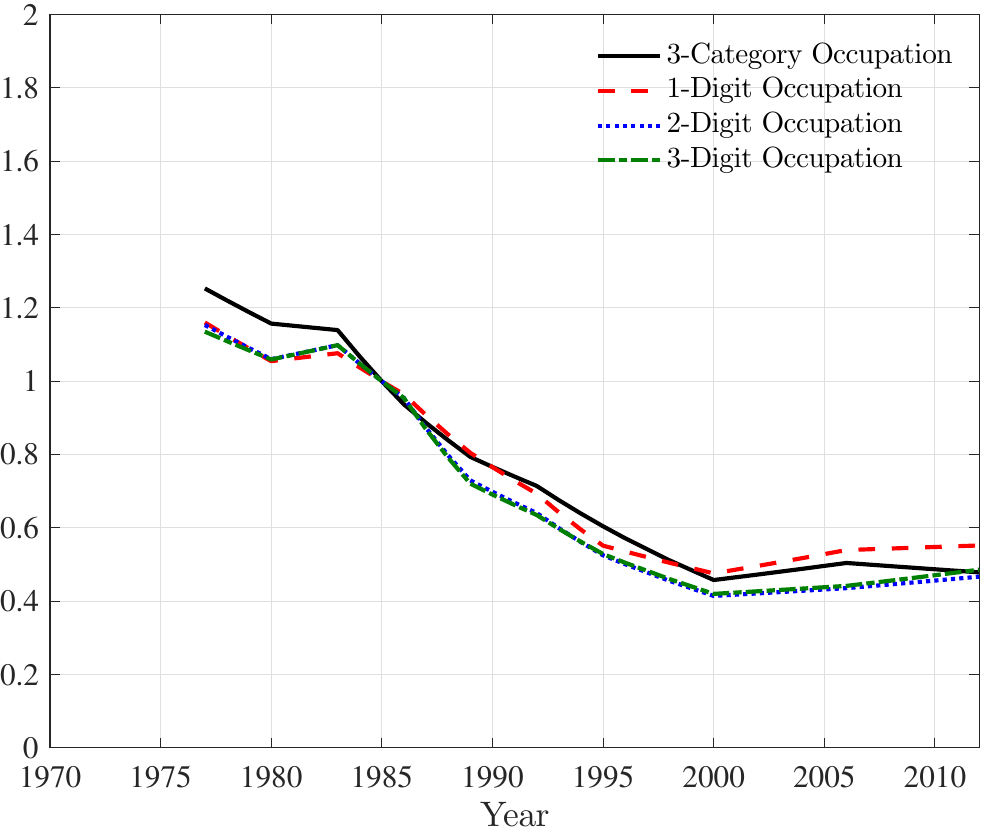}
    }\quad
    \subfloat[Constant Occupation between $t-2$ and $t$: College]{
      \includegraphics[width=0.42\columnwidth]{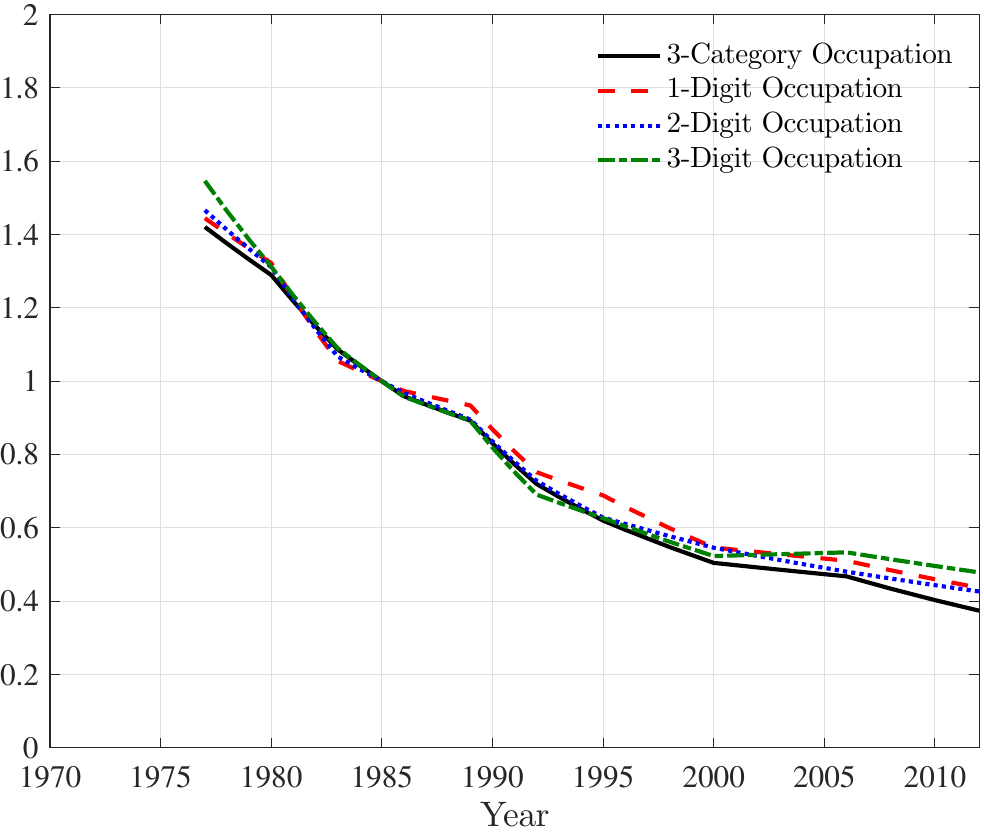}
    }\
\caption{$\mu_t$ Implied by 2SLS Estimates for Occupational Stayers from $t-2$ to $t$} 
\label{app fig: PSID returns occ}
\end{figure}

Figure~\ref{app fig: PSID returns occ} reports 2SLS estimates based on occupation-stayers in years $t-2$ to $t$ based on different occupation classifications.\footnote{These estimates are based on the same subperiods as reported in Tables~\ref{tab: 2SLS stayers 1} and \ref{tab: 2SLS stayers 2}.  The 3-category occupation estimates reflect those remaining within occupations classified as either cognitive, routine, or manual based on \cite{cortes_2016}.}  In all cases, $w_{i,t-8}$ and $w_{i,t-9}$ (or $w_{i,t-10}$) residuals are used as instruments.  Estimated skill return patterns are very similar regardless of how narrowly we define occupations. %\tcr{[Should we drop the following?:]} Figure~\ref{app fig: PSID returns init occ} further shows estimated skill returns conditioning on occupation in year $t-8$, matching the year used to create the instrument $w_{t-8}$. 

\begin{figure}[!htbp]
    \centering
    \subfloat[Same Occupation in $t'$ ($o_t=o_{t-2}=o_{t'}$)]{
    \includegraphics[width=0.45\columnwidth]{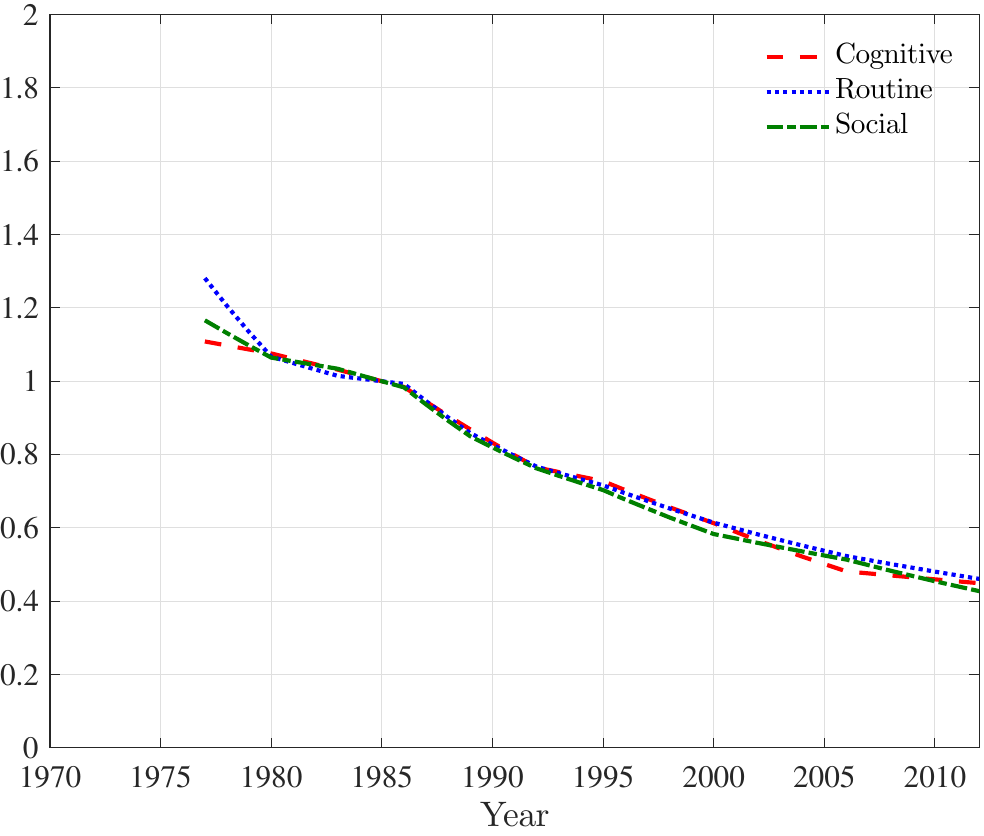}
    }\quad
    \subfloat[Different Occupation in $t'$ ($o_t=o_{t-2}\neq o_{t'}$)]{
    \includegraphics[width=0.45\columnwidth]{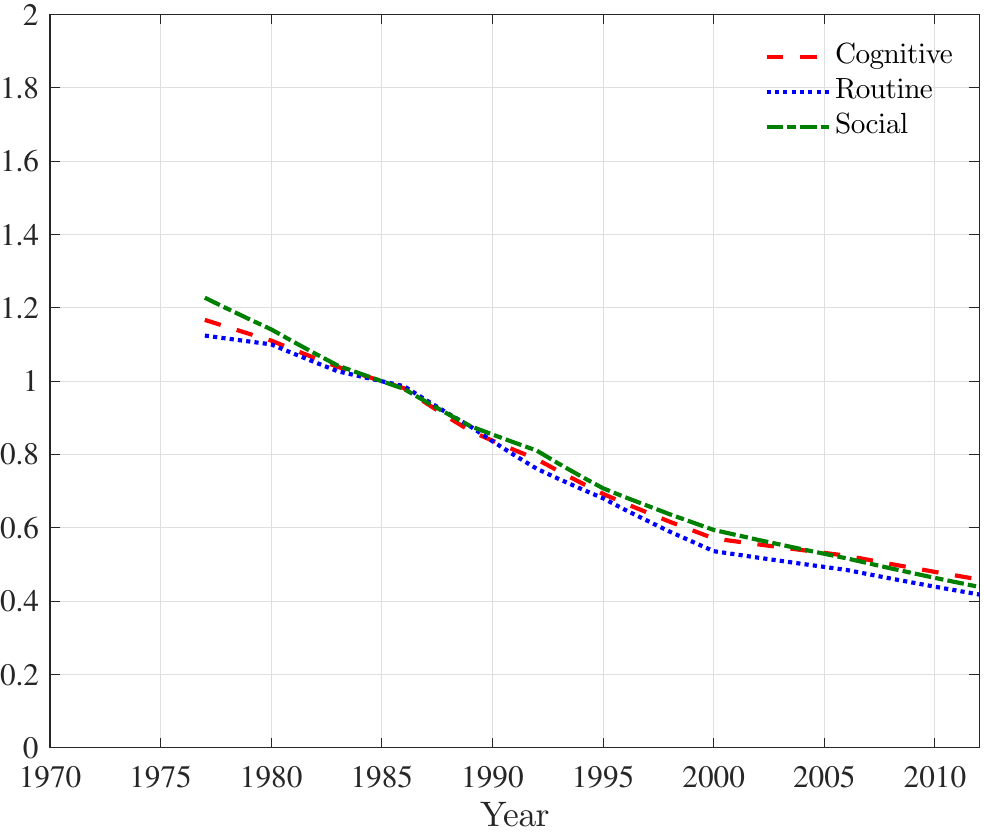}
    }    
\caption{$\mu_t^o / \mu_{1985}^o$ implied by 2SLS estimates for occupation-stayers: All experience levels}
\label{fig: PSID mu occ stayers by past occ 2SLS}
\end{figure}

\begin{figure}[!htbp]
    \centering
    \subfloat[Same Occupation in $t'$ ($o_t=o_{t-2}=o_{t'}$)]{
    \includegraphics[width=0.45\columnwidth]{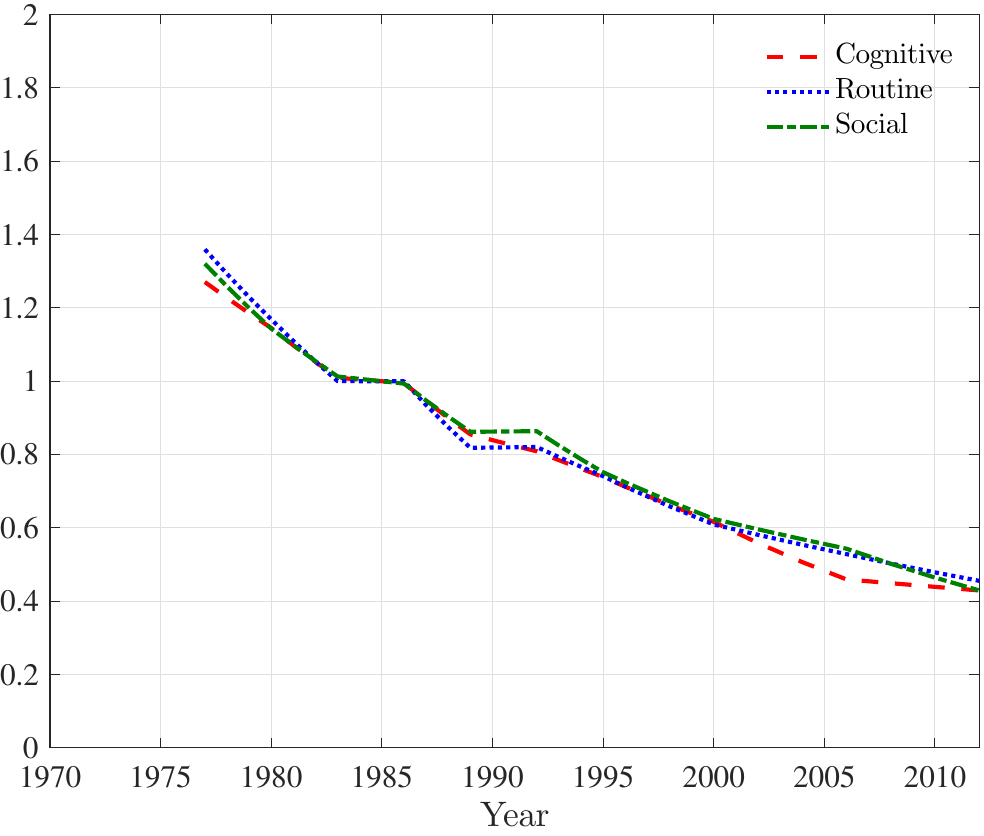}
    }\quad
    \subfloat[Different Occupation in $t'$ ($o_t=o_{t-2}\neq o_{t'}$)]{
    \includegraphics[width=0.45\columnwidth]{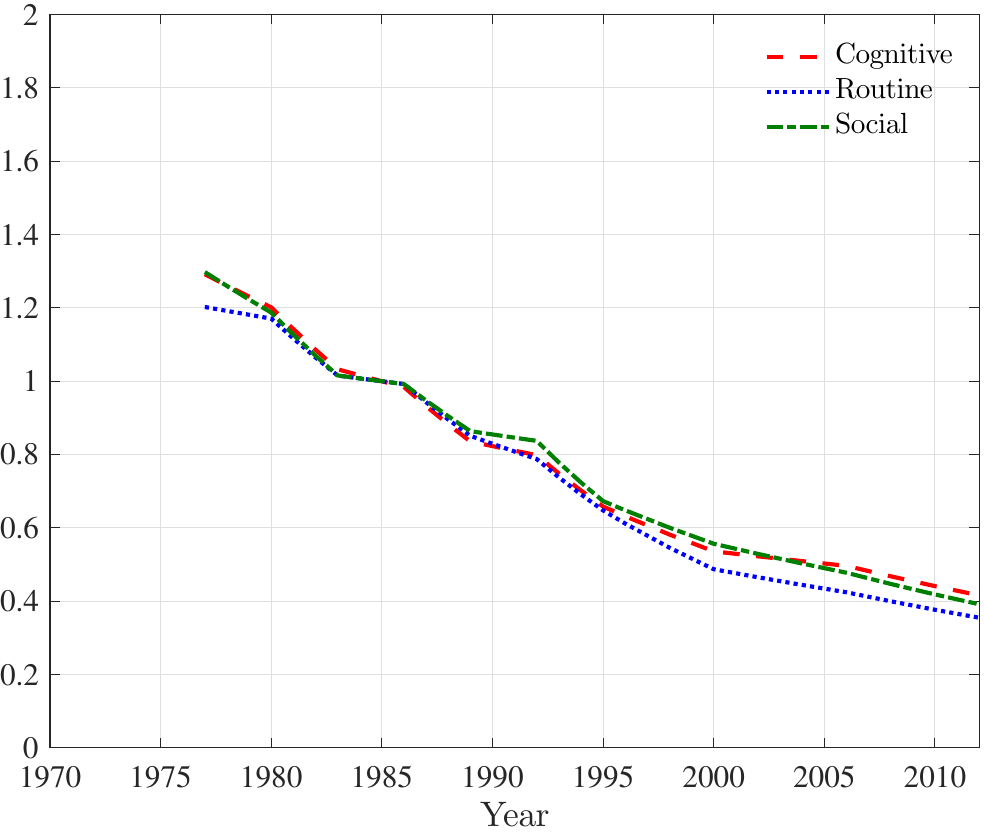}
    }    
\caption{$\mu_t^o / \mu_{1985}^o$ implied by 2SLS estimates for occupation-stayers: 21--40 years of experience}
\label{fig: PSID mu occ stayers by past occ old 2SLS}
\end{figure}

As noted in the text, estimated return growth for stayers in occupation $o_t=o_{t-1}=o$ should not depend on earlier occupation ($o_{t'}$) under Assumption~\ref{assum: mu ident occ}.
Estimates reported in Figures~\ref{fig: PSID mu occ stayers by past occ 2SLS} and \ref{fig: PSID mu occ stayers by past occ old 2SLS} indicate very similar estimated skill return profiles for occupation stayers with $o_{t'}=o$ vs.\ $o_{t'} \neq o$.

\begin{table}\small
\centering
\caption{2SLS Estimates for Occupational Stayers, 1979--1995}
\label{tab: 2SLS stayers 1}
{
\def\sym#1{\ifmmode^{#1}\else\(^{#1}\)\fi}
\begin{tabular}{l*{6}{c}}
\toprule
                    &\multicolumn{1}{c}{1979--1980}&\multicolumn{1}{c}{1981--1983}&\multicolumn{1}{c}{1984--1986}&\multicolumn{1}{c}{1987--1989}&\multicolumn{1}{c}{1990--1992}&\multicolumn{1}{c}{1993--1995}\\
\midrule
\multicolumn{7}{c}{\underline{A. Non-College Men}}\\
$\Delta_2 \mu_t/\mu_{t-2}$&      -0.045       &       0.024       &      -0.090       &      -0.171\sym{*}&      -0.081       &      -0.115       \\
                    &     (0.060)       &     (0.064)       &     (0.060)       &     (0.053)       &     (0.059)       &     (0.075)       \\
                    \addlinespace
Observations       &\multicolumn{1}{c}{509}       &\multicolumn{1}{c}{598}       &\multicolumn{1}{c}{548}       &\multicolumn{1}{c}{519}       &\multicolumn{1}{c}{511}       &\multicolumn{1}{c}{423}       \\
1st stage \(F\)-Statistic&      62.246       &      66.535       &      95.190       &      54.104       &      65.388       &      65.562       \\
\\
\midrule
\multicolumn{7}{c}{\underline{B. College Men}}\\
$\Delta_2 \mu_t/\mu_{t-2}$&      -0.104       &      -0.116       &      -0.083       &      -0.046       &      -0.157\sym{*}&      -0.063       \\
                    &     (0.058)       &     (0.065)       &     (0.058)       &     (0.055)       &     (0.046)       &     (0.064)       \\
                    \addlinespace                    
Observations       &\multicolumn{1}{c}{369}       &\multicolumn{1}{c}{511}       &\multicolumn{1}{c}{591}       &\multicolumn{1}{c}{694}       &\multicolumn{1}{c}{665}       &\multicolumn{1}{c}{731}       \\
1st stage \(F\)-Statistic&      88.830       &      76.805       &      86.170       &      55.780       &      95.572       &     132.245       \\
\\
\midrule
\multicolumn{7}{c}{\underline{C. Cognitive Occupations}}\\
$\Delta_2 \mu_t/\mu_{t-2}$&      -0.084       &      -0.075       &      -0.107\sym{*}&      -0.097       &      -0.136\sym{*}&      -0.102       \\
                    &     (0.071)       &     (0.062)       &     (0.053)       &     (0.050)       &     (0.043)       &     (0.056)       \\
                    \addlinespace
Observations        &\multicolumn{1}{c}{506}       &\multicolumn{1}{c}{776}       &\multicolumn{1}{c}{869}       &\multicolumn{1}{c}{905}       &\multicolumn{1}{c}{895}       &\multicolumn{1}{c}{914}       \\
1st stage \(F\)-Statistic&      92.258       &      75.839       &      37.259       &      74.934       &     123.756       &     144.648       \\
\\
\midrule
\multicolumn{7}{c}{\underline{D. Routine Occupations}}\\
$\Delta_2 \mu_t/\mu_{t-2}$&       0.020       &      -0.059       &      -0.129\sym{*}&      -0.073       &      -0.084       &      -0.107       \\
                    &     (0.070)       &     (0.049)       &     (0.047)       &     (0.048)       &     (0.065)       &     (0.056)       \\
                                        \addlinespace
Observations        &\multicolumn{1}{c}{648}       &\multicolumn{1}{c}{915}       &\multicolumn{1}{c}{908}       &\multicolumn{1}{c}{929}       &\multicolumn{1}{c}{899}       &\multicolumn{1}{c}{801}       \\
1st stage \(F\)-Statistic&      81.071       &      85.482       &      85.385       &     108.442       &      71.017       &     101.606       \\
\\
\midrule
\multicolumn{7}{c}{\underline{E. Social Occupations}}\\
$\Delta_2 \mu_t/\mu_{t-2}$&       0.062       &      -0.067       &      -0.108       &      -0.083       &      -0.202\sym{*}&      -0.048       \\
                    &     (0.097)       &     (0.069)       &     (0.068)       &     (0.068)       &     (0.054)       &     (0.085)       \\
                                        \addlinespace
Observations        &\multicolumn{1}{c}{374}       &\multicolumn{1}{c}{525}       &\multicolumn{1}{c}{573}       &\multicolumn{1}{c}{594}       &\multicolumn{1}{c}{598}       &\multicolumn{1}{c}{589}       \\
1st stage \(F\)-Statistic&      55.417       &      81.778       &      85.375       &      48.250       &      78.289       &      72.354       \\
\bottomrule
\multicolumn{7}{l}{Notes: Estimates from 2SLS regression of $\Delta_2 w_{i,t}$ on $w_{i,t-2}$ using instruments $(w_{i,t-8}, w_{i,t-9})$.  }\\
\multicolumn{7}{l}{$^*$ denotes significance at 0.05 level. }
\end{tabular}

}
\end{table}

\begin{table}\small
\centering
\caption{2SLS Estimates for Occupational Stayers, 1996--2012}
\label{tab: 2SLS stayers 2}
{
\def\sym#1{\ifmmode^{#1}\else\(^{#1}\)\fi}
\begin{tabular}{l*{3}{c}}
\toprule
                    &\multicolumn{1}{c}{1996--2000}&\multicolumn{1}{c}{2002--2006}&\multicolumn{1}{c}{2008--2012}\\
\midrule
\multicolumn{4}{c}{\underline{A. Non-College Men}}\\
$\Delta_2 \mu_t/\mu_{t-2}$&      -0.088       &       0.017       &       0.032       \\
                    &     (0.051)       &     (0.079)       &     (0.082)       \\
                    \addlinespace                    
Observations        &\multicolumn{1}{c}{407}       &\multicolumn{1}{c}{252}       &\multicolumn{1}{c}{325}       \\
1st stage \(F\)-Statistic&      68.156       &      30.257       &     105.090       \\
\\
\midrule
\multicolumn{4}{c}{\underline{B. College Men}}\\
$\Delta_2 \mu_t/\mu_{t-2}$&      -0.070       &       0.006       &      -0.035       \\
                    &     (0.045)       &     (0.049)       &     (0.042)       \\
                    \addlinespace                    
Observations        &\multicolumn{1}{c}{706}       &\multicolumn{1}{c}{443}       &\multicolumn{1}{c}{662}       \\
1st stage \(F\)-Statistic&     155.501       &      99.896       &     133.535       \\
\\
\midrule
\multicolumn{4}{c}{\underline{C. Cognitive Occupations}}\\
$\Delta_2 \mu_t/\mu_{t-2}$&      -0.110\sym{*}&      -0.002       &      -0.097\sym{*}\\
                    &     (0.038)       &     (0.052)       &     (0.041)       \\
                    \addlinespace                    
Observations        &\multicolumn{1}{c}{882}       &\multicolumn{1}{c}{836}       &\multicolumn{1}{c}{794}       \\
1st stage \(F\)-Statistic&     153.995       &     127.048       &     127.603       \\
\\
\midrule
\multicolumn{4}{c}{\underline{D. Routine Occuaptions}}\\
$\Delta_2 \mu_t/\mu_{t-2}$&      -0.068       &      -0.018       &      -0.078       \\
                    &     (0.049)       &     (0.039)       &     (0.054)       \\
                    \addlinespace                    
Observations        &\multicolumn{1}{c}{810}       &\multicolumn{1}{c}{801}       &\multicolumn{1}{c}{707}       \\
1st stage \(F\)-Statistic&     112.638       &     145.430       &     124.283       \\
\\
\midrule
\multicolumn{4}{c}{\underline{E. Social Occupations}}\\
$\Delta_2 \mu_t/\mu_{t-2}$&      -0.122\sym{*}&      -0.010       &      -0.038       \\
                    &     (0.049)       &     (0.043)       &     (0.042)       \\
                                        \addlinespace
Observations        &\multicolumn{1}{c}{580}       &\multicolumn{1}{c}{606}       &\multicolumn{1}{c}{587}       \\
1st stage \(F\)-Statistic&      94.614       &     123.053       &      93.845       \\
\bottomrule
\multicolumn{4}{l}{Notes: Estimates from 2SLS regression of $\Delta_2 w_{i,t}$ on}\\
\multicolumn{4}{l}{ $w_{i,t-2}$ using instruments $(w_{t-8},w_{t-9})$ for 1996--2000 and}\\
\multicolumn{4}{l}{ $(w_{t-8},w_{t-10})$ for 2002--2006 and 2008--2012. $^*$ denotes } \\
\multicolumn{4}{l}{significance at 0.05 level. }
\end{tabular}
}
\end{table}

\iffalse
\begin{figure}[h]
    \centering
      \includegraphics[width=0.42\columnwidth]{./figures/PSID_occupation/figures/mu_initial_occ}
\caption{$\mu_t$ Implied by 2SLS Estimates by Initial Occupation (in $t-8$)} 
\label{app fig: PSID returns init occ}
\end{figure}
\fi

\clearpage

%%%% Re-Number Tables and figures for the Appendix %%%%%%%%
\renewcommand{\thetable}{\thesection-\arabic{table}}
\setcounter{table}{0}

\renewcommand{\thefigure}{\thesection-\arabic{figure}}
\setcounter{figure}{0}

%%%%%%%%%%%%%%%%%%%%%%%%%%%%%%%%%%%%%%%%%%%%%%%%%%%%%%%%%%%%%%%%%%%%%%%%%%%%%%%%%%

\section{HRS Data and Results} \label{app: HRS}

\subsection{Data Description}

We use data from the Health and Retirement Study (HRS), a national U.S.\ panel survey of individuals over age 50 and their spouses.\footnote{More precisely, the sample does include some individuals age 50. For example, someone from the original cohort (born in 1931-1941) who was born late in 1941 may have been age 50 at the date of their first interview in 1992 if they were interviewed earlier in the calendar year.} We use data from six cohorts incorporated over time, beginning with the first cohort surveyed in 1992. (New cohorts of individuals were added in 1998, 2004, 2010, and 2016.)\footnote{The HRS sample was built up over time. The initial cohort consisted of persons born between 1931 and 1941 (aged 51 to 61 at first interview in 1992). The Asset and Health Dynamics Among the Oldest Old (AHEAD) cohort, born before 1924, was added in 1993.  Given the ages of respondents from this cohort (over 70 by 1994), it is excluded from our analysis. In 1998, two new cohorts were enrolled: the Children of the Depression (CODA) cohort, born 1924 to 1930, and the War Baby (WB) cohort, born 1942 to 1947. Early Baby Boomer (EBB, born 1948 to 1953) cohort was added in 2004, Mid Baby Boomer (MBB, born 1954 to 1959) cohort was added in 2010, and Late Baby Boomer (LBB, born 1960 to 1965) cohort was added in 2016. In addition to respondents from eligible birth years, the survey interviewed the spouses of married respondents or the partner of a respondent, regardless of age.} The survey has been fielded every two years since 1992, and it provides information about demographics, income, and cognition, making it ideal data for the purpose of our study. Because one of the cognitive tests (word recall) in 1992 and 1994 differs from that of later years, we use data collected from 1996 to 2018.\footnote{The word recall test contained a list of 20 words in 1992 and 1994, while it was reduced to 10 words in later years.}

The HRS records the respondent's and spouse's wage rates if they are working at the time of the interview. We use the hourly wage rate, deflating nominal values to 1996 dollars using the Consumer Price Index.\footnote{\url{https://www.bls.gov/cpi/research-series/home.htm\#CPI-U-RS20Data}} The HRS also provides various cognitive functioning measures.  We use word recall in our analysis, but report below on its correlation with two other measures available in several years: serial 7's and quantitative reasoning. Table \ref{tab:measure} provides a brief summary of these measures. The word recall test evaluates the memory of the respondents by reading a list of 10 words and asking them to recall immediately (immediate recall) and after a delay of about 5 minutes (delayed recall). We sum the number of words the respondent recalled in the two tasks and obtain a score of 21 different values. The serial 7's test asks the respondent to subtract 7 from the previous number, starting with 100 for five trials. This test score is the number of trials that the respondent answered correctly, and it has 6 different values. Quantitative reasoning consists of three simple arithmetic questions assessing the numeracy of the respondent. We construct a test score based on the answers and the resulting score ranges from 0 to 4.
Additional details about these measures are provided below.

\begin{table}[h]
%\small
\centering
\caption{Summary of Cognitive Measures}
\label{tab:measure}
\begin{tabular}{cccc}
\toprule
 & Meant to measure & Number of values & Available years \\
\midrule

Word recall & Memory & 21 (0--20) & 1996--2018\\
Serial 7's & Numeracy &  6 (0--5) & 1996--2018\\
Quantitative reasoning & Numeracy &  5 (0--4) & 2002--2018 \\
%Retrieval fluency & Fluency & 91 (0-90)  & 2010-2016\\

\bottomrule
\end{tabular}
\end{table}

Our sample is restricted to age-eligible (i.e.\ born in eligible years when first interviewed) men. We use observations when men are ages 50--70 if their potential labor market experience is between 30 and 50 years.\footnote{We use age recorded at the end of the interview (sometimes interviews occur over multiple dates). Potential experience is defined as age minus 6 minus years of schooling.} 
%We trim the top and bottom 1\% of all wages within year by college- vs.\ non-college-educated status and 10-year experience cells. [DIDN'T DO] 
In estimation, we use non-imputed wages and cognitive measures only. The sample contains 10,151 individuals and 43,096 person-year observations.

Our sample consists of 70\% white, 16\% black, 11\% Hispanic, and 4\% other races with an average age of 60 years.
We create five education categories based on years of completed schooling: 0-11 years (less than high school graduate), 12 years (high school graduate), 13-15 years (some college), 16 years (college graduate), and 17 or more years (above college). In our sample,  15\% had less than 12 years of schooling, 30\% had 12 years of schooling, 25\% had some college, 15\% completed college, and 15\% had more than 16 years of schooling. Table \ref{tab:statistics} shows the mean and the standard deviation of cognitive scores, the log hourly wage, and the log wage residual, along with correlations between these variables. The correlations between test scores range from nearly 0.29 to 0.48.  All three test scores exhibit similar positive correlations with log wages and log wage residuals. Most relevant to our analysis, word recall has a correlation of 0.21 with log wages and 0.07 with the log wage residual.

\begin{table}[!htbp]
\caption{Mean, standard deviation (S.D.), and correlations between cognitive scores and log wage residuals}
\label{tab:statistics}
\centering
\begin{tabular}{lccccccc}
\toprule
& Num.\ of obs. & Mean & S.D.  & \multicolumn{4}{c}{Correlations}\\
\cline{5-8}
& & & & WR & S7 & QR & $w$\\
\midrule
Word recall (WR)  & 39,222 & 10.33 & 3.15 & 1.00 \\
Serial 7's (S7)  & 39,865 & 3.91 & 1.46 & 0.29 & 1.00 \\
Quantitative reasoning (QR) & 17,828 & 2.03 & 1.26 & 0.33 & 0.48 & 1.00 \\
Log wage residual ($w$) & 23,027 & 0.00 & 0.75 & 0.07 & 0.05 & 0.08 & 1.00 \\
Log wage ($\ln W$) & 23,042 & 2.45 & 0.84 & 0.21 & 0.20 & 0.27 & 0.89 \\
\bottomrule
\end{tabular}    
\end{table}

\subsection{Detailed Description of Cognitive Measures}

\paragraph{Word recall.} The HRS contains two separate tasks to assess respondent's memory: immediate word recall and delayed word recall. During the interview, the interviewer read a list of 10 nouns to the respondent and asked the respondent to recall as many words as possible from the list in any order. After approximately 5 minutes of answering other survey questions, the respondent was asked to recall the nouns previously presented. We construct a single measure which is the sum of the number of nouns that the respondent recalled in the two tasks. This measure ranges from 0 to 20.

\paragraph{Serial 7's.} This test asks the respondent to subtract 7 from the prior number, beginning with 100 for five trials. Correct subtractions are based on the prior number given, so that even if one subtraction is incorrect subsequent trials are evaluated on the given (perhaps wrong) answer. This test score ranges from 0 to 5.

\paragraph{Quantitative reasoning.} In the 2002 wave of HRS, three questions were added to the core survey to assess respondents’ numerical ability:
\begin{enumerate}
\item ``Next I would like to ask you some questions which assess how people use numbers in everyday life. If the chance of getting a disease is 10 percent, how many people out of 1,000 would be expected to get the disease?"
\item ``If 5 people all have the winning numbers in the lottery and the prize is two million dollars, how much will each of them get?''
\item ``Let's say you have \$200 in a savings account. The account earns ten percent interest per year. How much would you have in the account at the end of two years?''
\end{enumerate}

We construct a single measure called quantitative reasoning using the answers from these three questions. For each of the first two questions, the respondent earns 1 point if the answer is correct and 0 otherwise. For the last question, the respondent earns 2 points if the answer is correct; 1 point if the respondent used 10\% as a simple interest rate rather than a compound interest rate (i.e., answered 240 instead of 242); otherwise,  he earns 0 points. The quantitative reasoning measure is the sum of points earned on all three questions, ranging from 0 to 4.

\clearpage

%%%%%%%%%%%%%%%%%%%%%%%%%%%%%%%%%%%%%%%%%%%%%%%%%%%%%%%%%%%%%%%%%%%%%%%%%%%%%%%%%%
%%%% Re-Number Tables and figures for the Appendix %%%%%%%%
\renewcommand{\thetable}{\thesection-\arabic{table}}
\setcounter{table}{0}

\renewcommand{\thefigure}{\thesection-\arabic{figure}}
\setcounter{figure}{0}

%%%%%%%%%%%%%%%%%%%%%%%%%%%%%%%%%%%%%%%%%%%%%%%%%%%%%%%%%%%%%%%%%%%%%%%%%%%%%%%%%%

\section{Survey of Income and Program Participation (SIPP) linked with  W-2 Forms} \label{app: GSF}

This appendix describes data from Internal Revenue Service (IRS)/Social Security Administration (SSA) W-2 Forms linked with the Survey of Income and Program Participation (SIPP), referred to as the Gold Standard File (GSF) by the Census Bureau \citep{ssb}.  These data include the full SSA history of annual earnings (i.e., wage and salary) for all linked respondents from 1951 to 2011.\footnote{This analysis was first performed using the SIPP Synthetic Beta (SSB), while final results were obtained by Census Bureau staff using the SIPP Completed Gold Standard Files.
See \cite{ssb_codebook} and \cite{benedetto_stanley_totty_2018} for additional details on the data.}  Because we use annual earnings from administrative records, annual hours of work and hourly wages are not available.

Our analysis is based on 16--69 year-old, US-born white men who could be linked to any of nine SIPP panels (1984, 1990, 1991, 1992, 1993, 1996, 2001, 2004, and 2008). The highest level of education achieved at the time of survey (asked only once in each panel) is available in 5 categories: no high school degree, high school degree, some college, college degree, and graduate degree. We map these categories to 10, 12, 14, 16, and 18 years of completed schooling in order to calculate potential experience (age - years of education - 6). Since some individuals were still young and unlikely to have completed their schooling at the time of survey, we exclude those who were under 30 years old or were enrolled in school when their education level was measured.

We focus mainly on results using Detailed Earnings Records (DER), which are uncapped and available from 1978 onward; however, we also take advantage of Summary Earnings Records (SER) available since 1951, which report earnings capped at the FICA taxable maximum.  We work with log earnings residuals constructed as with the PSID and restrict observations to years when individuals were no longer enrolled in school. We trim the top and bottom 1\% of DER-based earnings within year and college/non-college status by five-year experience cells, and residualize log DER-based earnings by regressing on experience indicators and interactions between education indicators and a third order polynomial in experience, separately by year and college/non-college status. Log SER-based earnings -- used only as instruments in our analysis -- are residualized by subtracting median values conditional on year, education, and five-year experience cells.

Based on a worker's primary job (i.e., the job with the highest earnings), the Census Bureau classified workers into 24 occupation categories each survey wave.  Table~\ref{app tab: GSF occupations} reports these occupation codes, along with our 3-category grouping of occupations (cognitive, manual, and routine). Since respondents can report different occupations in each of 3 survey waves each year, we define occupation stayers between two years as those who reported any occupation in both years.

\begin{table}[!htbp]
\centering
\caption{SIPP/W-2 Occupation Codes and 3-Category Grouping}
\label{app tab: GSF occupations}
\begin{tabular}{clc}
\toprule
Code & Occupation & 3-Category Grouping \\
\midrule
1 & Management & \multirow{10}{*}{Cognitive} \\
2 & Business and financial operations \\
3 & Computer and mathematical \\
4 & Architecture and engineering \\
5 & Life, physical, and social science \\
6 & Community and social service \\
7 & Legal \\
8 & Education, training, and libraries \\
9 & Arts, design, entertainment, sports, and media \\
10 & Healthcare practitioner and technical \\
\midrule
11 & Healthcare support & \multirow{5}{*}{Manual}\\
12 & Protective service \\
13 & Food prep and service \\
14 & Building and grounds cleaning and maintenance \\
15 & Personal care and service \\
\midrule
16 & Sales & \multirow{2}{*}{Routine} \\
17 & Office and administrative support \\
\midrule
18 & Farming, fishing, and forestry & Not classified \\
\midrule
19 & Construction and extraction & \multirow{5}{*}{Routine} \\
20 & Installation, maintenance, and repairs \\
21 & Production \\
22 & Transportation \\
23 & Material moving \\
\midrule
24 & Military & Not classified \\
\bottomrule
\end{tabular}    
\end{table}

\begin{figure}[!htbp]
  \centering
  \includegraphics[width=0.45\columnwidth]{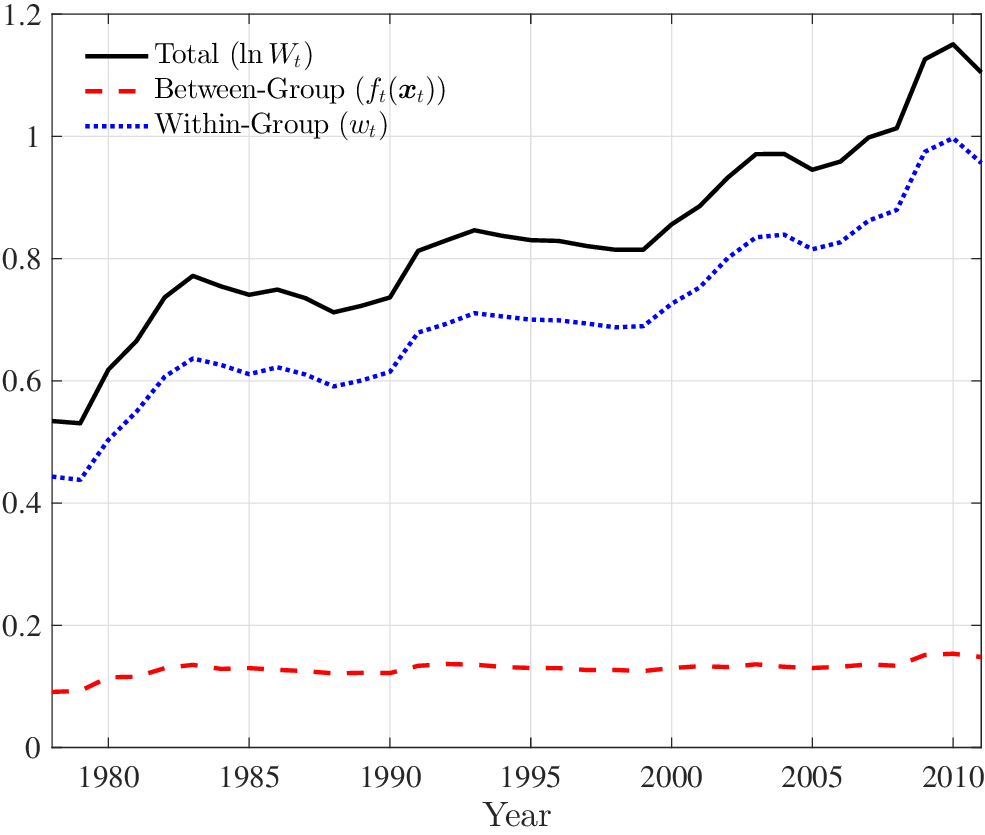}
\caption{Between- and within-group variances of log earnings, ages 16--64 with 5--40 years of experience (SIPP/W-2)}
    \label{fig: variance of earnings GSF}
\end{figure}

Figure~\ref{fig: variance of earnings GSF} reports log earnings inequality, along with between-group and within-group (residual) inequality, based on DER wage measures in the SIPP/W-2. The general trends are qualitatively similar to those for the PSID reported in Figure~\ref{fig: variance of hourly earnings}; although, the variance of total log earnings inequality and residual inequality is notably higher than their counterparts for log wages in the PSID. 
%Figure~\ref{fig: resid quart actual GSF} reports average log earnings by quartile each year.

\begin{figure}[!htbp]
    \centering
    \subfloat[Non-College]{
      \includegraphics[width=0.45\columnwidth]{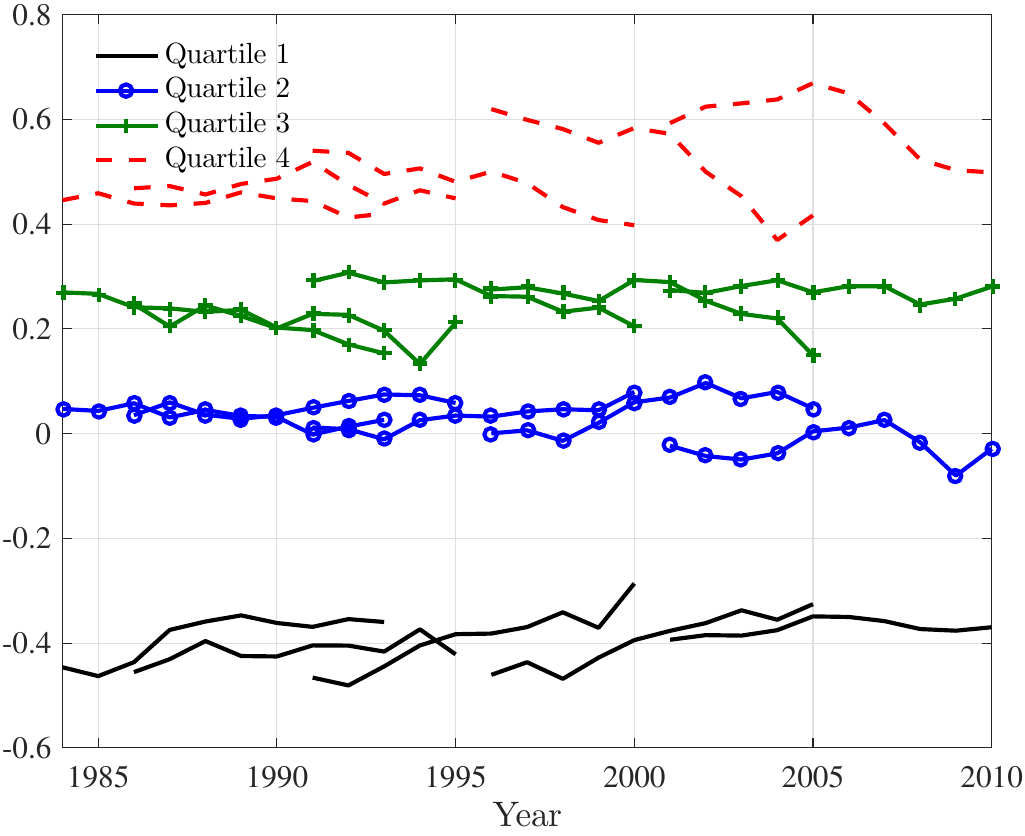}
}\quad
    \subfloat[College]{
      \includegraphics[width=0.45\columnwidth]{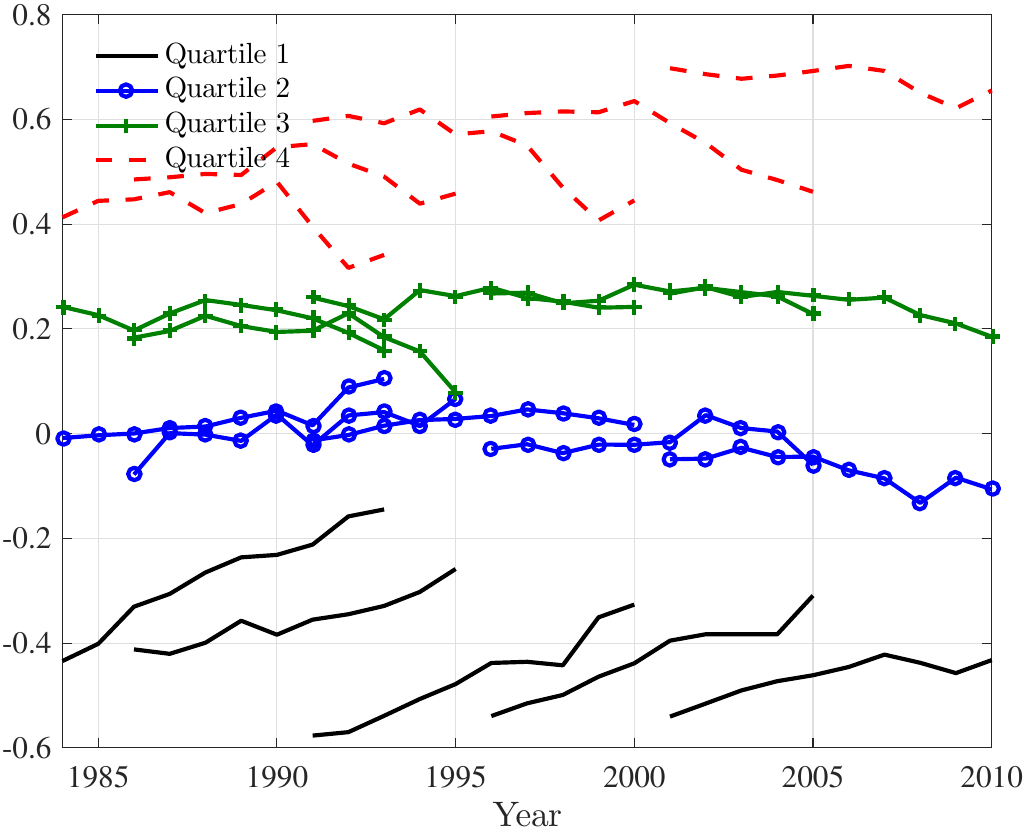}
}
\caption{Average predicted log earnings residuals by baseline residual quartile, 21--25 years of experience in base year (SIPP/W-2)}
    \label{fig: GSF pred w by educ}
\end{figure}

\begin{figure}[!htbp]
    \centering
    \subfloat[Non-College]{
      \includegraphics[width=0.45\columnwidth]{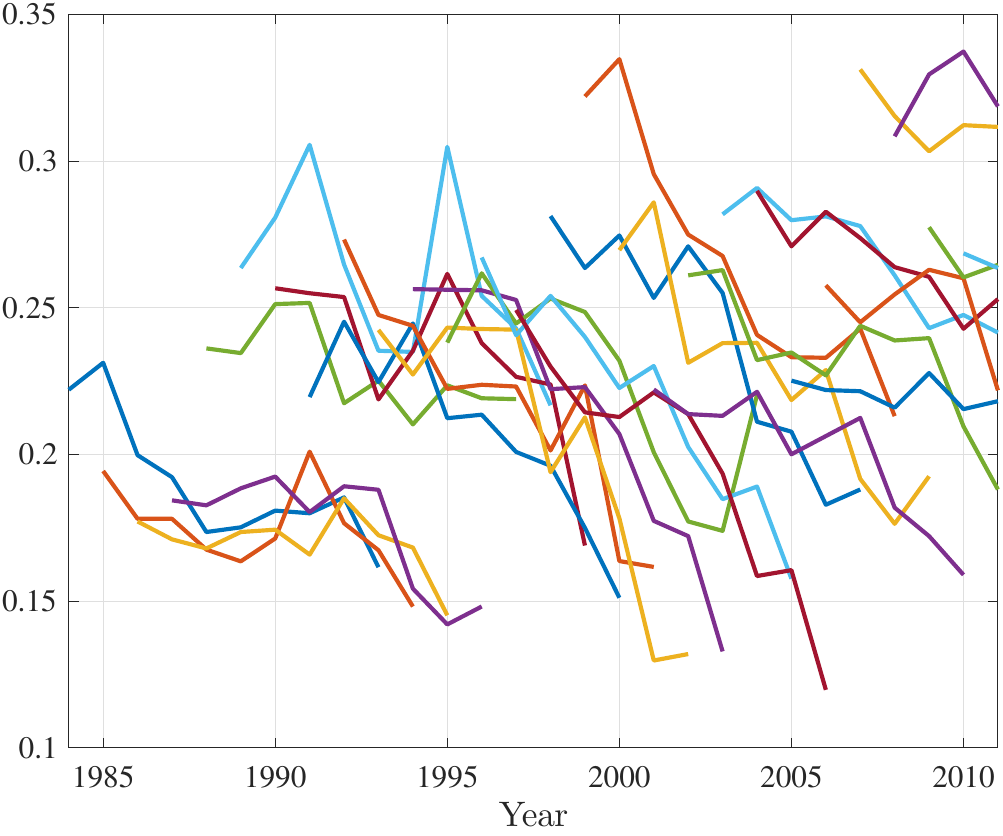}
}\quad
    \subfloat[College]{
      \includegraphics[width=0.45\columnwidth]{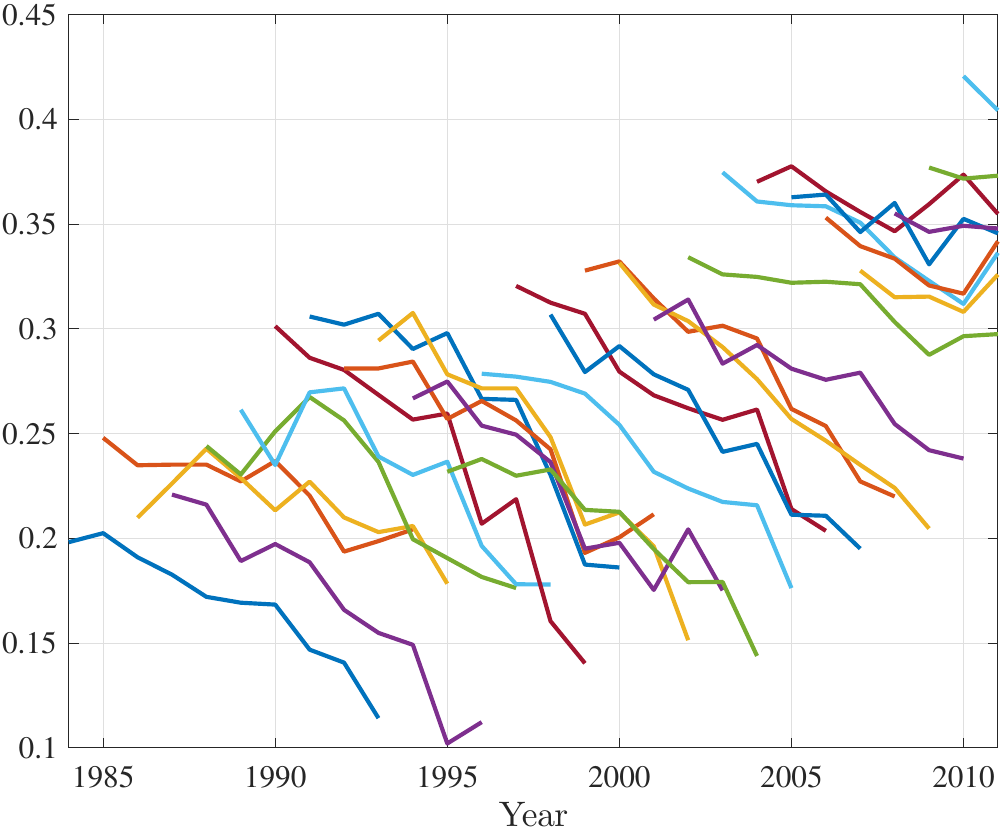}
}
\caption{Autocovariances for log earnings residuals, 21--25 years of experience in base year (SIPP/W-2)}
    \label{fig: GSF autocov by educ}
\end{figure}

Figure~\ref{fig: GSF pred w by educ} shows $\E\left[w_t|w_b \in Q_b^j\right]$ for different $t$ years where $Q_b^j$ reflects quartile $j$ in `base' year $b$, while Figure~\ref{fig: GSF autocov by educ} shows residual autocovariances $\cov(w_t,w_b)$ over years $t\geq b+6$ for fixed base year $b$.  Both figures are based on samples of non-college and college men with 21--25 years of experience in each base year, $b$.  Together, these indicate declines in the return to skills over the late-1980s and 1990s, consistent with our PSID-based results.

Tables~\ref{tab: IV SER GSF} and \ref{tab: IV DER GSF} report 2SLS estimates of skill return growth rates using SER- and DER-based lagged log earnings residuals ($w_{t-7}$), respectively, as instruments. (See Figure~\ref{fig: IV mu by educ GSF} in the paper.) Corresponding standard errors and sample sizes are also reported.

\if0
\begin{figure}[!htbp]
  \centering
  \includegraphics[width=0.45\columnwidth]{./figures/GSF_stage1/qtile_wage}
\caption{Average log earnings by quartile, ages 16--64 with 5--40 years of experience (SIPP/W-2) }
    \label{fig: resid quart actual GSF}
\end{figure}
\fi

\begin{table}[!htbp]
    \caption{2SLS estimates of $\Delta\mu_t/\mu_{t-1}$ (instrument: $w_{t-7}$ SER earnings),  32--40 years of experience in year $t$ (SIPP/W-2)}
    \label{tab: IV SER GSF}
    \centering\small
\begin{tabular}{ccccccc}
\toprule
\multirow{2}{*}{Year}&\multicolumn{3}{c}{Non-College}&\multicolumn{3}{c}{College}\\
\cmidrule(lr){2-4}\cmidrule(lr){5-7}
 & Estimate & Standard Error & Observations  & Estimate & Standard Error & Observations \\
\midrule
1979 & 0.024 & 0.059 & 3,600 & 0.122 & 0.075 & 1,900 \\
1980 & -0.037 & 0.043 & 3,700 & 0.047 & 0.063 & 2,100 \\
1981 & 0.053 & 0.051 & 3,800 & -0.135 & 0.077 & 2,200 \\
1982 & 0.077 & 0.045 & 3,800 & 0.015 & 0.063 & 2,400 \\
1983 & -0.044 & 0.043 & 3,700 & -0.002 & 0.059 & 2,500 \\
1984 & -0.024 & 0.041 & 3,700 & 0.010 & 0.069 & 2,600 \\
1985 & -0.068 & 0.039 & 3,600 & -0.024 & 0.058 & 2,700 \\
1986 & -0.039 & 0.040 & 3,600 & 0.010 & 0.050 & 2,900 \\
1987 & -0.139 & 0.037 & 3,700 & -0.041 & 0.032 & 3,000 \\
1988 & -0.043 & 0.038 & 3,800 & -0.101 & 0.030 & 3,100 \\
1989 & 0.020 & 0.033 & 3,800 & -0.043 & 0.043 & 3,100 \\
1990 & -0.078 & 0.033 & 3,800 & -0.009 & 0.044 & 3,200 \\
\bottomrule
\multicolumn{7}{l}{Notes: Reports coefficient estimates from 2SLS regression of $\Delta w_t$ on $w_{t-1}$ using $w_{t-7}$ as an}\\
\multicolumn{7}{l}{ instrument. The number of observations is rounded to the nearest 100 due to confidentiality.}
\end{tabular}
\end{table}

\begin{table}[!htbp]
    \caption{2SLS estimates of $\Delta \mu_t/\mu_{t-1}$ (instrument: $w_{t-7}$ DER earnings), 32--40 years of experience in year $t$ (SIPP/W-2)}
    \label{tab: IV DER GSF}    
\centering \small   
\begin{tabular}{ccccccc}
\toprule
\multirow{2}{*}{Year}&\multicolumn{3}{c}{Non-College}&\multicolumn{3}{c}{College}\\
\cmidrule(lr){2-4}\cmidrule(lr){5-7}
 & Estimate & Standard Error & Observations  & Estimate & Standard Error & Observations \\
\midrule
1985   & -0.060 & 0.029 & 3,800 & -0.026 & 0.032 & 2,900 \\
1986   & -0.072 & 0.030 & 3,800 & -0.101 & 0.026 & 3,100 \\
1987   & -0.074 & 0.030 & 3,900 & -0.049 & 0.027 & 3,200 \\
1988   & -0.055 & 0.037 & 4,000 & -0.084 & 0.028 & 3,400 \\
1989   & 0.016 & 0.030 & 3,900 & -0.025 & 0.028 & 3,400 \\
1990   & -0.065 & 0.029 & 4,000 & 0.012 & 0.030 & 3,500 \\
1991   & 0.010 & 0.032 & 4,000 & -0.009 & 0.028 & 3,600 \\
1992   & -0.061 & 0.026 & 4,100 & -0.096 & 0.025 & 3,800 \\
1993   & -0.073 & 0.023 & 4,100 & -0.040 & 0.024 & 3,900 \\
1994   & -0.025 & 0.025 & 4,200 & -0.055 & 0.027 & 4,200 \\
1995   & 0.012 & 0.036 & 4,300 & -0.045 & 0.023 & 4,400 \\
1996   & -0.018 & 0.028 & 4,300 & -0.094 & 0.022 & 4,700 \\
1997   & -0.058 & 0.024 & 4,400 & -0.020 & 0.021 & 5,000 \\
1998   & -0.088 & 0.019 & 4,400 & -0.073 & 0.020 & 5,500 \\
1999   & -0.057 & 0.023 & 4,400 & -0.084 & 0.020 & 6,000 \\
2000   & -0.107 & 0.028 & 4,400 & -0.048 & 0.022 & 6,600 \\
2001   & -0.102 & 0.029 & 4,400 & -0.039 & 0.018 & 7,100 \\
2002   & -0.045 & 0.029 & 4,300 & -0.034 & 0.021 & 7,700 \\
2003   & -0.085 & 0.028 & 4,400 & -0.045 & 0.018 & 8,300 \\
2004   & -0.010 & 0.029 & 4,800 & -0.019 & 0.016 & 8,900 \\
2005   & -0.052 & 0.024 & 5,000 & -0.064 & 0.015 & 9,400 \\
2006   & -0.002 & 0.024 & 5,100 & -0.011 & 0.014 & 9,800 \\
2007   & -0.031 & 0.023 & 5,400 & -0.023 & 0.015 & 10,000 \\
2008   & -0.011 & 0.022 & 5,600 & -0.014 & 0.015 & 10,500 \\
2009   & 0.006 & 0.024 & 5,700 & -0.042 & 0.015 & 10,500 \\
2010   & -0.001 & 0.024 & 5,700 & 0.004 & 0.014 & 10,000 \\
2011   & -0.048 & 0.023 & 5,700 & 0.003 & 0.013 & 9,900 \\
\bottomrule
\multicolumn{7}{l}{Notes: Reports coefficient estimates from 2SLS regression of $\Delta w_t$ on $w_{t-1}$ using $w_{t-7}$ as an}\\
\multicolumn{7}{l}{ instrument. The number of observations is rounded to the nearest 100 due to confidentiality.}
\end{tabular}
\end{table}

\end{document}